\documentclass[a4paper,11pt]{article}
\pdfoutput=1  

\usepackage{jheppub}  

\usepackage{tabularray}

\usepackage{chngpage}

\usepackage[T1]{fontenc} 

\usepackage{lipsum}  

\usepackage{multirow}  

\usepackage{tabularx}  

\usepackage{float}

\usepackage{xcolor}

\usepackage{accents}

\usepackage{comment}

\usepackage{enumitem}

\usepackage{nccmath}

\usepackage{dsfont}

\usepackage{bbm}

\usepackage{xcolor}

\usepackage{cancel}

\usepackage{pdflscape}

\usepackage{tikz}

\usetikzlibrary{shapes.arrows}

\usetikzlibrary{shapes.geometric, arrows}

\newcommand{\be}{\begin{equation}}
\newcommand{\ee}{\end{equation}}

\newcommand{\ba}{\begin{aligned}}
\newcommand{\ea}{\end{aligned}}

\newcommand{\nn}{\nonumber}

\newcommand{\hs}{\mathfrak{hs}^\text{s,s}}

\newcommand{\cA}{\mathcal{A}}
\newcommand{\cB}{\mathcal{B}}
\newcommand{\cC}{\mathcal{C}}
\newcommand{\cD}{\mathcal{D}}

\newcommand{\cF}{\mathcal{F}}
\newcommand{\cG}{\mathcal{G}}
\newcommand{\cH}{\mathcal{H}}
\newcommand{\cI}{\mathcal{I}}

\newcommand{\cK}{\mathcal{K}}
\newcommand{\cL}{\mathcal{L}}

\newcommand{\cN}{\mathcal{N}}
\newcommand{\cO}{\mathcal{O}}
\newcommand{\cP}{\mathcal{P}}
\newcommand{\cQ}{\mathcal{Q}}

\newcommand{\cV}{\mathcal{V}}
\newcommand{\cW}{\mathcal{W}}
\newcommand{\cX}{\mathcal{X}}
\newcommand{\cY}{\mathcal{Y}}
\newcommand{\cZ}{\mathcal{Z}}

\newcommand{\WW}{\mathcal{W}_{\infty}^\text{s,s}}

\newcommand*{\one}{\text{\usefont{U}{bbold}{m}{n}1}}

\newcommand{\bonetti}[1]{{\color{purple} #1}}

\title{\boldmath 
A universal W-algebra for $\mathcal N = 4$ super Yang-Mills
}

\author[a]{Federico Bonetti,} 
\author[b,c]{Carlo Meneghelli}

\affiliation[a]{Departamento de Electromagnetismo y Electr\'onica, Universidad de Murcia, Campus de Espinardo, 30100 Murcia, Spain}

\affiliation[b]{Dipartimento SMFI, Universit\`a di Parma, Viale G.P.~Usberti 7/A, 43100, Parma, Italy}

\affiliation[c]{INFN Gruppo Collegato di Parma}

\emailAdd{f.bonetti@um.es,   carlo.meneghelli@unipr.it}

\abstract{
Using bootstrap methods,
we provide evidence for the existence of a
non-linear W-algebra,
denoted 
$\mathcal W_\infty^\text{s,s}$,
which contains the \emph{small} $\mathcal{N} = 4$
super Virasoro
 algebra
and features 
an infinite tower of additional  
generators,
organized in \emph{short} supersymmetry multiplets.
The algebra $\mathcal {W}_\infty^\text{s,s}$
has one free parameter, its central charge $c$.
We claim that the simple quotient of
$\mathcal {W}_\infty^\text{s,s}$
for $c= -3(N^2-1)$ 
is isomorphic 
to the
vertex operator algebra associated to
 4d $\mathcal{N}=4$ $\mathfrak{su}(N)$ super Yang-Mills theory.
We define a filtration in 
$\mathcal {W}_\infty^\text{s,s}$ and 
provide evidence that it reproduces the $R$-filtration,
which is crucial to extract 4d SCFT data from the
vertex operator algebra.
Finally, we  present explicit formulae for all the (anti)commutators 
of the wedge algebra of $\mathcal W_\infty^\text{s,s}$.

}

\begin{document} 
 
\maketitle
\flushbottom

\section{Introduction and summary}

To any 4d
$\cN = 2$
superconformal field theory
(SCFT) one can associate a vertex 
operator algebra (VOA) \cite{Beem:2013sza}, 
which captures
a protected subsector of the 4d
SCFT.
This applies in particular to 
4d $\cN = 4$ $\mathfrak{su}(N)$ super Yang-Mills:
we denote its 
 associated VOA as  $\cV(A_{N-1})$.
Some properties of $\cV(A_{N-1})$  have been discussed in 
\cite{Bonetti:2018fqz}, which  proposes a
definition of this VOA in terms of a free field realization
based on the Weyl group  of $\mathfrak{su}(N)$.
This viewpoint allows to formulate   precise
conjectures on $\cV(A_{N-1})$, some of which
have been proven in \cite{Arakawa:2023cki}.

Motivated by the existence of the 1-parameter family of VOAs
$\cV(A_{N-1})$,
in this work we address the following question:
is there a universal W-algebra
for the sequence 
$\cV(A_{N-1})$, $N=2,3,4,\dots$?
In other words, we ask whether
all VOAs $\cV(A_{N-1})$ can be recovered as 
truncations of a single W-algebra,
which is expected to exist for generic values of the central charge and to admit an infinite
number of strong generators.\footnote{A subset $\mathcal S$ of operators in a W-algebra is
a set of strong generators if
all operators in the W-algebra can be written as finite linear combinations of
normal ordered products of operators in $\mathcal S$ and their derivatives.
}

Various examples of similar
universal W-algebras are available
in the literature. For instance, in the non-supersymmetric context, 
one can consider the sequence of VOAs usually denoted
$\cW_N$, $N=2,3,4,\dots$,
where the case $N=2$ is the Virasoro algebra, and 
the case $N=3$ is the 
$\cW_3$  algebra of Zamolodchikov
\cite{Zamolodchikov:1985wn}.
For each $N$, $\cW_N$ exists
for generic values of the central charge $c$
and has finitely
many strong generators.
(The VOA $\cW_N$
is obtained by Drinfeld-Sokolov
reduction \cite{Drinfeld:1984qv,Balog:1990dq,Bouwknegt:1992wg}
of the $\mathfrak{su}(N)$
affine Kac-Moody algebra.)
A universal W-algebra
for $\cW_N$ exists, usually denoted
$\cW_\infty[\mu]$,
discussed in 
\cite{Figueroa-OFarrill:1992uuf,Khesin:1993ru,Khesin:1993ww}
building on 
\cite{Pope:1989ew,Bakas:1991fs,
Bakas:1991gs,
Yamagishi:1991ax,
Figueroa-OFarrill:1991ynu,
Yu:1991ng,
Yu:1991bk}.
A mathematical proof of the existence
and properties of $\cW_\infty[\mu]$ can be found in
\cite{Linshaw:2017tvv}.
The free parameters of 
$\cW_\infty[\mu]$ are $\mu$ and the central charge $c$. 
For generic values of $\mu$,
$\cW_\infty[\mu]$ is a non-linear\footnote{Here non-linear refers to the fact that
the singular part of the OPE of
two strong generators  contains
normal ordered products of 
strong generators and their derivatives.
}
W-algebra with infinitely many strong generators.
In the limit $\mu \rightarrow N$, $N=2,3,4\dots$,
$\cW_\infty[\mu]$ develops an ideal of null states; modding them out, it reduces to 
$\cW_N$ at central charge $c$.
The existence and properties
of $\cW_\infty[\mu]$
are central to 
the minimal model
holography program, see the review
\cite{Gaberdiel:2012uj}
and references therein.
Even-spin version W-algebras
are also known \cite{Gaberdiel:2011nt,
Ahn:2011pv,
Candu:2012ne,
Kanade:2018qut,
Prochazka:2019yrm},
as well as 
supersymmetric 
versions 
\cite{
Creutzig:2011fe,
Candu:2012jq,
Candu:2012tr,
Beccaria:2013wqa,
Gaberdiel:2013vva,
Creutzig:2013tja,
Candu:2013fta,
Gaberdiel:2014yla,
Beccaria:2014jra,
Eberhardt:2018plx,
Ahn:2020rev,
Ahn:2022qex,
Ahn:2022orj,
Ahn:2023mdg}.

In this paper, we 
conjecture that a universal W-algebra for $\cV(A_{N-1})$ exists,
which we denote $\WW$ \cite{Leuv}.
We provide support
for this conjecture
based on 
operator product expansion
(OPE)
 bootstrap methods
(see e.g.~\cite{Blumenhagen:1990jv}).
We make extensive use 
of the \textit{Mathematica}
package
\texttt{OPEdefs}  \cite{Thielemans:1991uw,Thielemans:1994er}.
Our proposed
universal W-algebra $\WW$
has small $\mathcal N =4$
supersymmetry
but differs from other models
in the literature
with the same supersymmetry
\cite{Ahn:2020rev} 
because of the quantum numbers
of its strong generators
(besides those of the 
small $\cN = 4$ super Virasoro algebra).
Furthermore, we find that $\WW$
has no free parameters besides its central charge.
This sets it apart compared to other examples
mentioned in the previous paragraph.
Upon setting $c = -3 (N^2-1)$ to truncate
$\WW$ to $\cV(A_{N-1})$, no tunable parameter is left.
This is consistent with the expectation that
$\cV(A_{N-1})$ for $N \ge 3$
does not admit any continuous deformation,
and that its very existence relies on intricate patterns
of null states.

\noindent
\textbf{Note added.} While this paper was being finalized
\cite{Gaberdiel:2025eaf} appeared, which has some overlap 
with this work.

% including W-algebras
% with small $\mathcal N = 4$
% supersymmetry \cite{sdsds}.
% Their generators, however,
% have quantum numbers that are
% different from those required
% for a universal W-algebra
% for $\cV(A_{N-1})$.

% extensions and generalizations are also available in the literature,
% with $\cN = 2$  \cite{sdsd}, large $\cN = 4$ \cite{sdsds},
% and small $\cN = 4$ \cite{sdsd}
% supersymmetry.

% More precisely, our findings
% are summarized below.

% Thielemans papers:
% \cite{Thielemans:1991uw,Thielemans:1994er}

\subsection*{Summary of results}

\subsubsection*{The W-algebra $\WW$}

We propose that a non-linear W-algebra denoted $\WW$ exists,
with the following properties:
\begin{itemize}
\item[(1)] $\WW$ contains the small $\cN = 4$ super Virasoro algebra as a subalgebra.
\end{itemize}
Thus, the operator content of $\WW$ is organized in multiplets of
the superconformal algebra\footnote{In this work, otherwise stated, we work with  complexified Lie algebras.}
$\mathfrak{psl}(2|2)$, the global part of the small $\cN = 4$ super Virasoro algebra.
The bosonic subalgebra of 
$\mathfrak{psl}(2|2)$ is
$\mathfrak{sl}(2)_z \oplus \mathfrak{sl}(2)_y$, where the first summand is the conformal algebra in two dimensions, and the second factor is an R-symmetry. 

\begin{itemize}
\item[(2)] $\WW$ is strongly generated by 
the  operators
collected in the following table.
\begin{center}
\begingroup
\renewcommand{\arraystretch}{1.3}
 \begin{tabular}{ |c || c | c | c | c | c | c | c | c | } 
 \hline 
$\mathfrak{psl}(2|2)$ multiplets & \multicolumn{4}{c|}{$\mathbb J$}
 & \multicolumn{4}{c|}{$\mathbb W_p$ ($p=3,4,5,\dots$)}
 \\ \hline
operators & $J$ & $G$ & $\widetilde G$ & $T$
 & $W_p$ & $G_{W_p}$ & $\widetilde G_{W_p}$ &
 $T_{W_p}$ 
\\ \hline 
$\mathfrak{sl}(2)_z$ conformal weight $h$ & 
$1$ & 
$\frac 32$ &
$\frac 32$ &
$2$ & 
$\frac p2$ & 
$\frac p2 + \frac 12$ & 
$\frac p2 + \frac 12$ & 
$\frac p2 + 1$
\\ \hline 
$\mathfrak{sl}(2)_y$ spin $j$
& 
$1$ & 
$\frac 12$ & 
$\frac 12$ &
$0$ & 
$\frac p2$ & 
$\frac p2 - \frac 12 $ & 
$\frac p2 - \frac 12 $ & 
$\frac p2 - 1 $ \\ \hline 
\end{tabular}
\endgroup
\end{center}
The strong generators are organized in 
\emph{short}
$\mathfrak{psl}(2|2)$ multiplets.
(The labels s,s in $\WW$ stand for
``small'' and ``short.'')
The multiplet denoted $\mathbb J$
collects the generators of the small $\cN = 4$
super Virasoro algebra, with $T$ the stress tensor, $J$ the affine $\mathfrak{sl}(2)_y$ current,
and $G$, $\widetilde G$ the supercurrents.
Besides $\mathbb J$,
we have one additional $\mathfrak{psl}(2|2)$ multiplet $\mathbb W_p$ for each $p=3,4,5,\dots$.

\item[(3)] For each $p=3,4,5,\dots$
the operator
$W_p$ is a Grassmann even super Virasoro primary operator.

\item[(4)] The W-algebra $\WW$ exists for generic values of the central charge $c$.
$\WW$
has no free parameters besides $c$.
\end{itemize}

We argue for the existence and properties of $\WW$ by 
bootstrapping 
the OPEs of strong generators.
We leverage symmetry under the small $\cN = 4$ super Virasoro algebra, and the assumption that $W_p$ be a super Virasoro primary.
Thus, the non-trivial task is the determination of the OPEs $W_{p_1} \times W_{p_2}$.
We implement all these OPEs with
$p_1 + p_2 \le 10$.
Associativity allows us to fix all  coefficients in the 
$W_{p_1} \times W_{p_2}$ OPEs with
$p_1 + p_2 \le 9$;
for $p_1 + p_2=10$
we have partial results.
Our findings are summarized in Tables
\ref{tab_ansatz}-\ref{tab_OPEcoeffs_part8}
and equation \eqref{eq_best_normalization}, where
we make use of a parameter $\nu$ related to the central charge $c$ by $c=3(1 - \nu)$.

Next, we conjecture the following additional properties of $\WW$.
\begin{itemize}
\item [(5)] 
  If we tune the central charge to the values 
  \be 
  \label{cGOOD}
  c = - 3(N^2-1) \ , \qquad N=2,3,4,5,
\dots , 
  \ee 
  $\WW$ develops an ideal of null states. 
  In particular,
  all generators in the multiplets
  $\mathbb W_p$ with $p>N$ are null.
The simple quotient of $\WW$ is 
  isomorphic to 
 the VOA 
  $\cV(A_{N-1})$ associated to 4d $\cN = 4$ $\mathfrak{su}(N)$
  super Yang-Mills.
We stress that  the ideal of null states 
of $\WW$
contains states constructed with $\mathbb W_p$ with $p \le N$,
which are the VOA avatars of Higgs branch relations
of super Yang-Mills.

\end{itemize}

\begin{itemize}
\item [(6)]
The W-algebra $\WW$ admits an increasing weight based filtration
for generic values of the central charge.
The simple quotient of $\WW$ at $c = - 3 (N^2-1)$ inherits the  filtration, and thus per the conjecture above
gives a filtration on $\cV(A_{N-1})$. The latter is identified with the $R$-filtration.

\end{itemize}

We recall that accessing the quantum number $R$ in the VOA
$\cV(A_{N-1})$ is essential to being able to
invert the map from 4d operators to 2d operators and therefore
extract valuable information about 4d physics from the
VOA.

We provide various pieces of evidence for the 
connection between the W-algebra $\WW$ and 
4d $\cN = 4$ $\mathfrak{su}(N)$ super Yang-Mills.
First, we perform a detailed counting of states
in $\WW$ up to conformal weight $h=4$,
for various values of $N$, and check it against
predictions from the Macdonald index and Hall-Littlewood chiral ring of 4d $\cN = 4$ $\mathfrak{su}(N)$ super Yang-Mills.
Second, we test the counting of states in $\WW$ with the
BRST construction of \cite{Beem:2013sza} of the VOA $\cV(A_{N-1})$.
Finally, we also perform some checks against the
free field realization of $\cV(A_{N-1})$
\cite{Bonetti:2018fqz, Arakawa:2023cki}.
As a side product,
we exhibit the free-field realization
for $N=5$,
and the full set of OPE coefficients of
$\cV(A_4)$,
which had not appeared in the literature before.
They are collected in 
\eqref{free_su5_OPEs}
and Appendix \ref{app_free}.

\subsubsection*{The W-algebra $\WW$ and half-BPS correlators in 4d super Yang-Mills}

The information contained in the W-algebra $\WW$ 
can be used to compute 
a protected part of 
correlators of a class of 1/8-BPS operators
in 4d $\cN = 4$ $\mathfrak{su}(N)$ super Yang-Mills,
for any $N$. More precisely, the operators in question
are the Schur operators, reviewed briefly in Section \ref{sec_review_4d2dmap}.
The protected part of correlators of Schur operators
can in principle be computed in 4d by Wick contractions,
provided one knows the explicit forms of the operators.
The problem greatly simplifies for half-BPS operators, as
they do not participate in the mechanism of multiplet recombination.
Thus, the map between OPE data in $\WW$ and
4d correlators is transparent
for half-BPS operators.

% Conversely, this implies that
% the OPE data of 
% $\WW$  can in principle be extracted from Wick contractions in
% free 4d $\cN = 4$ $\mathfrak{su}(N)$ super Yang-Mills.
% For generic Schur operators, however, the map
% between 4d and 2d operators is 

% The W-algebra $\WW$ exhibits close ties
% to the half-BPS sector of 4d $\cN = 4$ $SU(N)$
% super Yang-Mills theory.
% In fact, we claim that an infinite number of OPE
% coefficients of $\WW$ can be computed \emph{exactly}
% as functions of the central charge $c$
% by performing Wick contractions in free 
% 4d $\cN = 4$ $SU(N)$
% super Yang-Mills theory.

To formulate precisely the correspondence
between $\WW$ and half-BPS operators
in 4d,
we use the notion of \emph{single-particle operator}
\cite{Aprile:2018efk, Aprile:2020uxk}. We recall that a basis of
the half-BPS chiral
ring of 4d $\cN = 4$ $SU(N)$
super Yang-Mills is provided by single-trace
operators $T_p = {\rm Tr}(\phi^p)$ 
and their products, the multi-trace operators.
(Here $\phi$ denotes schematically the six real
adjoint-valued scalars of 4d $\cN = 4$ super Yang-Mills.)
By definition, the single-particle operator
$\cO_p$ is the combination 
\be 
\cO_p = T_p + \text{(multi-trace corrections)} 
\ee 
that has vanishing two-point 
functions with \emph{any} multi-trace operator.
Some explicit examples of operators
$\cO_p$ can be found in \eqref{eq_single_particle_examples}.
The operator $\cO_p$   is identically zero for $p>N$.
(In contrast, for $p>N$ the single-trace
operator $T_p$ reduces to a complicated
combination of multi-trace operators via
trace relations.)

Now, the key observation is that
the two- and three-point
functions of half-BPS operators
are 
in 4d $\cN = 4$ $SU(N)$ super Yang-Mills
are 
protected
\cite{Howe:1995aq,
DHoker:1998vkc,
Howe:1998zi,
Intriligator:1998ig,
Gonzalez-Rey:1999ouj,
Intriligator:1999ff,
Eden:1999gh,
Skiba:1999im,
Penati:1999ba,
Dolan:2004mu}
and can  be computed in 
the
free-field  limit using Wick contractions.
We find it convenient to perform the rescaling
\be 
\widetilde \cO_p = \frac{1}{N^{(p-2)/2}} \cO_p \ .
\ee 
Then, under the chiral algebra map of \cite{Beem:2013sza},
the operator $\widetilde \cO_p$ maps to the strong generator~$W_p$,
\be 
 \widetilde \cO_p  \mapsto 
W_p   \ .
\ee 
We note that 2- and 3-point functions of 
$\widetilde \cO_p$ operators 
are rational functions of $N^2$
(and not merely rational functions of $N$).
We claim that, if we perform the
formal replacement
$N^2 \rightarrow 1-c/3$,
the $SU(N)$ theory correlators
$\langle \widetilde \cO_p
\widetilde \cO_p \rangle$, 
$\langle \widetilde \cO_{p_1}
\widetilde \cO_{p_2} 
\widetilde \cO_{p_3}
\rangle$
yield a subset of OPE coefficients of the W-algebra
$\WW$
as exact functions of $c$,
\be \label{eq_freefield_to_W}
\begin{array}{c}
\langle \widetilde \cO_p \widetilde \cO_p\rangle \ , \
\langle \widetilde \cO_{p_1} \widetilde \cO_{p_2}
\widetilde \cO_{p_3} \rangle  \\ 
\text{rational functions of $N^2$}
\end{array}
\;\; 
\xrightarrow{\quad N^2 \leadsto 1-c/3
\quad}
\;\;
\begin{array}{c}
\text{W-algebra OPE coefficients}
\\
g_p(c) \ , \
c_{p_1 p_2}{}^{p_3}(c)
\end{array}  \ .
\ee 
Here $g_p$ and $c_{p_1 p_2}{}^{p_3}$ are defined
by $W_p \times W_p \supset g_p \one$, 
$W_{p_1} \times W_{p_2} \supset 
c_{p_1 p_2}{}^{p_3} W_{p_3}$.
The precise formulation
of the correspondence
\eqref{eq_freefield_to_W}
can be found in equations
\eqref{eq_W_O_dictionary},
\eqref{eq_g_dictionary}, \eqref{eq_lambda_and_C}.
In stating
\eqref{eq_freefield_to_W}
we are implicitly choosing
a normalization for the $W_p$
generators on the W-algebra side. This is the normalization in which the results of equations
\eqref{eq_best_normalization}
 and Tables
\ref{tab_OPEcoeffs_part1}-\ref{tab_OPEcoeffs_part8}
are presented.

\subsubsection*{The wedge algebra of $\WW$}

We complement our OPE bootstrap
analysis of $\WW$
with a study of its wedge algebra, which we denote 
$\text{Wedge}(\WW)$.
Recall that, 
given a family of  W-algebras $\cW(c)$ with central charge $c$,
its wedge algebra $\text{Wedge}(\cW(c))$
is a Lie algebra obtained
from $\cW(c)$ in two steps
\cite{Bowcock:1991zk}
(see also \cite{Gaberdiel:2011wb} for a review).
(i) For each strong generator
(say $X$)
of the W-algebra, expand
$X$ in Laurent modes as
$X(z) = \sum_m z^{-m-h_X}$
and retain only the ``wedge modes'' $|m|<h_X$
(here $h_X$ is the conformal
weight of $X$).
(ii) Consider the limit 
of large central charge.
Heuristically, step (i)
eliminates all central terms in the W-algebra, and step (ii)
suppresses all non-linearities of $\cW(c)$, to yield an ordinary
Lie algebra.

The wedge algebra 
$\text{Wedge}(\WW)$
is an infinite-dimensional
higher-spin Lie superalgebra,
analogous
to the higher-spin algebra considered in the minimal model holography program
(see review \cite{Gaberdiel:2012uj}).
It has been considered in 
\cite{Costello:2018zrm} in the context
of twisted holography,
where it was denoted
$\mathfrak a_\infty$
and it was described as 
the superalgebra of global
symmetries of the large-$N$
VOA associated to
4d $\cN = 4$ $\mathfrak{su}(N)$ super Yang-Mills.
In this work, 
we recover this Lie superalgebra by means of a 
direct Lie algebra bootstrap.
In the process, we identify closed forms for all its
(anti)commutators.
More precisely, we seek a Lie superalgebra containg
$\mathfrak{psl}(2|2)$,
whose generators 
have the same quantum numbers
as the wedge modes of the strong generators of $\WW$.
Interestingly, these assumptions
are enough to completely single out an infinite dimensional
Lie superalgebra which we denote
$\hs$. Here hs stands for higher spin.
(The wedge algebra of the W-algebra of \cite{Ahn:2020rev}
has the same supersymmetry as
$\hs$, but a different spectrum.)
The generators of 
$\hs$ are listed in 
\eqref{eq_psl_table},
\eqref{eq_W_table}.
The (anti)commutators of
$\hs$ are given in closed
form in Appendix 
\ref{app_wedge}.\footnote{Even though we have closed expressions for all
(anti)commutators,
we have only
verified a finite subset of
Jacobi identities.}
The Lie superalgebra 
$\hs$ has no free parameter.
Covariance under
$\mathfrak{psl}(2|2)$
restricts heavily
its structure;
the  structure constants that are not fixed by $\mathfrak{psl}(2|2)$ symmetry
turn out to have a particularly simple form, see
\eqref{eq_wedge_algebra_coeffs}.

Having \emph{independently} boostrapped
the Lie superalgebra $\hs$,
we finally propose that indeed
\be 
\hs \cong \text{Wedge}(\WW) \ . 
\ee 
We perform explicit tests of this proposal using the large central charge limit of the
OPE coefficients reported  in Tables 
\ref{tab_OPEcoeffs_part1}-\ref{tab_OPEcoeffs_part8}.

% \subsection*{Organization of the paper}

% \bonetti{This will be updated in the end}

% The rest of this paper is organized as follows.
% Section \ref{sec_results} presents our main results
% from the OPE bootstrap analysis, while Section
% \ref{sec_methods} describes the 
% methods we have used to obtain those results.
% The truncation of $\WW$ to $\cV(A_{N-1})$
% is studied in Section \ref{sec_truncation}. Section 
% \ref{sec_wedge} is devoted to
% the wedge algebra of $\WW$. We conclude with some remarks on
% 4d $\cN = 4$ super Yang-Mills
% theory and the holography in Section \ref{sec_comments} and with an outlook of possible future
% directions in Section 
% \ref{sec_outlook}. Technical material
% is collected in several appendices.

% as the Lie superalgebra 
% $\hs$ is analogous
% to the higher-spin algebra considered in the minimal model holography program
% (see review \cite{Gaberdiel:2012uj}).
% We believe that
% $\hs$ might be of independent interest beyond applications to
% the W-algebra $\WW$.
% The generators of 
% $\hs$ are listed in 
% \eqref{eq_psl_table},
% \eqref{eq_W_table}.
% All (anti)commutators of
% $\hs$ are given in closed
% form in appendix 
% \ref{app_wedge}.\footnote{Even though we have closed expressions for all
% (anti)commutators,
% we cannot quite claim
% to have demonstrated rigorously
% the existence of the Lie superalgebra
% $\hs$, since we have only
% verified a finite subset of
% Jacobi identities. The expressions in appendix \ref{app_wedge}
% should then be regarded as a conjecture.}
% We find that the Lie superalgebra 
% $\hs$ has no free parameter.

\section{Bootstrapping $\WW$}
\label{sec_results}

In this section we 
summarize the assumptions and strategy underlying 
our bootstrap analysis of the W-algebra
$\WW$, 
and we present its main results.
Our key findings are the OPEs
listed in Tables \ref{tab_ansatz} to
\ref{tab_OPEcoeffs_part8}.
Further details of the methodology used to derive the results of this
section are described in Appendix \ref{sec_methods}.

\subsection{Preliminary: small $\mathcal N = 4$ super Virasoro algebra}

The bosonic
generators of the small $\mathcal N = 4$
super Virasoro algebra
are the stress tensor $T(z)$
(of conformal weight  2)
and a triplet $J^{IJ}(z)=J^{JI}(z)$  of affine
$\mathfrak{sl}(2)$ currents
(of weight 1).
The indices $I,J=1,2$
are fundamental indices of
$\mathfrak{sl}(2)$.
The fermionic generators 
are 
two doublets $G^I(z)$, $\widetilde G^{I}(z)$ 
of weight 3/2.
We find it convenient to adopt a
standard
index-free notation based on the introduction of an auxiliary variable $y$.
More precisely, we define
\be 
\ba 
J(z,y) = J^{IJ}(z) y_I y_J \ , \quad 
G(z,y) = G^I(z) y_I \ , \quad 
\widetilde G(z,y) = \widetilde G^I(z) y_I \ , \quad 
y_I = (1 , y) \ . 
\ea 
\ee 
The $J \times J$ and $T \times T$
OPEs  read 
\be 
\ba 
J(z_1,y_1) J(z_2,y_2) & =
\frac{-k y_{12}^2 \one}{z_{12}^2}  
+ \frac{2y_{12}
\big(
1 + \tfrac 12 y_{12} \partial_{y_2}
\big)
J(z_2,y_2)
}{z_{12}}   + \text{reg.} \ , \\
T(z_1) T(z_2)
& = 
\frac{\tfrac 12 c \one }{z_{12}^4} 
+ \frac{2 T(z_2)}{z_{12}^2}
+ \frac{\partial_{z_2} T(z_2)}{z_{12}}
+ \text{reg.} \ .
\ea 
\ee 
We have introduced the notation
$z_{12} = z_1 - z_2$,
$y_{12} = y_1 - y_2$.
The central charge $c$ is given in terms of the
affine Kac-Moody level $k$
as
\be 
c = 6k \ . 
\ee 
We record the full set
of OPEs of the small
$\mathcal N = 4$ super Virasoro 
algebra in Appendix \ref{app_notation}.

The small $\cN = 4$
super Virasoro algebra
is invariant under an
$SL(2)$ outer automorphism
under which $T$, $J$ are invariant and $G$, $\widetilde G$ rotate as a doublet.
We denote the Cartan
generator of this
$SL(2)$ automorphism
as $\mathfrak{gl}(1)_{r}$
and we assign charges $\pm \tfrac 12$ 
to $G$, $\widetilde G$
under  $\mathfrak{gl}(1)_{r}$,
respectively.

The global part of the small
$\cN=4$ super Virasoro algebra
is the superconformal
algebra
$\mathfrak{psl}(2|2)$.
Its bosonic subalgebra is
$\mathfrak{sl}(2)_z \oplus \mathfrak{sl}(2)_y$,
where $\mathfrak{sl}(2)_z$ denotes the
global conformal algebra on the Riemann sphere
(associated to the coordinate $z$)
and $\mathfrak{sl}(2)_y$ is the R-symmetry
algebra (associated to the auxiliary
variable $y$).

In this work we encounter two
kinds of $\mathfrak{psl}(2|2)$
superconformal multiplets.
Firstly, we have short multiplets,
whose content is of the form 
\be 
\text{short $\mathfrak{psl}(2|2)$ multiplet:}
\qquad 
\begin{array}{ccc}
& X \\
G_X := G^\downarrow X &&
\widetilde G_{X} := \widetilde G^\downarrow X \\
& T_X := \widetilde G^\downarrow G^\downarrow X 
\end{array} \ . 
\ee 
The operator $X$ transforms
as a primary under
$\mathfrak{sl}(2)_z \oplus
\mathfrak{sl}(2)_y$.
It has $\mathfrak{sl}(2)_y$
spin $j\in \{ 1,\frac 32, 2, \frac 52 , \dots\}$
and conformal weight $h=j$.
The notation $G^\downarrow X$
denotes the $\mathfrak{sl}(2)_z \oplus
\mathfrak{sl}(2)_y$
primary operator
of conformal weight
$h+\frac 12$ and 
spin $j-\frac 12$
that enters the order-one
pole in the singular OPE
of $G$ with $X$.
The notation
$\widetilde G^\downarrow$
is defined similarly,
so that $\widetilde G^\downarrow X$ is
an $\mathfrak{sl}(2)_z \oplus
\mathfrak{sl}(2)_y$ primary
with conformal weight
$h+\frac 12$ and spin $j-\frac 12$. Finally,
$\widetilde G^\downarrow G^\downarrow X$ is
an $\mathfrak{sl}(2)_z \oplus
\mathfrak{sl}(2)_y$ primary
with conformal weight
$h+1$ and spin $j-1$.
We refer to the operator $X$
as the $\mathfrak{psl}(2|2)$
primary in the multiplet.

For example,
the generators
$J$, $G$, $\widetilde G$,
$T$ of the 
small $\cN=4$
super Virasoro algebra
form a short $\mathfrak{psl}(2|2)$ multiplet, 
which we denote
\be 
\mathbb J = \left\{
\begin{array}{ccc}
& J \\
G = G^\downarrow J &&
\widetilde G = \widetilde G^\downarrow J \\
& T = \widetilde G^\downarrow G^\downarrow J 
\end{array} 
\right\} 
\ . 
\ee 

In this work we also
find 
long $\mathfrak{psl}(2|2)$ multiplets whose 
$\mathfrak{psl}(2|2)$ primary
has conformal weight $h$
and spin $j$ with $h<j$.
For generic $j$,
a long $\mathfrak{psl}(2|2)$
contains 16 
$\mathfrak{sl}(2)_z \oplus
\mathfrak{sl}(2)_y$ primary
operators, which are obtained
from the $\mathfrak{psl}(2|2)$ primary by repeated application of OPE with the
fermionic currents $G$, $\widetilde G$. We refer the reader to Appendix \ref{app_notation}
for further details.

We also need to recall the notion of super Virasoro
primary operator.
A super Virasoro primary operator $X$ is a
$\mathfrak{psl}(2|2)$
primary operator satisfying an additional requirement:
in the OPEs of the generators
$J$, $G$, $\widetilde G$, $T$
of the super Virasoro
algebra with $X$, we encounter
no poles of order 2 or higher,
with the exception of the
standard order-2 pole in the $TX$ OPE. We give 
more details on super Virasoro
primary operators in Appendices
\ref{app_notation}, \ref{app_superVir}.

% \be 
% \ba 
% J(z_1,y_1) J(z_2,y_2) & =
%  -k  \frac{y_{12}^2}{z_{12}^2} \one
% + 2 \frac{y_{12}}{z_{12}}
% \big(
% 1 + \tfrac 12 y_{12} \partial_{y_2}
% \big)
% J(z_2,y_2) + \text{reg.}
% \ , \\
% T(z_1, y_1) J(z_2,y_2)
% & = \frac{1}{z_{12}^2} 
% \big( 1 + z_{12} \partial_{z_2} \big)
% J(z_2,y_2)
% + \text{reg.} \ , 
% \\
% T(z_1) T(z_2)
% & = 
% \frac c2 \frac{1}{z_{12}^4} \one
% + 2 \frac{1}{z_{12}^2}
% \big( 1 
% + \tfrac 12 z_{12} \partial_{z_2} \big)
% T(z_2)
% + \text{reg.}
%  \ , 
% \\
% T(z_1, y_1) G(z_2,y_2)
% & = \frac 32 \frac{1}{z_{12}^2} 
% \big( 1 + \tfrac 23 z_{12} \partial_{z_2} \big)
% G(z_2,y_2)
% + \text{reg.} \ , 
% \\
% T(z_1, y_1) \widetilde G(z_2,y_2)
% & = \frac 32 \frac{1}{z_{12}^2} 
% \big( 1 + \tfrac 23 z_{12} \partial_{z_2} \big)
% \widetilde G(z_2,y_2)
% + \text{reg.} \ , 
% \\
% J(z_1,y_1) G(z_2,y_2)
% & = \frac{y_{12}}{z_{12}}
% \big(
% 1 +  y_{12} \partial_{y_2}
% \big)
% G(z_2,y_2) 
% + \text{reg.} \ , \\
% J(z_1,y_1) \widetilde G(z_2,y_2)
% & = \frac{y_{12}}{z_{12}}
% \big(
% 1 +  y_{12} \partial_{y_2}
% \big)
% \widetilde G(z_2,y_2) 
% + \text{reg.} 
% \ , \\
% G(z_1,y_1) G(z_2,y_2)
% & = \text{reg.} \ , \\
% \widetilde G(z_1,y_1) \widetilde G(z_2,y_2)
% & = \text{reg.} \ , \\
% G(z_1,y_1) \widetilde G(z_2,y_2)
% & = - 2 k \frac{y_{12}}{z_{12}^3} \one
% + 2 \frac{1}{z_{12}^2} 
% \big( 1 + \tfrac 12 z_{12}
% \partial_{z_2} \big)
% \big( 1 + \tfrac 12 y_{12}
% \partial_{y_2}
% \big)
% J(z_2,y_2)
% - \frac{y_{12}}{z_{12}}
% T(z_2,y_2) +\text{reg.} \ .
% \ea
% \ee

\subsection{Assumptions and strategy of the bootstrap analysis}

In this section we list the assumptions of our bootstrap
analysis and describe the strategy to implement it.

\subsubsection{Assumptions}
\label{subsec_assum}

Our goal is to set up and perform a bootstrap
analysis to gather evidence for the existence 
of a W-algebra $\WW$ with the following
properties.
\begin{itemize}
\item $\WW$ contains the small 
$\cN=4$ super Virasoro algebra as a subalgebra.
\item The set of strong generators of $\WW$ consists of:
\begin{itemize}
    \item the generators $J$, $G$, $\widetilde G$, $T$ of the small $\cN=4$ super-Virasoro algebra;
    \item for each $p=3,4,5,\dots$ 
    one short $\mathfrak{psl}(2|2)$ multiplet of operators 
\be 
\mathbb W_p = \left\{
\begin{array}{ccc}
& W_p \\
G_{W_p} := G^\downarrow W_p &&
\widetilde G_{W_p} = \widetilde G^\downarrow W_p \\
& T_{W_p} := \widetilde G^\downarrow G^\downarrow W_p 
\end{array} 
\right\} 
\ . 
\ee 
\end{itemize}
\item For each $p=3,4,5,\dots$
the operator $W_p$ is 
a Grassmann-even super Virasoro primary with conformal weight and spin given as $h=j=\frac 12p$.
\item The central charge $c$
of the small $\cN=4$ super Virasoro algebra is generic.
All operators $W_p$ are non-null
operators.
\end{itemize}
The fact that $W_p$ is not a null operator can be stated in terms of the two-point function
of $W_p$ with itself.
By $\mathfrak{sl}(2)_z \oplus \mathfrak{sl}(2)_y$ symmetry,
this quantity takes the form
\be 
\langle W_p(z_1,y_1) W_p(z_2,y_2) \rangle 
= g_{p} \frac{y_{12}^p}{z_{12}^p} \ .
\ee
The two-point function coefficient $g_p$ is assumed to be nonzero at generic values of the central charge $c$.
% We might set $g_p=1$ by rescaling $W_p$, but we prefer to leave $g_p$ arbitrary.

\subsubsection{Strategy} 

Our goal is to fix the singular parts of the OPEs among the strong generators of $\WW$. These OPEs are of three kinds,
\be 
\mathbb J \times \mathbb J \ , \qquad 
\mathbb J \times \mathbb W_p \ , \qquad 
\mathbb W_{p_1} \times \mathbb W_{p_2} \ . 
\ee 
We are using the following compact notation: if $\mathbb A$,
$\mathbb B$ denote
$\mathfrak{psl}(2|2)$ multiplets,
then 
$\mathbb A \times \mathbb B$
denotes collectively all OPEs
of the form
$A \times B$,
where $A$ is an $\mathfrak{sl}(2)_z \oplus \mathfrak{sl}(2)_y$
primary operator in the multiplet
$\mathbb A$,
and $B$ is an $\mathfrak{sl}(2)_z \oplus \mathfrak{sl}(2)_y$
primary operator in the multiplet~$\mathbb B$.

The $\mathbb J \times \mathbb J$ OPEs are the OPEs of generators
of the small $\cN = 4$ super Virasoro algebra, reported in 
\eqref{eq_sVir}.
The $\mathbb J \times \mathbb W_p$ OPEs are completely fixed by
super Virasoro covariance
and by the assumption that
$W_p$ be a super Virasoro primary.
We report the 
$\mathbb J \times \mathbb W_p$ OPEs in Appendix \ref{app_notation}.

Our main task is the determination of the OPEs
of the form
$\mathbb W_{p_1} \times \mathbb W_{p_2}$. 
We now describe our strategy to perform their computation.

\subsubsection*{Preliminary: OPE associativity }

A key tool in our bootstrap analysis
is associativity of the OPE.
This can be characterized as follows.
Let $A(z)$, $B(z)$ be two operators.
(Any dependence on the auxiliary variable $y$, if present, is suppressed for simplicity.)
The OPE $A \times B$ can be parameterized as 
\be 
A(z_1) B(z_2) = \sum_{n \in \mathbb Z} \frac{1}{z_{12}^n}  \{ AB \}_n(z_2) \ ,
\ee 
where $z_{12}=z_1 - z_2$
and $\{ AB \}_n(z_2)$ is identically
zero for sufficiently large $n$. 
With this notation,
the associativity of the OPE
is encoded in the identities
\cite{borcherds1986vertex,kac1998vertex,Thielemans:1994er}
\be  \label{eq_OPE_assoc}
\{ A \{ BC \}_p \}_q
= (-1)^{|A||B|} 
\{ B \{ AC \}_q \}_p
+ \sum_{\ell=1}^\infty 
\binom{q-1}{\ell-1} 
\{ \{ AB \}_\ell C\}_{p+q-\ell} \ ,
\qquad 
p,q \in \mathbb Z \ . 
\ee 
Here $|A| = 0,1$ mod 2 is the Grassmann parity of the operator $A$.

The identity \eqref{eq_OPE_assoc},
when $p$ or $q$ are non-positive,
encodes the rules to compute
OPEs of composite operators.
Throughout this work,
we assume that these rules hold.
We are then led to consider the identity
\eqref{eq_OPE_assoc}
in the case in which $A$, $B$, $C$
are strong generators of the algebra
and $p,q>0$.
Motivated by these considerations,
we introduce the following
notation.
Given operators $A$, $B$, $C$,
we define
\be 
\ba
\text{Jacobi}_{p,q}(A,B,C)
&:= \{ A \{ BC \}_p \}_q
- (-1)^{|A||B|} 
\{ B \{ AC \}_q \}_p
- \sum_{\ell=1}^\infty 
\binom{q-1}{\ell-1} 
\{ \{ AB \}_\ell C\}_{p+q-\ell} \ , 
\\
\text{Jacobi}(A,B,C) &:= 
\left\{ 
\text{Jacobi}_{p,q}(A,B,C) \ , 
\ p, q \in \mathbb N
\right\} \ . 
\ea 
\ee 
If $\mathbb A$, $\mathbb B$, $\mathbb C$ denote
$\mathfrak{psl}(2|2)$ multiplets,
we also introduce the compact
notation
\be 
\text{Jacobi}(\mathbb A , \mathbb B , \mathbb C) = \left\{
\text{Jacobi}(A,B,C) , A \in \mathbb A , B \in \mathbb B , C \in \mathbb C 
\right \} \ .  
\ee 
We note that
$\text{Jacobi}(A,B,C)$ is 
implemented as the command
\texttt{OPEJacobi[A,B,C]}
in the \textit{Mathematica}
package \texttt{OPEdefs} \cite{Thielemans:1994er}.

In order to ensure the
associativity of the OPE,
we have to impose that 
every element in the set
$\text{Jacobi}(A,B,C)$ is zero,
for any $A$, $B$, $C$ strong generators.
Even though the set 
$\text{Jacobi}(A,B,C)$ is indexed by
two positive integers $p$, $q$,
only a finite number of elements of
$\text{Jacobi}(A,B,C)$
can be potentially non-zero.
Therefore,
demanding that all elements of
$\text{Jacobi}(A,B,C)$ be zero
(for given $A$, $B$, $C$)
corresponds to a finite number of
constraints.
These are constraints on the OPE
coefficients in the singular parts of the OPEs among strong generators of the algebra.

In this paper we work
at generic values of the central charge $c$.
As a result, we expect that none
of the operators entering
$\text{Jacobi}(A,B,C)$ are null.
This simplifies our analysis:
we just have to set to zero
the coefficients of each
operator appearing in 
$\text{Jacobi}(A,B,C)$.

\subsubsection*{Workflow}

% As stated in section
% \ref{subsec_assum}, we seek
% a W-algebra $\WW$ with small
% $\cN=4$ supersymmetry
% that is generated by the 
% following $\mathfrak{psl}(2|2)$
% short multiplets
% of operators, 
% \be 
% \mathbb J = \{ J, G, \widetilde G, T\} \ , \qquad 
% \mathbb W_p = \{ W_p , G_{W_p} ,
% \widetilde G_{W_p} , T_{W_p}\} \ , \qquad 
% p=3,4,5,\dots \ .
% \ee 
% Specifying the vertex algebra
% amounts to specifying the 
% singular parts of all OPEs of the form
% $\mathbb J \times \mathbb J$,
% $\mathbb J \times \mathbb W_p$,
% $\mathbb W_{p_1} \times \mathbb W_{p_2}$ with $p_1 \le p_2$.
% Here the notation
% $\mathbb A \times \mathbb B$,
% where $\mathbb A$, $\mathbb B$ are
% $\mathfrak{psl}(2|2)$ multiplets,
% indicates the set of OPEs
% $\{ A \times B , A \in \mathbb A , B \in \mathbb B \}$.
% All OPEs $\mathbb J \times \mathbb J$
% and $\mathbb J \times \mathbb W_p$
% are fixed by our assumptions
% of small $\cN =4$ supersymmetry,
% and that $W_p$ is a super Virasoro
% primary operator, for each $p=3,4,5,
% \dots$.
% The OPEs $\mathbb J \times \mathbb J$
% were given  
% in \eqref{eq_sVir},
% while the $\mathbb J \times \mathbb W_p$ OPEs are given in appendix \ref{sdsd}.

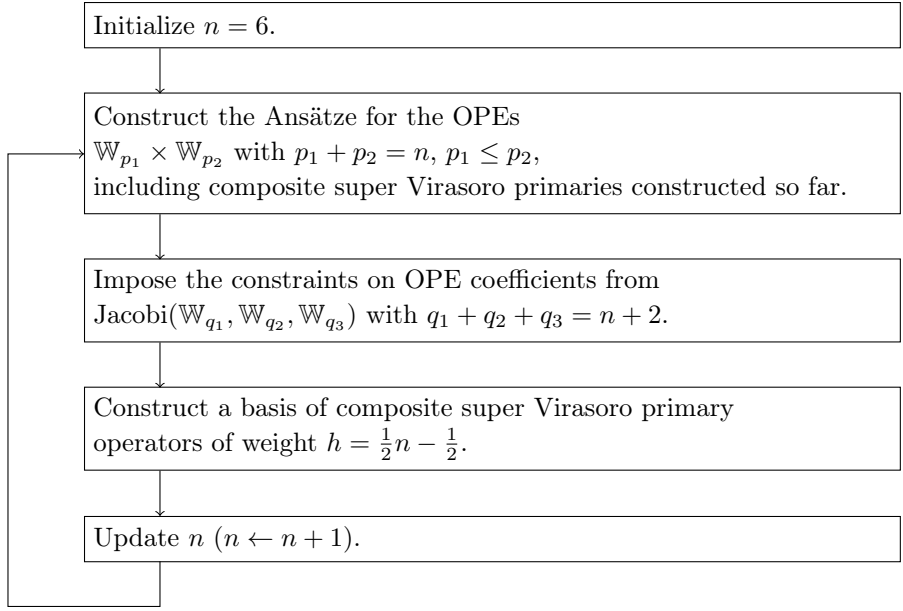
\begin{figure}
    \centering
\begin{tikzpicture}
\small

\node[draw, minimum width=10.5cm, 
minimum height=0.6cm,
anchor=south west, text width=10.5cm]  at (0,0) {Initialize $n=6$.};

\node[draw, minimum width=10.5cm,
minimum height=1.6cm,
anchor=south west, text width=10.5cm]  at (0,-1.6-0.6) {Construct the Ans\"atze
for the OPEs\\
$\mathbb W_{p_1} \times 
\mathbb W_{p_2}$ with
$p_1 + p_2 = n$, $p_1 \le p_2$,\\
including 
composite super Virasoro primaries
constructed so far.};

\node[draw, minimum width=10.5cm, 
minimum height=1.1cm, anchor=south west, text width=10.5cm]  at (0,-1.6-0.6-1.1-0.6) {Impose the constraints on OPE coefficients from\\
$\text{Jacobi}(\mathbb W_{q_1} , \mathbb W_{q_2}, \mathbb W_{q_3})$
with
$q_1 + q_2 + q_3 = n+2$.};

\node[draw, minimum width=10.5cm, anchor=south west,
minimum height=1.1cm,
text width=10.5cm]  at (0,-1.6-0.6-1.1-0.6-1.6-0.6+0.5) {Construct a basis of
composite super Virasoro primary\\ operators of weight
$h = \frac 12 n - \frac 12$.
};

\node[draw, minimum width=10.5cm, 
minimum height=0.6cm,
anchor=north west, text width=10.5cm]  at (0,-1.6-0.6-1.1-0.6-1.6-0.6-0.6+0.5) {Update $n$ ($n \leftarrow n+1$).};

\draw[->]   (1,0) -- (1,0-0.6);
\draw[->]   (1,-0.6-1.6) -- (1,-0.6-1.6-0.6);
\draw[->]   (1,-0.6-1.6-1.1-0.6) -- (1,-0.6-1.6-0.6-1.1-0.6);
\draw[->]   (1,-0.6-1.6-1.1-0.6-1.6-0.6+0.5) -- (1,-0.6-1.6-0.6-1.1-0.6-1.6-0.6+0.5);
\draw[->]   (1,-0.6-1.6-1.1-0.6-1.6-0.6-0.6-0.6+0.5) -- (1,-0.6-1.6-0.6-1.1-0.6-1.6-0.6-0.6-0.6+0.5)
-- (-1,-0.6-1.6-0.6-1.1-0.6-1.6-0.6-0.6-0.6+0.5)
-- (-1,-0.6-0.8)
-- (0,-0.6-0.8);

\end{tikzpicture}
    \caption{Flowchart describing our strategy to implement and constrain the OPEs
    $\mathbb W_{p_1} \times \mathbb W_{p_2}$ proceeding by  increasing values of $p_1 + p_2$.}
    \label{fig_flowchart}
\end{figure}

The central task of the bootstrap
analysis is the determination of the
$\mathbb W_{p_1} \times \mathbb W_{p_2}$ OPEs. Even though we have an infinite number of multiplets
$\mathbb W_p$, we can proceed
systematically by studying the
$\mathbb W_{p_1} \times \mathbb W_{p_2}$ OPEs for increasing values
of $p_1 + p_2$,
starting from $p_1 + p_2 = 6$.
Our strategy is summarized in the 
flowchart in figure \ref{fig_flowchart}.
In the following subsection
we describe how 
the Ansatz for $\mathbb W_{p_1} 
\times \mathbb W_{p_2}$ is constructed. 
For the construction of composite
super Virasoro primary operators
we refer the reader to Appendix
\ref{app_composites}.

It can be useful to unpack
the flowchart of figure 
\ref{fig_flowchart}
and list explicitly 
the steps
we have followed in implementing
our bootstrap analysis in this work:
\begin{enumerate}[noitemsep]
\item Construct the Ansatz for $\mathbb W_3 \times \mathbb W_3$.
\item Construct the Ansatz for $\mathbb W_3 \times \mathbb W_4$.
\item Impose the constraints on OPE coefficients from
$\text{Jacobi}(\mathbb W_3 , \mathbb W_3 , \mathbb W_3)$.
\item Construct the composite super Virasoro primary operators
$\cC_{3,1}^{W_3 W_3}$,
$\cC_{3,3}^{W_3 W_3}$.
\item Construct the Ans\"atze for $\mathbb W_3 \times \mathbb W_5$,
$\mathbb W_4 \times \mathbb W_4$.
\item Impose the constraints 
on OPE coefficients
from
$\text{Jacobi}(\mathbb W_3 , \mathbb W_3 , \mathbb W_4)$.
\item Construct the composite super Virasoro primary operators
$\cC_{\frac 72, \frac 12}^{W_3 W_4}$,
$\cC_{\frac 72, \frac 32}^{W_3 W_4}$,
$\cC_{\frac 72, \frac 52}^{W_3 W_4}$,
$\cC_{\frac 72, \frac 72}^{W_3 W_4}$.
\item Construct the Ans\"atze for $\mathbb W_3 \times \mathbb W_6$,
$\mathbb W_4 \times \mathbb W_5$.
\item Impose the constraints 
on OPE coefficients
from
$\text{Jacobi}(\mathbb W_3 , \mathbb W_3 , \mathbb W_5)$,
$\text{Jacobi}(\mathbb W_3 , \mathbb W_4 , \mathbb W_4)$.
\item Construct the composite super Virasoro primary operators
$\cC_{4,0}^{W_3 W_3}$,
$\cC_{4,0}^{W_4 W_4}$
$\cC_{4,1}^{W_3 W_5}$,
$\cC_{4,2}^{W_3 W_3}$,
$\cC_{4,2}^{W_3 W_5}$,
$\cC_{4,2}^{W_4 W_4}$
$\cC_{4,3}^{W_3 W_5}$,
$\cC_{4,4}^{W_3 W_5}$,
$\cC_{4,4}^{W_4 W_4}$.
\item Construct the Ans\"atze for $\mathbb W_3 \times \mathbb W_7$,
$\mathbb W_4 \times \mathbb W_6$,
$\mathbb W_5 \times \mathbb W_5$.
\item Impose the constraints 
on OPE coefficients
from
$\text{Jacobi}(\mathbb W_3 , \mathbb W_3 , \mathbb W_6)$,
$\text{Jacobi}(\mathbb W_3 , \mathbb W_4 , \mathbb W_5)$,
$\text{Jacobi}(\mathbb W_4 , \mathbb W_4 , \mathbb W_4)$.
\end{enumerate}
Clearly, this algorithm can in principle
be continued  indefinitely. Due to computational
limitations, in this work
we terminate the algorithm
at step 12.~above.

\subsubsection{Ansatz for the OPEs}

Let us now describe the salient features of our Ansatz
for the $\mathbb W_{p_1} \times \mathbb W_{p_2}$ OPEs.
Crucially, 
they are constrained by super Virasoro covariance (see Appendix \ref{app_superVir}).
Indeed,
all $\mathbb W_{p_1} \times \mathbb W_{p_2}$ OPEs can be reconstructed by knowledge of the OPE $W_{p_1} \times W_{p_2}$.
Secondly, 
the content of the RHS
of the $W_{p_1} \times W_{p_2}$
OPE can be organized into
contributions of super Virasoro primaries
and their descendants.
The possible super Virasoro primaries
that we can encounter on the RHS
of $W_{p_1} \times W_{p_2}$ are the identity operator
$\one$, the generators
$W_p$, as well as composite super Virasoro primary
operators.

The core of the bootstrap
analysis performed in this work
is the determination of the
$W_{p_1} \times W_{p_2}$ OPEs
with 
$p_1 \le p_2$ and 
$p_1 + p_2 \le 10$.
In the remainder of this section
we present our results for
these OPEs.

It is convenient to 
adopt a compact notation
in which, on the RHS of $W_{p_1} \times W_{p_2}$,
we only list the super Virasoro primary operators
that contribute to the singular
part of that OPE, together with
their OPE coefficients.
In particular, we do not show explicitly
any of their super Virasoro descendants
and we suppress
all $z$ and $y$ coordinates and derivatives.
This information can be unambiguously
reconstructed exploiting super Virasoro covariance,
as discussed in Appendix \ref{app_superVir}.

The OPEs $W_{p_1} \times W_{p_2}$
with $p_1 \le p_2$
and $p_1 + p_2 \le 10$
are collected in Table
\ref{tab_ansatz} in the compact
notation described in the previous paragraph.
The notation $\cC_{h,j}^{W_p W_q}$
indicates a composite super Virasoro primary operator
with conformal weight $h$
and spin $j$,
constructed starting from
a product of $W_p$ and $W_q$.
More explicitly,
the composite operators entering
the OPEs above are of the form
\be  \label{eq_composite_schematic}
\ba 
\cC_{3,1}^{W_3 W_3}  & = (W_3 W_3)_0^1 + \dots \ , 
&
\qquad \qquad 
\cC_{4,1}^{W_3 W_5}
& = (W_3 W_5)_0^1 + \dots \ , 
\\
\cC_{3,3}^{W_3 W_3}
 & = (W_3 W_3)_0^3 + \dots \ , 
& 
\cC_{4,2}^{W_3 W_3}
 & = (W_3 W_3)_{-1}^2 + \dots  \ , 
\\
\cC_{\frac 72 , \frac 12}^{W_3 W_4}
 & = 
(W_3 W_4)_0^{\frac 12} + \dots
&
\cC_{4,2}^{W_3 W_5}
 & = 
(W_3 W_5)^2_0 + \dots 
\\
\cC_{\frac 72 , \frac 32}^{W_3 W_4}
 & = 
(W_3 W_4)_0^{\frac 32} + \dots \ ,
&
\cC_{4,2}^{W_4 W_4}
 & = (W_4 W_4)_0^2 + \dots 
\\ 
\cC_{\frac 72 , \frac 52}^{W_3 W_4}
 & = 
(W_3 W_4)_0^{\frac 52} + \dots \ ,
&
\cC_{4,3}^{W_3 W_5}
 & = (W_3 W_5)_0^3 + \dots  \ ,
\\
\cC_{\frac 72 , \frac 72}^{W_3 W_4}
& = 
(W_3 W_4)_0^{\frac 72} + \dots \ ,
&
\cC_{4,4}^{W_3 W_5} 
& = (W_3 W_5)_0^4 + \dots 
 \ , 
\\
\cC_{4,0}^{W_3 W_3}
 & = (W_3 W_3)_{-1}^0
+ \dots 
&
\cC_{4,4}^{W_4 W_4}
  & = (W_4 W_4)_0^4 + \dots 
\\
\cC_{4,0}^{W_4 W_4}
& = (W_4 W_4)_0^0 + \dots  \ .
\ea 
\ee 
We are using the notation $(AB)^j_n$
defined in Appendix \ref{app_notation}.
Schematically,
if $A$, $B$
are $\mathfrak{sl}(2)_z \oplus \mathfrak{sl}(2)_y$ primary operators,
their OPE is organized into 
$\mathfrak{sl}(2)_z \oplus \mathfrak{sl}(2)_y$ primary operators;
$(AB)^j_n$ is the $\mathfrak{sl}(2)_z \oplus \mathfrak{sl}(2)_y$ primary operator
that enters  the $A \times B$ 
OPE in the pole in $z_{12}$ of order
$n$ (so values $n\le 0$ correspond
to regular terms in the OPE).
The omitted terms
in \eqref{eq_composite_schematic} 
are additional terms required to
have super Virasoro primary
operators.
Further details on the composite operators
listed in 
\eqref{eq_composite_schematic} are reported in 
Appendix~\ref{app_composites}.

We point out that the W-algebra also contains composite super Virasoro primary operators that are constructed with three or more $W_p$'s, for example
an operator of the form
\be 
\cC^{W_3W_3W_3}_{\frac 92, \frac 92}=(W_3(W_3 W_3)^3_0)^3_0 + \dots \ .
\ee 
The lightest such operators have weight $h=9/2$. They 
first appear in OPEs $W_{q_1} \times W_{q_2}$ with $q_1 + q_2 = 11$,
but in this work 
we study $q_1 + q_2 \le 10$.

\begin{table}[H]
    \centering
    \begin{tabular}{| l |}
\hline  
\rule[-3.5mm]{0mm}{9mm}
\begin{minipage}{1cm}
\small
\begin{fleqn}
\begin{equation*} 
\ba 
W_3 \times W_3 & = g_3 \one
+ c_{3 3}{}^{4} W_4 \ , 
\ea 
\end{equation*}
\end{fleqn}
\end{minipage}
\\
\hline
\rule[-3.5mm]{0mm}{9mm}
\begin{minipage}{1cm}
\small
\begin{fleqn}
\begin{equation*} 
\ba 
W_3 \times W_4 & = 
c_{3 4}{}^{3} W_3
+ c_{3 4}{}^{5} W_5 \ , 
\ea 
\end{equation*}
\end{fleqn}
\end{minipage}
\\
\hline
\rule[-9mm]{0mm}{19mm}
\begin{minipage}{1cm}
\small
\begin{fleqn}
\begin{equation*} 
\ba 
W_3 \times W_5
& =  c_{3 5}{}^{4} W_4
+ c_{3 5}{}^{6} W_6
 + c_{3 5}{}^{ \cC_{3,1}^{W_3 W_3} }
\cC_{3,1}^{W_3 W_3}  
+ c_{3 5}{}^{ \cC_{3,3}^{W_3 W_3} }
\cC_{3,3}^{W_3 W_3}   
\ , \\[2mm]
W_4 \times W_4
& = g_4 \one 
+ c_{4 4}{}^{4} W_4
+ c_{4 4}{}^{6} W_6
 + c_{4 4}{}^{   \cC_{3,1}^{W_3 W_3}  }
\cC_{3,1}^{W_3 W_3}     
+ c_{4 4}{}^{\cC_{3,3}^{W_3 W_3}}
\cC_{3,3}^{W_3 W_3}    
 \ , 
\ea 
\end{equation*}
\end{fleqn}
\end{minipage}
\\
\hline
\rule[-17mm]{0mm}{36mm}
\begin{minipage}{1cm}
\small
\begin{fleqn}
\begin{equation*} 
\ba 
W_3 \times W_6
& = c_{3 6}{}^{3} W_3
+ c_{3 6}{}^{5} W_5
+ c_{3 6}{}^{7} W_7
\\
& + c_{3 6}{}^{  \cC_{\frac 72, \frac 32}^{W_3 W_4}  } 
\cC_{\frac 72, \frac 32}^{W_3 W_4}  
+ c_{3 6}{}^{  \cC_{\frac 72 , \frac 52}^{W_3 W_4}  }
\cC_{\frac 72 , \frac 52}^{W_3 W_4}
 + c_{3 6}{}^{
\cC_{\frac 72, \frac 72}^{W_3 W_4}
} 
\cC_{\frac 72, \frac 72}^{W_3 W_4}
\ , \\[2mm]
W_4 \times W_5
& = c_{4 5}{}^{3} W_3
+ c_{4 5}{}^{5} W_5
+ c_{4 5}{}^{7} W_7
\\ 
& + c_{4 5}{}^{
\cC_{\frac 72 , \frac 12}^{W_3 W_4}
}
\cC_{\frac 72 , \frac 12}^{W_3 W_4}
+ c_{4 5}{}^{
\cC_{\frac 72 , \frac 32}^{W_3 W_4}
} 
\cC_{\frac 72 , \frac 32}^{W_3 W_4}
+ c_{4 5}{}^{
\cC_{\frac 72 , \frac 52}^{W_3 W_4}
} 
\cC_{\frac 72 , \frac 52}^{W_3 W_4}
+ c_{4 5}{}^{
\cC_{\frac 72 , \frac 72}^{W_3 W_4}
}
\cC_{\frac 72 , \frac 72}^{W_3 W_4}
 \ , 
\ea 
\end{equation*}
\end{fleqn}
\end{minipage}
\\
\hline
\rule[-34mm]{0mm}{70mm}
\begin{minipage}{14.5cm}
\small
\begin{fleqn}
\begin{equation*} 
\ba 
W_3 \times W_7
& = c_{3 7}{}^{4} W_4
+ c_{3 7}{}^{6} W_6
+ c_{3 7}{}^{8} W_8
+ c_{3 7}{}^{\cC_{3,3}^{W_3 W_3}}
\cC_{3,3}^{W_3 W_3}  
\\
&
+ c_{3 7}{}^{\cC_{4,2}^{W_3 W_3} }
\cC_{4,2}^{W_3 W_3}   
+ c_{3 7}{}^{\cC_{4,2}^{W_3 W_5}   }
\cC_{4,2}^{W_3 W_5}  
+ c_{3 7}{}^{ \cC_{4,2}^{W_4 W_4} }
\cC_{4,2}^{W_4 W_4}  
 \\ 
& + c_{3 7}{}^{\cC_{4,3}^{W_3 W_5} }
\cC_{4,3}^{W_3 W_5} 
+ c_{3 7}{}^{ \cC_{4,4}^{W_3 W_5} }
\cC_{4,4}^{W_3 W_5}  
+ c_{3 7}{}^{  \cC_{4,4}^{W_4 W_4}  }
\cC_{4,4}^{W_4 W_4}   
\ , \\[2mm]
W_4 \times W_6
& = c_{4 6}{}^{4} W_4
+ c_{4 6}{}^{6} W_6
+ c_{4 6}{}^{8} W_8
 + c_{4 6}{}^{ \cC_{3,1}^{W_3 W_3} }
\cC_{3,1}^{W_3 W_3}     
+ c_{4 6}{}^{ \cC_{3,3}^{W_3 W_3}  }
\cC_{3,3}^{W_3 W_3}   
\\
&
 + c_{4 6}{}^{ \cC_{4,1}^{W_3 W_5} }
\cC_{4,1}^{W_3 W_5}      
+ c_{4 6}{}^{  \cC_{4,2}^{W_3 W_3}   }
\cC_{4,2}^{W_3 W_3}   
+ c_{4 6}{}^{  \cC_{4,2}^{W_3 W_5}  }
\cC_{4,2}^{W_3 W_5}    
+ c_{4 6}{}^{  \cC_{4,2}^{W_4 W_4} }
\cC_{4,2}^{W_4 W_4}    
\\
&
 + c_{4 6}{}^{  \cC_{4,3}^{W_3 W_5}  }
\cC_{4,3}^{W_3 W_5}      
+ c_{4 6}{}^{  \cC_{4,4}^{W_3 W_5}  }
\cC_{4,4}^{W_3 W_5}    
+ c_{4 6}{}^{  \cC_{4,4}^{W_4 W_4}  }
\cC_{4,4}^{W_4 W_4}    
\ , \\[2mm] 
W_5 \times W_5
& = 
g_5 \one
+ c_{5 5}{}^{4} W_4
+ c_{5 5}{}^{6} W_6
+ c_{5 5}{}^{8} W_8
 + c_{5 5}{}^{  \cC_{3,1}^{W_3 W_3}  }
\cC_{3,1}^{W_3 W_3}    
+ c_{5 5}{}^{  \cC_{3,3}^{W_3 W_3}  }
\cC_{3,3}^{W_3 W_3}    
\\
&
+ c_{5 5}{}^{  \cC_{4,0}^{W_3 W_3}  }
\cC_{4,0}^{W_3 W_3}     
+ c_{5 5}{}^{  \cC_{4,0}^{W_4 W_4}  }
\cC_{4,0}^{W_4 W_4}    
+ c_{5 5}{}^{  \cC_{4,2}^{W_3 W_3}  }
\cC_{4,2}^{W_3 W_3}   
+ c_{5 5}{}^{  \cC_{4,2}^{W_3 W_5}  }
\cC_{4,2}^{W_3 W_5}    
\\
&
+ c_{5 5}{}^{  \cC_{4,2}^{W_4 W_4}  }
\cC_{4,2}^{W_4 W_4}      
+ c_{5 5}{}^{  \cC_{4,4}^{W_3 W_5}  }
\cC_{4,4}^{W_3 W_5}     
+ c_{5 5}{}^{  \cC_{4,4}^{W_4 W_4}  }
\cC_{4,4}^{W_4 W_4}    \ . 
\ea 
\end{equation*}
\end{fleqn}
\end{minipage}
\\
\hline
    \end{tabular}
    \caption{Ansatz for  $W_{p_1} \times W_{p_2}$ OPEs studied in this work. We adopt a schematic notation in which on the RHS of an OPE we list the super Virasoro primary operators that contribute to the singular part of the OPE with their OPE coefficients. We omit all super Virasoro descendants and suppress all $z$, $y$ coordinates and derivatives. 
    The notation $\cC_{h,j}^{W_p W_q}$ stands for a composite 
    super Virasoro primary operator
    of weight $h$ and spin $j$ constructed out of $W_p W_q$,
    see \eqref{eq_composite_schematic} and Appendix
    \ref{app_composites}.
    We group the OPEs based on increasing $p_1+p_2$.}
    \label{tab_ansatz}
\end{table}

\newpage

\newpage

%%%%%%%%%%%%%%%%

\subsection{Results of the bootstrap analysis}
\label{sec_results_nice}

Next, we discuss our results for the OPE coefficients
entering Table \ref{tab_ansatz}.
We find it convenient to write the central charge $c$ and the level $k$ in terms of a new parameter $\nu$ as
\be \label{eq_nu_def}
c = 6k = 3 (1 - \nu) \ . 
\ee 
We were able to fix all OPE coefficients in the OPEs
$W_3 \times W_3$, 
$W_3 \times W_4$, 
$W_3 \times W_5$, 
$W_4 \times W_4$,
$W_3 \times W_6$, 
and
$W_4 \times W_5$
in terms of $\nu$
and of the two-point function
constants $g_p$.
In contrast,
the OPE coefficients in the
OPEs 
$W_3 \times W_7$,
$W_4 \times W_6$, and
$W_5 \times W_5$ are only partially fixed.

% The findings of the bootstrap analysis are reported in
% Appendix \ref{sdsds} in tables \ref{ugly_tab_OPEcoeffs_part1}
% to \ref{ugly_tab_OPEcoeffs_part5}.
% We observe that 
% the OPE coefficients
% $c_{33}{}^4$,
% $c_{34}{}^5$,
% $c_{35}{}^6$,
% and $c_{45}{}^7$
% are determined up to a sign by the bootstrap equations.
% These signs, however,
% are not physical, as they can be reabsorbed by field redefinitions of the form $W_p'=s_p W_p$ with
% $s_p \in \{\pm 1\}$.

% We were able to fix some of the OPE
% coefficients in the OPEs
% $W_3 \times W_7$,
% $W_4 \times W_6$, and
% $W_5 \times W_5$.
% They are reported in tables
% \ref{ugly_tab_OPEcoeffs_part6} to 
% \ref{ugly_tab_OPEcoeffs_part8}
% in appendix \ref{app_results}.
% Finally, we have derived some constraints among the OPE coefficients
% that are not fully determined;
% they are collected in table
% \ref{tab_OPEcoeffs_part9}.

Based on our results,
our expectation is that
all OPE coefficients would be 
progressively
fixed, if we pushed the bootstrap
analysis to 
OPEs $W_{p_1} \times W_{p_2}$
with higher and higher values of
$p_1 + p_2$.

To present explicitly the results of the bootstrap analysis, we have to make a choice of generators $W_p$. More precisely, we have to address two possible sources of ambiguity:
\begin{itemize}
    \item[(i)] For $p\ge 6$, there exist composite super Virasoro primary operators constructed with $W_q$'s with $q<p$ that have the same quantum numbers as $W_p$. As a result, we can always redefine $W_p$ by shifting it by a linear combination of such composites. 
\item[(ii)] For $p\ge 3$ the overall normalization of $W_p$ is arbitrary.   
\end{itemize}
The first examples of 
shifts as in point (i) occur for $p=6,7$,
    \be \label{eq_redefiningW6W7}
    W_6' = W_6 + \mu_6 \cC^{W_3 W_3}_{3,3} \ , \qquad 
    W_7' = W_7 + \mu_7 \cC^{W_3 W_4}_{\frac 72 , \frac 72} \ , 
    \ee 
    where $\mu_6$, $\mu_7$ are arbitrary constants.
    
To deal with (i) we propose the following general prescription:
\be \label{eq_no_mixing_for_Wp}
\begin{array}{c}
\text{$W_p$ must have vanishing two-point function with}
\\
\text{any composite super Virasoro primary
built with $W_q$'s with $q<p$.}
\end{array}
\ee 
We expect that this prescription fixes any ambiguity of type (i).
We have checked this claim explicitly for $p=6,7$.

Next, we consider point (ii).
The freedom in the normalization of $W_p$ 
implies that the two-point function coefficients $g_p$ can be set to any desired value.
In this subsection we observe that a normalization choice exists in which the OPE coefficients of the W-algebra take a particularly simple form.
This choice is heavily inspired by half-BPS operators in 4d $SU(N)$ super Yang-Mills,
and in particular by the notion
of single-particle operators \cite{Aprile:2018efk, Aprile:2020uxk}.
The precise connection between the W-algebra and field theory 
is discussed in Section 
\ref{sec_halfBPS}.

To describe our choice of normalization for $g_p$ we need to introduce some notation.
For integer $p\ge 3$, 
we define the following function of $\nu$,
\be \label{eq_S_def}
S_p(\nu) = \frac{(\sqrt \nu +1)_{p-1}  - 
(\sqrt \nu -p+1)_{p-1}}{p(p-1)} \frac{\nu^{\lfloor \frac{p-2}{2} \rfloor }  }{ \nu^{\frac{p-2}{2}} } \ . 
\ee 
In the above expression
$(a)_n = \prod_{j=0}^{n-1}(a+j)$
is the (ascending) Pochhammer symbol.
Even if it is not apparent
from \eqref{eq_S_def},
 $S_p(\nu)$ is a monic polynomial in $\nu$
 of degree $\lfloor \frac{p-2}{2} \rfloor$
 with integer coefficients.
The explicit expressions of $S_p(\nu)$ for the first few values of $p$ are 
\be \label{eq_Sp_examples}
\ba
S_3(\nu) & = 1 \ , \\
S_4(\nu) & = \nu +1 \ , \\ 
S_5(\nu) & = \nu + 5  \ , \\ 
S_6 (\nu) & = \nu ^2+15 \nu +8
\ , \\ 
S_7(\nu) & = \nu ^2+35 \nu +84
\ , \\ 
S_8(\nu) & = \nu ^3+70 \nu ^2+469 \nu +180
 \ ,\\ 
S_9(\nu) & = \nu ^3+126 \nu ^2+1869 \nu +3044 \ .  
\ea
\ee

Our choice of normalization
for $W_p$ (up to a sign)
can then be stated as
declaring the two-point function coefficient $g_p$ to be given by
\be \label{eq_best_normalization}
g_p = 
p \nu   \frac{\prod_{r=1}^{p-1} (\nu-r^2)}{S_p(\nu) }
\frac{\nu^{\lfloor \frac{p-2}{2} \rfloor}}{\nu^{p-1}}
  \ . 
\ee 
This is a rational function of $\nu$.

We are now in a position to list the OPE coefficients extracted
from the bootstrap analysis 
 in the normalization 
\eqref{eq_best_normalization}.
This is done in Tables
\ref{tab_OPEcoeffs_part1}-\ref{tab_OPEcoeffs_part8}.
We refer the reader to Appendix
\ref{app_results} for the same results, in which the values of the two-point function coefficients $g_p$
are left arbitrary.

Some comments are in order:

\begin{itemize}
    \item All OPE coefficients are rational functions of $\nu$. 
\item In writing the OPE coefficients we have implicitly fixed some sign ambiguities in the generators $W_p$. (Clearly
setting \eqref{eq_best_normalization} 
still allows for a sign flip
$W_p \mapsto - W_p$.)

\item The bootstrap analysis does not fix the OPE coefficients $c_{36}{}^3$
and $c_{37}{}^4$.
This is expected because of ambiguities
of the form \eqref{eq_redefiningW6W7} in the definition of $W_6$, $W_7$.
Here we have fixed these OPE coefficients to zero,
\be \label{eq_gauge_fix}
c_{36}{}^3 = 0 \ , \qquad 
c_{37}{}^4 = 0 \  . 
\ee 
Indeed, 
this is equivalent to imposing 
orthogonality of $W_6$, $W_7$
with composite super Virasoro primary operators,
\be \label{eq_no_W6W7_mixing}
\langle W_6 \; \cC^{W_3 W_3}_{3,3} \rangle = 0 \ , \qquad 
\langle W_7 \; \cC^{W_3 W_4}_{\frac 72 , \frac 72}
\rangle = 0 \ . 
\ee 
in accordance with  the general prescription \eqref{eq_no_mixing_for_Wp}.

    \item We observe that
    some OPE coefficients are $\nu$-independent. In particular, 
    \be \label{eq_nice_OPE_coeffs}
    c_{q_1 q_2}{}^p = q_1 q_2 \ , \qquad \text{for $p \ge q_1$, $p \ge q_2$, $q_1 + q_2 = p+2$} \ . 
    \ee
    At this stage, we can regard this as an empirical observation about the results of the bootstrap analysis. In Section 
    \ref{sec_halfBPS} we will relate
    \eqref{eq_nice_OPE_coeffs}
to correlators of half-BPS operators in 4d $\cN = 4$ $SU(N)$ super Yang-Mills.

    \item We can interpret \eqref{eq_nice_OPE_coeffs} as follows.  
    Let us order the OPEs
    $W_{q_1} \times W_{q_2}$
    for increasing values of $q_1 + q_2$. The generator
    $W_p$ appears for the first time in the OPEs for $q_1 + q_2 = p+2$. 
   Thus, we can \emph{define} $W_p$ to be the super Virasoro primary operator with $h=j=p/2$ that enters $W_{q_1} \times W_{q_2}$ with the fixed coefficient $q_1 q_2$ and that satisfies 
   \eqref{eq_no_mixing_for_Wp}.
   
    If there are several pairs $(q_1, q_2)$ with $q_1 + q_2 = p+2$, it does not matter which one we pick to define $W_p$. This is a non-trivial feature of the W-algebra.
 
    \item The values of $c_{37}{}^8$, $c_{46}{}^8$, $c_{55}{}^8$ reported above are educated guesses, based on 
    \eqref{eq_nice_OPE_coeffs}. Our bootstrap analysis is not powerful enough to fix them independently. It provides the relations 
    \be 
    c_{46}{}^8 = \frac 87 c_{37}{}^8 \ , \qquad 
    c_{55}{}^8 = \frac{25}{21} c_{37}{}^8 \ .
    \ee 
    These are indeed compatible with our educated guesses.

\item 
The OPE coefficients $c_{55}{}^{\cC^{W_3 W_5}_{4,4}}$,
$c_{55}{}^{\cC^{W_4 W_4}_{4,4}}$ are not completely fixed at this
stage of the bootstrap, but we do expect that they would be fixed if we pushed the bootstrap analysis to higher weights.

%     \item The two-point factors $g_p$ as given by 
%     \eqref{eq_best_normalization}
%     are divergent for special values of $\nu$.
% These values have to be studied separately. 
% Some examples are reported in 
% Appendix \ref{app_special_nus}.

    \item The two-point factors $g_p$ as given by 
\eqref{eq_best_normalization} are zero for special values of $\nu$. This will be important when we discuss some truncations
of the W-algebra $\WW$ in Section \ref{sec_truncation}.

\item We observe that, in the limit of large $\nu$, we have 
\be \label{eq_large_nu_facts}
\nu \rightarrow \infty \; : 
\qquad 
g_p = p \nu  + \cO(\nu^0) \ , \qquad 
c_{q_1 q_2}{}^p = q_1 q_2
+ \cO(\nu^{-1})\ . 
\ee 
Crucially,  the second relation holds (conjecturally)
for \emph{any} $q_1$, $q_2$, $p$ in the large $\nu$ limit. 
It is straightforward to check that it is satisfied for all
the OPE coefficients reported in Tables \ref{tab_OPEcoeffs_part1}-\ref{tab_OPEcoeffs_part8}.
Further insights into 
\eqref{eq_large_nu_facts}
are provided by the connection
with correlators of half-BPS operators explained in
Section \ref{sec_halfBPS}.

The observations
\eqref{eq_large_nu_facts} will be important in Section \ref{sec_wedge} for our discussion of the wedge algebra
of $\WW$.

\end{itemize}

\newpage

%W3 W3 and W3 W4
\begin{table}[h!]
    \centering
    \begin{tabular}{| c |}
\hline  
\textbf{
\textit{
OPE coefficients in $W_3 \times W_3$ and
$W_3 \times W_4$
}
}
\\ 
\hline 
\rule[-16mm]{0mm}{34mm}
\begin{minipage}{9cm}
\begin{fleqn}
\begin{equation*} 
\ba 
c_{33}{}^4 & = 9 \ , \\[2mm]
c_{34}{}^3 & = 
\frac{12 (\nu -9)}{\nu +1}
\ , \\[2mm]
c_{34}{}^5 & = 12 \ .
\ea 
\end{equation*}
\end{fleqn}
\end{minipage}
\\
\hline
    \end{tabular}
    \caption{Values of the OPE coefficients in the $W_3 \times W_5$, $W_3 \times W_4$ OPEs
    assuming \eqref{eq_best_normalization}.
    \label{tab_OPEcoeffs_part1}
    }
\end{table}

%W3 W5
\begin{table}[h!]
    \centering
    \begin{tabular}{| c |}
\hline  
\textbf{
\textit{
OPE coefficients in $W_3 \times W_5$
}
}
\\ 
\hline 
\rule[-25mm]{0mm}{52mm}
\begin{minipage}{9cm}
\begin{fleqn}
\begin{equation*} 
\ba 
c_{35}{}^4 & = 
\frac{15 (\nu -16) (\nu +1)}{\nu  (\nu +5)}
\ , \\[2mm]
c_{35}{}^6 & = 15 \ , \\[2mm] 
c_{35}{}^{\cC^{W_3 W_3}_{3,1}} & = 
-\frac{10 (\nu -2)}{\nu  (\nu +5)}
\ , \\[2mm]
c_{35}{}^{\cC^{W_3 W_3}_{3,3}}
& = 
\frac{15 (\nu -16) (\nu -9) (\nu +3)}{\nu  (\nu +5) \left(\nu ^2+15 \nu +8\right)}
\ .   
\ea 
\end{equation*}
\end{fleqn}
\end{minipage}
\\
\hline
    \end{tabular}
    \caption{Values of the OPE coefficients in the $W_3 \times W_5$ OPE assuming \eqref{eq_best_normalization}.
    \label{tab_OPEcoeffs_part2}
    }
\end{table}

%W4 W4
\begin{table}[h!]
    \centering
    \begin{tabular}{| c |}
\hline  
\textbf{
\textit{
OPE coefficients in $W_4 \times W_4$
}
}
\\ 
\hline 
\rule[-23mm]{0mm}{49mm}
\begin{minipage}{9cm}
\begin{fleqn}
\begin{equation*} 
\ba 
c_{44}{}^4 & = 
\frac{16 \left(\nu ^2-20 \nu +9\right)}{\nu  (\nu +1)}
\ , \\[2mm]
c_{44}{}^6 & = 16 \ , \\[2mm]
c_{44}{}^{\cC^{W_3 W_3}_{3,1}} & = 
-\frac{16}{\nu }
\ , \\[2mm]
c_{44}{}^{\cC^{W_3 W_3}_{3,3}}
& = 
\frac{32 (\nu -9) (\nu -4)}{\nu  \left(\nu ^2+15 \nu +8\right)}
\ . 
\ea 
\end{equation*}
\end{fleqn}
\end{minipage}
\\
\hline
    \end{tabular}
    \caption{Values of the OPE coefficients in the $W_4 \times W_4$ OPE assuming \eqref{eq_best_normalization}.
    \label{tab_OPEcoeffs_part3}
    }
\end{table}

\newpage 

%W3 W6
\begin{table}[h!]
    \centering
    \begin{tabular}{| c |}
\hline  
\textbf{
\textit{
OPE coefficients in $W_3 \times W_6$
}
}
\\ 
\hline 
\rule[-35mm]{0mm}{72mm}
\begin{minipage}{10cm}
\begin{fleqn}
\begin{equation*} 
\ba 
c_{36}{}^{3} & = 0 \ , \\[2mm]
c_{36}{}^{5} & = 
\frac{18 (\nu -25) (\nu +5)}{\nu ^2+15 \nu +8}
\ , \\[2mm]
c_{36}{}^{7} & = 18 \ , \\[2mm]
c_{36}{}^{\cC^{W_3 W_4}_{\frac 72, \frac 32}} 
& = 
-\frac{21 (\nu -1)}{\nu ^2+15 \nu +8}
\ , \\[2mm]
c_{36}{}^{\cC^{W_3 W_4}_{\frac 72, \frac 52}} 
& = 
\frac{3 (\nu -25) (\nu -16)}{\nu  \left(\nu ^2+15 \nu +8\right)}
\ , \\[2mm]
c_{36}{}^{\cC^{W_3 W_4}_{\frac 72, \frac 72}} 
& = 
\frac{36 (\nu -25) (\nu -16) (\nu +7)}{\left(\nu ^2+15 \nu +8\right) \left(\nu ^2+35 \nu +84\right)}
\ .
\ea 
\end{equation*}
\end{fleqn}
\end{minipage}
\\
\hline
    \end{tabular}
    \caption{Values of the OPE coefficients in the $W_3 \times W_6$ OPE assuming \eqref{eq_best_normalization}.
    \label{tab_OPEcoeffs_part4}
    }
\end{table}

%W4 W5
\begin{table}[h!]
    \centering
    \begin{tabular}{| c |}
\hline  
\textbf{
\textit{
OPE coefficients in $W_4 \times W_5$
}
}
\\ 
\hline 
\rule[-45mm]{0mm}{92mm}
\begin{minipage}{10cm}
\begin{fleqn}
\begin{equation*} 
\ba 
c_{45}{}^{3} & = 
\frac{20 (\nu -16) (\nu -9)}{\nu  (\nu +5)}
\ , \\[2mm]
c_{45}{}^{5} & = 
\frac{20 \left(\nu ^3-32 \nu ^2-77 \nu +36\right)}{\nu  (\nu +1) (\nu +5)}
\ , \\[2mm]
c_{45}{}^{7} & = 20 \ , \\[2mm]
c_{45}{}^{\cC^{W_3 W_4}_{\frac 72, \frac 12}} 
& = 
-\frac{10 (\nu +12)}{\nu  (\nu +5)} 
\ , \\[2mm]
c_{45}{}^{\cC^{W_3 W_4}_{\frac 72, \frac 32}} 
& = 
-\frac{20 (2 \nu +3)}{\nu  (\nu +5)} 
\ , \\[2mm]
c_{45}{}^{\cC^{W_3 W_4}_{\frac 72, \frac 52}} 
& = 
\frac{10 (\nu -16)}{\nu  (\nu +5)}
\ , \\[2mm]
c_{45}{}^{\cC^{W_3 W_4}_{\frac 72, \frac 72}} 
& = 
\frac{40 (\nu -16) \left(2 \nu ^2-7 \nu -63\right)}{\nu  (\nu +5) \left(\nu ^2+35 \nu +84\right)}
\ .
\ea 
\end{equation*}
\end{fleqn}
\end{minipage}
\\
\hline
    \end{tabular}
    \caption{Values of the OPE coefficients in the $W_4 \times W_5$ OPE assuming \eqref{eq_best_normalization}.
    \label{tab_OPEcoeffs_part5}
    }
\end{table}
 
\newpage

%W3 W7
\begin{table}[H]
    \centering
    \begin{tabular}{| c |}
\hline  
\textbf{
\textit{
OPE coefficients in $W_3 \times W_7$
}
}
\\ 
\hline 
\rule[-59mm]{0mm}{118mm}
\begin{minipage}{12cm}
%\small
\begin{fleqn}
\begin{equation*} 
\ba 
c_{37}{}^4 & = 0 \ , \\[2mm]
c_{37}{}^6 & = 
\frac{21 (\nu -36) \left(\nu ^2+15 \nu +8\right)}{\nu  \left(\nu ^2+35 \nu +84\right)}
\ , \\[2mm]
c_{37}{}^8 & = 21  \ ,  \qquad 
\\[2mm]
c_{37}{}^{\cC^{W_3 W_3}_{3,3}}
& = 0 \ , \\[2mm]
c_{37}{}^{\cC^{W_3 W_3}_{4,2}}
& = 
-\frac{224 \left(\nu ^5+42 \nu ^4+264 \nu ^3-253 \nu ^2+17226 \nu +9720\right)}{3 \nu ^2 (\nu +5)
   \left(\nu ^2+15 \nu +8\right) \left(\nu ^2+35 \nu +84\right)}
\ , \\[2mm]   
c_{37}{}^{\cC^{W_3 W_5}_{4,2}}
& = 
-\frac{56 \left(2 \nu ^2+13 \nu -90\right)}{5 \nu  \left(\nu ^2+35 \nu +84\right)}
\ , \\[2mm]
c_{37}{}^{\cC^{W_4 W_4}_{4,2}}
& = 
-\frac{21 \left(\nu ^2-\nu +60\right)}{2 \nu  \left(\nu ^2+35 \nu +84\right)}
\ , \\[2mm]
c_{37}{}^{\cC^{W_3 W_5}_{4,3}}
& = 
\frac{28 (\nu -36) (\nu -25)}{5 \nu  \left(\nu ^2+35 \nu +84\right)} 
\ , \\[2mm]
c_{37}{}^{\cC^{W_3 W_5}_{4,4}}
& = 
-\frac{42 \left(\nu ^3-42 \nu ^2+29 \nu +396\right)}{\nu  (\nu +5) \left(\nu ^2+35 \nu +84\right)}
+ \frac{21}{25}
c_{55}{}^{\cC^{W_3 W_5}_{4,4}}
\ , \\[2mm]
c_{37}{}^{\cC^{W_4 W_4}_{4,4}}
& = 
-\frac{42 (\nu -16) (\nu -3)}{\nu  \left(\nu ^2+35 \nu +84\right)}
+ \frac{21}{25}
c_{55}{}^{\cC^{W_4 W_4}_{4,4}}
\ .
\ea 
\end{equation*}
\end{fleqn}
\end{minipage}
\\
\hline
    \end{tabular}
    \caption{Values of   OPE coefficients in the $W_3 \times W_7$ OPE assuming \eqref{eq_best_normalization}.
    \label{tab_OPEcoeffs_part6}
    }
\end{table}

\newpage

%W4 W6
\begin{table}[H]
    \centering
    \begin{tabular}{| c |}
\hline  
\textbf{
\textit{
OPE coefficients in $W_4 \times W_6$ 
}
}
\\ 
\hline 
\rule[-74mm]{0mm}{150mm}
\begin{minipage}{12cm}
%\small
\begin{fleqn}
\begin{equation*} 
\ba 
c_{46}{}^4 & = 
\frac{24 (\nu -25) (\nu -16) (\nu +1)}{\nu  \left(\nu ^2+15 \nu +8\right)}
\ , \\[2mm] 
c_{46}{}^6 & = 
\frac{12 (2 \nu -1) \left(\nu ^3-46 \nu ^2-387 \nu -240\right)}{\nu  (\nu +1) \left(\nu ^2+15 \nu
   +8\right)}
   \ , \\[2mm]  
c_{46}{}^8 & = 24 \ , \\[2mm] 
c_{46}{}^{\cC^{W_3 W_3}_{3,1}}
& =
-\frac{56 (\nu -25) (\nu -2)}{3 \nu  \left(\nu ^2+15 \nu +8\right)}
\ , \\[2mm] 
c_{46}{}^{\cC^{W_3 W_3}_{3,3}}
& = 
\frac{36 (\nu -25) (\nu -16) (\nu -9) (\nu +3)}{\nu  \left(\nu ^2+15 \nu +8\right)^2}
\ , \\[2mm] 
c_{46}{}^{\cC^{W_3 W_5}_{4,1}}
& = 
-\frac{84 (\nu +23)}{5 \left(\nu ^2+15 \nu +8\right)} \ , \\[2mm]
c_{46}{}^{\cC^{W_3 W_3}_{4,2}}
& =\scriptstyle 
\frac{16 \left(5 \nu ^7-499 \nu ^6-9539 \nu ^5+119535 \nu ^4+1843758 \nu ^3+1311460 \nu ^2-32542800
   \nu -18144000\right)}{35 (\nu -15) \nu ^2 (\nu +5) \left(\nu ^2+15 \nu +8\right)^2}
   \ , \\[2mm] 
c_{46}{}^{\cC^{W_3 W_5}_{4,2}}
& = 
-\frac{12 \left(19 \nu ^2+117 \nu +320\right)}{5 \nu  \left(\nu ^2+15 \nu +8\right)}
\ , \\[2mm] 
c_{46}{}^{\cC^{W_4 W_4}_{4,2}}
& = 
-\frac{24 (\nu -4) (\nu +5)}{\nu  \left(\nu ^2+15 \nu +8\right)}
\ , \\[2mm] 
c_{46}{}^{\cC^{W_3 W_5}_{4,3}}
& = 
\frac{96 (\nu -25) (\nu -2)}{5 \nu  \left(\nu ^2+15 \nu +8\right)}
\ , \\[2mm] 
c_{46}{}^{\cC^{W_3 W_5}_{4,4}}
& =  
\frac{672 (\nu -9) (\nu +2)}{\nu  (\nu +5) \left(\nu ^2+15 \nu +8\right)}
+ \frac{24}{25}
c_{55}{}^{\cC^{W_3 W_5}_{4,4}}
   \ , \\[2mm] 
c_{46}{}^{\cC^{W_4 W_4}_{4,4}}
& = 
-\frac{24 (\nu -16) (\nu +3)}{\nu  \left(\nu ^2+15 \nu +8\right)}
+ \frac{24}{25}
c_{55}{}^{\cC^{W_4 W_4}_{4,4}}
\ .
\ea 
\end{equation*}
\end{fleqn}
\end{minipage}
\\
\hline
    \end{tabular}
    \caption{Values  of the OPE coefficients in the $W_4 \times W_6$ OPE assuming \eqref{eq_best_normalization}.
    \label{tab_OPEcoeffs_part7}}
\end{table}

%W5 W5
\begin{table}[H]
    \centering
    \begin{tabular}{| c |}
\hline  
\textbf{
\textit{
OPE coefficients in $W_5 \times W_5$ 
}
}
\\ 
\hline 
\rule[-58mm]{0mm}{120mm}
\begin{minipage}{12cm}
%\small
\begin{fleqn}
\begin{equation*} 
\ba 
c_{55}{}^4 & =
\frac{25 (\nu -16) \left(\nu ^3-32 \nu ^2-77 \nu +36\right)}{\nu ^2 (\nu +5)^2}
\ , \\[2mm]
c_{55}{}^6 & = 
\frac{25 \left(\nu ^2-59 \nu +16\right)}{\nu  (\nu +5)}
\ , \\[2mm]
c_{55}{}^8 & = 25 \ , \\[2mm]
c_{55}{}^{\cC^{W_3 W_3}_{3,1}}
& = 
-\frac{50 \left(5 \nu ^3-103 \nu ^2-460 \nu -864\right)}{9 \nu ^2 (\nu +5)^2}
\ , \\[2mm] 
c_{55}{}^{\cC^{W_3 W_3}_{3,3}}
& = 
\frac{25 (\nu -16) (\nu -9) \left(3 \nu ^3-24 \nu ^2-115 \nu +64\right)}{\nu ^2 (\nu +5)^2 \left(\nu
   ^2+15 \nu +8\right)}
   \ , \\[2mm] 
c_{55}{}^{\cC^{W_3 W_3}_{4,0}}
& = 
-\frac{40 \left(\nu ^3-13 \nu ^2+210 \nu -558\right)}{3 (\nu -7) \nu ^2 (\nu +5)}
\ , \\[2mm] 
c_{55}{}^{\cC^{W_4 W_4}_{4,0}}
& =
-\frac{25}{\nu }
\ , \\[2mm] 
c_{55}{}^{\cC^{W_3 W_3}_{4,2}}
& =\scriptstyle 
\frac{40 \left(5 \nu ^7-226 \nu ^6-2904 \nu ^5+46954 \nu ^4+338011 \nu ^3+1286280 \nu ^2+7747320 \nu
   +3959280\right)}{21 (\nu -15) \nu ^2 (\nu +5)^3 \left(\nu ^2+15 \nu +8\right)}
\ , \\[2mm] 
c_{55}{}^{\cC^{W_3 W_5}_{4,2}}
& = -\frac{60 (\nu -2)}{\nu  (\nu +5)}
\ , \\[2mm] 
c_{55}{}^{\cC^{W_4 W_4}_{4,2}}
& =-\frac{25}{\nu }
\ .
\ea 
\end{equation*}
\end{fleqn}
\end{minipage}
\\
\hline
    \end{tabular}
    \caption{Values of the OPE coefficients in the $W_5 \times W_5$ OPE assuming \eqref{eq_best_normalization}.
    \label{tab_OPEcoeffs_part8}}
\end{table}

\newpage

\subsection{Filtration on $\WW$}
\label{sec_WW_filtration}

In this subsection we discuss a  
filtration structure on $\WW$, 
which we refer to as 
$\mathfrak R$-filtration.
Evidence for this filtration comes
from the 
OPE coefficients reported in Tables \ref{tab_OPEcoeffs_part1}-\ref{tab_OPEcoeffs_part8}.
The $\mathfrak R$-filtration is a  weight-based increasing 
filtration.
More explicitly, we assign weights  to the strong generators
of $\WW$, according to the following table,
\begin{center}
\begingroup
\renewcommand{\arraystretch}{1.3}
 \begin{tabular}{| c || c | c | c | c | c | c | c | c | } 
 \hline 
$\mathfrak{psl}(2|2)$ multiplets & \multicolumn{4}{c|}{$\mathbb J$}
 & \multicolumn{4}{c|}{$\mathbb W_p$ ($p=3,4,5,\dots$)}
 \\ \hline
operators & $J$ & $G$ & $\widetilde G$ & $T$
 & $W_p$ & $G_{W_p}$ & $\widetilde G_{W_p}$ &
 $T_{W_p}$ 
\\ \hline 
$\mathfrak R$-weight& 
$1$ & 
$1$ &
$1$ &
$1$ & 
$\frac p2$ & 
$\frac p2$ & 
$\frac p2$ & 
$\frac p2$
\\ \hline   
\end{tabular}
\endgroup
\end{center}
The identity operator $\one$ and the 
derivative $\partial_z$ are assigned weight zero.
The weight of a normal ordered product is the sum of the weights
of the operators that enter it.
The weight of a linear combination of terms is
the maximum of the weights of each summand.
We define 
\be 
\ba 
\cV_{h,j,r} & = \{ \text{$\mathfrak{sl}(2)_y$ primaries of definite $(h,j,r)$} \} \ , \\
\mathfrak F_{h,j,r, \mathfrak R} &= \{ \text{$\mathfrak{sl}(2)_y$ primaries of definite $(h,j,r)$ and weight $\mathfrak R -k$, $k\in \mathbb Z_{\ge0}$} \}  \ . 
\ea 
\ee 
(The states on the RHSs of the above relations must have definite $h$, but need not
be $\mathfrak{sl}(2)_z$ primaries.)
By construction, summing $\mathfrak F_{h,j,r, \mathfrak R}$
over $\mathfrak R \in \frac 12 \mathbb Z_{\ge 0}$
reproduces the full space
$\cV_{h,j,r}$.
The associated graded is 
\be 
\mathrm{gr} \cV_{h,j,r} = \bigoplus_{\mathfrak R \in \frac 12 \mathbb Z_{\ge 0}} \mathfrak G_{h,j,r, \mathfrak R} \ , \qquad 
\mathfrak G_{h,j,r, \mathfrak R} = \mathfrak F_{h,j,r, \mathfrak R}/\mathfrak F_{h,j,r, \mathfrak R-1} \ . 
\ee 
We also claim that the $\mathfrak R$-filtration of $\WW$
satisfies the following additional properties.
Suppose $\cO_1 \in \mathfrak F_{h_1,j_1,r_1, \mathfrak R_1}$
and $\cO_2 \in \mathfrak F_{h_2,j_2,r_2, \mathfrak R_2}$.
Parametrize the full 
$\cO_1  \times \cO_2$ OPE
as 
\be 
\cO_1(z_1, y_1) \cO_2(z_2,y_2) = \sum_{n \in \mathbb Z} 
\sum_j \frac{y_{12}^{j_1+j_2-j}}{z_{12}^n} \widehat \cD_{j_1, j_2;j}(y_{12},\partial_{y_2})\{ \cO_1 \cO_2\}^j_n(z_2,y_2) \ . 
\ee 
Here $z_{12}=z_1-z_2$,
$y_{12}=y_1=y_2$. On the RHS of the above expression we have decomposed
the resultin irreducible representations of $\mathfrak{sl}(2)_y$.
The sums over $j$ is over the usual spins
allowed by $\mathfrak{sl}(2)_y$ selection rules
when combining an irreducible representation of spin $j_1$ and one of spin $j_2$.
The quantity $\widehat \cD_{j_1, j_2;j}(y_{12},\partial_{y_2})$
is a differential operator, see \eqref{eq_sl2_diff_ops}
for its expression.
The operator $\{ \cO_1 \cO_2\}^j_n(z_2,y_2)$ is by construction
an $\mathfrak{sl}(2)_y$ primary, but needs not be
an $\mathfrak{sl}(2)_z$ primary.
With this notation, our claim about the
$\mathfrak R$ filtration is as follows,
\be 
\ba 
\{ \cO_1 \cO_2 \}^j_0 &\in 
\mathfrak F_{h_1 + h_2 , j , r_1 + r_2 , \mathfrak R_1 + \mathfrak R_2}
&
&\text{(normal ordered product)}
 \ ,   \\
\{ \cO_1 \cO_2 \}^j_1 &\in 
\mathfrak F_{h_1 + h_2 -1 , j  ,r_1 + r_2, \mathfrak R_1 + \mathfrak R_2 -1}
 &
&\text{(simple pole in the OPE)}\ .
\ea 
\ee 
While we do not have a proof of the above claims,
we have verified that they are compatible with the
explicit OPE data that we were able to fix
with the bootstrap analysis.

Another  property of the $\mathfrak R$
filtration is that all states in the same
$\mathfrak{psl}(2|2)$ multiplet
have the same $\mathfrak R$ degree.
This is apparent for the strong generators
listed in the table above,
and it can also be checked in some examples of $\mathfrak{psl}(2|2)$ multiplets of low conformal weight.

\paragraph{Tuning the central charge and simple quotients.}
The $\mathfrak R$-filtration described above is defined for generic values of the central charge $c$.
We may also consider the case in which 
 the central charge is tuned to a special value, 
and $\WW$ develops an ideal of null states.
In this case, the $\mathfrak R$-filtration descends
to the simple quotient.
Let us define 
\be 
\cN_{h,j,r} = \{ \text{null states in $\cV_{h,j,r}$}\} \ . 
\ee 
The quotient by null states is a map 
\be 
\pi : \cV_{h,j,r} \rightarrow \widetilde \cV_{h,j,r} := \cV_{h,j,r}/\cN_{h,j,r} \ , \qquad 
\pi: v \mapsto v + \cN_{h,j,r} \ . 
\ee 
The induced filtration is 
\be 
\widetilde {\mathfrak F}_{h,j,r, \mathfrak R} = \pi \Big( 
{\mathfrak F}_{h,j,r, \mathfrak R}
\Big)  \ . 
\ee 
In simpler terms,    
to determine the weight of a state in the quotient,
we  add arbitrary null states
and tune them to minimize the weight.

\section{Truncation: from $\WW$ to $\mathcal V(A_{N-1})$}
\label{sec_truncation}

The bootstrap analysis of the previous section provides strong
evidence for the existence of a
W-algebra $\WW$ whose only free parameter\footnote{Up to the choice of normalization of the generators and the freedom discussed around \eqref{eq_redefiningW6W7}} is the central charge $c$.
In this section we study
$\WW$
as the central charge approaches the values\footnote{One could ask the converse question namely for which value of the central charge $c$  all the strong generators in the multiplets $\mathbb W_p$ with $p>N$ are nulls.
While we do not have a complete answer to this question, we know that there are more values compared to the one given in \eqref{cGOOD}, for example $c=3$ for $\mathcal{W}(2,3)$ \cite{Bonetti:2018fqz}.}
\be 
c = - 3 (N^2-1) \quad
\text{or equivalently} \quad \nu = N^2 \ , \qquad N  = 2,3,4,\dots \ .
\ee 
For $\nu = N^2$, $\WW$ develops a large ideal of null states,
which we denote $\cI_N$. Recall that an ideal $\cI$
of a vertex algebra $\mathcal V$ is a vector subspace of states satisfying the requirements
\be
\partial^k A \in \cI \ , \quad 
\{ A X \}_n \in \cI \ , \quad
\{ X A \}_n \in \cI 
 \ , \quad 
\text{for any $A \in \cI$, $X\in \cV$, $n\in \mathbb Z$,
$k \in \mathbb Z_{\ge 0}$} 
\ . 
\ee 
More precisely, we conjecture that $\cI_N$ satisfies the following:
\begin{itemize}
    \item $\cI_N$ contains all strong generators $W_p$ with $p > N$.
    \item In fact, $\cI_N$ is generated by $W_p$ with $p>N$.
    \item The ideals $\cI_N$ satisfy strict inclusions as vector spaces\footnote{We refer the reader to \cite{Chang:2024zqi} for a related discussion associated to the space of BPS operators in 4d SCFTs.}
\be \label{eq_inclusion}
\cI_{N } \supsetneq \cI_{N+1} \ . 
\ee 
\end{itemize}
Clearly, the ideal generated by $W_{p>N}$
contains all states written as normal ordered product
of some (derivative of) $W_p$ with $p>N$
with any other (derivative of a) strong generator.
More interestingly, the ideal generated by 
$W_{p>N}$ also contains states that are found in the
singular part of the OPE of some $W_p$ with $p>N$ with any other
states. For example,  by looking at the right hand side of the OPEs $W_{q} \times W_{p>N}$ when $c = - 3 (N^2-1)$ we discover additional null states that are necessary for the existence of the truncated W-algebra. These null states are associated with VOA avatars of Higgs branch relations and can be matched against the free field realization of \cite{Bonetti:2018fqz}.

The choice of generators $W_p$ (and their normalizations)
encoded in \eqref{eq_no_mixing_for_Wp} and \eqref{eq_best_normalization}
is particularly well-suited to analyze the limit $
\nu \rightarrow N^2$ for $N=2,3,\dots$.
Indeed, we observe that,
in this limit
all two-point function coefficients $g_p$
in \eqref{eq_best_normalization} and all
OPE coefficients
in Tables 
\ref{tab_OPEcoeffs_part1}-\ref{tab_OPEcoeffs_part8} 
remain finite.\footnote{For each $p\ge 4$, the roots of the polynomial $S_p(\nu)$ are all
negative real numbers.
(This is a conjectural statement, which we have verified explicitly for $p\le 60$.)
}  
Moreover,  it is clear from the expression
\eqref{eq_best_normalization}
that $g_p$ goes to zero as $\nu \rightarrow N^2$ for $p > N$.

Besides the conjectures listed above, we also claim that 
\begin{itemize}
    \item The simple quotient of $\WW$ at $\nu = N^2$ by the ideal $\cI_N$ is a VOA 
 with small $\cN = 4$ supersymmetry and strong generators
    $\mathbb J$, $\{ \mathbb W_p \}_{p=3}^N$.
This VOA is isomorphic to 
 the VOA 
    $\mathcal V(A_{N-1})$
    associated to the Coxeter group
    $A_{N-1} = 
    \text{Weyl}(\mathfrak{a}_{N-1})=
    \text{Weyl}(\mathfrak{su}(N))$ according to the proposal
    of \cite{Bonetti:2018fqz}, proven in \cite{Arakawa:2023cki}.
    It is also isomorphic to the VOA associated to
    4d $\cN = 4$ $SU(N)$ super Yang-Mills by the map of 
    \cite{Beem:2013sza}.
\end{itemize}

% This fact, together with more checks discussed below, motivates us to formulate the following conjectures:
% \begin{enumerate}
%     \item In the limit $\nu \rightarrow N^2$, all generators $W_p$ with $p>N$ become null operators. 
% \item 
% After modding out by null operators, the W-algebra
%     $\WW$
%     reduces to a VOA 
%  with small $\cN = 4$ supersymmetry and strong generators
%     $\mathbb J$, $\{ \mathbb W_p \}_{p=3}^N$.
% This VOA is isomorphic to 
%  the VOA 
%     $\mathcal V(A_{N-1})$
%     associated to the Coxeter group
%     $A_{N-1} = 
%     \text{Weyl}(\mathfrak{a}_{N-1})=
%     \text{Weyl}(\mathfrak{su}(N))$ according to the proposal
%     of \cite{Bonetti:2018fqz}, proven in \cite{Arakawa:2023cki}.
% \end{enumerate}

While we do not have rigorous proofs 
of above  claims,
we have gathered evidence in their
favor by studying some explicit examples.
In particular, in Section \ref{sec_Wq_WpN_OPE}
we study some implication of the $W_q \times W_{p>N}$ OPEs
for the structure of the ideal $\cI_N$.
In Section \ref{sec_N_examples} we analyze 
explicitly the cases
$N=2,3,4,5$, and compare 
 with the
VOA $\cV(A_{N-1})$
and its the free-field realization
proposed in \cite{Bonetti:2018fqz}
and proven in \cite{Arakawa:2023cki}.

\paragraph{Remarks on inclusion.}
The conjectural inclusion 
\eqref{eq_inclusion} amounts to the statement that,
if a state is null for $\nu = N^2$, then
it is also null for $\nu = M^2$ with $M = 2,\dots,N-1$.
This is clear for the strong generators
$W_p$, since $\cI_N$ is (conjecturally) generated by $W_{p >N}$.
This property, however, is less obvious for 
states that are written as linear combinations of
normal ordered products of several strong generators and their derivatives. Nonetheless, we can verify that
\eqref{eq_inclusion} holds for all states in $\mathfrak{psl}(2|2)$
multiplets with $h \le 4$, by means of  explicit
computation of their norms.
We refer the reader to Appendix \ref{app_low_states}
for further details.

As an example of a nontrivial check of \eqref{eq_inclusion}, let us discuss here the two-point function coefficients of the
composite super Virasoro primaries
listed in \eqref{eq_composite_schematic},
which can be computed 
using their explicit expressions in Appendix \ref{app_composites}.
If we make use of \eqref{eq_best_normalization} for the two-point functions
coefficients $g_p$, the results are as follows:
\small 
\be \label{eq_norm_summary}
\ba 
(h,j) & = (3,1) \ , &
& \cC_{3,1}^{W_3 W_3}  \ , & 
&  \frac{81 (\nu -16) (\nu -9) (\nu -4) (\nu -1)^2}{5 (\nu -11) (\nu -2) \nu } \ , \\ 
(h,j) & = (3,3) \ , &
& \cC_{3,3}^{W_3 W_3}  \ , & 
&  
\frac{18 (\nu -4) (\nu -1)^2 S_6(\nu)}{\nu  (\nu +3) (\nu +7)}
\ , \\ 
(h,j) & = (\tfrac 72,\tfrac 12) \ , &
& \cC_{\frac 72, \frac 12}^{W_3 W_4}  \ , & 
&   
\frac{24 (\nu -16) (\nu -9) (\nu -4) (\nu -1)^2}{5 (\nu -6) \nu ^2}
\ , \\ 
(h,j) & = (\tfrac 72,\tfrac 32) \ , &
& \cC_{\frac 72, \frac 32}^{W_3 W_4}  \ , & 
&  
\frac{72 (\nu -25) (\nu -16) (\nu -9) (\nu -4) (\nu -1)}{5 (\nu -13) \nu ^2}
\ , \\ 
(h,j) & = (\tfrac 72,\tfrac 52) \ , &
& \cC_{\frac 72, \frac 52}^{W_3 W_4}  \ , & 
&  
\frac{144 (\nu -9) (\nu -4) (\nu -1)^2}{7 \nu  (\nu +2)}
\ , \\ 
(h,j) & = (\tfrac 72,\tfrac 72) \ , &
& \cC_{\frac 72, \frac 72}^{W_3 W_4}  \ , & 
&   
\frac{12 (\nu -9) (\nu -4) (\nu -1)^2 
S_7(\nu)
}{\nu ^2 (\nu +7) (\nu +9)}
\ , \\ 
(h,j) & = (4,0) \ , &
& \begin{array}{l}
\cC_{4,0}^{W_3 W_3} \\
\cC_{4,0}^{W_4 W_4}
\end{array}\ , & 
\det =&  
-\frac{162 (\nu -25) (\nu -16) (\nu -7) (\nu -4)^3 (\nu -1)^4}{(\nu -9) (\nu -6) \nu ^3 (3 \nu -11)}
\ , \\ 
(h,j) & = (4,1) \ , &
& \cC_{4,1}^{W_3 W_5}\ , & 
 & 
\frac{15 (\nu -25) (\nu -16) (\nu -9) (\nu -4) (\nu -1)^2}{2 (\nu -7) \nu ^3}
\ , \\ 
(h,j) & = (4,2) \ , &
& \begin{array}{l}
\cC_{4,2}^{W_3 W_3} \\ 
\cC_{4,2}^{W_3 W_5} \\ 
\cC_{4,2}^{W_4 W_4}
\end{array}\ , & 
\det = & 
-\frac{1749600 (\nu -36) (\nu -25) (\nu -16)^2 (\nu -9)^3 (\nu -4)^4 (\nu -1)^6 (\nu +5)}{7 \nu ^7 (3 \nu -1) \left(\nu ^2-12 \nu
   -57\right)^2}
\ , \\ 
(h,j) & = (4,3) \ , &
& \cC_{4,3}^{W_3 W_5}\ , & 
 & 
\frac{225 (\nu -16) (\nu -9) (\nu -4) (\nu -1)^2}{8 \nu ^2 (\nu +3)}
\ , \\ 
(h,j) & = (4,4) \ , &
& \begin{array}{l}
\cC_{4,3}^{W_3 W_5}  \\ 
\cC_{4,3}^{W_4 W_4} 
\end{array}\ , & 
\det = &  
\frac{480 (\nu -16) (\nu -9)^2 (\nu -4)^3 (\nu -1)^4 S_8(\nu)}{\nu ^5 (\nu +5) (\nu +9) (\nu +11)^2}
\ , \\ 
\ea 
\ee 
\normalsize
In the above list,
when there is a single composite super Virasoro primary operator, we have given its two-point function coefficients.
If there are two or more operators,
we have written the determinant of the matrix of
two-point function coefficients,
because the expressions of the full matrices are
quite cumbersome (the interested reader can find them in Appendix~\ref{app_composites}).

\subsection{Null states generated from the $W_{q}\times  W_{p>N}$ OPEs}
\label{sec_Wq_WpN_OPE}

We will now look more closely at the OPEs $W_{q}\times  W_{p>N}$ when $\nu=N^2$. The first interesting example is given by the OPE $W_3\times  W_{N+1}$. 
To analyze this case, let us first collect from Tables
\ref{tab_OPEcoeffs_part1}-\ref{tab_OPEcoeffs_part8}  the general form of the OPE
\begin{equation}
\label{W3withWpplus1generalOPE}
W_3\times  W_{p+1}=g_3\,\delta_{2,p}\,\mathbf{1}+
\,c_{3,p+1}{}^p\,W_{p}+
\,3(p+1)\,W_{p+2}+\\
\mu_p^+\,\mathcal{C}_{\frac{p}{2}+1,\frac{p}{2}+1}
+\mu_p^0\,\mathcal{C}_{\frac{p}{2}+1,\frac{p}{2}}
+
\mu_p^-\,\mathcal{C}_{\frac{p}{2}+1,\frac{p}{2}-1}
\,.
\end{equation}
In this expression $\mathcal{C}_{(h,j)}$ denote composite super-Virasoro primaries and the OPE coefficients are given by
\begin{equation}
c_{3,p+1}{}^p(\nu)=3 (p+1)(\nu-p^2)\,\frac{S_p(\nu)}{S_{p+1}(\nu)}\frac{1}{\nu^{1-(p\, \text{mod}2)}}\,,
\end{equation} 
and
\begin{equation}
\big{(}\mu^+_4(\nu),\mu^+_5(\nu),\dots\big{)}=
\big{(}\tfrac{15(\nu-16)(\nu-9)(\nu+3)}{\nu\,S_5(\nu)S_6(\nu)},\tfrac{36(\nu-25)(\nu-16)(\nu+7)}{S_6(\nu)S_7(\nu)},\dots\big{)}\,,
\end{equation}
\begin{equation}
\big{(}\mu^0_5(\nu),\mu^0_6(\nu),\dots\big{)}=
\big{(}\tfrac{3(\nu-25)(\nu-16)}{\nu\,S_6(\nu)},
\tfrac{28(\nu-36)(\nu-24)}{5\,\nu\,S_7(\nu)},
\dots\big{)}\,,
\end{equation}
\begin{equation}
\big{(}\mu^-_4(\nu),\mu^-_5(\nu),\dots\big{)}=
\big{(}\tfrac{-10(\nu-2)}{\nu\,S_5(\nu)},\tfrac{-21(\nu-1)}{S_6(\nu)},\dots\big{)}\,,
\end{equation}
with $S_p(\nu)$ is defined in \eqref{eq_S_def}. For low values of $p$ some of the composites super-Virasoro primaries $\mathcal{C}$ in \eqref{W3withWpplus1generalOPE} are absent, so we omit the corresponding $\mu$ coefficient.
We recall that, as above, we organized the right and side of the OPEs in super-Virasoso families. In particular, when $p=2$ the term $W_2$ is absent and is a descendant of the identity $\mathbf{1}$.

When $\nu=N^2$, the operators $W_{p>N}$ are null. So, when we quotient by these nulls, we should set to zero the result of the OPE of these nulls with any other operator. The simplest case is given by $W_3\times  W_{N+1}$, which is a specialization of \eqref{W3withWpplus1generalOPE} to $p=N$ and $\nu=N^2$. We observe that in this case $c_{3,N+1}{}^N(N^2)=0$. This is good news, since the generator $W_N$ is non zero. The generator $W_{N+2}$ on the other hand appears with non vanishing OPE coefficient, but this is not a problem since we set this generator to zero in the quotient.
The cases $N=2$ and $N=3$ should be treated with care since in these cases the OPE coefficient of the super-Virasoro descendants of the identity and $W_3$ respectively have poles when $\nu$ approaches $4$ and $9$. These poles cancel against the zero of $g_3$ and $c_{34}^3$ respectively to leave the operators
\begin{equation}
\label{NulllightN23}
\mathcal{N}_{(2,0)}^{\mathfrak{su}(2)}=
(JJ)^0_0 + \tfrac{1}{3} T \,,
\qquad
\mathcal{N}_{(\frac{5}{2},\frac{1}{2})}^{\mathfrak{su}(3)}=
(J W_3)^{\frac{1}{2}}_0+ \tfrac{1}{4} T_{W_3}
\end{equation}
This implies that these operators, whose norm is easily seen to vanish for $\nu=4$ and $\nu=9$ respectively, must be set to zero in the respective quotients. 

For $N\geq 4$ the super-Virasoro primaries $\mathcal{C}$ contributes to the OPEs $W_3\times  W_{N+1}$ as well.
The OPE coefficients $\mu^+_N(N^2)$, $\mu^0_N(N^2)$ vanish while $\mu^-_N(N^2)\neq 0$. This implies that the operator that multiplies $\mu^-_N(N^2)\neq 0$, namely
\begin{equation}
\label{NulllightNgen}
\mathcal{N}_{(\frac{N}{2}+1,\frac{N}{2}-1)}^{\mathfrak{su}(N)}
=\mathcal{C}_{(\frac{N}{2}+1,\frac{N}{2}-1)}\Big{|}_{\nu=N^2,W_{p>N}=0}\,,
\end{equation}
%where $\mathcal{C}_{(\frac{N}{2}+1,\frac{N}{2}-1)}$
must be set to zero in the quotient.
Also in this case, we can verify that the 
operator \eqref{NulllightNgen} has zero norm,
see \eqref{eq_norm_summary}.
The null operator $\mathcal{N}_{(\frac{N}{2}+1,\frac{N}{2}-1)}^{\mathfrak{su}(N)}$  given in \eqref{NulllightN23} for $N=2,3$ and in \eqref{NulllightNgen} for $N\geq 4$  is the VOA avatar of the lightest Higgs branch relations. 
We further elaborate of the interpretation of the nulls states from the four dimensional perspective in Section \ref{sec_connection_4d}.

For $N=2$ the state $\mathcal{N}_{(2,0)}^{\mathfrak{su}(2)}$ generates the maximal ideal (after we have set $W_{p\geq 3}$ to zero).
For higher values of $N$ there are more generators of the maximal ideal.
To discover them, lets consider the OPE $W_3\times W_{N+2}$.
Looking at \eqref{W3withWpplus1generalOPE} with $p=N+1$ and $\nu=N^2$ we see that on the right hand side $W_{N+1}$ and $W_{N+2}$ are zero.
The OPE coefficients $\mu^+_{N+1}(N^2)$, $\mu^0_{N+1}(N^2)$ vanish while $\mu^-_{N+1}(N^2)\neq 0$. This implies that the operator that multiplies $\mu^-_{N+1}(N^2)\neq 0$, namely
\begin{equation}
\label{NulllightNgen}
\mathcal{N}_{(\frac{N+3}{2},\frac{N+1}{2})}^{\mathfrak{su}(N)}
=\mathcal{C}_{(\frac{N+3}{2},\frac{N+1}{2})}\Big{|}_{\nu=N^2,W_{p>N}=0}\,,
\end{equation}
%where $\mathcal{C}_{(\frac{N}{2}+1,\frac{N}{2}-1)}$
must be set to zero in the quotient.
Also in this case, we can verify that the 
operator \eqref{NulllightNgen} has zero norm.
As conjectured in \cite{Bonetti:2018fqz} for $N=3$ the states 
$\mathcal{N}_{(\frac{5}{2},\frac{1}{2})}^{\mathfrak{su}(3)}$ and $\mathcal{N}_{(3,1)}^{\mathfrak{su}(3)}$ generate the maximal ideal, but for $N\geq 4$ we need more.
For low values of $N$ we conjecture that the following nulls generate the maximal ideal
\begin{equation}
\ba
N=2: &\quad \mathcal{N}_{(2,0)}\\
N=3: &\quad \mathcal{N}_{(\frac{5}{2},\frac{1}{2})};\,\,\,
\mathcal{N}_{(3,1)}\\
N=4: &\quad \mathcal{N}_{(3,1)}\,
;\,\,\,
\mathcal{N}_{(\frac{7}{2},\frac{3}{2})},\,\,\mathcal{N}_{(\frac{7}{2},\frac{1}{2})}
;\,\,\,
\mathcal{N}_{(4,2)},\,
\mathcal{N}_{(4,0)}
\\
N=5: &\quad  \mathcal{N}_{(\frac{7}{2},\frac{3}{2})}
;\,\,\,
\mathcal{N}_{(4,2)},\,
\mathcal{N}_{(4,1)},\,
\mathcal{N}_{(4,0)};\,\,\,
\mathcal{N}_{(\frac{9}{2},\frac{5}{2})},\,\,\mathcal{N}_{(\frac{9}{2},\frac{3}{2})},\,\,\mathcal{N}_{(\frac{9}{2},\frac{1}{2})}
\\
\ea
\end{equation}
We have already showed that the nulls in the first two columns are obtained by considering the OPEs $W_3\times W_{N+1}$ and  $W_3\times W_{N+2}$.
The OPE $W_{4} \times W_{N+1}$ for example gives the second and third column of nulls and so on.

\subsection{Analysis  for $N=2, 3,4, 5$}
\label{sec_N_examples}

\subsubsection{The case $N=2$}

The expression \eqref{eq_best_normalization} for $g_p$
shows that for $\nu \rightarrow 4$ \emph{all} generators
$W_p$ become null operators.
This can also be argued differently as follows.

The W-algebra $\WW$ contains the small $\cN=4$ super Virasoro algebra as a subalgebra. If we tune $\nu = 4$, equivalently $c=-9$, 
there is a non-trivial super Virasoro descendant of the identity operator that becomes super Virasoro primary
(hence, null). 
In turn, this restricts the allowed 
representations of the super Virasoro algebra.

More explicitly,
the following operator becomes super Virasoro primary,
\be 
\mathcal N = (JJ)^0_0 + \frac 13 T \ . 
\ee 
We   now consider the OPE $\cN \times W_p$. This is completely fixed by 
the super Virasoro algebra and the assumption that $W_p$ be a super Virasoro primary. Since $\cN$ is a null state, all operators appearing on the
RHS of the OPE $\cN \times W_p$ must be null. However, the operator $W_p$ itself
appears on the RHS of this OPE, with coefficient
\be 
\cN \times W_p \supset - \frac {p(p+1)}{6} W_p \ . 
\ee 
We see that, if we assume that $W_p$
is not null, we get a contradiction.
Hence we conclude that $W_p$ must be null.

 After modding out by null states,
the W-algebra $\WW$
reduces to a copy 
 of the 
small $\cN =4$ superconformal algebra.
This indeed agrees with the VOA
$\cV(A_1)$, which is the same as the
small $\cN = 4$ superconformal algebra
at central charge $c=-9$,
see also \cite{Adamovic:2014lra}.

\subsubsection{The case $N=3$}

It is straightforward to check that all OPE coefficient in Tables \ref{tab_OPEcoeffs_part1}-\ref{tab_OPEcoeffs_part8}
remain finite in the limit $\nu \rightarrow 9$. Moreover, we see from
\eqref{eq_best_normalization} that the operators $W_4$, $W_5$, etc.~become null.

Let us present an argument 
for the fact that $W_4$ must be null, which is independent on a choice of normalization for the $W_p$ generators.
The norm squared of the composite super Virasoro primary operator
$\cC^{W_3 W_3}_{3,1}$ 
is a function of $g_3$ and $\nu$, given by \eqref{eq_composite_norm_list_h3}, repeated here for convenience,\footnote{We use $\langle \cO \cO' \rangle$
to denote the coefficient of the identity operator
$\one$ in the OPE $\cO \times \cO'$.
}
\be \label{eq_C31_norm}
\langle \cC^{W_3  W_3}_{3,1} 
\cC^{W_3 W_3}_{3,1} 
\rangle  = 
\frac{9 (\nu -16) (\nu -9) \nu  }{5 (\nu -11) (\nu -4) (\nu
   -2)} g_3^2
\ . 
\ee
We observe that $\cC^{W_3 W_3}_{3,1}$ becomes null
for $\nu = 9$, even if $W_3$ is not null ($g_3 \neq 0$).
Moreover, we can compute the
OPE of $\cC^{W_3 W_3}_{3,1}$ with $W_4$. On the RHS, we encounter $W_4$ itself, with OPE coefficient
\be \label{eq_ideal_kills_W4}
\cC^{W_3 W_3}_{3,1} \times W_4 = 
-\frac{108 (\nu -16) (\nu +1) g_3 }{5 (\nu -11) (\nu -4) (\nu -2)}
  W_4 + \dots 
\ee 
The OPE coefficient 
has a finite, non-zero
limit for $\nu = 9$.
(Here we assume  $g_3 \neq 0$.)
Null states form an ideal.
For $\nu = 9$,
we see that a non-zero multiple of $W_4$ appears 
on the RHS of the OPE with the null state $\cC^{W_3 W_3}_{3,1}$ with something.
We conclude that $W_4$ must be null.
By a similar token, we  consider the OPE\footnote{Since we now know that $W_4$ is null, one might think to consider the OPE $W_3\times W_4$ to argue that $W_5$ is null.
However, the OPE coefficient
$c_{34}{}^5$ depends not only on $g_3$ but also on $g_4$ and $g_5$,
\be 
(c_{34}{}^5)^2 = \frac{60 (\nu -16) (\nu +1) g_3 g_4}{(\nu -4) (\nu -1) (\nu +5)
  g_5 } \ .  
\ee 
This is why we have to resort to the more complicated OPE $\cC^{W_3 W_3}_{3,1} \times W_5$,
in which $W_5$ appears with a coefficient that depends on $g_3$ and $\nu$ only.
}
\be 
\ba
\label{eq_ideal_kills_W5}
\cC^{W_3 W_3}_{3,1} \times W_5 &= 
-\frac{18 \sqrt{\frac{3}{5}} \sqrt{\nu -16} \sqrt{\nu -9} \sqrt{\nu } \sqrt{g_3}
   \sqrt{g_5}}{(\nu -11) (\nu -4) \sqrt{\nu +5}}
   W_3  
\\   
& - \frac{6 \left(5 \nu ^3-103 \nu ^2-460 \nu -864\right)  g_3}{(\nu -11) (\nu -4) (\nu
   -2) (\nu +5)}
W_5 + \dots
\ea
\ee 
The coefficient of $W_5$
is finite and non-zero for 
$\nu = 9$. We learn that $W_5$ must be null.
As a consistency check,
we see that the coefficient of $W_3$ goes to zero for $\nu = 9$ (it also has a factor $\sqrt{g_5}$).
This is consistent with $W_3$ being non-null for $\nu = 9$.
Next, we consider  
\be \label{eq_ideal_kills_W6}
\ba
\cC^{W_3 W_3}_{3,1} \times W_6 &= 
-\frac{42 \sqrt{6} \sqrt{\nu -25} \sqrt{\nu -16} \sqrt{\nu } \sqrt{\nu +1} g_3
   \sqrt{  g_6   }}{5 (\nu -11) (\nu -4) \sqrt{\nu ^2+15 \nu +8}
   \sqrt{  g_4    }}  W_4
   \\ 
 &   
 -\frac{18 \left(11 \nu ^4-292 \nu ^3-3223 \nu ^2-10304 \nu -2560\right) g_3}{5 (\nu -11) (\nu -4) (\nu
   -2) \left(\nu ^2+15 \nu +8\right)} W_6 + \dots 
\ea
\ee 
The coefficient of $W_6$ is finite and non-zero
for $\nu = 9$. Hence, $W_6$ must be null.
We expect that this line of reasoning can be extended to prove that $W_7$, $W_8$, etc.~are also null.

Let us now discuss other consistency checks of our proposal that $W_p$ with $p\ge 4$ is null.
As noted above, null states  form an ideal.
Thus, 
all OPE coefficients of the form
$c_{3p}{}^3$, $p\ge 4$,
and $c_{pq}{}^3$, $p,q \ge 4$
must go to zero as $\nu \rightarrow 9$.
Indeed, we verify that this is the case
for $c_{34}{}^3$, $c_{36}{}^3$, $c_{45}{}^3$. Now $c_{36}{}^3 =0$ by ``gauge fixing'' \eqref{eq_gauge_fix}.
In contrast, 
$c_{34}{}^3$,  $c_{45}{}^3$
are non-trivial functions of $\nu$ 
(see Tables \ref{tab_OPEcoeffs_part1} and
\ref{tab_OPEcoeffs_part5}) repeated here for convenience,
\be 
\ba
c_{34}{}^3 & = \frac{12 (\nu -9) \nu }{\nu +1} \ , &
c_{45}{}^{3} & = \frac{20 (\nu -16) (\nu -9) \nu }{\nu +5} \ ,
\ea
\ee 
They indeed have a zero
at $\nu=9$.

Upon modding out $W_p$'s with $p\ge 4$, the W-algebra $\WW$
reduces to a VOA 
with small $\cN = 4$ supersymmetry,
central charge $c=-24$,
and strong generators
$\mathbb J$, $\mathbb W_3$.
These features match those of the VOA
$\cV(A_2)$.
Another check is provided by the
composite super Virasoro primary
operators
$\cC_{3,3}^{W_3 W_3}$,
$\cC_{3,1}^{W_3 W_3}$.
In the VOA $\cV(A_2)$,
the composite $\cC_{3,1}^{W_3 W_3}$ is null, while 
$\cC_{3,3}^{W_3 W_3}$ is non-null.
(This follows from the free-field realization reviewed in Appendix
\ref{app_free}.)
We reproduce these features starting from $\WW$. Indeed,
using the explicit data from
the bootstrap analysis of $\WW$, we can compute
the 2-point function
coefficients of $\cC_{3,3}^{W_3 W_3}$,
$\cC_{3,1}^{W_3 W_3}$ with themselves. 
We have already reported 
the norm squared of $\cC^{W_3 W_3}_{3,1}$
in \eqref{eq_C31_norm}. That of $\cC^{W_3 W_3}_{3,3}$ reads
\be \label{eq_C33_norm}
\langle \cC^{W_3 W_3}_{3,3}
\cC^{W_3 W_3}_{3,3}
\rangle 
= 
\frac{2 \nu  \left(\nu ^2+15 \nu
   +8\right)}{(\nu -4) (\nu +3)
   (\nu +7)}
% \frac{2  (\epsilon +9) \left(\epsilon ^2+33 \epsilon +224\right)}{(\epsilon +5)
%    (\epsilon +12) (\epsilon +16)}
   g_3^2 \ . 
\ee 
As anticipated, this has a non-zero finite
limit for $\nu \rightarrow 9$.

\subsubsection{The case $N=4$}

It is straightforward to check that all OPE coefficient in Tables \ref{tab_OPEcoeffs_part1}-\ref{tab_OPEcoeffs_part8}
remain finite in the limit $\nu \rightarrow 16$. Moreover, we see from
\eqref{eq_best_normalization} that the operators $W_5$, $W_6$, etc.~become null.

We can also argue that $W_5$
must be null in a way independent on any choice of normalization.
To this end we consider again the composite operator
$\cC^{W_3 W_3}_{3,1}$ with norm squared
\eqref{eq_C31_norm}. We see that this operator is null for $\nu = 16$.
In the OPE 
\eqref{eq_ideal_kills_W4}, the coefficient of $W_4$ goes to zero for $\nu = 16$,
so we cannot draw any conclusion from this OPE.
This is consistent with $W_4$ being non-null.
In the OPE \eqref{eq_ideal_kills_W5},
the coefficient of 
$W_5$ is finite and non-zero
for $\nu = 16$. We learn that $W_5$ must be null.
As a sanity check, $W_3$ can be non-null, because its coefficient in 
\eqref{eq_ideal_kills_W5} is zero for $\nu = 16$.
In a similar way,
the coefficient of $W_6$ in
\eqref{eq_ideal_kills_W6} is finite and non-zero for
$\nu = 16$, demonstrating that $W_6$ must be null. As another sanity check,
the coefficient of $W_4$
in \eqref{eq_ideal_kills_W6} goes to zero
for $\nu = 16$, compatibly with $W_4$ being non-null. 
We expect that similar arguments can be used to infer that $W_7$, $W_8$, etc.~are null.

% We observe that the linear combination
% \be 
% \cC_{4,0} := \cC^{W_3 W_3}_{4,0} + 
% \frac{252   g_3 }{289  g_4}
% \cC^{W_4 W_4}_{4,0}
% \ee 
% becomes null for $\nu = 16$.
% This can be verified with the data given in appendix
% \ref{app_composites}.
% Moreover, we can compute 
% \be \label{eq_ideal_kills_W5_bis}
% \cC_{4,0} \times W_5
% = 
% \frac{3 g_3 P}{2890 (\nu -9)^3 (\nu -6) (\nu -4) (\nu -1) \nu  (3 \nu -11) \left(\nu ^2+15 \nu
%    +8\right) \left(\nu ^2+35 \nu +84\right)}
%    W_5 + \dots  \ , 
% \ee 
% where $P$ is the following
% polynomial in $\nu$,
% \begin{align}
% P & = \scriptstyle
% 42843 \nu ^{11}-2825712 \nu ^{10}+51241583 \nu ^9-79526856 \nu ^8-694946555 \nu ^7-11585595608 \nu
%    ^6
% \\
% & \scriptstyle 
% +370157028549 \nu ^5-1903159514848 \nu ^4-637643496660 \nu ^3+5921382552624 \nu ^2+4491490227840
%    \nu +301771008000 \ . \nn
% \end{align}
% In particular, $P$  has a finite non-zero value at $\nu = 16$.
% We then conclude that $W_5$ must be null.
% We expect that this method can also be used to show that $W_6$, $W_7$, and so on, must be null,
% but its complexity 
% prevents us from implementing it explicity.

We now turn to other consistency checks of our claim that $W_p$ with $p\ge 5$ is null. 
All OPE coefficients of the form
$c_{rp}{}^s$ ($p\ge 5$, $r,s\in \{3,4\}$),
and $c_{pq}{}^r$, ($p,q \ge 5$, $r \in \{3,4\}$)
must go to zero as $\nu \rightarrow 16$.
We  verify this
explicitly
for the OPE coefficients
$c_{35}{}^4$, 
$c_{45}{}^3$,
$c_{46}{}^4$,
$c_{55}{}^4$,
which are given by 
\be 
\ba
c_{35}{}^4 & = \frac{15 (\nu -16) (\nu +1)}{\nu +5} \ , & 
c_{45}{}^{3} & = \frac{20 (\nu -16) (\nu -9) \nu }{\nu +5} \ ,  \\
c_{46}{}^4 & = \frac{24 (\nu -25) (\nu -16) \nu  (\nu +1)}{\nu ^2+15 \nu +8} \ , & 
c_{55}{}^4 & = \frac{25 (\nu -16) \left(\nu ^3-32 \nu ^2-77 \nu +36\right)}{(\nu +5)^2} 
\ ,
\ea
\ee 
They all vanish for $\nu = 16$.

Upon modding out $W_p$'s with $p\ge 5$, the W-algebra $\WW$
reduces to a VOA 
with small $\cN = 4$ supersymmetry,
central charge $c=-45$,
and strong generators
$\mathbb J$, $\mathbb W_3$,
$\mathbb W_4$.
These features match those of the VOA
$\cV(A_3)$.
Another check is provided by the
composite super Virasoro primary
operators
$\cC_{3,3}^{W_3 W_3}$,
$\cC_{3,1}^{W_3 W_3}$.
In the VOA $\cV(A_3)$,
the composite $\cC_{3,1}^{W_3 W_3}$ is null, while 
$\cC_{3,1}^{W_3 W_3}$ is non-null.
(This follows from the free-field realization.)
We reproduce these features starting from $\WW$. Indeed,
we see from 
\eqref{eq_C31_norm}, \eqref{eq_C33_norm}
that, in the limit
$\nu \rightarrow 16$,
the composite 
$\cC_{3,1}^{W_3 W_3}$
becomes null, while 
$\cC_{3,3}^{W_3 W_3}$
remains non-null.

Next, we verify that
we can reproduce the OPE coefficients
of the VOA $\cV(A_3)$ starting from the OPE coefficients of the W-algebra
$\WW$.
The $W_{p_1} \times W_{p_2}$ OPEs
in the VOA $\cV(A_3)$
take the form 
\be 
\ba 
W_3 \times W_3 & = g_3 \one + c_{33}{}^4 W_4 \ , 
&
W_3 \times W_4  &= c_{34}{}^3 W_3 \ , 
\\
W_4 \times W_4 & = g_4 \one 
+ c_{44}{}^4 W_4 
+ c_{44}{}^{
\cC_{3,3}^{W_3 W_3}
}
\cC_{3,3}^{W_3 W_3}
\ ,
\ea 
\ee 
with OPE coefficients
\be \label{eq_VS4_OPEs} 
\ba 
c_{33}{}^4 & = \tfrac{4 \sqrt 7}{\sqrt{85}} \tfrac{g_3}{ \sqrt{g_4} }
 \ , & 
c_{34}{}^3 & =  \tfrac{4 \sqrt 7}{\sqrt{85}}
\sqrt{g_4} \ , \\
c_{44}{}^4 & = - \tfrac{11 \sqrt 5}{3 \sqrt{119}} \sqrt{g_4} \ , 
& 
c_{44}{}^{\cC_{3,3}^{W_3 W_3}}
& = \tfrac{17}{28} \tfrac{g_4}{g_3} \ . 
\ea 
\ee 
These are derived from the free-field
realization reviewed in Appendix
\ref{app_free}.

All these four OPE coefficients  are indeed exactly reproduced
from the OPE coefficients
of $\WW$,
as it can be checked using the explicit expressions
in Tables \ref{tab_OPEcoeffs_part1}-\ref{tab_OPEcoeffs_part8}
and $g_p$ in \eqref{eq_best_normalization}.

\subsubsection{The case $N=5$}

It is straightforward to check that all OPE coefficients in Tables \ref{tab_OPEcoeffs_part1}-\ref{tab_OPEcoeffs_part8}
remain finite in the limit $\nu \rightarrow 25$. Moreover, we see from
\eqref{eq_best_normalization} that the operators $W_6$, $W_7$, etc.~become null.

To argue that $W_p$, $p\ge 6$ is null in a way independent on any normalization choice,
we need an argument similar to those around 
\eqref{eq_ideal_kills_W4}, 
\eqref{eq_ideal_kills_W5}, 
\eqref{eq_ideal_kills_W6}.
First, we observe that the composite
$\cC^{W_3 W_4}_{\frac 72 , \frac 32}$
has norm squared
\be \label{eq_composite7232}
\langle
\cC^{W_3  W_4}_{\frac 72, \frac 32} 
\cC^{W_3  W_4}_{\frac 72, \frac 32} 
\rangle  = 
\frac{6 (\nu -25) (\nu -16) (\nu +1) }{5 (\nu -13)
   (\nu -4) (\nu -1)} g_3 g_4 \ .
\ee 
We see that it becomes null for $\nu = 25$.
Next, we consider the OPE
\be 
\ba
&\cC^{W_3 W_4}_{\frac 72, \frac 32} \times W_5
 =
 -\frac{24 \sqrt{\frac{3}{5}} (\nu -25) \sqrt{\nu -16} (\nu +1) (2 \nu +3)
 \sqrt{g_3} \sqrt{g_5}
}{(\nu -13) \sqrt{\nu -9} (\nu -4) (\nu -1) \sqrt{\nu } \sqrt{\nu +5}}
 W_4  
\\
& -\frac{4 \sqrt{\frac{3}{5}} \sqrt{\nu -16} (\nu +1) \left(13 \nu ^3-454 \nu ^2-2407 \nu -4352\right)
    \sqrt{    g_3     } g_4   }{\left( \nu -4\right)
   \left( \nu -1\right)    (\nu -13)
   \sqrt{\nu -9} \sqrt{\nu } (\nu +5)^{3/2} \sqrt{
   g_5
   }}
 W_6 + \dots   \ . 
 \ea
\ee 
The coefficient of $W_6$
has a finite non-zero value for $\nu = 25$.
We conclude that $W_6$ must be null.
In contrast, the coefficient of $W_4$ vanishes, consistently with the fact that $W_4$ is not a null operator.
We expect that similar arguments can be used to prove that $W_7$, $W_8$, etc.~are null.

% The first ingredient is a suitable composite operator 
% constructed with $W_3$, 
% $W_4$, $W_5$
% that becomes null
% for $\nu = 25$.
% The results of
% appendix 
% \ref{app_composites} 
% show that there are various 
% such composites with this property.
% One of them might suitable to run an argument as above.
% We do not pursue explicitly this strategy, however,
% because it is technically too challenging.

Null states should form an ideal.
Thus, 
as a consistency check,
all OPE  
coefficients of the form
$c_{rp}{}^s$ ($p\ge 6$, $r,s\in \{3,4,5\}$),
and $c_{pq}{}^r$, ($p,q \ge 6$, $r \in \{3,4,5\}$)
must go to zero as $\nu \rightarrow 25$.
The OPE coefficients $c_{36}{}^3$ and $c_{37}{}^4$
are of this form,
and have been set to zero in \eqref{eq_gauge_fix}.
A more interesting check is provided by the   OPE coefficient
\be 
c_{46}{}^4  = \frac{24 (\nu -25) (\nu -16) \nu  (\nu +1)}{\nu ^2+15 \nu +8} \ ,
\ee 
which indeed vanishes for $\nu=25$.

Upon modding out $W_p$'s with $p\ge 6$, the W-algebra $\WW$
reduces to a VOA 
with small $\cN = 4$ supersymmetry,
central charge $c=-72$,
and strong generators
$\mathbb J$, $\mathbb W_3$,
$\mathbb W_4$,
$\mathbb W_5$.
These features match those of the VOA
$\cV(A_4)$.
Another check is provided by the 
composite super Virasoro primary operator
$\cC_{\frac 72, \frac 32}^{W_3 W_4}$.
In the VOA $\cV(A_4)$,
this composite is null.
(This follows from the free-field
realization.)
We reproduce this feature
starting from $\WW$,
as already noted above, see
\eqref{eq_composite7232}.

% Indeed, the 2-point function
% coefficient of $\cC_{\frac 72, \frac 32}^{W_3 W_4}$ with itself
% in $\WW$ is computed to be 
% \be 
% % g_{\cC(\frac 72, \frac 32) ,  \cC(\frac 72, \frac 32)} = 
% g_{ \cC_{\frac 72 , \frac 32}^{W_3 W_4} \, 
% \cC_{\frac 72 , \frac 32}^{W_3 W_4}
% }
% =
% \frac{6 (\nu -25) (\nu -16) (\nu +1) }{5 (\nu -13)
%    (\nu -4) (\nu -1)} g_3 g_4
% \ , 
% \ee 
% which indeed vanishes as $\nu \rightarrow 25$.

Next, we verify that
we can reproduce 
the OPE coefficients
of the VOA $\cV(A_4)$ starting from the OPE coefficients of the W-algebra
$\WW$.

The $W_{p_1} \times W_{p_2}$ OPEs
of the VOA $\cV(A_4)$
are derived in Appendix
\ref{app_free} from the free-field realization. The result reads 
\small
\be \label{free_su5_OPEs}
\ba 
W_3 \times W_3 & = g_3 \one
+ c_{3 3}{}^{4} W_4 \ , 
\\
W_3 \times W_4 & = 
c_{3 4}{}^{3} W_3
+ c_{3 4}{}^{5} W_5 \ ,
\\
W_3 \times W_5
& =  c_{3 5}{}^{4} W_4 
+ c_{3 5}{}^{  \cC_{3,1}^{W_3 W_3}  }
\cC_{3,1}^{W_3 W_3}
+ c_{3 5}{}^{ \cC_{3,3}^{W_3 W_3} }
\cC_{3,3}^{W_3 W_3}
\ , \\
W_4 \times W_4
& = g_4 \one 
+ c_{4 4}{}^{4} W_4
 + c_{4 4}{}^{  \cC_{3,1}^{W_3 W_3}   }
\cC_{3,1}^{W_3 W_3}
+ c_{4 4}{}^{ \cC_{3,3}^{W_3 W_3}  }
\cC_{3,3}^{W_3 W_3} \ , 
\\
W_4 \times W_5
& = c_{4 5}{}^{3} W_3
+ c_{4 5}{}^{5} W_5
 + c_{4 5}{}^{ \cC_{\frac 72 , \frac 12}^{W_3 W_4}  } 
\cC_{\frac 72 , \frac 12}^{W_3 W_4}
+ c_{4 5}{}^{
\cC_{\frac 72 , \frac 52}^{W_3 W_4}
} 
\cC_{\frac 72 , \frac 52}^{W_3 W_4}
+ c_{4 5}{}^{
\cC_{\frac 72 , \frac 72}^{W_3 W_4}
} 
\cC_{\frac 72 , \frac 72}^{W_3 W_4}
 \ ,
\\
W_5 \times W_5
& = 
g_5 \one
+ c_{5 5}{}^{4} W_4
 + c_{5 5}{}^{ \cC_{3,1}^{W_3 W_3}   }
\cC_{3,1}^{W_3 W_3}
+ c_{5 5}{}^{  \cC_{3,3}^{W_3 W_3} }
\cC_{3,3}^{W_3 W_3}
+ \overline c_{5 5}{}^{ \cC_{4,0}^{W_4 W_4}   }
\cC_{4,0}^{W_4 W_4}
\\
& +  
\overline  c_{5 5}{}^{ \cC_{4,2}^{W_3 W_5}   }
\cC_{4,2}^{W_3 W_5}
+ \overline c_{5 5}{}^{ \cC_{4,2}^{W_4 W_4}  }
\cC_{4,2}^{W_4 W_4}
+ c_{5 5}{}^{  \cC_{4,4}^{W_3 W_5} }
\cC_{4,4}^{W_3 W_5}
+ c_{5 5}{}^{ \cC_{4,4}^{W_4 W_4}  }
\cC_{4,4}^{W_4 W_4}
\ ,
\ea 
\ee
\normalsize
where the OPE coefficients are given as 
\be \label{eq_su5_coeffs}
\ba 
c_{33}{}^4 & = \tfrac{10}{\sqrt{91}} 
\tfrac{g_3}{\sqrt {g_4}} \ ,
& 
c_{34}{}^4 & = 
\tfrac{10}{\sqrt{91}}
\sqrt{g_4} \ , 
&
c_{34}{}^5 
& = \tfrac{\sqrt{13}}{\sqrt{14}}
\tfrac{ \sqrt{g_3}  \sqrt{g_4} }{ \sqrt{g_5} }
 \ , 
\\
c_{35}{}^4 & = 
\tfrac{\sqrt{13}}{\sqrt{14}}
\tfrac{ \sqrt{g_3}  \sqrt{g_5} }{ \sqrt{g_4} }
 \ , 
 &
c_{35}{}^{ \cC_{3,1}^{W_3 W_3}  } & = - \tfrac{23}{30 \sqrt 2} \tfrac{\sqrt{g_5}}{\sqrt{g_3}}
\ , 
&
c_{35}{}^{ \cC_{3,3}^{W_3 W_3} } 
&= \tfrac{1}{5 \sqrt 2} \tfrac{\sqrt{g_5}}{\sqrt{g_3}} \ , 
\\
c_{44}{}^4  &= \tfrac{67}{15 \sqrt{91}} \sqrt{g_4} 
\ ,
& 
c_{44}{}^{  \cC_{3,1}^{W_3 W_3}  } 
& = - \tfrac{39}{50} \tfrac{g_4}{g_3}
\ , 
& 
c_{44}{}^{ \cC_{3,3}^{W_3 W_3}   } 
&=  \tfrac{13}{25} \tfrac{g_4}{g_3} \ ,
\\
c_{45}{}^3 & = 
\tfrac{\sqrt{13}}{\sqrt{14}}
\tfrac{ \sqrt{g_4}  \sqrt{g_5} }{ \sqrt{g_3} }
\ , 
&
c_{45}{}^5 & = 
- \tfrac{87}{10 \sqrt{91}}
\sqrt{g_4} \ ,
& 
c_{45}{}^{
\cC_{\frac 72 , \frac 12}^{W_3 W_4}
}
&= - \tfrac{37}{30 \sqrt 2} \tfrac{\sqrt{g_5}}{\sqrt{g_3}} \ , 
\\
c_{45}{}^{
\cC_{\frac 72 , \frac 52}^{W_3 W_4}
}
& = \tfrac{3}{10 \sqrt 2} 
\tfrac{\sqrt{g_5}}{\sqrt{g_3}} \ , 
&
c_{45}{}^{
\cC_{\frac 72, \frac 72}^{W_3 W_4}
}
&= \tfrac{23}{30 \sqrt 2} 
\tfrac{\sqrt{g_5}}{\sqrt{g_3}} \ , 
&
c_{55}{}^4 & = - \tfrac{87}{10 \sqrt{91}}
\tfrac{g_5}{ \sqrt{g_4} } \ , 
\\
c_{55}{}^{  \cC_{3,1}^{W_3 W_3}  }
&= - \tfrac{77}{1800} \tfrac{g_5}{g_3} \ , 
&
c_{55}{}^{ \cC_{3,3}^{W_3 W_3}  }
&= \tfrac{173}{300} \tfrac{g_5}{g_3}  \ , 
&
\overline c_{5 5}{}^{  \cC_{4,0}^{W_4 W_4} }
& = \tfrac{116}{351} \tfrac{g_5}{g_4} \ , 
\\
\overline c_{5 5}{}^{ \cC_{4,2}^{W_3 W_5} }
& = - \tfrac{3981}{17875 \sqrt 2} \tfrac{\sqrt{g_5}}{\sqrt{g_3}} 
 \ ,  
&
\overline c_{5 5}{}^{ \cC_{4,2}^{W_4 W_4}  }
& = - \tfrac{58012}{83655} \tfrac{g_5}{g_4} \ ,
&
c_{55}{}^{\cC_{4,4}^{W_3 W_5}}
& = - \tfrac{1}{\sqrt 2} \tfrac{\sqrt{g_5}}{ \sqrt{g_3} } \ , \\ 
c_{55}{}^{\cC_{4,4}^{W_4 W_4}}
&= \tfrac{25}{39} \tfrac{\sqrt{g_5}}{\sqrt{g_4}}  \ .
\ea 
\ee 
We notice the absence of
$\mathcal C_{\frac 72, \frac 32}^{W_3 W_4}$,
$\mathcal C_{4,0}^{W_3 W_3}$,
and 
$\mathcal C_{4,2}^{W_3 W_3}$.
This is because the following three
operators are null in
$\cV(A_4)$,
\be \label{eq_nulls_in_N5}
\cC_{\frac 72 , \frac 32}^{W_3 W_4}
\ , \quad 
\cC_{4,0}^{W_3 W_3}
+ \tfrac{25}{13} \tfrac{g_3}{g_4}
\cC_{4,0}^{W_4 W_4}
\ , \quad 
\cC_{4,2}^{W_3 W_3}
+ \tfrac{315 \sqrt 2}{143} \tfrac{\sqrt{g_3}}{\sqrt{g_5}} 
\cC_{4,2}^{W_3 W_5}
+ \tfrac{3500}{1859} \tfrac{g_3}{g_4} \cC_{4,2}^{W_4 W_4}
\ . 
\ee 
The fact that these operators
are null 
in $\WW$
for $\nu=25$ 
is confirmed from the 2-point functions of composite operators
recorded in Appendix
\ref{app_composites}.

% As mentioned above,
% we can reproduce the fact
% that
% $\cC_{\frac 72 , \frac 32}^{W_3 W_4}$ is null from 
% the OPEs of the W-algebra
% $\WW$.
% We expect that the same should be possible for the other
% two null operators in \eqref{eq_nulls_in_N5},
% but we have not performed this check. 

To compare the OPEs 
\eqref{eq_su5_coeffs}
of the VOA $\cV(A_4)$
to the OPEs  of $\WW$
in Table \ref{tab_ansatz},
we have to take into
account that the states in 
\eqref{eq_nulls_in_N5} are null.
More precisely,
we 
set the three quantities
in \eqref{eq_nulls_in_N5} to zero,
solve for 
$\cC_{\frac 72 , \frac 32}^{W_3 W_4}$,
$\cC_{4,0}^{W_3 W_3}$,
$\cC_{4,2}^{W_3 W_3}$,
and plug their expressions
in the OPEs in 
Table \ref{tab_ansatz}.
We obtain the same structure
as in \eqref{free_su5_OPEs}
with the identifications
\be \label{eq_cbar_identifications}
\ba
\overline c_{5 5}{}^{  \cC_{4,0}^{W_4 W_4} }
& = 
c_{5 5}{}^{  \cC_{4,0}^{W_4 W_4} }
- \tfrac{25}{13} \tfrac{g_3}{g_4}
c_{5 5}{}^{ \cC_{4,0}^{W_3 W_3} }
 \ , 
 \\
\overline c_{5 5}{}^{ \cC_{4,2}^{W_3 W_5} }
& = 
c_{5 5}{}^{ \cC_{4,2}^{W_3 W_5} }
- \tfrac{315 \sqrt 2}{143} \tfrac{\sqrt {g_3}}{\sqrt{g_5}} 
c_{5 5}{}^{ \cC_{4,2}^{W_3 W_3}   } 
 \ ,  
 \\
 \overline 
 c_{5 5}{}^{ \cC_{4,2}^{W_4 W_4}  }
 & = 
c_{5 5}{}^{ \cC_{4,2}^{W_4 W_4}  }
- \tfrac{3500}{1859} \tfrac{g_3}{g_4}
c_{5 5}{}^{ \cC_{4,2}^{W_3 W_3} }
\ . 
\ea 
\ee

We are now in a position
to verify that the OPE
coefficients of $\cV(A_4)$
in \eqref{eq_su5_coeffs} follow from
the OPE coefficients of $\WW$
taking the limit $\nu \rightarrow 25$. It is straightforward to 
take the OPE data in Tables
\ref{tab_OPEcoeffs_part1}-\ref{tab_OPEcoeffs_part8}
and verify that we indeed reproduce
all OPE coefficients 
in \eqref{eq_su5_coeffs}, with the exception of 
\be 
c_{55}{}^{\cC_{4,4}^{W_3 W_5}} \ , \qquad 
c_{55}{}^{\cC_{4,4}^{W_4 W_4}} \ . 
\ee 
This is because these OPE coefficients in $\WW$
have not been determined yet
at the current stage of the bootstrap analysis.
Our expectation is that they would also be reproduced, if we were to
push the bootstrap of $W_{p_1} \times W_{p_2}$ OPEs in $\WW$ to higher values of $p_1 + p_2$.

\section{Connection with 4d $\cN = 4$ $SU(N)$ super Yang-Mills}
\label{sec_connection_4d}

In this section we aim
to collect further 
evidence that, upon setting
$c = 3(1 - N^2)$, equivalently
$\nu = N^2$, the simple quotient
of $\WW$ is isomorphic
to the VOA associated to 4d $\cN = 4$
$SU(N)$ super Yang-Mills (SYM),
and that the $\mathfrak R$-filtration
is identified with the $R$-filtration
coming from 4d physics \cite{Beem:2013sza,Beem:2017ooy}.
To achieve this goal, we will perform 
a counting of $\mathfrak{psl}(2|2)$
primary in $\WW$ and compare it against:
\begin{itemize}
    \item the Macdonald index of 4d $\cN = 4$ $SU(N)$ SYM;
    \item the Hilbert series of the Hall-Littlewood chiral ring of  4d $\cN = 4$ $SU(N)$ SYM;
    \item the BRST construction 
    of the VOA associated to 4d $\cN = 4$ $SU(N)$ SYM.
\end{itemize}
We will test $\mathfrak{psl}(2|2)$ primaries with conformal weight $h \le 4$,
for any value of $N$.

Before going into the details of our tests, we briefly review the
map from 4d $\cN = 2$ SCFTs and VOAs and
the notion of $R$-filtration.

\subsection{Brief review of the 4d/2d map}
\label{sec_review_4d2dmap}

\subsubsection{Some general features}

The map of  \cite{Beem:2013sza} associates to a 4d $\cN = 2$ superconformal field theory
a VOA obtained by passing to the cohomology of a suitable nilpotent fermionic
generator $\mathbbmtt{Q}$ in the superconformal algebra.
The choice of fermionic generator $\mathbbmtt{Q}$
selects a 2d plane $w=\bar w = 0$
inside 4d spacetime with complex coordinates
$(z, \bar z , w , \bar w)$.
Non-trivial cohomology classes
of local operators inserted at the origin 
$(z, \bar z , w , \bar w)=(0,0,0,0)$
admit canonical
representatives given by Schur operators.
Let $(E, j_1, j_2, R, r)$ denote the  quantum numbers
of the 4d $\cN = 2$ superconformal algebra $\mathfrak{sl}(4|2)$.
Here $E$ is the conformal dimension, while
$j_1$, $j_2$, $R$, $r$ are the eigenvalues with respect to the Cartan generators
of $\mathfrak{sl}(2)_1 \oplus \mathfrak{sl}(2)_2 \oplus 
\mathfrak{sl}(2)_R \oplus \mathfrak{gl}(1)_r$, respectively.
A Schur operator is a  highest weight state of $\mathfrak{sl}(2)_1 \oplus \mathfrak{sl}(2)_2 \oplus 
\mathfrak{sl}(2)_R \oplus \mathfrak{gl}(1)_r$,
and its quantum numbers satisfy
\be 
E - (j_1 + j_2) - 2R = 0  \ , \qquad 
r + j_1 - j_2 = 0 \ . 
\ee 
A Schur operator inserted at the origin can 
be moved to any other point $(z,\bar z)$ on the VOA plane
$w=\bar w =0$ by means of a twisted translation,
see \cite{Beem:2013sza} for details. 
A twisted translated Schur operator 
transforms has
conformal weight 
\be \label{eq_h_for_Schur}
h = E-R = \tfrac 12 (E + j_1 + j_2) \ . 
\ee 
The OPE of twisted translated Schur operators
is defined by taking the 4d OPE and passing to the cohomology
of the supercharge $\mathbbmtt{Q}$. One obtains a meromorphic OPE
which endows the vector space of Schur operators with the 
structure of a VOA.

For the purpose of enumeration, it is sufficient to focus on those Schur operators that are $\mathfrak{sl}(2)_z$
primaries. 
In terms of (unitary) 4d $\cN = 2$ superconformal multiplets,
these Schur operators are found in short multiplets of
type $\widehat \cB_{R}$, $\mathcal D_{R(0,j_2)}$,
$\overline \cD_{R(j_1,0)}$, and
$\widehat \cC_{R(j_1,j_2)}$, in the notation of \cite{Dolan:2002zh}.
Table \ref{table_multiplets_give_Schur} summarizes how Schur operators enter such multiplets and how their quantum numbers are related to those of the superconformal primary operator in the multiplet.

\begin{table}
\begin{center}
\begingroup
\renewcommand{\arraystretch}{1.4}
\begin{tabular}{| c | c | c | c | c | }
\hline 
\multirow{2}{3.3cm}{4d $\cN = 2$ multiplet} & \multicolumn{4}{c|}{Schur operator} \\\cline{2-5}
 & structure & $h$ & $r$ & $R$ \\ \hline \hline 
$\widehat \cB_{R'}$ & $\Psi^{1\dots 1}$ & $R'$ & $0$ & $R'$  \\  \hline 
$\cD_{R'(0,j_2)}$ & $\widetilde \cQ_{\dot +}^1 \Psi^{1\dots 1}_{\dot + \dots \dot +}$ & $R'+j_2+1$ & $j_2 + \frac 12$ & $R' + \frac 12$ \\ \hline 
$\overline \cD_{R'(j_1,0)}$ & $ \cQ_{ +}^1 \Psi^{1\dots 1}_{ + \dots  +}$ & $R'+j_1+1$ & $-j_1 - \frac 12$ & $R' + \frac 12$ \\ \hline 
$\widehat \cC_{R'(j_1,j_2)}$ & $ \cQ_{ +}^1 \widetilde \cQ_{\dot +}^1 \Psi^{1\dots 1}_{ + \dots  +    \dot + \dots \dot +}$ & $R'+j_1 + j_2 +2$ & $j_2-j_1$ & $R' + 1$ \\
\hline
\end{tabular}
\endgroup
\end{center}
\caption{4d $\cN = 2$ superconformal multiplet containing a Schur operator.
For each multiplet we give the schematic form of the Schur operator in terms of the 
superconformal primary $\Psi$ of the multiplet.
We also report the quantum numbers $(h,r,R)$ of the Schur operator
in terms of the quantum numbers of the multiplet.
(We use $R'$ for the $\mathfrak{sl}(2)_R$ quantum number
of the multiplet.) 
    \label{table_multiplets_give_Schur}
    }
\end{table}

The OPE structure on the vector space of Schur operators
preserves the quantum numbers $h$ and $r$, but it does not preserve the
quantum number $R$. There exists, however, a filtration of the VOA
which encodes $R$ \cite{Beem:2013sza,Beem:2017ooy}. More precisely, let us define
\be 
\cV_{h,r,R} = \{ \text{Schur operators with definite $(h,r,R) \in \tfrac 12 \mathbb Z_{\ge 0} \times \tfrac 12 \mathbb Z \times \tfrac 12 \mathbb Z_{\ge 0}$} \} \ .
\ee 
We also introduce 
\be 
\cF_{h,r,R} = \bigoplus_{k \ge 0 }\cV_{h,r,R-k} \ . 
\ee 
The subspaces $\cF_{h,r,R}$ furnish an increasing filtration
of the space $\cV_{h,r}$ of Schur operators of definite $(h,r)$.
This filtration enjoys the following additional properties. 
The unit operator satisfies $\one \in \cF_{0,0,0}$.
Suppose $\cO_1 \in \cF_{h_1,r_1,R_1}$
and $\cO_2 \in \cF_{h_2,r_2,R_2}$.
Parametrize the full 
$\cO_1  \times \cO_2$ OPE
as 
\be 
\cO_1(z) \cO_2(0) = \sum_{n \in \mathbb Z} \frac{1}{z^n} \{ \cO_1 \cO_2\}_n(0) \ . 
\ee 
Then, one has
\be 
\ba 
\{ \cO_1 \cO_2 \}_0 &\in \cF_{h_1 + h_2 ,   r_1 + r_2, R_1 + R_2}  &
&\text{(normal ordered product)}
 \ ,   \\
\{ \cO_1 \cO_2 \}_1 &\in \cF_{h_1 + h_2 - 1, r_1 + r_2,  R_1 + R_2 -1} 
 &
&\text{(simple pole in the OPE)}\ .
\ea 
\ee 
Given the filtration $\cF_{h,r,R}$ of $\cV_{h,r}$, the associated graded
is defined as 
\be 
{\rm gr} \cV_{h,r} = \bigoplus_{R\in \frac 12 \mathbb Z_{\ge 0}}
\cG_{h,r,R} \ , \qquad 
\cG_{h,r,R} = \cF_{h,r,R}/\cF_{h, r,R-1} \ . 
\ee 
As a vector space, ${\rm gr} \cV_{h,r}$ is isomorphic to
$\cV_{h,r}$.

Without knowledge of the $R$-filtration
it is in general not possible to invert the map
from 4d $\cN = 2$ superconformal multiplets
to $\mathfrak{sl}(2)_z$ primaries in the VOA.
If we have access to the quantum number $R$, however,
we can reconstruct unambiguously the 4d origin of
a 2d $\mathfrak{sl}(2)_z$ primary: 
\be \label{eq_where_fromN2}
\begin{array}{l}
\text{$\mathfrak{sl}(2)_z$ primary} \\ 
\text{with $(h,R,r)$}
\end{array}
\qquad 
\text{comes from} \qquad
\widehat \cC_{R'(j_1, j_2)} \;\; \text{with} \;\; 
\begin{array}{rcl}
R' &=& R-1  \ , \\ 
2j_1 &=& h-R-r-1  \ ,\\ 
2j_2 &=& h-R+r-1  \ . 
\end{array}
\ee 
Formally, we can still apply \eqref{eq_where_fromN2}
in the the special cases in which 
$h=R+r$ and/or $h = R-r$,  
provided we make use of the identifications 
\cite{Dolan:2002zh} 
\be 
\ba 
\widehat \cC_{R'(j_1,- \frac 12)} & \cong \overline \cD_{R'+\frac 12 (j_1,0)} \ , \\ 
\widehat \cC_{R'(- \frac 12 ,j_2)} & \cong 
\cD_{R'+\frac 12 (0,j_2)} \ , \\
\widehat \cC_{R'(- \frac 12 , - \frac 12 )} & \cong 
\widehat \cB_{R'+1} \ .
\ea 
\ee

\subsubsection{The case of 4d $\cN = 4$ super Yang-Mills}
We are interested in studying Schur operators in 4d $\cN = 4$ $SU(N)$
super Yang-Mills theory. 
In this case, due to the additional supersymmetry
compared to a generic $\cN = 2$ theory,
Schur operators are
organized in $\mathfrak{psl}(2|2)$ multiplets.
We denote the quantum numbers of
the 4d $\cN = 4$ superconformal algebra as
$(E, j_1, j_2 , [q_1,p,q_2])$, 
with $E$, $j_1$, $j_2$ as in the $\cN = 2$ case,
and $[q_1, p ,q_2]$ Dynkin labels of $\mathfrak{sl}(4)$.
In Table \ref{table_N4multiplets_give_Schur} 
we list the (unitary) 4d $\cN = 4$ superconformal multiplets
that give rise to $\mathfrak{psl}(2|2)$ multiplets
of Schur operator.
We adopt the notation of \cite{Dolan:2002zh} for 4d multiplets.
For each $\mathfrak{psl}(2|2)$ multiplet
we give
the following quantum numbers of its $\mathfrak{psl}(2|2)$ primary:  
conformal weight $h$, spin $j$ under $\mathfrak{sl}(2)_y$,
$r$ and $R$ charges.
We refer the reader to Section 6.1.4 of \cite{Bonetti:2018fqz}
for further details.

If we know the quantum numbers $(h,j,r,R)$ of
a $\mathfrak{psl}(2|2)$ primary   Schur operator,
we can unambiguously reconstruct its origin in four dimensions.
In the generic case $h \neq R + r$, $h \neq R - r$, 
from Table \ref{table_N4multiplets_give_Schur}
we find that
\be \label{eq_where_from}
\begin{array}{l}
\text{$\mathfrak{psl}(2|2)$ primary} \\ 
\text{with $(h,j,r,R)$}
\end{array}
\qquad 
\text{comes from} \qquad
\cC_{[q_1, p, q_2](j_1, j_2)} \;\; \text{with} \;\; 
\begin{array}{rcl}
p &=& 2j  \ , \\ 
q_1 &=& R+r-j-1  \ , \\ 
q_2 &=& R-r-j-1  \ , \\ 
2j_1 &=& h-R-r-1  \ ,\\ 
2j_2 &=& h-R+r-1  \ . 
\end{array}
\ee 
Formally, we can still apply \eqref{eq_where_from}
in the the special cases in which 
$h=R+r$ and/or $h = R-r$,  
provided we make use of the identifications 
\cite{Dolan:2002zh} 
\be \label{eq_N4_multiplet_identifications}
\ba 
\cC_{[q_1, p, q_2](j_1,- \frac 12)} & \cong \overline \cD_{[q_1, p, q_2+1](j_1,0)} \ , \\ 
\cC_{[q_1, p, q_2](- \frac 12 ,j_2)} & \cong 
\cD_{[q_1+1, p, q_2](0,j_2)} \ , \\
\cC_{[q_1, p, q_2](- \frac 12 , - \frac 12 )} & \cong 
\cB_{[q_1+1, p, q_1+1](0,0)} \ .
\ea 
\ee 

We close this subsection with a remark on the $r$ charge. The small $\cN = 4$ superconformal algebra enjoys an 
$\mathfrak{sl}(2)_{\rm out}$ outer automorphism, under which the supercurrents
$G$, $\widetilde G$ transform as a doublet. The 4d quantum number $r$ is identified with the Cartan eigenvalue
of $\mathfrak{sl}(2)_{\rm out}$,
in conventions in which 
$G$, $\widetilde G$ have charge $+\frac 12$, $- \frac 12$ respectively.

\begin{table}
\begin{center}
\begingroup
\small
\renewcommand{\arraystretch}{1.4}
\begin{tabular}{| c | c | c | c | c | }
\hline 
\multirow{2}{3cm}{4d $\cN = 4$ multiplet} & \multicolumn{4}{c|}{$\mathfrak{psl}(2
|2)$ multiplet} \\\cline{2-5}
 & $h$ & $j$ & $r$ & $R$ \\ \hline \hline 
$\cB_{[q,p,q]}$ & $\frac 12 p+q$ & $\frac 12 p$ & $0$ & $\frac 12 p +q$  \\  \hline 
$\cD_{[q_1,p,q_2](0,j_2)}$ & $\frac 12 p + q_1$ & $\frac 12 p$ & $\frac 12 (q_1 - q_2-1)$ & $\frac 12 (p + q_1 + q_2 +1)$  \\  \hline 
$\overline \cD_{[q_1,p,q_2](j_1,0)}$ & $\frac 12 p + q_2$ & $\frac 12 p$ & $\frac 12(q_1 - q_2 +1)$ & $\frac 12 (p + q_1 + q_2 +1)$  \\  \hline 
$\cC_{[q_1,p,q_2](j_1,j_2)}$ & $\frac 12 (q + q_1 + q_2) +(j_1 + j_2) + 2$ & $\frac 12 p$ & $\frac 12(q_1 - q_2)$ & $\frac 12 (p + q_1 + q_2 + 2)$  \\  \hline 
\end{tabular}
\endgroup
\end{center}
\caption{Summary of 4d $\cN = 4$ superconformal multiplet 
yielding $\mathfrak{psl}(2|2)$ multiplets of Schur operators.
For each $\mathfrak{psl}(2|2)$ multiplet we
give the quantum numbers of its $\mathfrak{psl}(2|2)$
primary operator.
Recall that: a multiplet of type $\cD_{[q_1, p, q_2](0,j_2)}$
satisfies $q_1 - q_2 = 2(j_2+1)$;
a multiplet of type $\overline \cD_{[q_1, p, q_2](j_1,0)}$
satisfies $q_1 - q_2 = -2(j_1+1)$;
a multiplet of type $\cC_{[q_1, p, q_2](j_1,j_2)}$
satisfies $q_1 - q_2 = 2(j_2-j_1)$.
    \label{table_N4multiplets_give_Schur}
    }
\end{table}

\subsection{Low-lying $\mathfrak{psl}(2|2)$ primaries in $\WW$
}

In this subsection we present a detailed counting of
$\mathfrak{psl}(2|2)$ primaries 
with conformal weight $h\le 4$
in the W-algebra $\WW$.
Besides $h$, we keep track of the following quantum numbers:
the $\mathfrak{sl}(2)_y$ spin $j$,  the
$\mathfrak{gl}(1)_r$ charge $r$,
and the $\mathfrak R$ weight,
as defined in Section \ref{sec_WW_filtration}.

Our main results are collected in
Table \ref{table_summary_of_states}.
Some comments are in order.
For each value of $(h,j)$
we give the value of the quantity 
\be \label{eq_counting_poly}
\sum_{\mathfrak R  , r  } 
n_{h,j,r,\mathfrak R } \;
x^{2r} \xi^{\mathfrak R}  
\ , 
\ee 
where 
$n_{h,j,r,\mathfrak R }$ is the number of linearly independent 
$\mathfrak{psl}(2|2)$ primaries
with quantum numbers $(h,j, r,\mathfrak R)$.
Thus, here $\xi$ is a fugacity for the
$\mathfrak R$-weight,
and $x$ a fugacity for the $r$ charge.
In Table \ref{table_summary_of_states} we also make use of the shorthand notation
\be 
\mathfrak N = \xi^3 + \xi^2 + \xi^{7/2} \chi_1(x) \ , 
\ee 
where $\chi_1(x)$ is the character of the 2-dimesional irrep of $\mathfrak{sl}(2)_{\rm out}$,
\be 
\chi_1(x) = x + x^{-1} \ . 
\ee 
Under the column `generic $\nu$'
we give the value of
\eqref{eq_counting_poly} for a generic value of the central charge.
In the other columns  
we give the value of 
\eqref{eq_counting_poly} 
for the simple quotient of $\WW$
with $\nu = N^2$, equivalently
$c=3(1-N^2)$, for $N=2,3,\dots$.

Some additional remarks:
\begin{itemize}
    \item Up to $h=7/2$ all $\mathfrak{psl}(2|2)$ primaries are neutral under $\mathfrak{gl}(1)_r$.
    The lightest $\mathfrak{psl}(2|2)$ primaries with non-trivial $r$-charge appear at $(h,j) = (4,1)$ for generic $\nu$ or $N \ge 4$. (This is signaled by the presence of the quantity $\mathfrak N$, which contains $\chi_1(x)$.) These $\mathfrak{psl}(2|2)$ primaries 
    form a doublet of $\mathfrak{sl}(2)_{\rm out}$ and
    are Grassmann odd.
    
\item The entries of the column
`generic $\nu$' can be extracted easily
from a partition function, as explained 
below in Section \ref{sec_counting_generic}. In contrast, the entries of the columns
for $\nu = N^2$
require a careful analysis of 
the null states that appear in $\WW$
for those values of $\nu$.  
Some details are reported in Appendix
\ref{app_low_states}.

\item  For fixed $(h,j)$, considering 
$\nu = N^2$ for increasing values of $N$,
we observe that 
degeneracies $n_{h,j,r,\mathfrak R}$
are non-decreasing. From some value of $N$ onward, they stabilize and become equal to the
degeneracies for generic $\nu$.
We have used the color blue to highlight those entries that have stabilized.
We also observe that,
for $N \ge 2h$, all states with conformal weight $h$ have stabilized
(some stabilize even before, for values of $N$ smaller than $2h$).

\end{itemize}

\begin{table}
\footnotesize
\begin{adjustwidth}{-.8in}{-.8in}  
\begin{center}
\begingroup
\renewcommand{\arraystretch}{1.3}
\setlength{\tabcolsep}{2.1pt}
\begin{tabular}{| c c   | l 
      |  l | l | l | l | l | l | l   |}
\hline
 &  &  & \multicolumn{7}{c|}{$\nu = N^2$
 with $N$ equal to}  \\ \cline{4-10}
$h$ & $j$   & generic $\nu$  &
$2$   & $3$ & $4$ & $5$ & $6$ & $7$ & $8$, $9$, $\dots$  \\
\hline \hline
1 & 1  & $\xi$ &  
$\color{blue} \xi$  &
$\color{blue} \xi$  &
$\color{blue} \xi$  &
$\color{blue} \xi$  &
$\color{blue} \xi$  &
$\color{blue} \xi$  &
$\color{blue} \xi$  
   \\ \hline   \hline
   $\frac 32$ & $\frac 32$  & $\xi^{3/2}$ &  
0  &
$\color{blue} \xi^{3/2}$  &
$\color{blue} \xi^{3/2}$  &
$\color{blue} \xi^{3/2}$  &
$\color{blue} \xi^{3/2}$  &
$\color{blue} \xi^{3/2}$  &
$\color{blue} \xi^{3/2}$  
   \\ \hline  \hline 
$2$ & $2$  & $2 \xi^2$ &  
$\xi^2$  &
$\xi^2$  &
$\color{blue} 2\xi^2$  &
$\color{blue} 2\xi^2$  &
$\color{blue} 2\xi^2$  &
$\color{blue} 2\xi^2$  &
$\color{blue} 2\xi^2$  
   \\ \hline   
$2$ & $0$  & $ \xi^2$ &  
0  &
$\color{blue} \xi^2$  &
$\color{blue} \xi^2$  &
$\color{blue} \xi^2$  &
$\color{blue} \xi^2$  &
$\color{blue} \xi^2$  &
$\color{blue} \xi^2$  
   \\ \hline  \hline 
$\frac 52$ & $\frac 52$  & $2 \xi^{5/2}$ &  
0  &
$\xi^{5/2}$  &
$ \xi^{5/2}$  &
$\color{blue} 2 \xi^{5/2}$  &
$\color{blue} 2 \xi^{5/2}$  &
$\color{blue} 2 \xi^{5/2}$  &
$\color{blue} 2 \xi^{5/2}$  
   \\ \hline   
$\frac 52$ & $\frac 32$  & 
$\xi^{5/2}$ &  
0  &
$\color{blue} \xi^{5/2}$  &
$\color{blue} \xi^{5/2}$  &
$\color{blue} \xi^{5/2}$  &
$\color{blue} \xi^{5/2}$  &
$\color{blue} \xi^{5/2}$  &
$\color{blue} \xi^{5/2}$  
   \\ \hline   
$\frac 52$ & $\frac 12$  & 
$ \xi^{5/2}$ &  
0  &
0  &
$\color{blue} \xi^{5/2}$  &
$\color{blue} \xi^{5/2}$  &
$\color{blue} \xi^{5/2}$  &
$\color{blue} \xi^{5/2}$  &
$\color{blue} \xi^{5/2}$  
   \\ \hline   \hline 
$3$ & $3$  & 
$4 \xi^3$ &  
$\xi^3$  &
$2 \xi^3$  &
$3 \xi^3$  &
$3 \xi^3$  &
$\color{blue} 4 \xi^3$  &
$\color{blue} 4 \xi^3$  &
$\color{blue} 4 \xi^3$  
   \\ \hline  
$3$ & $2$  & 
$ \xi^3$ &  
0  &
0  &
$\color{blue} \xi^3$  &
$\color{blue} \xi^3$  &
$\color{blue} \xi^3$  &
$\color{blue} \xi^3$  &
$\color{blue} \xi^3$  
   \\ \hline 
$3$ & $1$  & 
$\xi^2 + 3 \xi^3$ &  
$\xi^2$  &
$\xi^2 + \xi^3$  &
$\xi^2 + 2 \xi^3$  &
$\color{blue} \xi^2 + 3 \xi^3$  &
$\color{blue} \xi^2 + 3 \xi^3$  &
$\color{blue} \xi^2 + 3 \xi^3$  &
$\color{blue} \xi^2 + 3 \xi^3$  
   \\ \hline  \hline 
$\frac 72$ & $\frac 72$  & 
$4 \xi^{7/2}$ &  
0  &
$\xi^{7/2}$  &
$2 \xi^{7/2}$  &
$3 \xi^{7/2}$  &
$3 \xi^{7/2}$  &
$\color{blue} 4 \xi^{7/2}$ &
$\color{blue} 4 \xi^{7/2}$  
   \\ \hline
$\frac 72$ & $\frac 52$  & 
$3 \xi^{7/2}$ &  
0  &
$\xi^{7/2}$  &
$2\xi^{7/2}$  &
$\color{blue} 3 \xi^{7/2}$  &
$\color{blue} 3 \xi^{7/2}$  &
$\color{blue} 3 \xi^{7/2}$  &
$\color{blue} 3 \xi^{7/2}$  
   \\ \hline 
$\frac 72$ & $\frac 32$  & 
$\xi^{5/2} + 4 \xi^{7/2}$ &  
0  &
$\xi^{5/2} + \xi^{7/2}$  &
$\xi^{5/2} + 2 \xi^{7/2}$  &
$\xi^{5/2} + 3 \xi^{7/2}$  &
$\color{blue} \xi^{5/2} + 4 \xi^{7/2}$  &
$\color{blue} \xi^{5/2} + 4 \xi^{7/2}$  &
$\color{blue} \xi^{5/2} + 4 \xi^{7/2}$  
   \\ \hline 
$\frac 72$ & $\frac 12$ & 
$\xi^{5/2}  + 2 \xi^{7/2}$ &  
0  &
$\xi^{5/2}$  &
$\xi^{5/2} +   \xi^{7/2}$  &
$\color{blue} \xi^{5/2} + 2 \xi^{7/2}$  &
$\color{blue} \xi^{5/2} + 2 \xi^{7/2}$  &
$\color{blue} \xi^{5/2} + 2 \xi^{7/2}$  &
$\color{blue} \xi^{5/2} + 2 \xi^{7/2}$
   \\ \hline \hline 
$4$ & $4$  & 
$7 \xi^4$ &  
$\xi^4$  &
$2 \xi^4$  &
$4\xi^4$  &
$5\xi^4$  &
$6\xi^4$  &
$6\xi^4$  &
$\color{blue} 7\xi^4$  
   \\ \hline 
$4$ & $3$  & 
$4 \xi^4$ &  
0  &
$\xi^4$  &
$2\xi^4$  &
$3 \xi^4$  &
$\color{blue} 4 \xi^4$  &
$\color{blue} 4 \xi^4$  &
$\color{blue} 4 \xi^4$  
   \\ \hline 
$4$ & $2$  & 
$3 \xi^3 + 8 \xi^4$ &  
$\xi^3$  &
$2 \xi^3 +   \xi^4$  &
$3 \xi^3 + 4 \xi^4$  &
$3 \xi^3 + 6 \xi^4$  &
$3 \xi^3 + 7 \xi^4$  &
$\color{blue} 3 \xi^3 + 8 \xi^4$   &
$\color{blue} 3 \xi^3 + 8 \xi^4$
   \\ \hline 
$4$ & $1$  & 
$\xi^3 + 2 \xi^4 + \mathfrak N$ &  
0  &
$\xi^3$  &
$\xi^3 + \mathfrak N$  &
$\xi^3 + \xi^4 + \mathfrak N$  &
$\color{blue}  \xi^3 + 2\xi^4 + \mathfrak N$  &
$\color{blue}  \xi^3 + 2\xi^4 + \mathfrak N$   &
$\color{blue}  \xi^3 + 2\xi^4 + \mathfrak N$
   \\ \hline 
$4$ & $0$  & 
$\xi^2 + 2 \xi^3 + 4 \xi^4$ &  
$\xi^2$  &
$\xi^2 + \xi^3 + \xi^4$  &
$\xi^2 + 2 \xi^3 + 2 \xi^4$  &
$\xi^2 + 2 \xi^3 + 3 \xi^4$  &
$\color{blue} \xi^2 + 2 \xi^3 + 4 \xi^4$  &
$\color{blue} \xi^2 + 2 \xi^3 + 4 \xi^4$
&
$\color{blue} \xi^2 + 2 \xi^3 + 4 \xi^4$
   \\ \hline 
\end{tabular}
\endgroup
\normalsize
\end{center}
\end{adjustwidth}
\caption{Counting $\mathfrak{psl}(2|2)$ primaries with definite quantum numbers
$(h,j,r,\mathfrak R)$ in $\WW$,
up to $h=4$.
We make use of the shorthand notation
$\mathfrak{N} = \xi^3 + \xi^2 + \xi^{7/2} \chi_1(x)$,
$\chi_1(x) = x+x^{-1}$.
The variable $\xi$ is a fugacity for the quantum number $\mathfrak R$,
and the variable $x$ for the quantum number $r$. The entries in blue are those that are equal to the entries in  the column `generic $\nu$'.
    \label{table_summary_of_states}
    }
\end{table}

\subsubsection{Counting states in $\WW$ at generic central charge}
\label{sec_counting_generic}

We can define the following partition function for $\WW$,
\be \label{eq_WW_partition_function}
Z_{\WW} (q,a,x,\xi)={\rm Tr} \; q^h a^j x^{2r} \xi^{\mathfrak R} \ ,
\ee 
where $q$ is the fugacity of the conformal weight $h$,
$a$ is the fugacity for the Cartan generator of $\mathfrak{sl}(2)_y$,
$x$ is the fugacity for the $\mathfrak{gl}(1)_r$ charge, and $\xi$ is the fugacity for the $\mathfrak R$ weight.
By slight abuse of notation
we use the same symbol 
\eqref{eq_WW_partition_function}
for the partition function at generic values of the central charge,
and for $\nu = N^2$. In the latter case,  the trace is taken over the simple quotient, i.e.~modding out by the states that become null
for $\nu = N^2$.

If we start from the partition function
\eqref{eq_WW_partition_function} and specialize the $x$ fugacity to 
\be \label{eq_specialize_x}
x = - \xi^{1/2} \ , 
\ee 
we obtain the supercharacter of $\WW$,
\be \label{eq_supercharacter_WW}
\chi_{\WW}(q,a,\xi) = {\rm Tr} (-1)^F q^h a^j   \xi^{\mathfrak R + r} \ .
\ee 
Here $F$ is fermion number, and we have used the fact that in $\WW$ we have a
relation between Grassmann parity and $r$ quantum number,
\be 
(-1)^{2r} = (-1)^F \ . 
\ee

The partition function \eqref{eq_WW_partition_function}
can be expanded into contributions of long and short
$\mathfrak{psl}(2|2)$ multiplets,
\be 
Z_{\WW} (q,a,x,\xi) = \sum_{\substack{h,j,r,\mathfrak R \\ h>j}} n_{h,j,r,\mathfrak R} \;\mathfrak L_{h,j}(q,a,x) \; x^{2r} \xi^{\mathfrak R} 
+ \sum_{j,r,\mathfrak R} n_{j,j,r,\mathfrak R} \;\mathfrak S_{j}(q,a,x) \; x^{2r} \xi^{\mathfrak R} 
\ .
\ee 
Here $\mathfrak L_{h,j}(q,a,x)$ encodes the contribution of a long
$\mathfrak{psl}(2|2)$ multiplet 
and $\mathfrak S_{j}(q,a,x)$ that of a short multiplet with $h=j$.
The nonnegative integer
$n_{h,j,r,\mathfrak R}$ is the number of
linearly independent $\mathfrak{psl}(2|2)$ multiplets
with quantum numbers $(h,j,r,\mathfrak R)$, as in \eqref{eq_counting_poly}.
More explicitly, the quantities
$\mathfrak L_{h,j}(q,a,x)$,
$\mathfrak S_{j}(q,a,x)$ are defined as follows,
\be \label{eq_long_and_short_contributions}
\ba 
\mathfrak L_{h,j}(q, a, x)
& = \frac{1 }{1-q} \bigg[
q^h \chi_{2j}(a)
+ q^{h+\frac 12} \chi_{2j-1}(a) \chi_1(x)
+ q^{h+\frac 12} \chi_{2j+1}(a) \chi_1(x)
\\
& + q^{h+1} \chi_{2j}(a) \Big( 1+ \chi_2(x) \Big)
+ q^{h+1} \chi_{2j-2}(a)
+ q^{h+1} \chi_{2j+2}(a)
\\
& + q^{h+\frac 32} \chi_{2j-1}(a) \chi_1(x)
+ q^{h+\frac 32} \chi_{2j+1}(a) \chi_1(x)
 + q^{h+2} \chi_{2j}(a)
\bigg] \ , \\
\mathfrak S_{j}(q, a, x) &=\frac{ 1 }{1-q} \bigg[
q^j \chi_{2j}(a) + q^{j + \frac 12} \chi_{2j-1}(a) \chi_1(x)
+ q^{j+1} \chi_{2j-2}(a)
\bigg] \ .
\ea 
\ee  
In the above expressions, $\chi_{2j}(a)$,
$\chi_{2r}(a)$ are characters of $\mathfrak{sl}(2)_y$, $\mathfrak{sl}(2)_{\rm out}$ irrep, respectively,
\be 
\chi_{2j}(a) = a^{-2j} + a^{-2j+2} + \dots + a^{2j} 
\ ,  \qquad 
\chi_{2r}(x) = x^{-2r} + x^{-2r+2} + \dots + x^{2r}  
 \ . 
\ee 
For a long multiplet, in the case $j=1/2$, the states that would have negative
$\mathfrak{sl}(2)_y$ spin are absent.
In the case $j=0$, the states that would have negative
$\mathfrak{sl}(2)_y$ spin are absent,
and we also lose one of the states with $h+1$ and $j=0$.

The computation of \eqref{eq_WW_partition_function} for generic values of the central charge
is straightforward, due to the absence of non-trivial null state. 
We can use the following formula,
\be \label{eq_W_alg_Z} 
Z_{\WW} (q,a,x,\xi) = \prod_{p=2}^\infty \prod_{n=0}^\infty 
\frac{
\bigg[ \displaystyle \prod_{m=-\frac p2+\frac 12}^{\frac p2- \frac 12}(1 + x \xi^{\frac p2} q^{\frac p2+\frac 12 +n} a^{2m}) \Big]
\Big[ \prod_{m=-\frac p2 + \frac 12 }^{\frac p2 - \frac 12 }(1 + x^{-1} \xi^{\frac p2} q^{\frac p2 +\frac 12 +n} a^{2m}) \Big]
}{ \Big[ \displaystyle 
\prod_{m=-\frac p2 }^{\frac p2}(1 - \xi^{\frac p2} q^{\frac p2+n} a^{2m}) \Big]
\Big[ \prod_{m=-\frac p2+1}^{\frac p2-1}(1 - \xi^{\frac p2} q^{\frac p2+1+n} a^{2m}) \Big]} 
\ . 
\ee 
The product over $p$ encodes the contribution of the strong generators, with the term $p=2$ associated to the generators
$J$, $G$, $\widetilde G$, $T$ of the small $\cN = 2$ SCA
and the terms with $p\ge 3$ associated to
$W_p$, $G_{W_p}$, $\widetilde G_{W_p}$, $T_{W_p}$.
In the denominator we find the contributions of the bosonic generators $W_p$, $T_{W_p}$, whose conformal weight and spin are $(\frac p2, \frac p2)$, $(\frac p2 +1, \frac p2-1)$, respectively. In the numerator we find the contributions of the fermionic generators $G_{W_p}$, $\widetilde G_{W_p}$,
which have both conformal weight and spin 
$(\frac p2 + \frac 12, \frac p2 - \frac 12)$, but opposite
charges under the Cartan generator of $\mathfrak{sl}(2)_{\rm outer}$.

% Finally, we have introduced a book-keeping parameter $\Xi_p$ for each $p 
% \ge 2$.

% The fugacity $q$ keeps track of the conformal dimension $h$.
% The fugacity $a$ keeps track of the eigenvalue under the Cartan generator of $\mathfrak{sl}(2)_y$.
% The fugacity $x$ keeps track of the eigenvalue under the Cartan generator of $\mathfrak{sl}(2)_{\rm outer}$.
% The parameter $\alpha$ is introduced to keep track of bosonic/fermionic statistic. 

It is not difficult to
compute  \eqref{eq_W_alg_Z}
to low orders in a $q$-expansion
and reorganize the results
in terms of 
$\mathfrak L_{h,j}(q,a,x)$,
$\mathfrak S_{j}(q,a,x)$.
Working up to $q^4$, we confirm indeed the counting of $\mathfrak{psl}(2|2)$ primaries
for generic value of the central charge
reported
in   Table \ref{table_summary_of_states}.

\subsection{Comparison with
Macdonald index and Hall-Littlewood Hilbert series}

The counting of state reported above can be tested against some protected quantities that can be explicitly evaluated in 
4d $\cN = 4$ $SU(N)$ SYM:
the Macdonald limit of the superconformal index and the Hilbert series of the Hall-Littlewood chiral ring.

\subsubsection{Connection with Macdonald index}

We recall that superconformal index \cite{Romelsberger:2005eg,Kinney:2005ej} 
of a  4d $\cN = 2$ SCFTs
in the conventions of \cite{Gadde:2011uv}
is given as
\be 
\cI(\rho,\sigma,\tau)=
{\rm Tr}(-1)^F \rho^{\frac 12 (E - 2j_1 - 2R - r)}
 \sigma^{\frac 12 (E + 2j_1 - 2R - r)}
\tau^{\frac 12(E + 2j_2 + 2R + r)}
(e^{-\beta})^{\frac 12(E - 2 j_2 - 2R + r)} \ ,
\ee 
where $F$ is fermion number.
We consider the limit $\rho \rightarrow 0$ (which is one realization of the Macdonald limit, the other equivalent one being $\sigma \rightarrow 0$).
For $\rho \rightarrow 0$ the only states that contribute are those for which the exponents of $\rho$ and $e^{-\beta}$ are both zero, which is equivalent to
the conditions that define the subspace of Schur operators.
If we set 
\be 
\sigma = \frac q \tau \ , \qquad 
\tau^2 = q \xi \ ,  
\ee 
we get the expression for the Macdonald index.
In the case of 
4d $\cN = 4$ SYM, we can write
\be \label{eq_def_part_func2}
\cI_{\rm Macdonald}^{\text{4d $SU(N)$ SYM}}(q,a,\xi) = {\rm Tr}_{\cH_{\rm Schur}}
(-1)^F q^{E-R} \xi^{R +r} a^j   \ , 
\ee 
where
we have included an additional 
fugacity $a$ associated to the Cartan of
the $\mathfrak{sl}(2)_y$ 
symmetry (commutant of the
$\cN = 2$ superconformal algebra inside
the $\cN =4$ one).
Recall from \eqref{eq_h_for_Schur} that
$E-R = h$ for a Schur operator.

It is known that the Macdonald index for 4d $\cN = 4$ $SU(N)$ SYM can be computed 
from free fields. 
Using for example Table 2 in \cite{Gadde:2011uv}, we can
easily assemble the Macdonald
index of a free 4d $\cN = 4$ vector multiplet, combining the Macdonald indices of a free 4d $\cN =2$ vector multiplet and a free 4d $\cN = 2$ half hypermultiplet. The result is
\be \label{eq_single_letter_N4_Mac}
f^{V}_{\rm Macdonald}(q,a, \xi) =  -\frac{q (\xi +1)}{1-q}   + \chi_1(a)\frac{
\sqrt q \sqrt \xi
}{1-q}  \ .    
\ee 
The Macdonald index of 4d $\cN = 4$ $SU(N)$ SYM is then obtained by taking a plethystic exponential and projecting onto gauge singlets in the usual way,
\be \label{eq_SYM_Mac}
\cI_{\rm Macdonald}^{\text{4d $SU(N)$ SYM}}(q,a,\xi)= \int[db] {\rm PE} \bigg[ 
f^{V}_{\rm Macdonald}(q,a, \xi) \chi_\text{adj of $\mathfrak{su}(N)$}(b)
\bigg]  \ . 
\ee 
Here $b$ is a fugacity for $SU(N)$ gauge symmetry and the integral is over the Cartan torus of $SU(N)$.
The plethystic exponential PE is defined by
\be 
{\rm PE}[g(y_1, y_2, \dots )]
= \exp \sum_{k=1}^\infty \frac 1k g(y_1^k, y_2^k, \dots) \ ,
\ee 
where $g$ stands for any function
of any number of arguments $y_1$, $y_2$, \dots.

After these preliminaries, let us 
go back to discussing $\WW$.
We claim that the simple quotient of $\WW$ at $\nu = N^2$ is isomorphic to the VOA associated to 4d $\cN = 4$ $SU(N)$ SYM. This claim implies in particular that the supercharacter \eqref{eq_supercharacter_WW}
must match the Macdonald index \eqref{eq_SYM_Mac}. As it is already implicit in our choice of notation for fugacities on the two sides of the identification,
we also claim that $\mathfrak R$ on the $\WW$ side is identified with the 4d quantum number $R$ on the Macdonald index side.

On the $\WW$ side, Table \ref{table_summary_of_states}
contains all necessary data to compute the supercharacter of the simple quotient of $\WW$ up to order $q^4$ for any $N = 2,3,\dots$. All we have to do is use the counting of $\mathfrak{psl}(2|2)$ primaries in Table \ref{table_summary_of_states}
and the expressions \eqref{eq_long_and_short_contributions}
for the contrinutions of $\mathfrak{psl}(2|2)$ primaries. Furthermore, we have 
to specialize 
the fugacity $x$  as in \eqref{eq_specialize_x}.

On the SYM side, we have computed the Macdonald index up to order $q^5$
for $N=2,3,\dots, 10$. The results
are reported in Appendix \ref{all_macdonald}.

We indeed find a perfect match between
the supercharacter of the simple quotient of $\WW$ for $\nu = N^2$
and the Macdonald index of 4d $\cN = 4$ SYM,
up to order $q^4$.

Let us emphasize that Table 
\eqref{table_summary_of_states}
contains strictly more information
than the Macdonald index.
This is due to the presence of
fermionic $\mathfrak{psl}(2|2)$ multiplets
for $(h,j) = (4,1)$, as signaled by the quantity $\mathfrak N$ in 
Table \eqref{table_summary_of_states}.
In the Macdonald index, the contributions of these fermionic 
$\mathfrak{psl}(2|2)$ multiplets cancel against those of bosonic
$\mathfrak{psl}(2|2)$ multiplets
with the same quantum numbers.
This is seen explicitly from 
the fact that
\be 
\text{for $x = - \xi^{1/2}$,} \qquad   \mathfrak N := \xi^3 + \xi^2 + \xi^{7/2}\chi_1(x) =0  \ .
\ee

\subsubsection{Connection with Hall-Littlewood chiral ring}

The Hall-Littlewood (HL) chiral ring 
of a 4d $\cN = 2$ SCFT
is a consistent truncation of the usual $\cN = 1$ chiral ring
obtained by restricting to Schur operators.
More explicitly, the HL chiral ring 
captures the contributions of Schur operators
satisfying the additional condition
\be 
h = R+r \ . 
\ee 
In the case of 4d $\cN = 4$ $SU(N)$ SYM, the HL chiral ring is 
identified with the ring of functions on
\cite{Cachazo:2002ry,Kinney:2005ej} 
\be \label{eq_for_HL}
\frac{\mathbb C^{(2|1)} \times \mathbb R^{N-1}}{S_N} \ , 
\ee 
where we notice that $N-1$ is the rank of $SU(N)$ and
$S_N$ its Weyl group.
We recall that the HL chiral ring admits an alternative presentation
as a certain BRST cohomology problem, which is a truncation of the full
BRST problem yielding the VOA associated to the 4d theory \cite{Beem:2013sza},
see also Section \ref{sec_BRST} below. 
Non-trivially, the BRST presentation
reproduces the HL chiral ring and ring of functions on \eqref{eq_for_HL}, see the main conjecture in \cite{berest2015representationhomologyliealgebra}.

The Hilbert series of the
Hall-Littlewood chiral ring of 4d $\cN = 4$ $SU(N)$ SYM
can be written as
\be \label{eq_Hilbert_series_HL}
{\rm HS}_{\rm HL}(\tau,a,x) = {\rm Tr}_{\rm HL} \tau^{2h} a^j x^{2r} \  .
\ee 
Crucially, this quantity
can be computed via the Molien formula based on \eqref{eq_for_HL},
for any values of the fugacities $\tau$, $a$, $x$,
see \cite{Kinney:2005ej} and Section 6.4 of \cite{Bonetti:2018fqz} for details.
We stress the refinement by $x$, which captures the quantum
number $r$.
If we specialize to $x=-1$, we recover the HL limit of the superconformal index,
\be 
{\rm HS}_{\rm HL}(\tau,a,x=-1) = \cI_{\rm HL}^{\text{4d $SU(N)$ SYM}}(\tau ,a) \  .
\ee
We recall that the RHS is related to the Macdonald index as
\be 
\cI_{\rm HL}^{\text{4d $SU(N)$ SYM}}(\tau ,a) = \lim_{q \rightarrow 0} \cI_{\rm Macdonald}^{\text{4d $SU(N)$ SYM}}(q , a, \xi = \tau^2/q ) \ . 
\ee 
We notice that the original
Hilbert series 
\eqref{eq_Hilbert_series_HL} is sensitive to the 
fermionic $\mathfrak{psl}(2|2)$ primaries at $(h,j) = (4,1)$
with $R=7/2$, $r=1/2$. In contrast, this information is washed away
when we specialize to $x=-1$.
Thus, if we compute 
\eqref{eq_Hilbert_series_HL} for 4d $\cN = 4$ $SU(N)$
up to $\tau^8$, we can perform a non-trivial check against
the counting of $\mathfrak{psl}(2|2)$ primaries in $\WW$ in Table
\ref{table_summary_of_states}.
We compute \eqref{eq_Hilbert_series_HL}
for $N=2,3,\dots,8$ in Appendix \ref{app_HL},
where we also give the contribution to 
\eqref{eq_Hilbert_series_HL} of long and short multiplets.
Our findings show a perfect agreement between the Hilbert series of the HL chiral ring and the data of Table
\ref{table_summary_of_states}.

\subsection{Remarks on recombinations}

In this section we briefly discuss some aspects of the 
comparison between the the partition
function $Z_{\WW} (q,a,x,\xi)$ of $\WW$
and the partition function of \emph{free} 4d $\cN = 4$ $SU(N)$ SYM.

As explained in Section \ref{sec_counting_generic},
it is not hard to compute the partition
function $Z_{\WW} (q,a,x,\xi)$ of $\WW$
at generic values of the central charge,
using formula  \eqref{eq_W_alg_Z}.
It is harder to compute the partition
function of the simple quotient of $\WW$
at $\nu = N^2$, but the data of Table \ref{table_summary_of_states}
allows us to do it up to order $q^4$.

On the 4d side, 
if we consider free 
4d $\cN = 4$ $SU(N)$ SYM
we can compute not only the superconformal index,
but also more refined unprotected partition functions.
In particular, we can enrich
the Macdonald index \eqref{eq_def_part_func2}
with an extra fugacity $x$ to keep track of the charge $r$,
\be \label{sec_free_partition_function}
Z_{\text{free 4d $SU(N)$ SYM}} (q,a,x,\xi) =  {\rm Tr}_{\cH_{\rm Schur}}
 q^{E-R} \xi^{R } x^{2r} a^j   \ . 
\ee 
This quantity reduces to \eqref{eq_def_part_func2}
for $x=- \xi^{1/2}$. 
Computing \eqref{sec_free_partition_function} is not harder than computing \eqref{eq_def_part_func2}. The starting point is the
contribution to \eqref{sec_free_partition_function} of a free 4d $\cN = 2$ vector multiplet,
\be \label{eq_single_letter_N4}
f^{V}(q,a, \xi, x) =   \frac{q \sqrt \xi}{1-q} \chi_1(x)  + \chi_1(a)\frac{
\sqrt q \sqrt \xi
}{1-q}  \ . 
\ee 
This can be derived, for example, using 
Table 2 in \cite{Gadde:2011uv}
and assembling the 4d $\cN = 4$ vector multiplet
from a 4d $\cN = 2$ vector multiplet and a 
4d $\cN =2$ half hypermultiplet.
The quantity \eqref{sec_free_partition_function} is then given as 
\be \label{eq_gauge_inv_words}
Z_{\text{free 4d $SU(N)$ SYM}} (q,a,x,\xi)= \int[db] {\rm PE} \bigg[ 
f^{V}(q,a, \xi, x, \alpha) \chi_\text{adj of $\mathfrak{su}(N)$}(b)
\bigg]  \ ,
\ee 
in close analogy with \eqref{eq_SYM_Mac}.

Now, the quantity $Z_{\text{free 4d $SU(N)$ SYM}} (q,a,x,\xi)$
simply counts gauge invariant words
that can be written down using the Lagrangian letters of
4d $\cN = 4$ $SU(N)$ SYM. In the interacting theory,
some of this words recombine
and drop away from the partition function.
On the other hand, 
we  claim that the simple
quotient of $\WW$ at $\nu = N^2$ is isomorphic to the VOA associated to (interacting) 4d $\cN = 4$ $SU(N)$ SYM.
As a result, comparing  
$Z_{\WW} (q,a,x,\xi)$
against $Z_{\text{free 4d $SU(N)$ SYM}} (q,a,x,\xi)$
allows us to extract information about the states that recombine 
in the transition from the free to the interacting theory.

The states that can recombine are collections of short multiplets
whose combined content equals that of a long multiplet at threshold. For a 4d $\cN = 4$ SCFT, the possible recombinations can be described collectively as \cite{Dolan:2002zh} 
\be \label{sec_N4_recombinations}
\ba 
& \cA^{p + q_1 + q_2 + j_1 + j_2 + 2}_{[q_1, p, q_2]} \bigg|_{q_1 - q_2 = 2(j_2 - j_1)}
 \cong \\
 & \cong
\cC_{[q_1, p, q_2](j_1, j_2)}
\oplus
\cC_{[q_1 +1, p, q_2](j_1- \frac 12 , j_2)}
\oplus
\cC_{[q_1, p, q_2 +1](j_1, j_2- \frac 12)}
\oplus 
\cC_{[q_1+1, p, q_2 +1](j_1- \frac 12, j_2- \frac 12)}
\ .
\ea 
\ee 
We can formally apply the above also to the cases in which $j_1$
and/or $j_2$ are zero, if we make use of the identifications
\eqref{eq_N4_multiplet_identifications}.
We can traslate the multiplets on the RHS of
\eqref{sec_N4_recombinations} into $\mathfrak{psl}(2|2)$
multiplets using Table \ref{table_N4multiplets_give_Schur}.
We are thus led to define the following expressions,
\be \label{eq_A_quantities_recom}
\ba 
A_{[q_1,p,q_1]} (q,a,x,\xi)& =
(1 + \xi)\xi^{\frac p2 + q_1 + 1} \mathfrak L_{\frac p2 + q_1 + 2, \frac p2 } (q,a,x)
\\
& + \xi^{\frac p2 + q_1 + \frac 32} \chi_1(x)
\mathfrak L_{\frac p2 + q_1 + 2, \frac p2   }(q,a,x)
\ ,  \\ 
A_{[q_1,p,q_2](j_1,j_2)} (q,a,x,\xi)  & =    (1 + \xi) \xi^{\frac{p + q_1 + q_2}{2}+1}
x^{2(j_2 -j_1)}
\mathfrak L_{\frac{p + q_1 + q_2}{2} + j_1 + j_2 + 2 , \frac p2}(q,a,x)
\\
& + 
\xi^{\frac{p+q_1+q_2}{2} + \frac 32}
x^{2(j_2 -j_1)}
\chi_1(x)
\mathfrak L_{\frac{p + q_1 + q_2}{2} + j_1 + j_2 + 2 , \frac p2 ; j_2-j_1 + \frac 12 }(q,a,x) \ . 
\ea
\ee 
On the RHSs we encounter the quantity 
$\mathfrak L_{h,j}(q,a,x)$ of \eqref{eq_long_and_short_contributions}  that captures the contribution
of a long $\mathfrak{psl}(2|2)$ multiplet to the partition function.
The expressions \eqref{eq_A_quantities_recom} capture the contributions
of 4d long multiplets at threshold. As a small sanity check,
$A_{[q_1,p,q_1]}$ and $A_{[q_1,p,q_2](j_1,j_2)}$
vanish identically if we set $x = - \xi^{1/2}$.

After these preliminaries, we can finally present
our main result concerning the comparison
of the W-algebra partition function
and the partition function of free SYM.
We work up to order $q^4$. The result quoted below is valid for
$N = 8,9,\dots$ This is the ``stable'' regime for $N$,
in the sense that the counting of states with weight $h\le 4$
on either side becomes independent of $N$ for $N\ge 8$.
We find
\be 
\ba 
Z_{\WW} & = Z_{\text{free 4d $SU(N\ge 8)$ SYM}}
- \bigg[  
A_{[0,0,0]} 
+ A_{[0,1,0]} 
+ 3 A_{[0,2,0]} 
+ A_{[1,2,1]} 
\\
& + 2 A_{[0,1,0](\frac 12 , \frac 12)}
+ A_{[0,1,1](\frac 12 , 0)}
+ A_{[1,1,0](0,\frac 12 )}
+ 4 A_{[0,3,0]}
+ 3 A_{[1,1,1]}
\\ 
& + A_{[0,0,0](1,1)}
+ A_{[0,0,2](1,0)}
+ A_{[2,0,0](0,1)}
+ 3 A_{[0,2,0](\frac 12 , \frac 12)}
\\
& + 4 A_{[0,2,1](\frac 12 ,0)}
+ 4 A_{[1,2,0](0,\frac 12 )}
+ 6 A_{[1,0,1](\frac 12 , \frac 12)}
\\
& + A_{[1,0,2](\frac 12 ,0)}
+ A_{[2,0,1](0,\frac 12 )}
+ 8 A_{[0,4,0]}
+ 7 A_{[1,2,1]}
+ 7 A_{[2,0,2]}
\bigg] 
+ \mathcal O(q^{\frac 92}) \ . 
\ea 
\ee 
The above holds for arbitrary values of the fugacities
$a$, $\xi$, $x$.
This result is derived using Table \ref{table_summary_of_states} for the W-algebra side. For the free SYM side,
we have used the results collected in Appendix \ref{app_free_SYM_counting}.

It can also be interesting to 
study the difference between 
$Z_{\WW}$ and $Z_{\text{free 4d $SU(N)$ SYM}}$
for $N = 2,3,\dots,7$ (as always, we can do it only up to $q^4$
with the information we have on $\WW$).
Combining Table \ref{table_summary_of_states} and Appendix \ref{app_free_SYM_counting},
this can be achieved in a straightforward way, if desired.
We refrain, however, from a detailed discussion.

% \be 
% \ba 
% \mathsf A_{[q_1,p,q_1]} (q,a,x,\xi)& =
% (1 + \xi)\xi^{\frac p2 + q_1 + 1} \mathfrak L_{\frac p2 + q_1 + 2, \frac p2 ; 0} (q,a,x)
% \\
% & + \xi^{\frac p2 + q_1 + \frac 32}
% \Big[ 
% \mathfrak L_{\frac p2 + q_1 + 2, \frac p2 ; + \frac 12 }(q,a,x)
% + \mathfrak L_{\frac p2 + q_1 + 2, \frac p2 ; -\frac 12 }(q,a,x)
% \Big]
% \ ,  \\ 
% \mathsf A_{[q_1,p,q_2](j_1,j_2)} (q,a,x,\xi)  & =    (1 + \xi) \xi^{\frac{p + q_1 + q_2}{2}+1}
% \mathfrak L_{\frac{p + q_1 + q_2}{2} + j_1 + j_2 + 2 , \frac p2 ; j_2-j_1}(q,a,x)
% \\
% & + 
% \xi^{\frac{p+q_1+q_2}{2} + \frac 32}
% \Big[
% \mathfrak L_{\frac{p + q_1 + q_2}{2} + j_1 + j_2 + 2 , \frac p2 ; j_2-j_1 + \frac 12 }(q,a,x)
% \\
% &
% + \mathfrak L_{\frac{p + q_1 + q_2}{2} + j_1 + j_2 + 2 , \frac p2 ; j_2-j_1 -\frac 12 } (q,a,x)
% \Big] \ . 
% \ea
% \ee 

\subsection{Comparison with the BRST construction}
\label{sec_BRST}

The VOA associated to a 4d $\cN = 2$ SCFTs that admits a presentation as a gauge theory can be computed systematically
using the BRST construction introduced in \cite{Beem:2013sza}.
Clearly, this applies in particular to 4d $\cN= 4$ SYM.
We review the BRST construction briefly below.
While computationally intensive, the BRST construction has the
advantage of facilitating the identification of the 4d quantum number $R$ in the VOA. Thus, we can use it to provide further evidence that the proposed $\mathfrak R$-filtration in $\WW$
corresponds to the $R$-filtration coming from 4d physics.

\subsubsection{Brief review of the BRST construction}

The BRST construction of the VOA associated to 4d $\cN = 4$ $SU(N)$ SYM starts from 
a collection of  free fields,
more precisely $N^2-1$ copies of a symplectic boson and $N^2-1$
copies of a $bc$ system,
\be \label{eq_free_fields_of_BRTS}
\phi(z,y)^I{}_J \ , \qquad 
b(z)^I{}_J \ , \qquad 
c(z)^I{}_J \ .
\ee 
In the above expression $I,J = 1,\dots,N$ 
 are fundamental indices of $SU(N)$ and all letters are traceless matrices,
\be 
\phi(x,y)^I{}_I = b(z)^I{}_I =  c(z)^I{}_I  = 0 \ . 
\ee 
The field $\phi(z,y)^I{}_J$ is Grassmann even and   a polynomial in $y$ of degree 1,
\be 
\phi(z,y)^I{}_J = \phi_0(z)^I{}_J + y \phi_1(z)^I{}_J  \ . 
\ee 
The fields $b$, $c$ are Grassmann odd and $y$-independent.
The non-trivial OPEs among free letters read 
\be 
\ba 
\phi(y_1,z_1)^I{}_J  \phi(y_2,z_2)^K{}_L &=  \frac{y_{12}}{z_{12}} \bigg( \delta^I{}_L 
\delta^{K}{}_J
- \frac 1N \delta^I{}_J \delta^K{}_L
\bigg) + \text{reg.} \ , \\ 
b(z_1)^I{}_J  c(z_2)^K{}_L &=  \frac{1}{z_{12}} \bigg( \delta^I{}_L 
\delta^{K}{}_J
- \frac 1N \delta^I{}_J \delta^K{}_L
\bigg) + \text{reg.} \ .
\ea 
\ee 

Next, we construct the BRST current
\be 
\ba 
J_{\rm B} &= (\phi_0)^{I_1}{}_{I_2} (\phi_1)^{I_2}{}_{I_3}  c^{I_3}{}_{I_1}
- (\phi_0)^{I_1}{}_{I_2} (\phi_1)^{I_3}{}_{I_1}  c^{I_2}{}_{I_3}
 +   b^{I_1}{}_{I_2} c^{I_2}{}_{I_3}  c^{I_3}{}_{I_1}
\ .
\ea 
\ee 
The associated BRST charge is defined by acting with
$J_{\rm B}$ via the order-one pole in the OPE,
\be \label{eq_def_QB}
Q_{\rm B} \, X := \{ J_{\rm B}  X \}_1 \ , 
\ee 
where $X$ stands for any expression built from the free letters.\footnote{In this paper, when we work with free fields,
we always use the nested normal order product 
\be
X_1 X_2 \dots X_n := \{ X_1 \{ X_2 \{ \dots \{ X_{n-1}, X_n\}_0 \dots \}_0 \}_0 \}_0  \ ,  \nonumber 
\ee 
which is associative and graded commutative in a free VOA.
}
The charge $Q_{\rm B}$ is nilpotent.
The VOA associated to 4d $\cN = 4$ $SU(N)$ SYM
is obtained by passing to the cohomology of
$Q_{\rm B}$ relative to the zeromodes of the fields
$b(z)^I{}_J$. In practice, it means that
we consider states that are 
$Q_{\rm B}$-closed and that do not contain
any undifferentiated $c(z)^I{}_J$,
and we consider them modulo 
$Q_{\rm B}$-exact and that also do not contain
any undifferentiated $c(z)^I{}_J$.

We know that the sought-for VOA always
contains the generators of the small $\cN = 4$ super Virasoro algebra. Their explicit realizations in terms of the free fields
\eqref{eq_free_fields_of_BRTS} reads 
\be \label{eq_SCA_in_BRST}
\ba 
J(y) & = \tfrac 12 \phi(y)^I{}_J \phi(y)^J{}_I \ , \\ 
G(y) & = \phi(y)^I{}_J b^J{}_I \ , \\ 
\widetilde G(y) & = \phi(y)^I{}_J \partial_z c^J{}_I \ , \\ 
T & = - b^I{}_J \partial_z c^I{}_J - \tfrac 12 (\phi_0)^I{}_J \partial_z (\phi_1)^J{}_I
+ \tfrac 12 \partial_z  (\phi_0)^I{}_J (\phi_1)^J{}_I \ . 
\ea 
\ee 
In particular, the expressions for $J$ and $T$ allow
us to verify that the current $J_{\rm B}$ is a  $\mathfrak{sl}(2)_y \oplus \mathfrak{sl}(2)_z$ primary with weight $h=1$ and spin $j=0$. We also notice that  it has $\mathfrak{gl}(1)_r$ charge $-1/2$.
In passing, 
we observe that $J_{\rm B}$ is not a $\mathfrak{psl}(2|2)$ primary,
but rather a  descendant of the $\mathfrak{psl}(2|2)$ primary
\be 
\phi(y)^{I_1}{}_{I_2} c^{I_2}{}_{I_3} c^{I_3}{}_{I_1}  \ .
\ee

\paragraph{Further obervations on $Q_{\rm B}$.}
Let $X(y)$ be any operator. From the Borcherds identity 
\eqref{eq_OPE_assoc}, the definition 
\eqref{eq_def_QB}, and the $Q_{\rm B}$-closure of
$J$, $G$, $\widetilde G$, $T$,
we derive 
\be 
\ba 
 Q_{\rm B}  \{ J(y_1) X(y_2) \}_n   & = +   \{ J(y_1)  \; Q_{\rm B} X(y_2)  \}_n \ , \\
 Q_{\rm B}  \{ G(y_1) X(y_2) \}_n  & = -   \{ G(y_1) \; Q_{\rm B} X(y_2)   \}_n \ , \\ 
Q_{\rm B}  \{ \widetilde G(y_1) X(y_2) \}_n   & = -   \{ \widetilde G(y_1)  \; Q_{\rm B} X(y_2)   \}_n \ , \\ 
Q_{\rm B}  \{ T X(y_2) \}_n   & = +   \{ T  \; Q_{\rm B} X(y_2)  \}_n \ ,
\ea 
\ee 
for any $n\in \mathbb Z$.
These relations 
encode the compatibility between the action of $Q_{\rm B}$
and~$\mathfrak{psl}(2|2)$. 

Next, we observe that an operator $X$
does not contain zeromodes of $c$ if and only if
\be 
\{ b^I{}_J X(y_2) \}_1 = 0  \ . 
\ee  
We observe the property
\be 
\{ b^I{}_J J_{\rm B} \}_1 = K^I{}_J  \ , 
\ee 
where $K^I{}_J$ is the $SU(N)$ symmetry current, satisfying
$K^I{}_I = 0$ and given explicitly by
\be 
K^I{}_J = 
(\phi_0)^I{}_K (\phi_1)^K{}_J
- (\phi_0)^K{}_J (\phi_1)^I{}_K
+ b^I{}_K c^K{}_J - b^K{}_J c^I{}_K \ . 
\ee 
From the above remarks and the associativity identity
\eqref{eq_OPE_assoc}
we can check the following claim.
Suppose the operator $X(y_2)$ is a gauge invariant operator in the small algebra, namely
\be 
\{ b^I{}_J X(y_2) \}_1 = 0 \ , \qquad 
\{ K^I{}_J X(y_2) \}_1 = 0 \ . 
\ee 
Then, the operator $Y(y_2) := Q_{\rm B} X(y_2)$
is gauge invariant and lies in the small algebra, namely
\be 
\{ b^I{}_J Y(y_2) \}_1 = 0 \ , \qquad 
\{ K^I{}_J Y(y_2) \}_1 = 0 \ . 
\ee 
In checking the second equation, it is useful to observe that
$\{ K^I{}_J J_{\rm B} \}_1 = 0$.

\subsubsection{BRST analysis}

Within the BRST construction reviewed above,
we have tackled the following task.
For each $(h,j)$, with $h\le 4$:
\begin{itemize}
    \item Identify a set of linearly independent states constructed from the free letters 
    \eqref{eq_free_fields_of_BRTS}~that:  
    \begin{itemize}
        \item contain no $c$ zeromode;
        \item are singlets of $SU(N)$;
        \item are $\mathfrak{psl}(2|2)$ primaries 
        of definite $(h,j)$ with respect to $J$, $G$, $\widetilde G$, $T$ given in \eqref{eq_SCA_in_BRST}.
    \end{itemize}
 \item Determine the action of $Q_{\rm B}$ on the   states found above.
 \item Compile a list of $Q_{\rm B}$-closed states modulo 
 $Q_{\rm B}$-exact states, keeping track, besides $(h,j)$, of the quantum numbers $r$ and $R$.
\end{itemize}
Some comments on the last bullet point are in order.
In the vertex algebra spanned by the free letters
\eqref{eq_free_fields_of_BRTS},
the quantum number $r$ is identified with the
ghost number in the $bc$ system, in conventions
in which we assign ghost number $+1/2$ to $b$ and $-1/2$ to $c$.

For the quantum number $R$, we observe that 
all the free letters 
\eqref{eq_free_fields_of_BRTS} originate from 4d Schur operators that have
$R = 1/2$. Thus, 
in the vertex algebra spanned by the free letters,
we are led to consider a filtration based on length:
a normal ordered product of $n$ free fields
is assigned degree $n/2$. (The derivatives $\partial_z$ 
do not contribute.)
As usual, the degree of a linear combination of normal ordered products is the maximum of the degree of all summands.

The filtration described in the previous paragraph
is defined for any object constructed 
with the free fields \eqref{eq_free_fields_of_BRTS}.
We are interested in studying it
for states that are $Q_{\rm B}$-closed and have no $c$ zeromode, modulo states that are $Q_{\rm B}$-exact and have no $c$ zeromode.
To identify the correct degree of an expression,
we have to consider adding 
$Q_{\rm B}$-exact pieces and tuning them in such a way as to achieve the minimal degree possible.

We have performed the computations outlined above
for any value of $N$, for states with $h \le 4$.
In fact, for such states, the case $N \ge 8$
is easier to treat, as one can safely ignore any
trace relations. Indeed, states with $h=4$
are comprised at most of 8 free letters,
and if we consider $SU(N)$ with $N \ge 8$,
there are no trace relations of length $\le 8$.
For $N = 2,\dots,7$, on the contrary,
it is crucial to take trace relations into account in order
to perform the counting correctly.

The outcome of our computations
gives a perfect match with the counting of states in the simple
quotient of $\WW$ at $\nu = N^2$, see Table
\ref{table_summary_of_states}. In particular, the match is valid keeping track of all quantum numbers $(h,j,r,R)$ and thus provides further
evidence of the identification between the $\mathfrak R$-filtration
of $\WW$ and the $R$-filtration from 4d physics.

In closing this section, we notice that we can also make sense of the study of $Q_{\rm B}$ on $\mathfrak{psl}(2|2)$ primaries
if we formally extend $N$ from an integer to an arbitrary parameter. See \cite{Budzik:2023vtr} for a similar discussion in the context of the one-loop dilation operator
in 4d $\cN = 4$ SYM.
The key idea is that the matrix that represents
the action of $Q_{\rm B}$ on the basis of 
$\mathfrak{psl}(2|2)$ primaries with some definite $(h,j)$ is
a finite-dimensional matrix whose entries are
rational functions of $N$. As such, this matrix can be extended to arbitrary non-integer values of $N$.

\section{$\WW$ and half-BPS operators in 4d $\mathcal N = 4$ super Yang-Mills}
\label{sec_halfBPS}

% In previous sections,
% we have gathered evidence for the existence of the W-algebra $\WW$. We have also argued that $\WW$ truncates
% to the VOA $\cV(A_{N-1})$ labeled by
% the Weyl group of $\mathfrak{su}(N)$
% when the central charge $c$ approaches
% $c= -3(N^2-1)$.
% We recall that the VOA $\cV(A_{N-1})$ can be defined
% as the VOA associated to the Coxeter group
% ${\rm Weyl}(\mathfrak{su}(N))$,
% as in the proposal of \cite{Bonetti:2018fqz}
% proven in \cite{Arakawa:2023cki}.
% However, $\cV(A_{N-1})$ is also expected to be the VOA associated to 4d $\mathcal N=4$ $SU(N)$ SYM theory
% by the map of \cite{Beem:2013sza}.
% As a result, we expect
% that the W-algebra 
% $\WW$ encodes non-trivial information 
% on protected operators of 
% 4d $\mathcal N=4$ $SU(N)$ SYM,
% for any value of $N$.

In this section, 
we establish a precise connection between the
W-algebra $\WW$
and half-BPS operators in 4d
$\cN = 4$ $SU(N)$ super Yang-Mills.
The correspondence is tested by 
comparing 
the
OPE coefficients of $\WW$ from the bootstrap
analysis with known results on suitable correlators of half-BPS operators. Crucial to this analysis is the notion of
single-particle operator,
defined in 
\cite{Aprile:2018efk, Aprile:2020uxk}
building on \cite{Arutyunov:1999en, DHoker:1999jke, Rastelli:2017udc}.

The main point of this section can be summarized as follows:
\begin{itemize}
    \item  
      Two- and three-point functions of half-BPS single-particle operators in 4d $\mathcal N = 4$ $SU(N)$ super Yang-Mills
are independent on the Yang-Mills coupling \cite{Howe:1995aq,
DHoker:1998vkc,
Howe:1998zi,
Intriligator:1998ig,
Gonzalez-Rey:1999ouj,
Intriligator:1999ff,
Eden:1999gh,
Skiba:1999im,
Penati:1999ba,
Dolan:2004mu}
and can be computed for any integer $N\ge 2$ using \emph{Wick contractions}.
Choosing a convenient normalization for single-particle operators,
their two- and three-point functions are
rational functions of $N^2$.
\item The W-algebra data $c_{p_1 p_2}{}^{p_3}$, $g_p$
at \emph{generic} values of the central charge 
 can be extracted from 
field theory correlators
in which we formally perform the
replacement $N^2 \leadsto  \nu$.  
\end{itemize}
This claim entails an infinite number of conjectures regarding OPE data of $\WW$
as exact functions of the central charge. Clearly, our finite bootstrap data only allows us to check a finite subset of them.

In the rest of this subsection we first review relevant properties of half-BPS operators in 4d $\cN = 4$ $SU(N)$ super Yang-Mills
and their correlators. We then describe in detail the relation between correlators and W-algebra data.

\subsection{Review of half-BPS operators and their correlators}
\label{sec_SPOs}

\subsubsection{Single-particle operators}

Let us consider
4d $\cN = 4$ $\mathfrak{su}(N)$ SYM. 
Its  spectrum contains half-BPS
operators that are the superconformal
primaries of multiplets of type
$\cB^{\frac 12 , \frac 12}_{[0,p,0](0,0)}$, $p=2,3,4,\dots$,
in the notation of \cite{Dolan:2002zh}.
We recall that these half-BPS operators
are absolutely protected
and form the half-BPS chiral ring.

A customary basis
of half-BPS operators is furnished by single-trace operators (and their products).
Let $\phi(x,Y)$ denote the six real adjoint-valued scalars in 4d $\cN = 4$ super Yang-Mills. Here
$x$ are coordinates in 4d spacetime, while $Y$ is an
auxiliary $SO(6)$ null vector used as a proxy for R-symmetry indices.
Single-trace operators are of the form
\be 
T_p(x,Y) = {\rm Tr} \Big( 
\phi(x,Y)^p  
\Big) \ ,
\ee 
for integer $p\ge 2$.
The trace is in the fundamental representation of $\mathfrak{su}(N)$.
The operator $T_p$
transforms in the $[0,p,0]$ representation of the R-symmetry $\mathfrak{su}(4) \cong \mathfrak{so}(6)$
and has conformal dimension $\Delta = p$.
We refer to $T_p$ as having charge $p$, for brevity.
Multi-trace operators
are simply products of single-trace operators,
\be 
T_{p_1 , p_2 , \dots  ,p_n} = T_{p_1} T_{p_2} \dots T_{p_n} \ , 
\ee 
with charge $p_1 + \dots + p_n$.

Since correlators of half-BPS operators
are independent of the Yang-Mills coupling
$g_{\rm YM}$, they can be computed in the free field limit
$g_{\rm YM} \rightarrow 0$
 using Wick contractions.
The basic contraction reads
\be 
\langle \phi_a{}^b (x_1,Y_1)
\phi_c{}^d (x_2,Y_2)
\rangle 
= \bigg( 
\delta^b_c \delta^d_a - \frac 1N \delta^b_a \delta^d_c
\bigg)
\frac{Y_1 \cdot Y_2}{(x_1-x_2)^2} \ . 
\ee 
The indices $a$, $b$, $c$, $d =1, \dots N$
are fundamental indices of
$\mathfrak{su}(N)$.

For later applications
to the W-algebra $\WW$,
a different basis of half-BPS operators is best suited. This basis is comprised of so-called single-particle operators (SPOs) (and their products).  

By definition, an SPO is a half-BPS operator
with vanishing two-point function
with all multi-trace operator,
\be \label{eq_SPO_orthogonal}
\langle \cO_p  T_{q_1 ,  q_2  , \dots ,  q_n} \rangle = 0 \ , 
\ee 
with $n\ge 2$.
We 
 use the notation
$\cO_p$ for a single-particle
operator of charge $p$.
Crucially, an SPO is generically not
a single-trace operator,
but rather an admixture of
single-trace and multi-trace
operators.
Some concrete low-charge
examples are (in the $SU(N)$ theory)
\begin{align}
\label{eq_single_particle_examples}
\cO_2 & = T_2 \nn \ , 
\\ 
\cO_3 & = T_3  \nn \ , 
\\ 
\cO_4 & = T_4 - \frac{2N^2-3}{N(N^2 + 1)} 
T_{2,2}
\ , \\
\cO_5 & = T_5
- \frac{5(N^2-2)}{N(N^2 +5)}
T_{2,3} \  ,  \nn \\ 
\cO_6 &= T_6 
- \frac{3N^4 - 11 N^2 + 80}{N(N^4 + 15 N^2 + 8)} T_{3,3}
- \frac{6(N^2-4)(N^2+5)}{N(N^4 + 15 N^2 + 8)} T_{4,2}
+ \frac{7(N^2-7)}{N^4 + 15 N^2 + 8} T_{2,2,2} \nn \ . 
\end{align} 
In general, we fix the normalization of $\cO_p$
by requiring it to have the form
\be \label{eq_Op_normalization}
\cO_p = T_p + \text{multi-trace corrections} \ , 
\ee 
with the corrective terms uniquely determined as rational functions of $N$
by imposing 
\eqref{eq_SPO_orthogonal}.
Closed formulae for the corrective terms are reported in \cite{Aprile:2020uxk}.

SPOs enjoy remarkable properties, including the following
\cite{Aprile:2018efk, Aprile:2020uxk}:
\begin{itemize}
\item  If the total charge $p$ of an SPO is greater than $N$, 
the operator vanishes automatically.
In contrast, if we consider a single-trace operator of charge $p>N$,
it is not zero, but rather a complicated
combination of multi-trace operators.

\item In the context of AdS/CFT,
single-particle operators
are dual to single-particle supergravity states in Type IIB on $AdS_5 \times S^5$. In particular, they interpolate between point-like gravitons and
giant gravitons (i.e.~D3-branes wrapping $S^3 \subset S^5$).
In the strict large $N$ limit
($N=\infty$)
single-particle operators
and single-trace operators can be identified,
but they generically differ by $1/N$ corrections as soon as $N$ is finite. This is apparent in the examples 
\eqref{eq_single_particle_examples}.
\end{itemize}

\subsubsection{Some closed expressions for correlators}

Several exact in $N$ results for the correlators of single-particle
operators are derived in 
\cite{Aprile:2020uxk}. 
Here we review some of them.

As a preliminary, we recall that two- and three-point functions 
of $\cO_p$ operators
are fixed by
superconformal symmetry up to
an overall constant. We write
\be 
\ba 
\langle \cO_p(x_1, Y_1)
\cO_p(x_2,Y_2) \rangle 
& = \langle \cO_p \cO_p \rangle (Z_{12})^p  \  , 
\\
\langle \cO_{p_1}(x_1,Y_1)
\cO_{p_2}(x_2,Y_2)
\cO_{p_3}(x_3,Y_3)
\rangle 
& = \langle \cO_{p_1} \cO_{p_2} \cO_{p_3} \rangle
(Z_{12})^{p_{12}}
(Z_{23})^{p_{23}}
(Z_{13})^{p_{13}} \ ,
\ea
\ee 
where 
$\langle \cO_p \cO_p \rangle$, 
$\langle \cO_{p_1} \cO_{p_2} \cO_{p_3} \rangle$ (without $x$, $Y$ dependence)
denote the constant  two- and three-point functions coefficients and 
we have made use of the shorthand notation
\be 
Z_{ij} = \frac{Y_i \cdot Y_j}{(x_i-x_j)^2} \ , \quad 
p_{12} = \frac{p_1 + p_2 - p_3 }{2}   \ , \quad 
p_{23} = \frac{p_2 + p_3 - p_1 }{2} \ ,
\quad 
p_{13} = \frac{p_1 + p_3 - p_2 }{2} \ .
\ee 
A non-zero three-point function requires $p_1 + p_2 + p_3$ even
(due to R-symmetry selection rules,
or equivalently analyticity in the $Y$ vectors).
We observe that $\langle \cO_{p_1} \cO_{p_2} \cO_{p_3} \rangle$ is totally symmetric in its $p_i$ labels, due to the Bose symmetry of the $\cO_{p_i}$
operators.

Let us now state some results of
\cite{Aprile:2020uxk} regarding the two- and three-point function
coefficients 
$\langle \cO_p \cO_p \rangle$
and
$\langle \cO_{p_1} \cO_{p_2} \cO_{p_3} \rangle$.

We assume that $\cO_p$
is normalized as in 
\eqref{eq_Op_normalization}.
Then one has
\be \label{eq_Rp_expr1}
\langle \cO_p \cO_p \rangle = p^2(p-1) \frac{(N - p+1)_{p-1} (N +1)_{p-1}   }{
(N +1)_{p-1} - (N - p+1)_{p-1}
} \ . 
\ee 
Here $(a)_n$ is the ascending Pochhammer symbol.

Next, let us consider a three-point function of single-particle operators,
\be \label{eq_three_SPOs}
\langle \cO_{p} \cO_{q_1} \cO_{q_2} \rangle
%\cC_{p q_1 q_2}
\qquad 
\text{where $p$ is the largest charge: } 
p \ge q_{1}  \ , 
p \ge q_{2}\ .
\ee 
With this notation, the following holds.
\begin{itemize}
\item If $p \ge q_1 + q_2$, the correlator vanishes:
\be \label{eq_SPO_zero}
\langle \cO_{p} \cO_{q_1} \cO_{q_2} \rangle
%\cC_{p q_1 q_2} 
= 0 \ . 
\ee 
\item If $p= q_1 + q_2 -2$,
\be \label{eq_SPO_2}
\langle \cO_{p} \cO_{q_1} \cO_{q_2} \rangle
%\cC_{p q_1 q_2} 
= q_1 q_2 \langle \cO_p \cO_p \rangle \ .
\ee 
\item If $p= q_1 + q_2 - 4$, 
\begin{align} \label{eq_SPO_4}
&
\langle \cO_{p} \cO_{q_1} \cO_{q_2} \rangle
%\cC_{p q_1 q_2} 
  =  \langle \cO_p \cO_p \rangle
 \bigg\{ 
q_1 q_2 \bigg[ 
N - \frac{(q_1-1)(q_2-1)}{N}
\bigg]
+    f(q_1, q_2, N)
+    f(q_2, q_1, N)
\bigg \}
\ ,  
\\
& f(q_1, q_2,N)  = 
\frac{q_1 q_2 (q_2 -1)}{2N (q_1-2)} 
\bigg[ 
2N^2
+ N(q_1-1)_2
+ 2 (q_1-2)_2
+ \frac{2N (q_1-1) _2(N)_{q_1}}{
(N-q_1+1)_{q_1}
-
(N)_{q_1}
}
\bigg] \ .   \nn
\end{align}
\end{itemize}

It is also useful to review some results for correlators
in the planar limit,
i.e.~as $N \rightarrow \infty$.
In this limit, the difference between the single-particle operator $\cO_p$
and the single-trace operator
$T_p$ is subleading.
The correlators of single trace operators in the planar limit
are known for any $p_1$, $p_2$, $p_3$ \cite{Lee:1998bxa}.
We have
\be 
\text{at large $N$:}
\qquad 
\frac{  
\langle \cO_{p_1} \cO_{p_2} \cO_{p_3} \rangle
}{\sqrt{  
\langle \cO_{p_1} \cO_{p_1} \rangle 
\langle \cO_{p_2} \cO_{p_2} \rangle 
\langle \cO_{p_3} \cO_{p_3} \rangle 
}}
= \frac{\sqrt{p_1 p_2 p_3}}{N} 
+ \text{subleading} \ . 
\ee 
Making use of 
\eqref{eq_Rp_expr1} at large $N$,
$\langle \cO_p \cO_p \rangle \approx pN^p$,
we obtain 
\be \label{eq_largeN_YM}
\text{at large $N$:}
\qquad 
\langle \cO_{p_1} \cO_{p_2} \cO_{p_3}
\rangle 
= p_1 p_2 p_3 N^{\frac{p_1 + p_2 + p_3}{2} -1} + \text{subleading} \ .
\ee

\subsubsection{A useful rescaling}

Let us define a rescaled version of $\cO_p$,
\be 
\widetilde \cO_p = \frac{1}{N^{(p-2)/2}}
\cO_p \  .
\ee 
The two- and three-point functions coefficients for
the rescaled $\cO_p$ operators
are given by
\be
\langle \widetilde \cO_p 
 \widetilde \cO_p
 \rangle 
 = N^{2-p} \langle  \cO_p 
  \cO_p
 \rangle  \ , \qquad 
\langle 
\widetilde \cO_{p_1}
\widetilde \cO_{p_2}
\widetilde \cO_{p_3}
\rangle 
= N^{3 - \frac{p_1 + p_2 + p_3}{2}}
\langle
 \cO_{p_1}
 \cO_{p_2}
 \cO_{p_3}
\rangle 
\ . 
\ee 
This rescaling is useful because 
$\langle \widetilde \cO_p 
 \widetilde \cO_p
 \rangle $
and $\langle 
\widetilde \cO_{p_1}
\widetilde \cO_{p_2}
\widetilde \cO_{p_3}
\rangle $
are rational functions of $N^2$
(as opposed to $N$).  
This can be checked explicitly for $\langle \widetilde \cO_p 
 \widetilde \cO_p
 \rangle $ using
\eqref{eq_Rp_expr1} and for $\langle 
\widetilde \cO_{p_1}
\widetilde \cO_{p_2}
\widetilde \cO_{p_3}
\rangle $ in the cases reported above,
\eqref{eq_SPO_zero}, \eqref{eq_SPO_2}, 
\eqref{eq_SPO_4}.\footnote{
A general argument is as follows.
The quantity $\langle 
 \cO_{p_1}
 \cO_{p_2}
 \cO_{p_3}
\rangle $ is 
rational function of $N$
but its large-$N$ expansion
is actually a series in $1/N^2$
(this is a general feature of large-$N$ pertubation
theory involving only fields in the adjoint representation).
Schematically,
$\langle 
 \cO_{p_1}
 \cO_{p_2}
 \cO_{p_3}
\rangle  = N^\alpha (a_0 + a_1 N^{-2} + a_2 N^{-4} + \dots )$,
where $\alpha = (p_1 + p_2 + p_3)/2-1$ as can be seen from
\eqref{eq_largeN_YM}.
After rescaling, $\langle 
\widetilde \cO_{p_1}
\widetilde \cO_{p_2}
\widetilde \cO_{p_3}
\rangle$ is still a rational function of $N$
and it has the form
$\langle 
\widetilde \cO_{p_1}
\widetilde \cO_{p_2}
\widetilde \cO_{p_3}
\rangle = N^2 (a_0 + a_1 N^{-2} + a_2 N^{-4} + \dots )$.
Then, it must be a rational function of $N^2$.
}

In what follows, we make use of the notation
\be 
\langle \widetilde \cO_p 
 \widetilde \cO_p
 \rangle
 (N^2 \leadsto \nu) \ , \qquad 
\langle 
\widetilde \cO_{p_1}
\widetilde \cO_{p_2}
\widetilde \cO_{p_3}
\rangle
(N^2 \leadsto \nu) \ . 
\ee 
This means that 
$\langle \widetilde \cO_p 
 \widetilde \cO_p
 \rangle$, $\langle 
\widetilde \cO_{p_1}
\widetilde \cO_{p_2}
\widetilde \cO_{p_3}
\rangle$ are first computed at integer $N$ and expressed as rational functions of $N^2$, and then the replacement of $N^2$ with $ \nu$ is made.

% We close this subsection by stressing that all two-point functions coefficients $R_p$
% and all three-point function coefficients $\cC_{p_1 p_2 p_3}$
% are rational functions of $N$.
% They are computed by Wick contractions for integer $N \ge 2$, but the final results can be  continued to generic  values of $N$. We use the notation
% \be 
% R_p(N \leadsto \sqrt \nu ) \ , \qquad 
% \cC_{p_1 p_2 p_3}(N \leadsto \sqrt \nu ) \ ,
% \ee 
% in order to indicate that
% $R_p$, $\cC_{p_1 p_2 p_3}$ are first computed at integer $N$ and expressed as rational functions, and then the replacement of $N$ with $\sqrt \nu$ is made.

\subsection{From half-BPS correlators to W-algebra data}

We are now in a position to 
describe precisely the relation
between half-BPS correlators and W-algebra data.
The correspondence  
is as follows:
\begin{itemize}
\item  \emph{The generator $W_p$
is identified with the SPO $\cO_p$, for $p\ge 3$, up to an $N$-dependent normalization factor}
\be \label{eq_W_O_dictionary}
W_p \leftrightarrow \widetilde \cO_p = 
\frac{1}{N^{(p-2)/2}} \cO_p \ .
\ee 
Here $W_p$ is assumed to satisfy 
\eqref{eq_no_mixing_for_Wp}. This is the W-algebra analog of the condition 
\eqref{eq_SPO_orthogonal}
in Yang-Mills.
\item \emph{
In accordance with
\eqref{eq_W_O_dictionary}, the two-point function coefficient $g_p$ is given by}
\be \label{eq_g_dictionary}
g_p(\nu) =\langle
\widetilde \cO_p
\widetilde \cO_p
\rangle 
(N^2 \leadsto  \nu ) \ .
\ee 
% \be \label{eq_g_dictionary}
% g_p(\nu) =\frac{1}{\nu^{\frac{p-2}{2}}} R_p(N \leadsto \sqrt \nu ) \ .
% \ee 
The factor $N^{(p-2)/2}$
in \eqref{eq_W_O_dictionary}
is engineered to  ensure that 
\eqref{eq_g_dictionary} matches with
our normalization choice
\eqref{eq_best_normalization}.

\item \emph{In accordance with
\eqref{eq_W_O_dictionary}, 
all OPE coefficients of the form
$c_{p_1 p_2}{}^{p_3}$
are given by
}
\be \label{eq_lambda_and_C}
c_{p_1 p_2}{}^{p_3} (\nu)= 
\frac{
\langle
\widetilde \cO_{p_1}
\widetilde \cO_{p_2}
\widetilde \cO_{p_3}
\rangle (N^2 \leadsto \nu)
}{
\langle
\widetilde \cO_{p_3}
\widetilde \cO_{p_3}
\rangle(N^2 \leadsto \nu)
} \ .
\ee 
% \be \label{eq_lambda_and_C}
% c_{p_1 p_2}{}^{p_3} (\nu)= 
% \nu^{1-\frac{p_1 + p_2 - p_3}{2}} \;
% \frac{\cC_{p_1 p_2 p_3}(N \leadsto \sqrt \nu)}{R_{p_3}(N \leadsto \sqrt \nu)} \ .
% \ee 
% We observe that $p_1 + p_2 - p_3$ has to be even
% for $W_{p_3}$ to enter the 
% $W_{p_1} \times W_{p_2}$ OPE.

\end{itemize}
The relation 
\eqref{eq_lambda_and_C} amounts to an infinite number of predictions about W-algebra OPE coefficients as exact functions of $\nu$.
We can test it against the explicit bootstrap data
collected in Tables \ref{tab_OPEcoeffs_part1}-\ref{tab_OPEcoeffs_part8}.
We have indeed performed this check 
in the following cases (we order the entries of triples in descending order):
\begin{align}
&\text{$p= q_1 + q_2$} \ : &
&(p ,q_1, q_2) \in \Big\{ (6,3,3), (7,4,3)  \Big\} \ , \nn \\
&\text{$p= q_1 + q_2-2$} \ : &
&(p ,q_1, q_2) \in \Big\{ 
(4,3,3), (5,4,3), (6,5,3), (7,6,3),
(6,4,4), (7,5,4)
\Big\} \ , \nn \\
&\text{$p= q_1 + q_2-4$} \ : &
&(p ,q_1, q_2) \in \Big\{ 
(4,4,4),
(5,5,4),
(6,6,4),
(6,5,5)
\Big\} \ . 
\end{align}
We can also consider
\eqref{eq_lambda_and_C} at large $\nu$, 
equivalent to large $N$.
Using \eqref{eq_largeN_YM}
and $\langle \cO_p \cO_p
\rangle 
\approx pN^p$,
we reproduce 
\be 
\text{at large $\nu$:}
\qquad 
c_{p_1 p_2}{}^{p_3} (\nu) = p_1   p_2  + \text{subleading} \ . 
\ee

We emphasize that the
correspondence   outlined in the three bullet points above holds for generic values of $\nu$. For special values of $\nu$, the quantity $g_p(\nu)$
as defined by 
\eqref{eq_g_dictionary}  develops divergences
and a special analysis is needed.
Some examples are reported
in Appendix \ref{app_special_nus}.

\section{Wedge algebra of $\WW$}
\label{sec_wedge}

In this section we recall the notion
of wedge algebra and 
give a fully explicit
presentation of  the wedge algebra of the W-algebra $\WW$. 
This wedge algebra has already appeared in~\cite{Costello:2018zrm}.

\subsection{Reminder on wedge algebras}

The notion of wedge algebra associated to a  W-algebra can be described
as follows \cite{Bowcock:1991zk}.
Let $X(z)$ denote a strong
generator of the W-algebra,
with conformal weight $h_X$.
The Laurent modes of $X(z)$
are defined as usual via
\be 
X(z) = \sum_{m} z^{-m-h_X} X_m \  .
\ee 
The sum is over  integers
(resp.~half integers)
if $h_X$ is integer
(resp.~half integer).
The modes satisfying
\be 
|m| \le h_X - 1
\ee 
are referred to as vacuum preserving modes, as they preserve the
$\mathfrak{sl}(2)_z$ invariant vacuum.
The main idea is to seek a truncation
of the full (super) W-algebra to the vacuum
preserving modes of its generators,
in order to obtain a  Lie (super)algebra.
The restriction to the vacuum preserving modes kills all central terms in the W-algebra (terms with the identity operator $\one$).
A na\"ive truncation, however, fails in general
to give a well-defined Lie algebra,
due to the non-linealities in the OPEs
of the W-algebra.
Let us suppose that the W-algebra under examination exists for generic values of the central charge $c$. Then, by considering the large $c$ limit, all non-linearities are suppressed. Under mild conditions on the W-algebra \cite{Bowcock:1991zk},
one obtains a Lie algebra
whose generators are the vacuum
preserving modes of the generators of the original W-algebra.
This is precisely the sought-for wedge algebra.

Let us consider two simple examples.
The wedge algebra of the Virasoro algebra is $\mathfrak{sl}(2)$.
This is readily seen 
from the familiar commutators 
of the Virasoro algebra\footnote{It is customary to denote the Laurent modes of $T(z)$ as $L_m$, but we prefer to write them as $T_m$
for uniformity of notation
with modes of other operators, introduced below. 
},
\be  \label{eq_Vir_reminder}
T(z) = \sum_{m\in \mathbb Z} z^{-m-2} T_m \ , \qquad 
[T_m, T_n] = (m-n) T_{m+n} + \frac{c}{12}(m^3-m) \delta_{m+n,0} \ . 
\ee 
The vacuum preserving modes are $T_{m}$ with $m=0,\pm 1$. For this values of $m$ the central term drops out and we recover the Lie algebra 
$\mathfrak{sl}(2)$.

Our second example is the small 
$\cN = 4$ super Virasoro algebra
\eqref{eq_sVir}. Its wedge algebra
is $\mathfrak{psl}(2|2)$.
The generators of this
(finite dimensional)
 Lie superalgebra
are the vacuum preserving modes of $T$, $J$, $G$, $\widetilde G$.
The Laurent modes of $T$
have already been defined in 
\eqref{eq_Vir_reminder}.
Those of $J$, $G$, $\widetilde G$ are
\be \label{eq_superVir_modes}
\ba 
J(z,y) &= \sum_{m\in \mathbb Z} z^{-m-1} J_m(y)  \ , \\
G(z,y) &= \sum_{m \in \frac 12 + \mathbb Z} z^{-m-\frac 32} G_m(y) \ , &
\widetilde G(z,y) &= \sum_{m \in \frac 12 + \mathbb Z} z^{-m-\frac 32} \widetilde G_m(y) \ . 
\ea 
\ee 
We find it convenient to keep the auxiliary variable $y$ in performing the Laurent expansion in the coordinate $z$.
The vacuum preserving modes are
\be \label{eq_superVir_vac_modes}
\{T_m\}_{m=-1,0,1} \ , 
\quad 
\{ J_m(y) \}_{m=0} \ , \quad 
\{ G_m(y) \}_{m=-\frac 12, \frac 12} \ , \quad 
\{ \widetilde G_m(y) \}_{m=-\frac 12, \frac 12} \ . 
\ee 
The OPEs reported in 
\eqref{eq_sVir}
are readily translated into 
(anti)commutation
relations for the modes 
in \eqref{eq_superVir_modes}.
We refer the reader to Appendix
\ref{app_wedge} for details.
In particular,
if we restrict to the vacuum
preserving modes \eqref{eq_superVir_vac_modes},
the central terms in the $[T,T]$,
$[J,J]$, and $[ G, \widetilde G ]$
(anti)commutators all drop out.
We are left with the
(anti)commutators of
$\mathfrak{psl}(2|2)$.

\subsection{Properties of the wedge algebra of $\WW$}

Our knowledge of the OPE
coefficients of the W-algebra
$\WW$ is partial.
For this reason, instead of trying
to construct the wedge algebra
of $\WW$ directly, we will proceed indirectly. 

In a first step, using Lie algebra bootstrap
methods, we will 
determine explicitly the structure  
of an infinite-dimensional
Lie superalgebra, which we denote $\hs$.
Here hs stands 
for higher-spin,
in analogy with the higher-spin algebras of minimal
model holography
(see \cite{Gaberdiel:2012uj} for a review).
The generators of $\hs$ fall into short
multiplets of $\mathfrak{psl}(2|2)$.
This distinguishes
$\hs$ from other higher spin
algebras with the same amount of
supersymmetry considered in the literature, such as
the wedge algebra of the W-algebra of \cite{Ahn:2020rev}.
In fact, the Lie superalgebra
$\hs$ has already appeared
in \cite{Costello:2018zrm},
where it was denoted
$\mathfrak{a}_\infty$
and was referred to as the
algebra of global symmetries of the VOA associated to large-$N$ 4d $\cN = 4$ $\mathfrak{su}(N)$ super Yang-Mills.

% All (anti)commutators of the Lie superalgebra 
% $\hs$
% are given in closed form
% in appendix \ref{sdsd},
% see equations
% XXXXX.
% We notice that the existence and properties of $\hs$
% might be of interest in their own
% right, independently on the W-algebra
% $\WW$.

In a second step, we 
propose that $\hs$ is indeed isomorphic to the wedge algebra
of~$\WW$,
\be 
\hs \cong \text{Wedge}(\WW) \ . 
\ee 
We perform some partial checks of this claim, using the explicit
OPE coefficients of $\WW$
collected in Tables
\ref{tab_OPEcoeffs_part1}-\ref{tab_OPEcoeffs_part8}.

\subsubsection*{Bootstrapping the   Lie superalgebra $\hs$}

We  seek a  Lie superalgebra
$\hs$ with the following properties.
\begin{itemize}
    \item $\hs$ contains
    $\mathfrak{psl}(2|2)$ as a subalgebra; all generators of
    $\hs$ fall into $\mathfrak{psl}(2|2)$ multiplets.
   \item Besides the generators of $\mathfrak{psl}(2|2)$, the generators of $\hs$ consist
   of one short $\mathfrak{psl}(2|2)$ multiplet with $h=j=p/2$ for each $p\in \{ 3,4,5,\dots\}$.
\end{itemize}
We describe the generators
of $\hs$ with a notation that makes
contact with the notation we have used
for the W-algebra $\WW$.
The generators of
$\mathfrak{psl}(2|2)$ are 
summarized in the table below,
together with their $\mathfrak{sl}(2)_z$ weight $h$ and
$\mathfrak{sl}(2)_y$ spin $j$,
\be \label{eq_psl_table}
\begingroup
\renewcommand{\arraystretch}{1.2}
\begin{array}{c| | c | c | c}
& \;\; h\;\;  & \;\; j\;\; & \text{allowed values of $m$}
\\
\hline 
J_m(y) & 1 & 1 & 0 \\ 
G_m(y) & \tfrac 32 & \tfrac 12 & - \tfrac 12, \tfrac 12 \\ 
\widetilde G_m(y) & \tfrac 32 & \tfrac 12 & - \tfrac 12, \tfrac 12 \\ 
T_m & 2 & 0 & -1,0,1
\end{array}
\endgroup
\ee 
We remind the reader that an object of spin $j$ is a polynomial
in the auxiliary variable $y$
of degree $2j$.
The additional generators
of $\hs$ are collected in the table below
(there is one such multiplet for each $p=3,4,5,\dots$)
\be \label{eq_W_table}
\begingroup
\renewcommand{\arraystretch}{1.2}
\begin{array}{c| | c | c | c}
& \;\;\;\; h\;\;\;\;  & \;\;\;\; j\;\;\;\; & \text{allowed values of $m$}
\\
\hline 
(W_p)_m(y) & \frac p2 & \frac p2 & - \frac p2 +1 , - \frac p2 +2, \dots, \frac p2 -1 \\ 
(G_{W_p})_m(y) & \tfrac p2 + \frac 12 & \tfrac p2 - \frac 12 & 
- \frac p2 + \frac 12 \ , 
- \frac p2 + \frac 32 , \dots ,
\frac p2 - \frac 12
\\ 
( \widetilde G_{W_p})_m(y) & \tfrac p2 + \frac 12 & \tfrac p2 - \frac 12 & 
- \frac p2 + \frac 12 \ , 
- \frac p2 + \frac 32 , \dots ,
\frac p2 - \frac 12 \\ 
(T_{W_p})_m(y) & \frac p2 + 1 & 
\frac p2 -1 &
- \frac p2  , - \frac p2 +1 , \dots,
\frac p2
\end{array}
\endgroup
\ee 
Clearly, tables
\eqref{eq_psl_table} and \eqref{eq_W_table}
are nothing but a list of the vacuum
preserving modes of the strong generators of $\WW$.

The bootstrap construction of
$\hs$ is reported in Appendix \ref{app_wedge}.
We are able to determine
all (anti)commutators
of $\hs$
in closed form.
They are encoded in
equations
\eqref{eq_psl22_compact},
\eqref{eq_JW_comm},
\eqref{eq_Wcomm},
\eqref{eq_GWcomm},
\eqref{eq_HWcomm},
\eqref{eq_TWcomm},
\eqref{eq_c_result}.
These expressions
make use of a compact notation for (anti)commutators, explained in Appendix 
\ref{app_wedge}.

Let us discuss here some
salient features of
$\hs$.
The (anti) commutators
of $\hs$
fall into three classes,
\be 
[\mathbb J , \mathbb J] \ , \qquad 
[\mathbb J , \mathbb W_p] \ , \qquad 
[\mathbb W_{p_1} , \mathbb W_{p_2}] \ . 
\ee 
Here $[\mathbb J, \mathbb J]$
is a shorthand notation
for all (anti)commutators
among $J_m(y)$,
$G_m(y)$, $\widetilde G_m(y)$,
$T_m$. Similar remarks apply to
the notation $[\mathbb J , \mathbb W_p]$ and $[\mathbb W_{p_1} , \mathbb W_{p_2}]$.

Now, the (anti)commutators 
$[\mathbb J , \mathbb J]$
are those of $\mathfrak{psl}(2|2)$,
and are therefore completely fixed,
see \eqref{eq_psl22_compact}.
Next,
the (anti)commutators 
$[\mathbb J , \mathbb W_p]$
are completely fixed by demanding that
$W_p$ be a $\mathfrak{psl}(2|2)$
primary with $h=j=p/2$.
These (anti)commutators are
reported in \eqref{eq_JW_comm}.
The only non-trivial
(anti)commutators
are those of the form 
$[\mathbb W_{p_1} , \mathbb W_{p_2}]$.
Even though we have an infinite
number of $\mathfrak{psl}(2|2)$
multiplets $\mathbb W_p$,
we can study these (anti)commutators
systematically,
by organizing them according to increasing values of $p_1 + p_2$.
This is completely analogous to the
W-algebra bootstrap problem
for $\WW$.
In particular,
in any given commutator
$[W_{p_1}, W_{p_2}]$,
only a finite number of terms
appear on the RHS.

Due to $\mathfrak{psl}(2|2)$
covariance,
all (anti)commutators of the form
$[\mathbb W_{p_1}, \mathbb W_{p_2}]$
can be reconstructed from knowledge
of the commutator
$[W_{p_1}, W_{p_2}]$.
Moreover, this commutator receives contributions organized in terms of $\mathfrak{psl}(2|2)$ primaries
and their descendants.
The possible $\mathfrak{psl}(2|2)$ primaries we can encounter on the RHS
of $[W_{p_1}, W_{p_2}]$
are $J$ (if $p_1=p_2$)
and $W_q$. We also encounter the $\mathfrak{psl}(2|2)$ descendants $T$ and $T_{W_q}$.
Schematically, 
\be \label{eq_WW_comm_shorthand}
[W_{p_1} , W_{p_2}]
 =  \delta_{p_1,p_2} \gamma_{p_1}    \Big( J 
- \tfrac 16
T \Big)
+ \sum_q 
\kappa_{p_1 p_2}{}^{q}
\Big(
W_q -\tfrac{\left(p_1-p_2+q\right) \left(-p_1+p_2+q\right)}{4 (q-1) q (q+1)}
T_{W_q}
 \Big) \ .
 \ee 
The range of the summation over $q$ is 
\be 
\max(3, |p_1-p_2|+2) \le q \le p_1+p_2-2 \ , 
\qquad 
q - p_1 -p_2 \in 2 \mathbb Z \ . 
\ee 
In \eqref{eq_WW_comm_shorthand} we are using a shorthand notation.
We are omitting the Laurent mode label $m$ on $(W_p)_m$, $J_m$, and so on, as well as
all dependence on the auxiliary
coordinate $y$.
The relative factors between $T$ and $J$,
and between $T_{W_q}$
and $W_q$
are fixed by 
$\mathfrak{psl}(2|2)$ covariance.
We refer the reader to Appendix
\ref{app_wedge} for further details
on the compact notation used in
\eqref{eq_WW_comm_shorthand}.

The quantities $\gamma_{p}$,
$\kappa_{p_1 p_2}{}^q$ are 
the only piece of information 
in \eqref{eq_WW_comm_shorthand}  that   is not fixed by symmetry considerations.
The aim of the Lie algebra bootstrap is precisely to determine $\gamma_p$,
$\kappa_{p_1 p_2}{}^q$.
They turn out to be given by very simple expressions:
\be
\label{eq_wedge_algebra_coeffs}
\gamma_p = 2p^2 \ , \qquad 
\kappa_{p_1 p_2}{}^q = p_1 p_2 \ . 
\ee
In writing these expressions,
we have implicitly made a convenient choice of
normalization for the  $W_p$
generators.

We have checked explicitly that,
if we assume \eqref{eq_WW_comm_shorthand},
\eqref{eq_wedge_algebra_coeffs},
then all
(super) Jacobi identities 
of the form
$[\mathbb W_{p_1},[ \mathbb W_{p_2} , \mathbb W_{p_3}]] \pm \text{cyclic}=0$
with $p_1+p_2+p_3 \le 14$
are satisfied.

% Based on this evidence,
% we conjecture that
% \eqref{eq_wedge_algebra}, \eqref{eq_wedge_algebra_coeffs}
% specify a well-defined infinite
% dimensional   Lie superalgebra with
% $\mathfrak{psl}(2|2)$ symmetry.
% Given that all (anti)commutators
% are known in closed form
% (see appendix \ref{app_wedge}),
% it might be possible to
% furnish a general proof that
% all Jacobi identities hold, but we leave this endeavor to future work.

\subsubsection*{Some checks against the OPE coefficients of $\WW$}

Having determined the structure and properties of
$\hs$, we now proceed to 
make a stronger case
that 
$\hs \cong \text{Wedge}(\WW)$, by examining some
of the OPE coefficients
of $\WW$
in the large $c$ regime.
The match between
 $\hs$ and $\text{Wedge}(\WW)$ is encoded in the relations
\be \label{eq_hs_vs_wedge}
\begin{array}{rcl}
\text{$\hs$ side} &&
\text{$\WW$ side} \\
\hline
\gamma_p & = & 
\displaystyle
\lim_{c\rightarrow \infty} \bigg(
-\frac{6p}{c} g_p \bigg)
\rule[-6mm]{0mm}{14mm}
\\[2mm] 
\kappa_{p_1 p_2}{}^q & = &
\displaystyle
\lim_{c\rightarrow \infty}
c_{p_1 p_2}{}^q
\end{array}
\ee 
The appearance of the quantity
$- 6p g_p/c$ in the first relation is motivated as follows.
 The
contribution of the identity operator $\one$
and its super Virasoro descendants to the OPE
$W_p \times W_p$ takes the form
\cite{Bonetti:2018fqz}
  \be \label{eq_block_of_one}
W_p \times W_p \supset  
g_p \bigg[ 
\one -\tfrac{6p}{c} (J -\tfrac{1}{6} T)
+\tfrac{18 p(p-1)}{c(c-6)}\, (JJ)^2_0
-\tfrac{9p(p+1)}{c(c+9)}\,((JJ)^0_0+ \tfrac 13 T)+
\dots
\bigg] \ . 
  \ee  
In particular, we 
observe that $-6p g_p/c$
is the coefficient in front of $J$.

We recall that the normalization of the generators of the W-algebra is not fixed a priori.
In writing 
\eqref{eq_hs_vs_wedge} we assume that a normalization is used,
such that the limits exist finite. In other words, the limit of large central charge
is understood as
\be \label{eq_large_c_limit}
c \rightarrow \infty \ , \qquad 
g_p \rightarrow \infty \ , \qquad 
\frac{g_p}{c} \text{ fixed} \ . 
\ee

Luckily, the normalization 
\eqref{eq_best_normalization} that we have used in Section 
\ref{sec_results} does indeed
possess this property.
In fact, it yields a perfect match with 
the $\hs$ coefficients
\eqref{eq_wedge_algebra_coeffs}.
Recalling $c=3(1-\nu)$, we have
\be 
\lim_{c \rightarrow \infty}
\left( - \frac{6p}{c}g_p\right)
= \lim_{\nu \rightarrow \infty}
\left(  \frac{2p}{\nu-1} g_p \right)
= \lim_{\nu \rightarrow \infty}
\frac{2p^2 \nu}{\nu -1}
   \frac{\prod_{r=1}^{p-1} (\nu-r^2)}{S_p(\nu) }
\frac{\nu^{\lfloor \frac{p-2}{2} \rfloor}}{\nu^{p-1}} 
= 2p^2 \ . 
\ee 
We have already pointed out in 
\eqref{eq_large_nu_facts} that 
\be 
\lim_{\nu \rightarrow \infty}
c_{p_1 p_2}{}^q = p_1 p_2
\ .
\ee 
By comparing these results
with the values of $\gamma_p$, $\kappa_{p_1 p_2}{}^q$
reported in \eqref{eq_wedge_algebra_coeffs},
we confirm the validity of \eqref{eq_hs_vs_wedge}.

In addition to \eqref{eq_hs_vs_wedge}, we have verified explicitly in a few examples that, in the large $c$ limit,
all composite operators
are suppressed in all $W_{p_1} \times W_{p_2}$ OPEs. 
For instance,
this phenomenon is clearly visible in
\eqref{eq_block_of_one}:
inside the square brackets,
the composite super Virasoro descendants of $\one$
are suppressed by at least one additional power of $c$,
compared to the non-composite
descendants $J$, $T$.

Finally, let us comment on the
interplay between field redefinitions
of the form 
\eqref{eq_redefiningW6W7},
and the wedge algebra limit.
To this end, let us focus for definiteness
on the redefinition \eqref{eq_redefiningW6W7} of $W_6$,
repeated here for convenience,
\be \label{eq_W6_redef_again}
W_6' = W_6 + \mu_6  \, 
\cC_{3,3}^{W_3 W_3} \ .
\ee 
As reported in Appendix
\ref{app_redef},  
\eqref{eq_W6_redef_again} has the effect of
modifying some of the OPE
coefficients of the form $c_{p6}{}^q$.
If we work in the normalization 
\eqref{eq_best_normalization},
\be 
\ba 
(c_{36}{}^{3})'   & = 
c_{36}{}^{3}
+ \frac{6   (\nu -1) \left(\nu ^2+15 \nu +8\right)}{(\nu +3) (\nu +7)} 
\mu_6 \ , 
   \\
(c_{36}{}^{5})'  & = c_{36}{}^{5}
+ \frac{54  (\nu -1)}{\nu +7}
\mu_6 \ , 
   \\
(c_{36}{}^{7})'    & = c_{36}{}^{7} \ ,  
   \\
   (  c_{46}{}^4)'
   & = 
   c_{46}{}^4
+ \frac{144   (\nu -4) (\nu -1) (\nu +1)}{\nu  (\nu +3) (\nu +7)}
\mu_6
    \ , 
   \\ 
   (  c_{46}{}^6)'
   & = 
   c_{46}{}^6
  + \frac{108   (\nu -1)}{\nu +7}
  \mu_6
    \ ,  
   \\
   (c_{46}{}^8)'
   & = 
   c_{46}{}^8 \ . 
\ea 
\ee 
We are interested in the limit of large $\nu$.
We want to avoid divergences
in the above transformations in this limit.
The relation for $(c_{36}{}^3)'$
shows that we must take $\mu_6$ to zero,
scaling as $1/\nu$. With this scaling,
we observe that all other transformations become trivial in the limit
$\nu \rightarrow \infty$, e.g.~$(c_{36}{}^5)' = c_{36}{}^5$, and so on. The only OPE coefficient in the above list with a non-trivial
transformation in the large-$\nu$ limit is therefore
$c_{36}{}^3$, but
it does not enter the wedge algebra
(anti)commutators
(because we only have those $c_{p_1 p_2}{}^q$
where $q \ge |p_1-p_2|+2$).

Similar remarks apply to the redefinition
of $W_7$ as in \eqref{eq_redefiningW6W7}. The results of Appendix \ref{app_redef} show that,
in order to have finite expressions in the limit
\eqref{eq_large_c_limit}, we must take the 
parameter $\mu_7$ to zero, scaling like $1/\nu$.
Then, in the limit \eqref{eq_large_c_limit}
all OPE coefficients of the form
$c_{p7}{}^q$ become invariant, with the exception of $c_{37}{}^4$
(we only consider here the OPE coefficients that appear in $W_{p_1}\times W_{p_2}$
with $p_1+p_2\le 10$, as  in Table \ref{tab_ansatz}). The coefficient $c_{37}{}^4$, however,
does not appear in the wedge algebra,
since it does not satisfy
$q \ge |p_1 - p_2|+2$.

In conclusion,
even though in the full W-algebra
we have the freedom to perform
redefinitions such as
\eqref{eq_W6_redef},  \eqref{eq_W7_redef},
in the limit \eqref{eq_large_c_limit}
we can nonetheless 
 extract unambiguously the relevant OPE
coefficients $c_{p_1 p_2}{}^q$
and match them with  \eqref{eq_wedge_algebra_coeffs}.

\section{Outlook}
\label{sec_outlook}

There are several possible directions for future investigation. We discuss some of them below.

In this paper we 
have investigated the universal W-algebra for the VOAs $\cV(A_{N-1})$
associated to 
4d $\cN = 4$ $\mathfrak{su}(N)$ super Yang-Mills,
or alternatively to the $A$-series of Weyl groups
$A_{N-1} = \text{Weyl}(\mathfrak{a}_{N-1})$.
A natural direction for the future is the study
of the series
$\text{Weyl}(\mathfrak{b}_{N}) \cong
\text{Weyl}(\mathfrak{c}_{N})$,
and possibly $\text{Weyl}(\mathfrak{d}_{N})$.
A further generalization is provided by VOAs
associated to complex reflection groups,
which have $\cN = 2$ supersymmetry \cite{Bonetti:2018fqz}.
It would be interesting to explore universal W-algebra for series of complex reflection groups
with unbounded rank.
For instance, table 1 of \cite{Bonetti:2018fqz}
contains several examples of series of crystallographic
complex reflections groups, many of which correspond to known
4d $\cN = 3$ SCFTs.
Similarly, one could consider 
a universal vertex operator algebra for the 
 rank $N$ 
Deligne-Cvitanovi\'c theories, building on \cite{Beem:2019snk}.
Analyzing those examples could provide generalizations of the holographic dual of the VOA associated to 4d $\cN = 4$ 
$\mathfrak{su}(N)$
super Yang-Mills
to setups with lower supersymmetry.

Moreover, it would be interesting to study  
modules and characters
of the W-algebra $\WW$.
For instance, one may wonder
if characters of 
 $\WW$ exhibit 
properties analogous to the rich
structures found
in the non-supersymmetric setting in characters of
$\cW_\infty[\mu]$
and of the related W-algebra
$\cW_{1+\infty}$,
see e.g.~\cite{Gaberdiel:2012ku, Prochazka:2015deb}.
Moreover, the study of modules of
$\WW$
might provide inroads into
studying a class of supersymmetric surface defects in  4d $\mathcal \cN =4$
super Yang-Mills, building on the SCFT/VOA correspondence of \cite{Beem:2013sza} and on 
\cite{Cordova:2017mhb,
Pan:2017zie}.

In this work we have emphasized the role of the small $\cN = 4$ super Virasoro symmetry and we have worked in a basis of super Virasoro primary operators. It would be interesting to explore whether relaxing this condition can yield to a different basis of generators, in which the structure of OPEs simplifies,
along the lines of the quadratic basis of 
$\cW_\infty[\mu]$ and $\cW_{1 + \infty}$ \cite{Figueroa-OFarrill:1992uuf,Khesin:1994ey,
Lukyanov,  Prochazka:2014gqa}.

% Another interesting problem  is
% to analyze possible
% truncations of
% $\WW$.
% This might include
% truncations to
% known W-algebras with infinitely many strong generators and 
%  lower supersymmetry,
% but also truncations
% to W-algebras with finitely many strong generators. 
% More broadly, 
% the OPE coefficients in tables 
% \ref{tab_OPEcoeffs_part1}-\ref{tab_OPEcoeffs_part9}
% are rational functions of
% the central charge $c$
% (recall $c=3(1-\nu)$),
% and it would be interesting to explore the significance of
% the zeros in their denominators.

The notion of wedge algebra allows us
to start from a W-algebra,
and extract a Lie (super)algebra.
Under favorable circumstances, this 
procedure can be inverted:
the W-algebra can be recovered from the wedge algebra by means of a 
Drinfeld-Sokolov reduction 
\cite{Drinfeld:1984qv,Balog:1990dq,Bouwknegt:1992wg}.
It would be intersting to epxlore further
if a strategy based on 
Drinfeld-Sokolov reduction  could be useful is reconstructing 
the non-linear W-algebra 
$\WW$ from the Lie (super)algebra
$\hs$.

Finally, 
the universal W-algebra $\WW$ can be 
regarded   as
an analytic continuation
of $\cV(A_{N-1})$,
in which the discrete label $N$ is replaced by the
continuum label $c$,
see \cite{Zeng:2025pss} for related work.
From this point of view,
we can say that the W-algebra
$\WW$
captures an analytic
continuation
of the protected Schur sector
of 4d $\cN = 4$ super Yang-Mills. 
Similar continuations have been studied in \cite{Binder:2019zqc}.
This point of view can be useful in the study of the properties
of 4d protected operators
in the large $N$ limit.
Indeed, the relevance
of the universal W-algebra
of $\cV(A_{N-1})$ for
the study of $\tfrac 18$-BPS
operators in 4d $\cN = 4$ super Yang-Mills has  been anticipated e.g.~in~\cite{Chang:2023ywj}.

The 
holographic dual of the
VOA associated to
4d $\cN = 4$ super Yang-Mills
with gauge algebra $\mathfrak{su}(N)$
has been studied in \cite{Bonetti:2016nma}
with supersymmetric localization methods.
It has also been analyzed
in \cite{Costello:2018zrm} within the 
twisted holography program.
In \cite{Bonetti:2016nma}
it was argued that
the holographic dual
of the VOA associated to
4d $\cN = 4$ $\mathfrak{su}(N)$
super Yang-Mills
is a higher-spin Chern-Simons theory in $AdS_3$, 
whose gauge algebra consists of two copies (left-moving and right-moving)
of a suitable infinite-dimensional Lie superalgebra.
The latter
is the Lie superalgebra $\mathfrak{a}_\infty$
of \cite{Costello:2018zrm},
which we have recovered in this work with the name
$\hs$.
Our explicit construction of
the universal W-algebra $\WW$ (at least
for low-lying operators)
completes this circle of
ideas and might facilitate further investigations.

\section*{Acknowledgements}
We are grateful to
Christopher Beem,
Tom\'a\v{s} Proch\'azka,
Jingxiang Wu
for useful discussions
and correspondence. 
FB is grateful to the organizers of the Pollica Summer Workshop supported by the Regione Campania, Universit\`a degli Studi di Salerno, Universit\`a degli Studi di Napoli ``Federico II'', the Physics Department ``Ettore Pancini'' and ``E.R. Caianiello'', and Istituto Nazionale di Fisica Nucleare.
FB is supported by 
the Program ``Saavedra Fajardo'' 22400/SF/23.
 The work of CM is supported in part by the Italian Ministero
dell’Universit`a e della Ricerca (MIUR), and by Istituto Nazionale di Fisica Nucleare (INFN) through
the “Gauge and String Theory” (GAST) research project.

\newpage

\newpage

\appendix
\addtocontents{toc}{\protect\setcounter{tocdepth}{1}}

\section{Notation, conventions, useful formulae}

\subsection{Notation and conventions for OPEs}
\label{app_notation}

In this appendix we summarize
conventions and useful formulae
for OPEs and supersymmetry multiplets.

\subsubsection{OPEs of the small
$\mathcal N = 4$ super Virasoro algebra}
The OPEs
among the generators
$T(z)$, $J(z,y)$, $G(z,y)$,
$\widetilde G(z,y)$ 
of the small
$\mathcal N = 4$ super Virasoro algebra
read 
\small
\be \label{eq_sVir}
\ba 
J(z_1,y_1) J(z_2,y_2) & =
\frac{-k y_{12}^2 \one}{z_{12}^2}  
+ \frac{2y_{12}
\big(
1 + \tfrac 12 y_{12} \partial_{y_2}
\big)
J(z_2,y_2)
}{z_{12}}   + \text{reg.}
\ , \\
T(z_1) J(z_2,y_2)
& = \frac{J(z_2,y_2)}{z_{12}^2}
+ \frac{\partial_{z_2} J(z_2,y_2)}{z_{12}}
+ \text{reg.} \ , 
\\
T(z_1) T(z_2)
& = 
\frac{\tfrac 12 c \one }{z_{12}^4} 
+ \frac{2 T(z_2)}{z_{12}^2}
+ \frac{\partial_{z_2} T(z_2)}{z_{12}}
+ \text{reg.}
 \ , 
\\
T(z_1) G(z_2,y_2)
& = 
\frac{\frac 32 G(z_2,y_2)}{z_{12}^2}
+ \frac{\partial_{z_2} G(z_2,y_2)}{z_{12}}
+ \text{reg.} \ , 
\\
T(z_1) \widetilde G(z_2,y_2)
& =
\frac{\frac 32 \widetilde G(z_2,y_2)}{z_{12}^2}
+ \frac{\partial_{z_2} \widetilde G(z_2,y_2)}{z_{12}}
+ \text{reg.}  \ , 
\\
J(z_1,y_1) G(z_2,y_2)
& = \frac{y_{12}\big(
1 +  y_{12} \partial_{y_2}
\big)
G(z_2,y_2) }{z_{12}}
+ \text{reg.} \ , \\
J(z_1,y_1) \widetilde G(z_2,y_2)
& = \frac{y_{12}\big(
1 +  y_{12} \partial_{y_2}
\big)
\widetilde G(z_2,y_2) }{z_{12}}
+ \text{reg.} 
\ , \\
G(z_1,y_1) G(z_2,y_2)
& = \text{reg.} \ , \\
\widetilde G(z_1,y_1) \widetilde G(z_2,y_2)
& = \text{reg.} \ , \\
G(z_1,y_1) \widetilde G(z_2,y_2)
& =  \frac{-2ky_{12}\one }{z_{12}^3}  
+ \frac{2\big( 1 + \tfrac 12 y_{12}
\partial_{y_2}
\big)
J(z_2,y_2)}{z_{12}^2}
+ \frac{ \big( 1 + \tfrac 12 y_{12}
\partial_{y_2}
\big)
\partial_{z_2}J(z_2,y_2)
- y_{12}T(z_2)}{z_{12}}
 +\text{reg.} \ .
\ea
\ee
\normalsize
We have introduced the notation
$z_{12} = z_1 - z_2$,
$y_{12} = y_1 - y_2$.
The central charge $c$ is given in terms of the
affine Kac-Moody level $k$
as
\be 
c = 6k \ . 
\ee 

\subsubsection{Notation for operators on the RHS of an OPE}

Let $A(z)$, $B(z)$ be two
operators. (Any possible dependence on the  auxiliary variable $y$ is suppressed for simplicity.)
We parametrize the OPE of $A\times B$
as
\be
A(z_1) B(z_2) = \sum_{n \in \mathbb Z} \frac{1}{z_{12}^n} \{ AB \}_n(z_2) \ , 
\ee 
where $z_{12} = z_1-z_2$.
The operator $\{ AB \}_n(z)$
is identically zero for $n$ sufficiently
large (the OPE contains only a finite number of singular terms).
The operator $\{AB\}_0(z)$
is the conformal normal ordered product
of $A$ and $B$. 
The operators $\{ AB\}_n(z)$ with
$n <0$ can all be expressed
as a normal ordered product
of derivatives of $A$ and  $B$
(see e.g.~\cite{Thielemans:1994er}),
\be 
\{ AB \}_{-n}(z) = \frac{1}{n!}
\{ \partial_z^n A \, B \}_0(z) \ , \qquad 
n >0 \ . 
\ee

% We introduce
% a special notation for it,
% \be 
% {\rm NO}[A,B](z) := \{ AB \}_0(z) \ . 
% \ee 

\subsubsection{$\mathfrak{sl}(2)_z \oplus \mathfrak{sl}(2)_y$ primary operators
and their OPEs}

The global part of the
small $\cN = 4$ superconformal
algebra is $\mathfrak{psl}(2|2)$,
whose bosonic subalgebra is
$\mathfrak{sl}(2)_z \oplus \mathfrak{sl}(2)_y$.
We describe operators that
transform as primaries under
$\mathfrak{sl}(2)_z \oplus \mathfrak{sl}(2)_y$
as functions of the variable $z$
(coordinate on the Riemann sphere)
and of an auxiliary variable $y$.
More precisely,
let $A(z,y)$
be an $\mathfrak{sl}(2)_z \oplus \mathfrak{sl}(2)_y$
primary of conformal weight
$h_A$ and spin $j_A \in \{ 0, \tfrac 12, 1, \tfrac 32, \dots\}$.
Then $A(z,y)$
is a polynomial of degree $2j_A$ in $y$.
If $T$ denotes the stress energy tensor and $J$ to affine Kac-Moody $\mathfrak{sl}(2)_y$
current, the OPEs of $T$ and $J$ with $A$
can be parametrized as
\be 
\ba 
T(z_1) A(z_2, y_2) & =\sum_{n \in \mathbb Z}
\frac{1}{z_{12}^n} \{ T A(y_2) \}_n(z_2) \ , \\
J(z_1, y_1) A(z_2,y_2) & = \sum_{n\in \mathbb Z}
\frac{1}{z_{12}^n} \{ J(y_1) A(y_2)\}_n(z_2) \ .
\ea
\ee 
The conditions that $A$ be an 
$\mathfrak{sl}(2)_z \oplus \mathfrak{sl}(2)_y$
primary
of weight $h_A$ and spin $j_A$ are
\be 
\ba 
\{ T A(y_2) \}_3(z_2) & = 0  \ , \\
\{ T A(y_2) \}_2(z_2) &= h_A A(z_2,y_2) \ , \\ 
\{ T A(y_2) \}_1(z_2) &= \partial_{z_2}A(z_2,y_2)  \ , \\ 
\{ J(y_1) A(y_2) \}_1(z_2)  &=
y_{12}(2 j_A + y_{12} \partial_{y_2}) A(z_2,y_2) \ .
\ea 
\ee

% \begin{gather}
% \{ T A(y_2) \}_3(z_2) = 0 \ , \quad 
% \{ T A(y_2) \}_2(z_2) = h_A A(z_2,y_2) \ , \quad
% \{ T A(y_2) \}_1(z_2) = \partial_{z_2}A(z_2,y_2)  \ , \nn \\ 
% \{ J(y_1) A(y_2) \}_1(z_2)  =
% y_{12}(2 j_A + y_{12} \partial_{y_2}) A(z_2,y_2) \ . 
% \end{gather}

Let $A$, $B$ be
$\mathfrak{sl}(2)_z \oplus \mathfrak{sl}(2)_y$ primary
operators with
conformal weights and spins
$(h_A, j_A)$, $(h_B, j_B)$, respectively.
The OPE $A \times B$ can be organized
into contributions of
$\mathfrak{sl}(2)_z \oplus \mathfrak{sl}(2)_y$ primary operators and their descendants,
\be \label{eq_covariant_OPE}
A(z_1, y_1) B(z_2,y_2) = \sum_C 
c_{AB}{}^C \frac{y_{12}^{j_A + j_B - j_C}}{z_{12}^{h_A + h_B - h_C}}
\cD_{h_A, h_B; h_C}(z_{12},\partial_{z_2})
\widehat \cD_{j_A, j_B;j_C}(y_{12},\partial_{y_2}) C(z_2, y_2) \ . 
\ee 
The sum runs over
$\mathfrak{sl}(2)_z \oplus \mathfrak{sl}(2)_y$ primary
operators $C$ of definite 
conformal weight $h_C$ and spin $j_C$. 
The constant $c_{AB}{}^C$
is the OPE coefficient.
The standard $\mathfrak{sl}(2)_y$ selection rules imply that 
$c_{AB}{}^C$ can only be non-zero for
\be 
j_C \in \{ |j_A - j_B| , 
|j_A - j_B| + 1 , \dots , j_A + j_B 
\} \ . 
\ee 
The differential operators 
$\cD_{h_A, h_B; h_C}(z_{12},\partial_{z_2})$
and 
$\widehat \cD_{j_A, j_B;j_C}(y_{12},\partial_{y_2})$
are defined as follows
(see e.g.~\cite{Thielemans:1994er})
\be \label{eq_sl2_diff_ops}
\ba 
\cD_{h_A, h_B; h_C}(z_{12},\partial_{z_2})
& = \sum_{k=0}^\infty 
\frac{(h_C + h_A - h_B)_k}{k! (2h_C)_k} z_{12}^k \partial_{z_2}^k \ , \\ 
\widehat \cD_{j_A, j_B;j_C}(y_{12},\partial_{y_2})
& = \sum_{k=0}^\infty 
\frac{(-j_C - j_A + j_B)_k}{k! (-2j_C)_k} z_{12}^k \partial_{z_2}^k
 \ , 
\ea 
\ee
where $(x)_k$ is the ascending
Pochhammer symbol,
\be 
(x)_k = \prod_{i=0}^{k-1}(x+i) \ ,
\qquad (x)_0 = 1 \ . 
\ee 

Thanks to \eqref{eq_covariant_OPE},
we can adopt a compact 
notation for OPEs, in which
the operators on the LHS
are the $\mathfrak{sl}(2)_z \oplus \mathfrak{sl}(2)_y$ primary operators $A$ and $B$,
and on the RHS
we simply list
the $\mathfrak{sl}(2)_z \oplus \mathfrak{sl}(2)_y$ primary operators $C$ that contribute to
the singular part of the $A \times B$ OPE, with their OPE
coefficients
(omitting the $y_{12}$, $z_{12}$
powers and the differential
operators $\cD_{h_A, h_B;h_C}$,
$\widehat \cD_{j_A, j_B;j_C}$).
As an example,
the OPEs of the small
$\cN = 4$ superconformal
algebra in compact notation
read
\be \label{eq_SCA_compact}
\ba 
J \times J & = - k \one +  2J \ , &
T \times T & = \tfrac c2 \one + 2T \ , & 
T \times J & = J \ , \\
J \times G & = G \ , & 
J \times \widetilde G & = \widetilde G \ , \\
T \times G & = \tfrac 32 G \ , &
T \times \widetilde G & = \tfrac 32 \widetilde G \ , \\ 
G \times G & = 0 \ , & 
\widetilde G \times \widetilde G & = 0 \ , & 
G \times \widetilde G
& = - 2 k \one + 2 J - T \ .
\ea 
\ee 
where $c = 6k$, cfr.~the explicit
expressions \eqref{eq_sVir}.

\subsubsection{The $\mathfrak{sl}(2)_z \oplus \mathfrak{sl}(2)_y$
primary operators  
$(AB)^j_n$
}
Suppose $A$, $B$
are $\mathfrak{sl}(2)_z \oplus \mathfrak{sl}(2)_y$ primary
operators with weights
and spins $(h_A,j_A)$,
$(h_B,j_B)$ respectively.
Recall that $A$ is a polynomial
in $y$ of degree $2j_A$, and similarly for $B$. Let us write
\be 
A(z,y) =\sum_{ \alpha =0}^{2j_A} A_\alpha (z) y^\alpha \ , \qquad 
B(z,y) =\sum_{ \alpha =0}^{2j_B} B_\alpha (z) y^\alpha  \ . 
\ee 
Let us fix an integer $n$
and a value 
\be 
j \in \{ |j_A - j_B| , |j_A - j_B| +1 , \dots , j_A + j_B \} \ . 
\ee 
We can construct
a new $\mathfrak{sl}(2)_z \oplus \mathfrak{sl}(2)_y$
primary operator
of weight $h_A + h_B -n$
and spin $j$ out of the quantities
$\{ A_\alpha B_\beta\}_n(z)$.
It is denoted $(AB)_n^j$
and is defined as follows,
\be \label{eq_def_slP}
(AB)_n^j(z,y) = \sum_{\alpha = 0}^{2j_A} \sum_{\beta = 0}^{2j_B}
\cC_{j_A, j_B, j, \alpha,\beta}
y^{j-j_A - j_B + \alpha + \beta }
\sum_{p=0}^\infty \cK_{h_A, h_B, n, p} \partial_z^p \{ A_\alpha B_\beta\}_{n+p}(z) \ ,
\ee 
where the coefficients 
$\cC_{j_A, j_B, j, \alpha,\beta}$,
$\cK_{h_A, h_B, n, p}$
are given as
\be 
\ba 
\cC_{j_A, j_B, j, \alpha, \beta}
& = \sum_{\ell = 0}^{j_A + j_B - j}
\frac{(-1)^\ell (j_A - j_B + j - \ell )^\downarrow_\ell }{\ell! 
(j_A + j_B - j -\ell)!
(2j + \ell +1)^\downarrow_\ell
}
\sum_{s=0}^\ell \binom{\ell}{s}
(\alpha)^\downarrow_{j_A + j_B - j - s} (\beta)^\downarrow_s 
 \ , \\
\cK_{h_A, h_B, n, p}
& = \frac{(-1)^p  (2h_A - n -p)_p}{p! (2h_A + 2h_B - 2n - p - 1)_p} \ . 
\ea 
\ee 
We have introduced the descending Pochhammer symbol
\be 
(x)^\downarrow_n = \prod_{i=0}^{n-1} (x-i) \ , \qquad 
(x)^\downarrow_0 = 1
\ . 
\ee 
The overall normalization of
$(AB)^j_n(z,y)$ has been chosen in such a way that
the $A\times B$ OPE
takes the form 
\be 
A(z_1, y_1) B(z_2,y_2) = \sum_{n,j} 
 \frac{y_{12}^{j_A + j_B - j}}{z_{12}^{n}}
\cD_{h_A, h_B; h_A + h_B -n}(z_{12},\partial_{z_2})
\widehat \cD_{j_A, j_B;j}(y_{12},\partial_{y_2}) 
(AB)^j_n(z_2,y_2)
\ . 
\ee

\subsubsection{The operations
$G^\uparrow$, $G^\downarrow$,
$\widetilde G^\uparrow$, $\widetilde G^\downarrow$}

Let $A$ be an $\mathfrak{sl}(2)_z \oplus \mathfrak{sl}(2)_y$
primary operator of weight $h$
and spin $j$.
We introduce the following notation,
\be 
\ba
(G^\uparrow A)(z,y) &= (GA)^{j + \frac 12}_{1}(z,y) \ , & 
(G^\downarrow A)(z,y) &= (GA)^{j - \frac 12}_{1}(z,y) \ , \\
(\widetilde G^\uparrow A)(z,y) &= (\widetilde GA)^{j + \frac 12}_{1}(z,y) \ , & 
(\widetilde G^\downarrow A)(z,y) &= (\widetilde GA)^{j - \frac 12}_{1}(z,y) \ .
\ea 
\ee 
In other words,
$G^\uparrow A$ is the 
$\mathfrak{sl}(2)_z \oplus \mathfrak{sl}(2)_y$
primary operator that appears
in the first order pole in the
OPE $G \times A$, and similarly for $G^\downarrow$, $\widetilde G^\uparrow$, $\widetilde G^\downarrow$.

\subsubsection{$\mathfrak{psl}(2|2)$
primary operators
and $\mathfrak{psl}(2|2)$ multiplets}

An operator $X$ is a
$\mathfrak{psl}(2|2)$ primary operator
of weight $h_X$ and spin $j_X$ 
if
\be 
\ba 
\{ T X(y_2) \}_3(z_2) & = 0 \ , \\ 
\{ T X(y_2) \}_2(z_2) &= h_X X(z_2,y_2) \ , \\
\{ T X(y_2) \}_1(z_2) &= \partial_{z_2}X(z_2,y_2)  \ ,   \\ 
\{ J(y_1) X(y_2) \}_1(z_2)  &=
y_{12}(2 j_X + y_{12} \partial_{y_2}) X(z_2,y_2) \  , \\ 
\{ G(y_1) X(y_2) \}_2(z_2) &= 0 \ , \\
\{ \widetilde G(y_1) X(y_2) \}_2(z_2) &= 0 \ . 
\ea  
\ee 
% \begin{gather}
% \{ T X(y_2) \}_3(z_2) = 0 \ , \quad 
% \{ T X(y_2) \}_2(z_2) = h_X X(z_2,y_2) \ , \quad
% \{ T X(y_2) \}_1(z_2) = \partial_{z_2}X(z_2,y_2)  \ , \nn \\ 
% \{ J(y_1) X(y_2) \}_1(z_2)  =
% y_{12}(2 j_X + y_{12} \partial_{y_2}) X(z_2,y_2) \  , \\ 
% \{ G(y_1) X(y_2) \}_2(z_2) = 0 \ , \quad 
% \{ \widetilde G(y_1) X(y_2) \}_2(z_2) = 0 
%  \ . \nn
% \end{gather}
In particular, $X$ is an 
 $\mathfrak{sl}(2)_z \oplus \mathfrak{sl}(2)_y$
primary operator.
The vanishing of the order-two
pole in the $G \times X$ OPE is
equivalent to demanding that $X$ be
annihilated by the $S$-type
supercharge associated to the current $G$, and similarly for
$\widetilde G$.

Starting from the $\mathfrak{psl}(2|2)$ primary operator $X$
we can act with the operations
$G^\uparrow$, $G^\downarrow$,
$\widetilde G^\uparrow$, $\widetilde G^\downarrow$
and generate the entire
$\mathfrak{psl}(2|2)$  multiplet
of $X$. 
In this work we encounter the following types of 
$\mathfrak{psl}(2|2)$  multiplets.

\begin{itemize}
\item Short multiplet: $h_X = j_X$,
$j_X \in \{ 1, \tfrac 32, 2, \dots \}$.
The $\mathfrak{psl}(2|2)$ 
multiplet of $X$ contains
four 
$\mathfrak{sl}(2)_z \oplus \mathfrak{sl}(2)_y$
primary operators,
with the following quantum numbers,
\begin{center}
\renewcommand{\arraystretch}{1.2}
\begin{tabular}{c | | c | c | c | c}
& $X$ & $G^\downarrow X \equiv G_X$ & $\widetilde G^\downarrow X \equiv \widetilde G_X$  &
$\widetilde G^\downarrow G^\downarrow X \equiv T_X$ \\
\hline
$h$ & $j_X$ & $j_X + \tfrac 12$ &
$j_X + \tfrac 12$ & 
$j_X + 1 $\\
$j$ & $j_X$ & $j_X - \tfrac 12$ & 
$j_X - \tfrac 12$ & 
$j_X -1 $\\
$r$ & $r_X$ & $r_X + \tfrac 12$ & 
$r_X - \tfrac 12 $ & 
$r_X$
\end{tabular} 
\end{center}
We have included the charge $r$ under the $\mathfrak{u}(1)_r$ Cartan
generator of the $\mathfrak{sl}(2)_{\rm out}$ outer automorphism
of the small $\cN = 4$ superconformal algebra.
We assume that the 
 $\mathfrak{psl}(2|2)$ 
primary operator $X$ has definite $r$-charge $r_X$.

\item Long multiplet: $h_X > j_X$,
$j_X \in \{ 1, \tfrac 32, 2, \dots\}$.
This multiplet contains 16 
$\mathfrak{sl}(2)_z \oplus \mathfrak{sl}(2)_y$
primary operators,
with the following quantum numbers,
\\[4mm]
\small 
{
\hspace*{1cm}
\renewcommand{\arraystretch}{1.2}
\begin{tabular}{c | | c | c | c | c | c}
& 
$X$ & 
$
\begin{array}{c}
G^\downarrow X
\\[-2mm]
\equiv X_{[\frac 12, - \frac 12, \frac 12]}
\end{array}$ & 
$
\begin{array}{c}
\widetilde G^\downarrow X
\\[-2mm]
\equiv 
X_{[\frac 12, - \frac 12, - \frac 12]}
\end{array}
$ &
$\begin{array}{c}
G^\uparrow X 
\\[-2mm]
\equiv X_{[\frac 12,  \frac 12 , \frac 12]}
\end{array}
$ & 
$\begin{array}{c}
\widetilde G^\uparrow X
\\[-2mm]
\equiv 
X_{[\frac 12,  \frac 12, - \frac 12]}
\end{array}$ 
 \\
\hline
$h$ & $h_X$ & $h_X + \tfrac 12$ &
$h_X + \tfrac 12$ & 
$h_X + \tfrac 12  $ & 
$h_X + \tfrac 12  $ 
\\
$j$ & $j_X$ & $j_X - \tfrac 12$ & 
$j_X - \tfrac 12$ & 
$j_X  + \tfrac 12$ & 
$j_X  + \tfrac 12$ 
\\
$r$ & $r_X$ & $r_X + \tfrac 12$ & 
$r_X - \tfrac 12 $ & 
$r_X + \tfrac 12 $ & 
$r_X - \tfrac 12 $ 
\end{tabular} 
}
\\[4mm]
{
\hspace*{1cm}
\renewcommand{\arraystretch}{1.2}
\begin{tabular}{c | | c | c | c | c | c | c }
& 
$\begin{array}{c}
G^\downarrow \widetilde G^\downarrow X 
\\[-2mm]
\equiv X_{[1,-1,0]}
\end{array}$ &
$\begin{array}{c}
G^\downarrow G^\uparrow X
\\[-2mm] \equiv 
X_{[1,0,1]}
\end{array}$ & 
$\begin{array}{c}
\widetilde G^\downarrow \widetilde G^\uparrow X
\\[-2mm]
\equiv 
X_{[1,0-1]}
\end{array}$ &
$\begin{array}{c}
G^\downarrow \widetilde G^\uparrow X
\\[-2mm]
\equiv X_{[1,0,0]}
\end{array}$ &
$\begin{array}{c}
\widetilde G^\downarrow G^\uparrow X
\\[-2mm]
\equiv X'_{[1,0,0]}
\end{array}$
&
$\begin{array}{c}
G^\uparrow \widetilde G^\uparrow X
\\[-2mm]
\equiv X_{[1,1,0]}
\end{array}
$
 \\
\hline
$h$ & 
$h_X + 1$ &
$h_X + 1$ &
$h_X + 1$ & 
$h_X +1$ & 
$h_X +1$ & 
$h_X + 1$
\\
$j$ &
$j_X  -1$ & 
$j_X$ &
$j_X$ & 
$j_X$ & 
$j_X$ & 
$j_X + 1$
\\
$r$ &  
$r_X$ & 
$r_X + 1$ & 
$r_X -1$ & 
$r_X$ & 
$r_X$ & 
$r_X$
\end{tabular} 
}
\\[4mm]
{
\hspace*{1cm}
\renewcommand{\arraystretch}{1.2}
\begin{tabular}{c | | c | c | c | c | c }
&
$\begin{array}{c}
G^\downarrow \widetilde G^\downarrow \widetilde G^\uparrow X
\\[-2mm]
\equiv X_{[\frac 32, - \frac 12, - \frac 12]}
\end{array}$
& 
$\begin{array}{c}
\widetilde G^\downarrow G^\downarrow G^\uparrow X
\\[-2mm]
\equiv X_{[\frac 32, - \frac 12 , \frac 12]}
\end{array}$
&
$\begin{array}{c}
G^\uparrow \widetilde G^\downarrow \widetilde G^\uparrow X
\\[-2mm]
\equiv X_{[\frac 32 , \frac 12 , - \frac 12]}
\end{array}$
& 
$\begin{array}{c}
\widetilde G^\uparrow G^\downarrow G^\uparrow X
\\[-2mm]
\equiv X_{[\frac 32 , \frac 12 , \frac 12]}
\end{array}$
&
$\begin{array}{c}
G^\downarrow \widetilde G^\downarrow
G^\uparrow \widetilde G^\uparrow X
\\[-2mm]
\equiv X_{[2,0,0]}
\end{array}$
 \\
\hline
$h$ & $h_X +\tfrac 32$ & 
$h_X +\tfrac 32$ & 
$h_X +\tfrac 32$ & 
$h_X +\tfrac 32$ & 
$h_X + 2$
\\
$j$ & $j_X - \tfrac 12$ & 
$j_X - \tfrac 12$ & 
$j_X + \tfrac 12$ & 
$j_X + \tfrac 12$ & 
$j_X$
\\
$r$ & $r_X - \tfrac 12$ &
$r_X + \tfrac 12$ & 
$r_X - \tfrac 12$ & 
$r_X + \tfrac 12$ & 
$r_X$
\end{tabular} 
}
\normalsize

We have introduced a shorthand notation $X_{[\Delta h, \Delta j, \Delta r]}$ for the
$\mathfrak{psl}(2|2)$ descendants of
$X$, based on the amounts $\Delta h$,
$\Delta j$, $\Delta r$ 
that their weights, spins, and $\mathfrak {u}(1)_r$ charges
differ from those of $X$.
Since $G^\downarrow \widetilde G^\uparrow X$ and 
$\widetilde G^\downarrow  G^\uparrow X$ share the same quantum numbers,
we use a prime to distiguish them.

\item Long multiplet: $h_X > j_X$,
$j_X = \tfrac 12 $.
This multiplet contains 15
$\mathfrak{sl}(2)_z \oplus \mathfrak{sl}(2)_y$
primary operators.
Compared to a long multiplet
with generic $j_X$, in the case
$j_X=\tfrac 12$ the operator
$G^\downarrow X \widetilde G^\downarrow X$
is absent (it would have spin $-\tfrac 12$).

\item Long multiplet: $h_X > j_X$,
$j_X =0$.
This multiplet contains 10
$\mathfrak{sl}(2)_z \oplus \mathfrak{sl}(2)_y$
primary operators.
Compared to a long multiplet
with generic $j_X$, in the case
$j_X=0$ the operators
$G^\downarrow X$,
$\widetilde G^\downarrow X$,
$G^\downarrow \widetilde G^\downarrow X$,
$G^\downarrow \widetilde G^\downarrow \widetilde G^\uparrow X$,
$\widetilde G^\downarrow  G^\downarrow  G^\uparrow X$
are absent (they would have negative spin).
Moreover, the two operators
$G^\downarrow \widetilde G^\uparrow X$
and $\widetilde G^\downarrow G^\uparrow X$
become equal,
$G^\downarrow \widetilde G^\uparrow X
=\widetilde G^\downarrow G^\uparrow X$.
\end{itemize}

\subsubsection{Super Virasoro primary
operators}

An operator $X$ is a
super Virasoro primary operator
of weight $h_X$ and spin $j_X$ if
\be
\ba 
\{ T X(y_2) \}_{n \ge 3}(z_2) &= 0 \ , 
\\
\{ T X(y_2) \}_2(z_2) & = h_X X(z_2,y_2) \ , \\
\{ T X(y_2) \}_1(z_2) &= \partial_{z_2}X(z_2,y_2)  \ ,  \\ 
\{ J(y_1) X(y_2) \}_{n \ge 2}(z_2) & = 0 \ , 
\\
\{ J(y_1) X(y_2) \}_1(z_2) & =
y_{12}(2 j_X + y_{12} \partial_{y_2}) X(z_2,y_2) \  , \\ 
\{ G(y_1) X(y_2) \}_{n \ge 2}(z_2) & = 0 \ , \\
\{ \widetilde G(y_1) X(y_2) \}_{n \ge 2}(z_2) &  = 0
\ea 
\ee 
% \begin{gather}
% \{ T X(y_2) \}_{n \ge 3}(z_2) = 0 \ , \quad 
% \{ T X(y_2) \}_2(z_2) = h_X X(z_2,y_2) \ , \quad
% \{ T X(y_2) \}_1(z_2) = \partial_{z_2}X(z_2,y_2)  \ , \nn \\ 
% \{ J(y_1) X(y_2) \}_{n \ge 2}(z_2) = 0 \ ,\quad 
% \{ J(y_1) X(y_2) \}_1(z_2)  =
% y_{12}(2 j_X + y_{12} \partial_{y_2}) X(z_2,y_2) \  , \\ 
% \{ G(y_1) X(y_2) \}_{n \ge 2}(z_2) = 0 \ , \quad 
% \{ \widetilde G(y_1) X(y_2) \}_{n \ge 2}(z_2) = 0 
%  \ . \nn
% \end{gather}
In particular, $X$ is a 
 $\mathfrak{psl}(2|2)$
primary operator.

\subsubsection{OPEs
$\mathbb J \times  \mathbb X$
for $X$ a short super Virasoro 
primary operator}

The generators of the superconformal
algebra form the short $\mathfrak{psl}(2|2)$
multiplet
\be 
\mathbb J = \{ J , G , \widetilde G, T \} \ . 
\ee 
Let $X$ be a super Virasoro primary operator
of weight and spin $h_X=j_X \in \{1, \tfrac 32, 2, \dots\}$.
We denote the short
$\mathfrak{psl}(2|2)$ multiplet of $X$ as
\be 
\mathbb X = \{ X, G_{X}, \widetilde G_{X}, T_{X}
\}  \ . 
\ee 
Let us record
all OPEs of the form
$\mathbb J \times \mathbb X$
using the same compact notation as in 
\eqref{eq_SCA_compact},
\be \label{eq_J_short_OPEs} 
\ba 
J \times X & = 2j_X X \ , 
&
J \times \widetilde G_X & =
(2j_X-1) \widetilde G_X
\ , 
\\
G \times X & = G_X   \ , 
&
 G \times \widetilde G_X & =
2j_X X - T_X 
\ , 
\\
\widetilde G \times X & =\widetilde G_X   \ , 
&
\widetilde G \times \widetilde G_X & = 0
\ , 
\\
T \times X & = 
j_X X\ , 
&
T \times \widetilde G_X & = 
\left( j_X + \frac 12 \right)
\widetilde G_X
\ , 
\\[2mm]
J \times G_X & = (2j_X-1) G_X \ , 
&
J \times  T_X & = 
(2j_X-1) X
+(2j_X-2) T_X
\ , 
\\
G \times G_X & = 0   \ , 
&
 G \times T_X & = 
 \left( 2j_X - \frac{1}{2j_X}\right) G_X
\ , 
\\
\widetilde G \times G_X & =
- 2 j_X X + T_X \ , 
&
\widetilde G \times  T_X & =
 \left( 2j_X - \frac{1}{2j_X}\right) \widetilde G_X
\ , 
\\
T \times G_X & =
\left( j_X + \frac 12
\right) G_X\ , 
&
T \times  T_X & = 
( j_X + 1) T_X
\ , 
\\
\ea 
\ee 
These OPEs are consistent with the vanishing of all
$\text{Jacobi}(\mathbb J , \mathbb J , \mathbb X)$ Jacobiators.

\subsubsection{OPEs
$\mathbb J \times  \mathbb X$
for $X$ a long super Virasoro 
primary operator}

Let $X$ be a super Virasoro primary operator
of weight $h_X$
and spin $j_X$,
with $h_X>j_X$ and $j_X \in \{1, \tfrac 32, 2, \dots\}$.
We denote the long
$\mathfrak{psl}(2|2)$ multiplet of $X$ as
\be 
\mathbb X = \{ X, 
X_{[\frac 12 , - \frac 12 , \frac 12]} ,
X_{[\frac 12 , - \frac 12 ,- \frac 12]}\dots ,
X_{[2,0,0]}
\}  \ . 
\ee 
Let us record
all $\mathbb J \times \mathbb X$ OPEs in compact form.
These OPEs have been obtained 
via trial and error
exploiting the vanishing of 
$\text{Jacobi}(\mathbb J , \mathbb J , \mathbb X)$.

The OPEs $\mathbb J \times \mathbb X$
in the case $j_X = \tfrac 12$ or $0$
are obtained with minimal
modifications, by removing
the appropriate descendants of $X$ that
are absent compared to the case
$j_X \ge 1$.

\begin{flalign}
J \times X & = 2 j_X X  \ , & \nonumber \\
G \times X & = X_{[\frac 12 , - \frac 12 , \frac 12]}
+ X_{[\frac 12 ,  \frac 12 , \frac 12]}  \ , 
\nonumber
\\
\widetilde G \times X & = X_{[\frac 12 , - \frac 12 , - \frac 12]}+X_{[\frac 12 , \frac 12 , - \frac 12]}  \ ,
\nonumber \\
T \times X & = h_X X  \ , 
\end{flalign}

\begin{flalign}
J \times X_{[\frac 12 , - \frac 12 , \frac 12]} & =
\textstyle
2  \left(j_X-\frac{1}{2}\right)  X_{[\frac 12 , - \frac 12 , \frac 12]} \ , & 
\nonumber  \\
G \times X_{[\frac 12 , - \frac 12 , \frac 12]} & = 
\textstyle
-\frac{2  j_X}{2 j_X+1} 
X_{[1,0,1]} \ , 
\nonumber 
\\
\widetilde G \times X_{[\frac 12 , - \frac 12 , \frac 12]} & = \textstyle
-   X_{[1,-1,0]}-X_{[1,0,0]}+\frac{1}{2 j_X+1} X'_{[1,1,0]} -\frac{2 j_X 
   (h_X+j_X+1)}{2 j_X+1} X \ , 
   \nonumber 
   \\
T \times X_{[\frac 12 , - \frac 12 , \frac 12]} & =  
 \textstyle
 \left(h_X+\frac{1}{2}\right) X_{[\frac 12 , - \frac 12 , \frac 12]}  \ ,  
\end{flalign}

\begin{flalign}
J \times X_{[\frac 12 , \frac 12 , \frac 12]} & =  \textstyle
2  \left(j_X+\frac{1}{2}\right) X_{[\frac 12 , \frac 12 , \frac 12]} \ , & 
\nonumber \\
G \times X_{[\frac 12 , \frac 12 , \frac 12]} & =
 \textstyle
 X_{[1,0,1]}  \ ,
 \nonumber
 \\
\widetilde G \times X_{[\frac 12 , \frac 12 , \frac 12]} & =
 \textstyle
 -X_{[1,1,0]}+X'_{[1,1,0]}+ (h_X-j_X) X \ ,  \nonumber
 \\
T \times X_{[\frac 12 , \frac 12 , \frac 12]} & =  
 \textstyle
 \left(h_X+\frac{1}{2}\right) X_{[\frac 12 , \frac 12 , \frac 12]}  \ , 
\end{flalign}

\begin{flalign}
J \times X_{[\frac 12 , - \frac 12 , - \frac 12]} & =
 \textstyle
 2 \left(j_X-\frac{1}{2}\right)
 X_{[\frac 12 , - \frac 12 , - \frac 12]} \ , & 
 \nonumber \\
G \times X_{[\frac 12 , - \frac 12 , - \frac 12]} & =
 \textstyle
 X_{[1, -1, 0]}+\frac{1}{2 j_X+1} X_{[1,0,0]} 
-X'_{[1,1,0]}+\frac{2 j_X 
   (h_X+j_X+1)}{2 j_X+1} X \ , 
    \nonumber \\
\widetilde G \times X_{[\frac 12 , - \frac 12 , - \frac 12]} & =
 \textstyle
 -\frac{2  j_X}{2 j_X+1} X_{[1,0,-1]} \ , 
  \nonumber
\\
T \times X_{[\frac 12 , - \frac 12 , - \frac 12]} & = 
 \textstyle
 \left(h_X+\frac{1}{2}\right) 
X_{[\frac 12 , - \frac 12 , - \frac 12]}  \ ,  
\end{flalign}

\begin{flalign}
J \times X_{[\frac 12 , \frac 12 , - \frac 12]} & = 
 \textstyle
 2 \left(j_X+\frac{1}{2}\right) 
X_{[\frac 12 , \frac 12 , - \frac 12]}  \ , & \nonumber   \\
G \times X_{[\frac 12 , \frac 12 , - \frac 12]} & = 
 \textstyle
 X_{[1,0,0]}+X_{[1,1,0]}+ (j_X-h_X) X  \ , 
 \nonumber 
 \\
\widetilde G \times X_{[\frac 12 , \frac 12 , - \frac 12]} & = 
 \textstyle
 X_{[1,0,-1]}  \ ,
 \nonumber 
 \\
T \times X_{[\frac 12 , \frac 12 , - \frac 12]} & =
 \textstyle
\left(h_X+\frac{1}{2}\right)
X_{[\frac 12 , \frac 12 , - \frac 12]} 
\end{flalign}

\begin{flalign}
J \times X_{[1,-1,0]}& = 
 \textstyle
2 (j_X-1) X_{[1,-1,0]}
-\frac{(2 j_X-1)  (h_X+j_X+1)}{2
   j_X+1} 
   X \ ,  & \nonumber
   \\
G \times X_{[1,-1,0]}& =  
 \textstyle
 -\frac{ (2 j_X-1) (h_X+j_X+1)}{2
   j_X}
   X_{[\frac 12 , - \frac 12 , \frac 12]}
   -\frac{ 2 j_X-1 }{2 j_X} X_{[\frac 32 , - \frac 12 , \frac 12]}  \ ,
   \nonumber
   \\
\widetilde G \times X_{[1,-1,0]}& =
 \textstyle
 \frac{  2 j_X-1 }{2 j_X} X_{[\frac 32 , - \frac 12 , -\frac 12]} 
-\frac{(2 j_X-1)
   (h_X+j_X+1)}{2 j_X} 
   X_{[\frac 12 , - \frac 12 , - \frac 12]}  \ , 
   \nonumber
   \\
T \times X_{[1,-1,0]}& = 
 \textstyle
 (h_X+1) X_{[1,-1,0]}  \ ,
\end{flalign}

\begin{flalign}
J \times X_{[1,0,1]} & =
 \textstyle
2  j_X X_{[1,0,1]}  \ , & 
\nonumber  \\
G \times X_{[1,0,1]} & =
 \textstyle
 0  \ , 
 \nonumber 
 \\
\widetilde G \times X_{[1,0,1]} & = 
 \textstyle
 (j_X-h_X)  X_{[\frac 12 , - \frac 12 , \frac 12]} 
-   (h_X+j_X+1) X_{[\frac 12 , \frac 12 , \frac 12]} +X_{[\frac 32 , - \frac 12 , \frac 12]}+X_{[\frac 32 , \frac 12 , \frac 12]} \  , 
\nonumber 
\\
T \times X_{[1,0,1]} & =  
 \textstyle
 (h_X+1) X_{[1,0,1]}  \ , 
\end{flalign}

\begin{flalign}
J \times X_{[1,0,0]} & =
 \textstyle
 2  j_X X_{[1,0,0]} 
+\frac{j_X  (h_X-j_X)
   (h_X+j_X+1)}{h_X (j_X+1)}
   X  \ , \nonumber  
   \\
G \times X_{[1,0,0]} & =
 \textstyle
 \frac{ (2 h_X+1) (h_X-j_X)}{2 h_X}  X_{[\frac 12 , - \frac 12 , \frac 12]} 
+\frac{ 
   (h_X+j_X+1) (h_X-j_X)}{2 h_X
   (j_X+1)}
X_{[\frac 12 , \frac 12 , \frac 12]}  
   -\frac{  2 j_X+1 }{2 (j_X+1)} 
   X_{[\frac 32 , \frac 12 , \frac 12]} \ , \nonumber  
   \\
\widetilde G \times X_{[1,0,0]} & =
 \textstyle
 -X_{[\frac 32 , - \frac 12 , -\frac 12]}-\frac{1 }{2 (j_X+1)} 
X_{[\frac 32 , \frac 12 ,- \frac 12]}
+\frac{
   h_X-j_X}{2 h_X}
   X_{[\frac 12 , - \frac 12 , - \frac 12]}
   -\frac{ (h_X+j_X+1) (2
   h_X j_X+h_X+j_X)}{2 h_X (j_X+1)} 
 X_{[\frac 12 , \frac 12 , - \frac 12]}  
 \ , \nonumber  
 \\
T \times X_{[1,0,0]} & =  (h_X+1) 
X_{[1,0,0]} \ , 
\end{flalign}

\begin{flalign}
J \times X'_{[1,1,0]}
& =
\textstyle
2  j_X X'_{[1,1,0]} 
-\frac{j_X  (h_X-j_X)
   (h_X+j_X+1)}{h_X (j_X+1)} X  \ , 
   & 
   \nonumber   \\
G \times X'_{[1,1,0]}
& = 
 \textstyle
 -\frac{  (h_X-j_X)}{2 h_X}
X_{[\frac 12 , - \frac 12 , \frac 12]}
+\frac{ 
   (h_X+j_X+1) (2 h_X j_X+h_X+j_X)}{2 h_X
   (j_X+1)}
X_{[\frac 12 , \frac 12 , \frac 12]}  
   -X_{[\frac 32 , - \frac 12 , \frac 12]}
   -\frac{1}{2 (j_X+1)} X_{[\frac 32 , \frac 12 , \frac 12]} \ ,
     \nonumber
     \\
\widetilde G \times X'_{[1,1,0]}
& = 
 \textstyle
 -\frac{  2 j_X+1 }{2 (j_X+1)}
X_{[\frac 32 , \frac 12 ,- \frac 12]} 
-\frac{ (2 h_X+1)
   (h_X-j_X)}{2 h_X}
   X_{[\frac 12 , - \frac 12 , - \frac 12]}
   -\frac{ (h_X+j_X+1)
   (h_X-j_X)}{2 h_X (j_X+1)}
X_{[\frac 12 , \frac 12 , - \frac 12]}   
   \ , 
     \nonumber
     \\
T \times X'_{[1,1,0]}
& = 
 \textstyle
 (h_X+1) X'_{[1,1,0]} \ ,
\end{flalign}

\begin{flalign}
J \times X_{[1,0,-1]}
& =
 \textstyle
2  j_X X_{[1,0,-1]} \ , &
\nonumber  \\
G \times X_{[1,0,-1]}
& = X_{[\frac 32 , - \frac 12 , -\frac 12]}+X_{[\frac 32 , \frac 12 ,- \frac 12]} + (h_X-j_X)
X_{[\frac 12 , - \frac 12 , - \frac 12]}+
   (h_X+j_X+1)
   X_{[\frac 12 , \frac 12 , - \frac 12]} \ ,
   \nonumber
   \\
\widetilde G \times X_{[1,0,-1]}
& = 0 \ , 
\nonumber
\\
T \times X_{[1,0,-1]}
& =  (h_X+1) X_{[1,0,-1]} \ , 
\end{flalign}

\begin{flalign}
J \times X_{[1,1,0]}
& =
2  (j_X+1) X_{[1,1,0]} 
+ (j_X-h_X) X \ ,  & \nonumber 
\\
G \times X_{[1,1,0]}
& =   (h_X-j_X)
X_{[\frac 12 , \frac 12 ,\frac 12]}
+X_{[\frac 32 , \frac 12 , \frac 12]} \ , 
\nonumber 
\\
\widetilde G \times X_{[1,1,0]}
& =  (h_X-j_X) 
X_{[\frac 12 , \frac 12 , - \frac 12]}
-X_{[\frac 32 , \frac 12 ,- \frac 12]}  \ ,  
\nonumber 
\\
T \times X_{[1,1,0]}
& =  (h_X+1) X_{[1,1,0]} 
 \ , 
\end{flalign}

\begin{flalign}
J \times X_{[\frac 32 , - \frac 12 , -\frac 12]}
& = \textstyle
2  \left(j_X-\frac{1}{2}\right)
X_{[\frac 32 , - \frac 12 , -\frac 12]}
   -\frac{2  j_X (h_X+j_X+1)}{2 j_X+1}
   X_{[\frac 12 , \frac 12 , - \frac 12]} \  & \nonumber   \\
& \textstyle
-\frac{2  (2 j_X-1)
   (h_X-j_X) (h_X+j_X+1)}{(2 h_X+1) (2
   j_X+1)}
   X_{[\frac 12 , - \frac 12 , - \frac 12]}
    \ , 
\nonumber \\   
G \times X_{[\frac 32 , - \frac 12 , -\frac 12]}
& =  \textstyle
-\frac{2 (h_X+1) (h_X-j_X)}{2 h_X+1}
X_{[1,-1,0]}
+\frac{2
    j_X}{2 j_X+1}
    X_{[2,0,0]}
    -\frac{2 
   (h_X+j_X+1) (2 h_X j_X+h_X+j_X)}{(2 h_X+1)
   (2 j_X+1)}
   X_{[1,0,0]}
\nonumber    \\
  &   \textstyle  +\frac{2  (h_X+j_X+1)
   (h_X-j_X)}{(2 h_X+1) (2 j_X+1)}
   X'_{[1,1,0]}
   -\frac{4 j_X 
   (h_X+j_X+1) (h_X-j_X)}{(2 h_X+1) (2 j_X+1)} X  \ , 
   \nonumber 
   \\
\widetilde G \times X_{[\frac 32 , - \frac 12 , -\frac 12]}
& =  \textstyle
-\frac{4  (h_X+1) j_X (h_X+j_X+1)}{(2 h_X+1)
   (2 j_X+1)}
   X_{[1,0,-1]} \ ,
  \nonumber 
  \\
T \times X_{[\frac 32 , - \frac 12 , -\frac 12]}
& =  \textstyle
\left(h_X+\frac{3}{2}\right) 
X_{[\frac 32 , - \frac 12 , -\frac 12]}  \  , 
\end{flalign}

\begin{flalign}
J \times X_{[\frac 32 , - \frac 12 , \frac 12]}
& = \textstyle
2  \left(j_X-\frac{1}{2}\right)
X_{[\frac 32 , - \frac 12 , \frac 12]}  
   +\frac{2   j_X (h_X+j_X+1)}{2
   j_X+1}
    X_{[\frac 12 ,  \frac 12 , \frac 12]}  
    & 
   \nonumber 
   \\
& \textstyle 
+\frac{2   (2 j_X-1) (h_X-j_X) (h_X+j_X+1)}{(2
   h_X+1) (2 j_X+1)}
 X_{[\frac 12 , - \frac 12 , \frac 12]}
  \ , 
 \nonumber 
 \\ 
G \times X_{[\frac 32 , - \frac 12 , \frac 12]}
& =  \textstyle
\frac{4  (h_X+1) j_X (h_X+j_X+1)}{(2 h_X+1)
   (2 j_X+1)} 
   X_{[1,0,1]}
  \ ,    \nonumber 
  \\
\widetilde G \times X_{[\frac 32 , - \frac 12 , \frac 12]}
& =  \textstyle
-\frac{2 (h_X+1) (h_X-j_X)}{2 h_X+1}
X_{[1,-1,0]}
+\frac{2
    j_X}{2 j_X+1}
    X_{[2,0,0]}
    -\frac{2 
   (h_X+j_X+1) (h_X-j_X)}{(2 h_X+1) (2
   j_X+1)}
   X_{[1,0,0]}
   \nonumber 
   \\
  &   \textstyle
  +\frac{2  (h_X+j_X+1) (2 h_X
   j_X+h_X+j_X)}{(2 h_X+1) (2 j_X+1)}
   X'_{[1,1,0]}
   -\frac{4 j_X
    (h_X+j_X+1) (h_X-j_X)}{(2 h_X+1) (2
   j_X+1)}  X \ , 
      \nonumber 
      \\
T \times X_{[\frac 32 , - \frac 12 , \frac 12]}
& =  \textstyle
\left(h_X+\frac{3}{2}\right)
X_{[\frac 32 , - \frac 12 , \frac 12]}  \  , 
\end{flalign}

\begin{flalign}
J \times X_{[\frac 32 , \frac 12 ,- \frac 12]} 
&=  \textstyle
2   \left(j_X+\frac{1}{2}\right)
X_{[\frac 32 , \frac 12 ,- \frac 12]}
+    (h_X-j_X)
X_{[\frac 12 , - \frac 12 , - \frac 12]}
+\frac{2  (h_X+j_X+1)
   (h_X-j_X)}{2 h_X+1}
 X_{[\frac 12 , \frac 12 , - \frac 12]}  
  \ ,  & 
  \nonumber \\
G \times X_{[\frac 32 , \frac 12 ,- \frac 12]} 
&=  \textstyle
-X_{[2,0,0]}+\frac{(h_X-j_X)}{2 h_X+1}
X_{[1,0,0]} 
-\frac{2
    (h_X+1) (h_X+j_X+1)}{2 h_X+1}
    X_{[1,1,0]}
   \nonumber   \\
  &  \textstyle
  +
   (h_X-j_X)
   X'_{[1,1,0]}
   +\frac{2  (h_X-j_X)
   (h_X+j_X+1)}{2 h_X+1}
   X
  \ ,  \nonumber   \\
\widetilde G \times X_{[\frac 32 , \frac 12 ,- \frac 12]} 
&=  \textstyle
\frac{2  (h_X+1) (h_X-j_X)}{2 h_X+1} 
X_{[1,0,-1]} \ , 
  \nonumber 
\\
T \times X_{[\frac 32 , \frac 12 ,- \frac 12]} 
&=    \textstyle
\left(h_X+\frac{3}{2}\right)
X_{[\frac 32 , \frac 12 ,- \frac 12]}  \ , 
\end{flalign}

\begin{flalign}
J \times X_{[\frac 32 , \frac 12 , \frac 12]}
&=  \textstyle
 (j_X-h_X)
 X_{[\frac 12 , - \frac 12 , \frac 12]}
 -\frac{2   (h_X-j_X)
   (h_X+j_X+1)}{2 h_X+1}
 X_{[\frac 12 , \frac 12 , \frac 12]} 
   +2 
   \left(j_X+\frac{1}{2}\right)
   X_{[\frac 32 , \frac 12 , \frac 12]}
    \ , & 
   \nonumber 
    \\
G \times X_{[\frac 32 , \frac 12 , \frac 12]}
&=  \textstyle
-\frac{2  (h_X+1) (h_X-j_X)}{2 h_X+1}
X_{[1,0,1]}  \ , 
   \nonumber 
   \\
\widetilde G \times X_{[\frac 32 , \frac 12 , \frac 12]}
&=  \textstyle
-X_{[2,0,0]}
+ (j_X-h_X)
X_{[1,0,0]}
-\frac{2  (h_X+1)
   (h_X+j_X+1)}{2 h_X+1}
   X_{[1,1,0]}
    \nonumber 
    \\
   &  \textstyle
   -\frac{  h_X-j_X }{2
   h_X+1}
   X'_{[1,1,0]}
   +\frac{2  (h_X-j_X) (h_X+j_X+1)}{2
   h_X+1}
   X \ , 
    \nonumber 
    \\
T \times X_{[\frac 32 , \frac 12 , \frac 12]}
&=   \textstyle
\left(h_X+\frac{3}{2}\right)
X_{[\frac 32 , \frac 12 , \frac 12]}  \ ,
\end{flalign}

\begin{flalign}
J \times X_{[2,0,0]}
&=  \textstyle
2  j_X X_{[2,0,0]}
+ (j_X-h_X)
X_{[1,-1,0]}
-\frac{
   (h_X-j_X) (h_X+j_X+1)}{h_X+1}
   X_{[1,0,0]} 
  & \nonumber  \\
&   \textstyle   -
   ( h_X + j_X + 1)
   X_{[1,1,0]}
+\frac{(h_X-j_X)
   (h_X+j_X+1)}{h_X+1}
   X'_{[1,1,0]} 
   +\frac{2 j_X 
   (h_X-j_X) (h_X+j_X+1)}{(h_X+1) (2 h_X+1)} 
   X \ , 
 \nonumber   \\
G \times X_{[2,0,0]}
&=  \textstyle
\frac{2   (h_X+j_X+1) (h_X-j_X)}{2
   h_X+1}
 X_{[\frac 12 , - \frac 12 , \frac 12]}  
   +\frac{2   (h_X+j_X+1) (h_X-j_X)}{2
   h_X+1}
    X_{[\frac 12 ,  \frac 12 , \frac 12]} 
\\
& \textstyle 
-\frac{ (2 h_X+3) (h_X-j_X)}{2
   (h_X+1)}
   X_{[\frac 32 , - \frac 12 , \frac 12]}
 -\frac{ (2 h_X+3) (h_X+j_X+1)}{2
   (h_X+1)} 
   X_{[\frac 32 , \frac 12 , \frac 12]} \ ,  
 \nonumber   \\
\widetilde G \times X_{[2,0,0]}
&=  \textstyle
\frac{ (2 h_X+3) (h_X-j_X)}{2
   (h_X+1)}
   X_{[\frac 32 , - \frac 12 , -\frac 12]}
   +\frac{ (2 h_X+3) (h_X+j_X+1)}{2
   (h_X+1)}
   X_{[\frac 32 , \frac 12 ,- \frac 12]} 
   \nonumber
\\
& \textstyle
   +\frac{2  (h_X+j_X+1) (h_X-j_X)}{2
   h_X+1}
   X_{[\frac 12 , - \frac 12 , - \frac 12]}
   +\frac{2  (h_X+j_X+1) (h_X-j_X)}{2
   h_X+1}
   X_{[\frac 12 , \frac 12 , - \frac 12]}  \ , \nonumber  \\
T \times X_{[2,0,0]}
&=  \textstyle  (h_X+2)
X_{[2,0,0]}
-\frac{3  (h_X-j_X)
   (h_X+j_X+1)}{2 h_X+1}
   X \ , 
\end{flalign}

\subsubsection{$\mathfrak{psl}(2|2)$ covariance and two-point function coefficients}

Let $A$, $B$ be $\mathfrak{sl}(2)_z \oplus 
\mathfrak{sl}(2)_y$ primaries with definite
$r$ charges.
Their two-point function can only be nonzero
for $h_A = h_B = h$, $j_A = j_B = j$,
and $r_A = - r_B = r$,
and takes the form 
\be 
\langle A(z_1,y_1) B(z_2, y_2) \rangle 
= \langle AB \rangle \frac{y_{12}^{2j}}{z_{12}^{2h}} \ .
\ee 
We refer to the number $\langle AB\rangle$
as two-point function coefficient.

Now, suppose $X$, $Y$ are 
$\mathfrak{psl}(2|2)$ primaries
with $h_X = h_Y = h$, $j_X = j_Y = j$,
$r_X = - r_Y = r$. 
We are interested in expressing the two-point function coefficients of 
$\mathfrak{psl}(2|2)$ descendants of $X$, $Y$ in terms of the two-point function coefficient $\langle XY \rangle$.
The operators $X$, $Y$ can be both Grassmann even or both Grassmann odd.

\paragraph{Short primaries.}

Suppose $X$ and $Y$ are short, hence $h=j$.
The non-zero
two-point function coefficients of
$\mathfrak{psl}(2|2)$ descendants of $X$, $Y$ are 
\be 
\ba 
\langle G_X \widetilde G_Y \rangle 
& = 2h (-1)^{|X|}\langle X Y \rangle \ , \\ 
\langle \widetilde G_X G_Y \rangle 
& = - 2 h (-1)^{|X|}\langle XY \rangle \ ,
\\
\langle T_X T_Y \rangle & = - (2h-1)(2h+1) \langle X Y \rangle \ .
\ea 
\ee 
Here $|X|\in \{0,1\}$ is the Grassmann parity of $X$. 
(In this work, all short $\mathfrak{psl}(2|2)$
primaries are Grassmann even.)

\paragraph{Long primaries.}
The generic case is $j \ge 1$.
The non-zero
two-point function coefficients of
$\mathfrak{psl}(2|2)$ descendants of $X$, $Y$ are 
\small
\be 
\ba 
\langle 
X_{[\frac 12, - \frac 12, +\frac 12]}
\; 
Y_{[\frac 12, - \frac 12, -\frac 12]}
\rangle & = \frac{2j(h+j+1 )}{2j+1} 
(-1)^{|X|}
\langle XY \rangle \ , \\
\langle 
X_{[\frac 12, - \frac 12, -\frac 12]}
 \; 
 Y_{[\frac 12, - \frac 12, +\frac 12]}
 \rangle & =  
- \frac{2j(h+j+1 )}{2j+1}
(-1)^{|X|}
\langle XY \rangle \ , \\
\langle 
X_{[\frac 12, \frac 12, + \frac 12]}
\; 
Y_{[\frac 12, \frac 12, - \frac 12]}
\rangle & = (h-j )
(-1)^{|X|}
\langle XY \rangle \ , \\
\langle 
X_{[\frac 12, \frac 12, - \frac 12]}
\; 
Y_{[\frac 12, \frac 12, +\frac 12]}
\rangle & = -(h-j )
(-1)^{|X|}
\langle XY \rangle \ , \\
\langle 
X_{[1,-1,0]}  \;
Y_{[1,-1,0]}
\rangle 
& = 
- \frac{(2j-1)(h+j+1)^2 }{2j+1}
\langle XY \rangle 
\ , \\
\langle 
X_{[1,1,0]} \; 
Y_{[1,1,0]}
\rangle 
& = 
- (h-j)^2
\langle XY \rangle 
\ , \\
\langle 
X_{[1,0,+1]} \; 
Y_{[1,0,-1]} 
\rangle 
& = 
- (h-j)(h+j+1)
\langle XY \rangle 
 \ , \\ 
 \langle 
X_{[1,0,-1]} \; 
Y_{[1,0,+1]} 
\rangle 
& = 
 - (h-j )(h+j+1 )
\langle XY \rangle 
 \ , \\ 
\begin{pmatrix}
  \langle 
  X_{[1,0,0]} \;
     Y_{[1,0,0]}
  \rangle  &  \langle
    X_{[1,0,0]} \;
     Y'_{[1,0,0]}
  \rangle 
  \\
  \langle
  X'_{[1,0,0]} \;
      Y_{[1,0,0]}
  \rangle  &  
  \langle
    X'_{[1,0,0]} \;
    Y'_{[1,0,0]}
  \rangle 
\end{pmatrix}
& = \begin{pmatrix}
    M_{11} & M_{12} \\
    M_{21}& M_{22}
\end{pmatrix}
\langle XY \rangle 
\ , \\ 
\langle 
X_{[\frac 32, - \frac 12, - \frac 12]}
\; 
Y_{[\frac 32, - \frac 12, + \frac 12]}
 \rangle 
&  = 
- \frac{4 j(h+1) (h-j) (h+j+1)^2  }{(2h+1)(2j+1)}
(-1)^{|X|}
\langle XY \rangle  
\ , \\ 
\langle 
X_{[\frac 32, - \frac 12, + \frac 12]}
\; 
Y_{[\frac 32, - \frac 12, - \frac 12]}
\rangle 
&  = 
 \frac{4 j(h+1) (h-j) (h+j+1)^2 }{(2h+1)(2j+1)}
 (-1)^{|X|}
\langle XY \rangle  
\ , \\ 
\langle 
X_{[\frac 32, \frac 12, +\frac 12]} 
\; 
Y_{[\frac 32, \frac 12, -\frac 12]} 
\rangle
& = 
 \frac{2 (h+1) (h+j+1 ) (h-j)^2 }{2h+1}
 (-1)^{|X|}
\langle XY \rangle 
\ , \\
\langle 
X_{[\frac 32, \frac 12, -\frac 12]} 
\; 
Y_{[\frac 32, \frac 12, +\frac 12]} 
\rangle
& = 
-   
\frac{2 (h+1) (h+j+1 ) (h-j)^2}{2h+1}
(-1)^{|X|}
\langle XY \rangle 
\ , \\
\langle 
X_{[2,0,0]}
\; 
Y_{[2,0,0]}
\rangle 
& = 
\frac{(2h+3)(h-j)^2(h+j+1)^2}{2h+1}
\langle XY \rangle \ ,
\ea 
\ee 
where 
\be 
\ba 
M_{11} & = M_{22} = \frac{(h+j+1)(h-j)^2}{2h(j+1)} \ , \\
M_{12} & = M_{21} = \frac{(h-j)(h+j+1)(h+j+2 hj)}{2h(j+1)} \  .
\ea 
\ee 
\normalsize
Here $|X|\in \{0,1\}$ is the Grassmann parity of $X$. 
In this work, we do encounter examples of long $\mathfrak{psl}(2|2)$
primaries that are Grassmann odd,
for example 
a pair of super Virasoro descendants of
$W_3$ with $(h,j,r) = (4,1, \pm \frac 12)$.

The cases $j=\frac 12$, $j=0$
can be deduced from the general case
by omitting the relevant
$\mathfrak{psl}(2|2)$ descendants
and specializing the above expressions to
these values of $j$.

\subsection{More details on the strategy for bootstrapping $\WW$}
\label{sec_methods}

In this section we present some more technical points related to the strategy for the OPE bootstrap of Section \ref{sec_results} and its implementation.

\subsubsection{Construction of 
the Ansatz for the $\mathbb W_{p_1} \times \mathbb W_{p_2}$ OPEs}
\label{sec_construct_ansatz}
 
The construction of the 
$\mathbb W_{p_1} \times \mathbb W_{p_2}$ OPEs proceeds in two steps.

Firstly, we need to identify which
super Virasoro primary operators can contribute to the singular part of the $\mathbb W_{p_1} \times \mathbb W_{p_2}$ OPEs. Let us distinguish two cases: $p_1 < p_2$ and $p_1 = p_2$.
If $p_1 < p_2$, the following super Virasoro primary operators can contribute:
\begin{itemize}
\item A short super Virasoro primary operator $(h=j)$ with 
\be 
\tfrac 12 (p_2 - p_1) \le h \le
\tfrac 12 (p_1+p_2)-1 \ , \qquad 
h - \tfrac 12 (p_1 + p_2) =0 \mod \mathbb Z \ .
\ee 
In particular, this includes
the generators $W_q$ 
with $3 \le q \le p_1+p_2 -2$
and $q - p_1 - p_2=0$ mod 2.

\item A long super Virasoro 
primary operator $(h>j)$
with 
\be 
\ba 
 h &\le \tfrac 12 (p_1 + p_2)-1 \ , &
h - \tfrac 12 (p_1 + p_2) &=0 \mod \mathbb Z \ , \\
\tfrac 12 (p_2 - p_1) \le j &\le 
\tfrac 12 (p_1 + p_2) -1 \ , & 
j - \tfrac 12 (p_1 + p_2) &=0 \mod \mathbb Z \ . 
\ea 
\ee 
\end{itemize}
If $p_1 = p_2$ we have to also take into account the Bose symmetry of the
$\mathbb W_{p_1} \times \mathbb W_{p_2}$ OPE. As a result,
the super Virasoro primary operators that can contribute are:
\begin{itemize}
\item A short super Virasoro primary operator $(h=j)$ with 
\be 
0 \le h \le \tfrac 12 (p_1 + p_2) -1 \ , \qquad 
h - \tfrac 12 (p_1 + p_2) = 0 \mod \mathbb Z \ . 
\ee 
In particular, this includes the identity operator
$\one$
as well as 
the generators $W_q$ 
with $3 \le q \le p_1+p_2 -2$
and $q - p_1 - p_2=0$ mod 2.
\item A long super Virasoro primary operator ($h>j$) with 
\be 
\ba 
 h &\le \tfrac 12 (p_1 + p_2)-1 \ , &
h - \tfrac 12 (p_1 + p_2) &=0 \mod \mathbb Z \ , \\
\tfrac 12 (p_2 - p_1) \le j &\le 
\tfrac 12 (p_1 + p_2) -1 \ , & 
j - \tfrac 12 (p_1 + p_2) &=0 \mod \mathbb Z \ , \\
&&
h-j &= 0  \mod 2 \ . 
\ea 
\ee 
\end{itemize}

Once we have identified all
super Virasoro primary operators
that can contribute to the singular
parts of the $\mathbb W_{p_1} \times \mathbb W_{p_2}$ OPEs,
we consider them 
one by one and 
construct the explicit Ansatz
for each of the 
$\mathbb W_{p_1} \times \mathbb W_{p_2}$ OPEs,
namely
$W_{p_1} \times W_{p_2}$,
$W_{p_1} \times G_{W_{p_2}}$,
$W_{p_1} \times \widetilde G_{W_{p_2}}$, \dots,
$T_{W_{p_1}} \times T_{W_{p_2}}$.
We refer the reader
to Appendix
\ref{app_superVir}
for further details
and explicit examples.

\subsubsection{Redefinitions of the generators $W_p$}
\label{sec_redef_W}

We have already pointed out in section
\ref{sec_results_nice} that, for $p\ge 6$,
redefinitions of the form
\eqref{eq_redefiningW6W7} are possible, 
repeated here for convenience,
\begin{align}
    \label{eq_W6_redef}
W_6' = W_6 + \mu_6  \, 
\cC_{3,3}^{W_3 W_3} \ , \\ 
\label{eq_W7_redef}
W_7' =  W_7 + \mu_7 \,
\cC_{\frac 72 , \frac 72}^{W_3 W_4}
 \ . 
\end{align}

% Starting from weight $h=3$,
% the algebra contains short composite
% super Virasoro primary operators
% with the same quantum numbers as one of the generators $W_q$.
% For instance, the composite operator
% $\cC_{3,3}^{W_3 W_3} = (W_3 W_3)_0^3 + \dots $
%  and the generator $W_6$
% both have $h=j=3$.
% As a result, we are free to perform
% a redefinition of the generator $W_6$, of the form
% \be  \label{eq_W6_redef}
% W_6' = W_6 + \mu_6  \, 
% \cC_{3,3}^{W_3 W_3}
%  \ , 
% \ee 
% with $\mu_6$ a constant parameter.
% By a similar token, we can consider the redefinition
% \be \label{eq_W7_redef}
% W_7' =  W_7 + \mu_7 \,
% \cC_{\frac 72 , \frac 72}^{W_3 W_4}
%  \ . 
% \ee   

We may regard redefinitions
such as \eqref{eq_W6_redef}, \eqref{eq_W7_redef}
as a ``gauge redundancy''
in our bootstrap analysis.
They affect the OPE coefficients in
$\mathbb W_{p_1} \times \mathbb W_{p_2}$ OPEs.
For example, the redefinition
\eqref{eq_W6_redef} affects
directly 
the OPEs $\mathbb W_{6} \times \mathbb W_{p_2}$, but it also
affects other $\mathbb W_{p_1} \times \mathbb W_{p_2}$ OPEs in which
$W_6$ appears on the RHS.
All constraints on OPE coefficients
deriving from $\text{Jacobi}(\mathbb W_{q_1} , \mathbb W_{q_2} , \mathbb W_{q_3})$ must be compatible with the ``gauge redundancy''
associated to redefinitions such as
\eqref{eq_W6_redef},
\eqref{eq_W7_redef}
We have verified this claim explicitly;  it provides a self-consistency check for our results. We refer the reader to Appendix
\ref{app_redef}
for explicit expressions
of the effect of the redefinitions
\eqref{eq_W6_redef}, \eqref{eq_W7_redef} on  OPE coefficients.

In practice it is convenient to ``gauge fix'' the ``gauge redundancy''
\eqref{eq_W6_redef},
\eqref{eq_W7_redef} 
by prescribing the values of suitable OPE coefficients.
A particularly convenient choice is
\be \label{eq_gauge_fix_App}
c_{36}{}^3 = 0 \ , \qquad 
c_{37}{}^4 = 0 \  . 
\ee 
This corresponds to
$W_6$, $W_7$ being orthogonal
to composite super Virasoro primaries, see
\eqref{eq_no_W6W7_mixing}.
All results of section
\ref{sec_results} are presented
with this ``gauge fixing'' condition.

% \bonetti{I've added here a small speculation
% for general $W_p$. It will link to the single particle operator story later}

% More generally, building on 
% \eqref{eq_gauge_fix}, we propose the following
% prescription to ``gauge fix''   redefinition ambiguities of the strong
% generators $W_p$:
% each $W_p$ should have vanishing 2-point functions with all composite super Virasoro primary operators constructed out of 
% $W_q$'s with $q < p$.
% We elaborate further on this prescription
% in section 
% \ref{sec_SPOs}.

\subsubsection{Permutation symmetry of $\langle W_{p_1} W_{p_2} W_{p_3} \rangle$ three-point functions}

The three-point function of three
(not necessarily distinct)
$W_p$ operators is fixed by
$\mathfrak{sl}(2)_z \oplus \mathfrak{sl}(2)_y$
covariance to take the form
\be \label{eq_threepoint}
\langle W_{p_1}(z_1,y_1)
W_{p_2}(z_2,y_2)
W_{p_3}(z_3,y_3)
\rangle 
= \lambda_{p_1 p_2 p_3} 
\frac{
y_{12}^{\frac 12 (p_1 + p_2 - p_3)}
y_{23}^{\frac 12 (p_2 + p_3 - p_1)}
y_{13}^{\frac 12 (p_1 + p_3 - p_2)}
}{
z_{12}^{\frac 12 (p_1 + p_2 - p_3)}
z_{23}^{\frac 12 (p_2 + p_3 - p_1)}
z_{13}^{\frac 12 (p_1 + p_3 - p_2)}
}  \ , 
\ee 
where $z_{ij}=z_i - z_j$,
$y_{ij} = y_i - y_j$ and $\lambda_{p_1 p_2 p_3}$ is a constant.
The latter is related to the OPE coefficients in the $W_{p_1} \times W_{p_2}$ OPE as
\be 
\lambda_{p_1 p_2 p_3} = \sum_{\cO} c_{p_1 p_2}{}^\cO g_{\cO p_3} \ , 
\ee 
where 
$\cO$ is an $\mathfrak{sl}(2)_z \oplus \mathfrak{sl}(2)_y$ primary
operator and 
$g_{\cO p_3}$ is the two-point function coefficient defined by
\be 
\langle \cO(z_1, y_1) W_{p_2}(z_2, y_2) 
\rangle = 
g_{\cO p_3} \frac{y_{12}^{ p_3}}{z_{12}^{p_3}} \ . 
\ee 
Clearly, $g_{\cO p_3}$ can only be non-zero if $\cO$ and $W_{p_3}$ have the same weight and spin,
$h=j=\tfrac 12 p_3$.

The Bose symmetry of the
three-point function 
\eqref{eq_threepoint} guarantees that the
coefficients $\lambda_{p_1 p_2 p_3}$
are totally symmetric in the
permutation of the $p_i$ labels.
In turn, this imposes some constraints on OPE coefficients.

For example, let us consider
$(p_1,p_2,p_3) = (3,3,4)$.
On the one hand, we have the relations
\be 
\sum_{\cO} c_{33}{}^\cO g_{\cO 4}
= \sum_{\cO'} c_{34}{}^{\cO'} g_{\cO' 3} \ . 
\ee 
On the other hand,
the only $\mathfrak{sl}(2)_z \oplus \mathfrak{sl}(2)_y$ primary with a non-zero two-point function
with $W_4$ is $W_4$ itself,
$g_{\cO 4} = g_{44} \equiv g_4$,
and similarly for
$g_{\cO' 3}$. We then obtain
the relation
\be  \label{eq_c334}
c_{33}{}^4  g_4 = c_{34}{}^3 g_3 \ . 
\ee   

Relations such as
\eqref{eq_c334}
are not independent from the constraints deriving from
the vanishing of 
$\text{Jacobi}(\mathbb W_{q_1}, \mathbb W_{q_2}, \mathbb W_{q_3})$.
Using the permutation symmetry of 
$\langle W_{p_1} W_{p_2} W_{p_3} \rangle$ three-point functions,
however, can be a 
much more efficient way of
deriving such constraints,
compared to the analysis of 
$\text{Jacobi}(\mathbb W_{q_1}, \mathbb W_{q_2}, \mathbb W_{q_3})$.
This is illustrated by the examples
$(p_1,p_2,p_3)=(3,6,5)$
and $(p_1,p_2,p_3)=(3,7,6)$.
Permutation symmetry
yields the relations
\be \label{eq_from_bose}
\sum_\cO c_{36}{}^{\cO} g_{\cO 5}
= \sum_{\cO'} c_{35}{}^{\cO} g_{\cO 6} 
\ , \qquad 
\sum_{\cO} c_{37}{}^\cO g_{\cO 6}
= \sum_{\cO'} c_{36}{}^\cO g_{\cO 7} \ . 
\ee 
In the first relation,
the only $\mathfrak{sl}(2)_z \oplus \mathfrak{sl}(2)_y$ primary
that can have a non-zero two-point
function with $W_5$ is $W_5$ itself,
$g_{\cO 5} = g_{55} \equiv g_5$.
On the RHS, however,
we could have a two-point function mixing between $W_6$ and the composite
operator 
$\cC_{3,3}^{W_3 W_3}$.
Here is useful to
make use of the ``gauge fixing'' condition
$c_{36}{}^3 = 0$, see \eqref{eq_gauge_fix}.
Indeed, setting 
$c_{36}{}^3$ to zero
also guarantees that the two-point function between $W_6$ and the composite 
$\cC_{3,3}^{W_3 W_3}$
 vanishes.
As a result, 
the first relation in \eqref{eq_from_bose}
implies 
\be \label{eq_c365}
c_{36}{}^5 g_5 = c_{35}{}^6 g_6 \ . 
\ee 
Let us now turn to the second relation
in \eqref{eq_from_bose}. We have already argued
that there is no two-point function
mixing involving $W_6$.
A priori, however, there is  a two-point function mixing involving
$W_7$ and the composite
$\cC_{\frac 72 , \frac 72}^{W_3 W_4}$.
Once again, we resort to the
gauge fixing condition
$c_{37}{}^4 =0$
in \eqref{eq_gauge_fix} to kill
the two-point function between
$W_7$ and 
$\cC_{\frac 72  , \frac 72}^{W_3 W_4}$. We then have simply
\be \label{eq_c376}
c_{37}{}^6 g_6 = c_{36}{}^7 g_7 \ . 
\ee 

The relations \eqref{eq_c365},
\eqref{eq_c376} could be alternatively
derived studying Jacobiators of the form
$\text{Jacobi}(\mathbb W_{q_1}, \mathbb W_{q_2}, \mathbb W_{q_3})$,
but that would necessarily require
pushing the bootstrap analysis
all the way to OPEs $\mathbb W_{p_1} \times \mathbb W_{p_2}$ with
$p_1+p_2 = 14$, in order to capture
the two-point function coefficient
$g_7$ (which governs how the unit 
operator $\one$ enters in the $W_7 \times W_7$ OPE).
This demonstrates how the permutation symmetry of 
$\langle W_{p_1} W_{p_2} W_{p_3} \rangle$ three-point functions
enhances 
the analysis based only on Jacobiators.
The results presented in Section
\ref{sec_results} were indeed obtained
making use of \eqref{eq_c365},
\eqref{eq_c376}---even though
we have only implemented
OPEs $\mathbb W_{p_1} \times \mathbb W_{p_2}$ with
$p_1+p_2 \le 10$.

%%%%%%%%%%%%%%%%%%%

\subsubsection{Results for OPE coefficients without fixing $g_p$}
\label{app_results}

We report here the results of the bootstrap analysis without fixing the normalization of $W_p$, i.e.~not
imposing \eqref{eq_best_normalization} and leaving instead the two-point function coefficients $g_p$ arbitrary.
We still impose $c_{36}{}^3 =0$,
$c_{37}{}^4= 0$ to ensure the orthogonality between $W_6$, $W_7$ and composite super Virasoro primary operators.

\newpage

%W3 W3 and W3 W4
\begin{table}[h!]
    \centering
    \begin{tabular}{| c |}
\hline  
\textbf{
\textit{
OPE coefficients in $W_3 \times W_3$ and
$W_3 \times W_4$
}
}
\\ 
\hline 
\rule[-18mm]{0mm}{38mm}
\begin{minipage}{9cm}
\begin{fleqn}
\begin{equation*} 
\ba 
(c_{3 3}{}^{4})^2 
& = 
\frac{36 (\nu -9) \nu }{(\nu -4) (\nu -1) (\nu +1)}
\frac{(g_3)^2}{g_4} 
 \ , \\[2mm]
 c_{3 4}{}^{3} & = \frac{g_4}{g_3}
c_{3 3}{}^{4}
\ , 
\\[2mm]
(c_{3 4}{}^{5})^2
& = 
\frac{60 (\nu -16) (\nu +1)}{(\nu -4) (\nu -1) (\nu +5)}
\frac{g_3 g_4}{g_5} 
\ea 
\end{equation*}
\end{fleqn}
\end{minipage}
\\
\hline
    \end{tabular}
    \caption{Values of the OPE coefficients in the $W_3 \times W_5$, $W_3 \times W_4$ OPEs.}
    \label{ugly_tab_OPEcoeffs_part1}
\end{table}

%W3 W5
\begin{table}[h!]
    \centering
    \begin{tabular}{| c |}
\hline  
\textbf{
\textit{
OPE coefficients in $W_3 \times W_5$
}
}
\\ 
\hline 
\rule[-25mm]{0mm}{52mm}
\begin{minipage}{9cm}
\begin{fleqn}
\begin{equation*} 
\ba 
c_{3 5}{}^{4} & = 
\frac{g_5}{g_4} c_{3 4}{}^{5}
\ , 
\\[2mm]
(c_{3 5}{}^{6})^2
& =
\frac{90 (\nu -25) \nu  (\nu +5)}{(\nu -4) (\nu -1) \left(\nu ^2+15 \nu +8\right)}
\frac{g_3 g_5}{g_6} \ ,
\\[2mm]
%c_{3 5}{}^{\cC(3,1)} 
c_{35}{}^{ \cC_{3,1}^{W_3 W_3}  }
& =   
-\frac{ (\nu -4) (\nu -2) (\nu -1) }{6
   (\nu -16) (\nu -9) \nu  }
\frac{c_{3 3}{}^{4} c_{3 4}{}^{5}g_5}{(g_3)^2}   
 \ , \\[2mm]
%c_{3 5}{}^{\cC(3,3)}
c_{35}{}^{ \cC_{3,3}^{W_3 W_3} }
& = 
\frac{ (\nu -4) (\nu -1) (\nu +3) }{4 \nu
    \left(\nu ^2+15 \nu +8\right)  }
\frac{c_{3 3}{}^{4} c_{3 4}{}^{5} g_5}{(g_3)^2}  
\ea 
\end{equation*}
\end{fleqn}
\end{minipage}
\\
\hline
    \end{tabular}
    \caption{Values of the OPE coefficients in the $W_3 \times W_5$ OPE. The values of $c_{33}{}^4$, $c_{34}{}^5$ are given in Table \ref{ugly_tab_OPEcoeffs_part1}.}
    \label{ugly_tab_OPEcoeffs_part2}
\end{table}

%W4 W4
\begin{table}[h!]
    \centering
    \begin{tabular}{| c |}
\hline  
\textbf{
\textit{
OPE coefficients in $W_4 \times W_4$
}
}
\\ 
\hline 
\rule[-25mm]{0mm}{52mm}
\begin{minipage}{9cm}
\begin{fleqn}
\begin{equation*} 
\ba 
c_{4 4}{}^{4}
& = 
\frac{4  \left(\nu ^2-20 \nu +9\right) }{3 (\nu -9) \nu 
   }
\frac{c_{3 3}{}^{4} g_4}{g_3}   
\ , \\[2mm]
c_{4 4}{}^{6} & = 
\frac{(\nu -4) (\nu -1) (\nu +1)
   }{45 (\nu -9) \nu  }
\frac{c_{3 3}{}^{4} c_{3 4}{}^{5} c_{3 5}{}^{6}g_4 }{(g_3)^2}
\ , 
\\[2mm] 
%c_{4 4}{}^{\cC(3,1)} 
c_{4,4}{}^{ \cC_{3,1}^{W_3 W_3} }
& =
-\frac{12 (\nu +1) }{(\nu -9) \nu  }
\frac{g_4}{g_3}
\ , 
\\[2mm]
% c_{4 4}{}^{\cC(3,3)}
c_{4,4}{}^{ \cC_{3,3}^{W_3 W_3} }
& = 
\frac{24 (\nu -4) (\nu +1) }{\nu  \left(\nu ^2+15 \nu +8\right) }
\frac{g_4}{g_3}  
\ea 
\end{equation*}
\end{fleqn}
\end{minipage}
\\
\hline
    \end{tabular}
    \caption{Values of the OPE coefficients in the $W_4 \times W_4$ OPE.
    The values of $c_{33}{}^4$, $c_{34}{}^5$ are given in table \ref{ugly_tab_OPEcoeffs_part1}.
    The value of $c_{35}{}^6$ is given in Table \ref{ugly_tab_OPEcoeffs_part2}.}
    \label{ugly_tab_OPEcoeffs_part3}
\end{table}

\newpage 

%W4 W5
\begin{table}[h!]
    \centering
    \begin{tabular}{| c |}
\hline  
\textbf{
\textit{
OPE coefficients in $W_4 \times W_5$
}
}
\\ 
\hline 
\rule[-45mm]{0mm}{92mm}
\begin{minipage}{10cm}
\begin{fleqn}
\begin{equation*} 
\ba 
c_{4 5}{}^{3} & = 
\frac{g_5}{g_3}
c_{3 4}{}^{5} \ , \\[2mm] 
c_{4 5}{}^{5} & = 
\frac{5  \left(\nu ^3-32 \nu ^2-77 \nu +36\right) }{3 (\nu -9)
   \nu  (\nu +5) }
\frac{c_{3 3}{}^{4}
g_4
}{
g_3
}    \ , \\[2mm]
(c_{4 5}{}^{7})^2
& = 
\frac{140 (\nu -36) (\nu -25) (\nu +1) (\nu +5) }{(\nu -9) (\nu
   -4) (\nu -1) \left(\nu ^2+35 \nu +84\right) }
\frac{g_4 g_5}{g_7}   
\ , 
\\[2mm]
%c_{4 5}{}^{\cC(\frac 72 , \frac 12)}
c_{45}{}^{  \cC_{\frac 72, \frac 12}^{W_3 W_4} }
& = 
-\frac{ (\nu -4) (\nu -1) (\nu +12) }{6
   (\nu -16) (\nu -9) \nu  }
\frac{c_{3 3}{}^{4} c_{3 4}{}^{5}
g_5}{
(g_3)^2
}    \ , \\[2mm] 
% c_{4 5}{}^{\cC(\frac 72 , \frac 32)}
c_{45}{}^{  \cC_{\frac 72, \frac 32}^{W_3 W_4} }
& = 
-\frac{ (\nu -4) (\nu -1) (2 \nu +3) }{3
   (\nu -16) (\nu -9) \nu  }
\frac{c_{3 3}{}^{4} c_{3 4}{}^{5}
g_5}{
(g_3)^2
}   
\ , \\[2mm] 
%c_{4 5}{}^{\cC(\frac 72 , \frac 52)}
c_{45}{}^{ \cC_{\frac 72 , \frac 52}^{W_3 W_4}  }
& = 
\frac{ (\nu -4) (\nu -1) }{6 (\nu -9) \nu
    }
\frac{c_{3 3}{}^{4} c_{3 4}{}^{5}
g_5}{
(g_3)^2
}     \ , \\[2mm]
%c_{4 5}{}^{\cC(\frac 72 , \frac 72)}
c_{45}{}^{  \cC_{\frac 72 , \frac 72}^{W_3 W_4}  }
& = 
\frac{2 (\nu -4) (\nu -1) \left(2 \nu ^2-7 \nu -63\right)
  }{3 (\nu -9) \nu  \left(\nu ^2+35 \nu +84\right) }
\frac{c_{3 3}{}^{4} c_{3 4}{}^{5}  g_5}{
(g_3)^2
}   
\ea 
\end{equation*}
\end{fleqn}
\end{minipage}
\\
\hline
    \end{tabular}
    \caption{Values of the OPE coefficients in the $W_4 \times W_5$ OPE. The values of $c_{33}{}^4$, $c_{34}{}^5$ are given in Table \ref{ugly_tab_OPEcoeffs_part1}.}
    \label{ugly_tab_OPEcoeffs_part4}
\end{table}

%W3 W6
\begin{table}[h!]
    \centering
    \begin{tabular}{| c |}
\hline  
\textbf{
\textit{
OPE coefficients in $W_3 \times W_6$
}
}
\\ 
\hline 
\rule[-35mm]{0mm}{72mm}
\begin{minipage}{10cm}
\begin{fleqn}
\begin{equation*} 
\ba 
c_{3 6}{}^{3} & = 0 \ , \\[2mm]
c_{3 6}{}^{5} & = 
\frac{g_6}{g_5}
c_{3 5}{}^{6} \ , \\[2mm] 
c_{3 6}{}^{7}
& = 
\frac{ (\nu -4) (\nu -1) \left(\nu
   ^2+15 \nu +8\right) }{60 (\nu -25) \nu  (\nu +5) }
\frac{c_{3 3}{}^{4} c_{3 5}{}^{6} c_{4 5}{}^{7}
g_6}{
g_3
   g_5
}   
\ , 
\\[2mm]
% c_{3 6}{}^{\cC(\frac 72 , \frac 32)}
c_{36}{}^{  \cC_{\frac 72 , \frac 32}^{W_3 W_4} }
& = 
-\frac{7  (\nu -4) (\nu -1)^2 }{30 (\nu
   -25) (\nu -16) (\nu +1) }
\frac{c_{3 4}{}^{5} c_{3 5}{}^{6}
g_6
}{g_3 g_4}
\ , \\[2mm]
% c_{3 6}{}^{\cC(\frac 72 , \frac 52)}
c_{36}{}^{ \cC_{\frac 72 , \frac 52}^{W_3 W_4}  }
& = 
\frac{(\nu -4) (\nu -1) }{30 \nu  (\nu
   +1) }
\frac{c_{3 4}{}^{5} c_{3 5}{}^{6} 
g_6}{
g_3 g_4
}  \ , \\[2mm]
% c_{3 6}{}^{\cC(\frac 72 , \frac 72)}
c_{36}{}^{ \cC_{\frac 72, \frac 72}^{W_3 W_4} }
& = 
\frac{2 (\nu -4) (\nu -1) (\nu +7) }{5
   (\nu +1) \left(\nu ^2+35 \nu +84\right) }
\frac{c_{3 4}{}^{5} c_{3 5}{}^{6}
g_6
}{
g_3 g_4
} 
\ea 
\end{equation*}
\end{fleqn}
\end{minipage}
\\
\hline
    \end{tabular}
    \caption{Values of the OPE coefficients in the $W_3 \times W_6$ OPE.
    The values of $c_{33}{}^4$, $c_{34}{}^5$ are given in Table \ref{ugly_tab_OPEcoeffs_part1}.
    The value of $c_{35}{}^6$ is given in Table \ref{ugly_tab_OPEcoeffs_part2}.
    The value of $c_{45}{}^7$ is given in Table \ref{ugly_tab_OPEcoeffs_part4}
    \label{ugly_tab_OPEcoeffs_part5}.}
\end{table}

\newpage

%W3 W7
\begin{table}[H]
    \centering
    \begin{tabular}{| c |}
\hline  
\textbf{
\textit{
OPE coefficients in $W_3 \times W_7$  (partial)
}
}
\\ 
\hline 
\rule[-39mm]{0mm}{79mm}
\begin{minipage}{14cm}
\small
\begin{fleqn}
\begin{equation*} 
\ba 
c_{3 7}{}^{4} & = 0 
\ , \\[2mm]
c_{3 7}{}^{6}
& = 
\frac{ (\nu -4) (\nu -1) \left(\nu
   ^2+15 \nu +8\right) }{60 (\nu -25) \nu  (\nu +5) }
\frac{c_{3 3}{}^{4} c_{3 5}{}^{6} c_{4 5}{}^{7}
g_7}{
g_3
   g_5
}
\ , \\[2mm]
% c_{3 7}{}^{\cC(3,3)} 
c_{37}{}^{ \cC_{3,3}^{W_3 W_3}  }
& = 0 
\ , \\[2mm]
% c_{3 7}{}^{\cC(4,2)_\text{I}}
c_{37}{}^{ \cC_{4,2}^{W_3 W_3}  }
& = 
-\frac{8  (\nu -4) (\nu -1) \left(\nu ^5+42 \nu ^4+264 \nu
   ^3-253 \nu ^2+17226 \nu +9720\right) }{15 (\nu -36) (\nu -25) (\nu -16) \nu  (\nu
   +1) (\nu +5) \left(\nu ^2+15 \nu +8\right)}
\frac{
c_{3 4}{}^{5} c_{4 5}{}^{7}
g_7
}{
 g_3 g_4
}   
\ , \\[2mm]
% c_{3 7}{}^{\cC(4,2)_\text{II}}
c_{37}{}^{ \cC_{4,2}^{W_3 W_5} }
& =
-\frac{2  (\nu -4) (\nu -1) \left(2 \nu ^2+13 \nu -90\right)
   g_7}{15 (\nu -36) (\nu -25) \nu  (\nu +5) }
\frac{
c_{3 3}{}^{4} c_{4 5}{}^{7}
}{
g_3 g_5
}   
\ , \\[2mm]
% c_{3 7}{}^{\cC(4,2)_\text{III}}
c_{37}{}^{  \cC_{4,2}^{W_4 W_4}  }
& = 
-\frac{ (\nu -9) (\nu -4) (\nu -1) \left(\nu ^2-\nu
   +60\right) }{10 (\nu -36) (\nu -25) (\nu -16) (\nu +1)^2 }
\frac{
c_{3 4}{}^{5} c_{4 5}{}^{7}
g_7
}{
(g_4)^2
}    
\ , \\[2mm]
% c_{3 7}{}^{\cC(4,3)}
c_{37}{}^{ \cC_{4,3}^{W_3 W_5} }
&=
\frac{ (\nu -4) (\nu -1) }{15 \nu  (\nu
   +5) }
\frac{
c_{3 3}{}^{4} c_{4 5}{}^{7}
g_7
}{
g_3 g_5
}   
\ea 
\end{equation*}
\end{fleqn}
\end{minipage}
\\
\hline
    \end{tabular}
    \caption{Values of the some of the OPE coefficients in the $W_3 \times W_7$ OPE.
    The values of $c_{33}{}^4$, $c_{34}{}^5$ are given in Table \ref{ugly_tab_OPEcoeffs_part1}.
    The value of $c_{35}{}^6$ is given in Table \ref{ugly_tab_OPEcoeffs_part2}.
    The value of $c_{45}{}^7$ is given in Table~\ref{ugly_tab_OPEcoeffs_part4}.
    \label{ugly_tab_OPEcoeffs_part6}}
\end{table}

\newpage

%W4 W6
\begin{table}[H]
    \centering
    \begin{tabular}{| c |}
\hline  
\textbf{
\textit{
OPE coefficients in $W_4 \times W_6$  (partial)
}
}
\\ 
\hline 
\rule[-53mm]{0mm}{108mm}
\begin{minipage}{14cm}
\small
\begin{fleqn}
\begin{equation*} 
\ba 
c_{4 6}{}^{4}
& = 
\frac{ (\nu -4) (\nu -1) (\nu +1)
   }{45 (\nu -9) \nu  }
\frac{
c_{3 3}{}^{4} c_{3 4}{}^{5} c_{3 5}{}^{6}
g_6
}{
(g_3)^2
}   
\ , \\[2mm]
c_{4 6}{}^{6}
& = 
\frac{ (2 \nu -1) \left(\nu ^3-46 \nu ^2-387 \nu -240\right)
   }{(\nu -9) \nu  \left(\nu ^2+15 \nu +8\right) }
\frac{
c_{3 3}{}^{4}
g_4
}{
g_3
}   
\ , \\[2mm]
% c_{4 6}{}^{\cC(3,1)}
c_{46}{}^{ \cC_{3,1}^{W_3 W_3}  }
& = 
-\frac{7  (\nu -4) (\nu -2) (\nu -1) }{45
   (\nu -16) (\nu -9) \nu  }
\frac{
c_{3 4}{}^{5} c_{3 5}{}^{6}
g_6
}{
(g_3)^2
}   
\ , 
\\[2mm]
% c_{4 6}{}^{\cC(3,3)}
c_{46}{}^{ \cC_{3,3}^{W_3 W_3} }
& = 
\frac{3  (\nu -4) (\nu -1) (\nu +3) }{10
   \nu  \left(\nu ^2+15 \nu +8\right) }
\frac{
c_{3 4}{}^{5} c_{3 5}{}^{6}
g_6
}{
(g_3)^2
}    
\ , \\[2mm]
% c_{4 6}{}^{\cC(4,1)}
c_{46}{}^{ \cC_{4,1}^{W_3 W_5} }
& = 
-\frac{7  (\nu -4) (\nu -1) (\nu +1) (\nu +23)
}{30 (\nu -25) (\nu -9) \nu  (\nu +5) }
\frac{
c_{3 3}{}^{4} c_{3 5}{}^{6}
   g_4 g_6
}{
(g_3)^2
   g_5
}   
\ , 
\\[2mm]
% c_{4 6}{}^{\cC(4,2)_\text{I}}
c_{46}{}^{  \cC_{4,2}^{W_3 W_3} }
& = 
{\scriptstyle
\frac{2  (\nu -4) (\nu -1) \left(5 \nu ^7-499 \nu ^6-9539 \nu
   ^5+119535 \nu ^4+1843758 \nu ^3+1311460 \nu ^2-32542800 \nu -18144000\right) }{525
   (\nu -25) (\nu -16) (\nu -15) (\nu -9) \nu ^2 (\nu +5) \left(\nu ^2+15 \nu +8\right)
  }
\frac{
c_{3 4}{}^{5} c_{3 5}{}^{6}
g_6
}{
( g_3 )^2
}   
}
\ , 
\\[2mm]
% c_{4 6}{}^{\cC(4,2)_\text{II}}
c_{46}{}^{ \cC_{4,2}^{W_3 W_5} }
& = 
-\frac{ (\nu -4) (\nu -1) (\nu +1) \left(19 \nu ^2+117 \nu
   +320\right) }{30 (\nu -25) (\nu -9) \nu ^2 (\nu +5)
}
\frac{
c_{3 3}{}^{4} c_{3 5}{}^{6}
g_4 g_6
}{
  ( g_3 )^2 g_5
}   
\ , \\[2mm]
% c_{4 6}{}^{\cC(4,2)_\text{III}}
c_{46}{}^{ \cC_{4,2}^{W_4 W_4} }
& = 
-\frac{4  (\nu -4)^2 (\nu -1) (\nu +5)
 }{15 (\nu -25) (\nu -16) \nu  (\nu +1) }
\frac{
c_{3 4}{}^{5} c_{3 5}{}^{6}
  g_6
}{
g_3 g_4
}   
\ , \\[2mm]
% c_{4 6}{}^{\cC(4,3)}
c_{46}{}^{ \cC_{4,3}^{W_3 W_5} }
& = 
\frac{4  (\nu -4) (\nu -2) (\nu -1) (\nu +1)
}{15 (\nu -9) \nu ^2 (\nu +5) }
\frac{
c_{3 3}{}^{4} c_{3 5}{}^{6}
   g_4 g_6
}{
( g_3 )^2
   g_5
}   
\ea 
\end{equation*}
\end{fleqn}
\end{minipage}
\\
\hline
    \end{tabular}
    \caption{Values of the some of the OPE coefficients in the $W_4 \times W_6$ OPE.
    The values of $c_{33}{}^4$, $c_{34}{}^5$ are given in Table \ref{ugly_tab_OPEcoeffs_part1}.
    The value of $c_{35}{}^6$ is given in Table \ref{ugly_tab_OPEcoeffs_part2}.
    The value of $c_{45}{}^7$ is given in Table~\ref{ugly_tab_OPEcoeffs_part4}.
    \label{ugly_tab_OPEcoeffs_part7}}
\end{table}

%W5 W5
\begin{table}[H]
    \centering
    \begin{tabular}{| c |}
\hline  
\textbf{
\textit{
OPE coefficients in $W_5 \times W_5$  (partial)
}
}
\\ 
\hline 
\rule[-53mm]{0mm}{108mm}
\begin{minipage}{12cm}
\small
\begin{fleqn}
\begin{equation*} 
\ba 
c_{5 5}{}^{4}
& = 
\frac{5  \left(\nu ^3-32 \nu ^2-77 \nu +36\right) }{3 (\nu -9)
   \nu  (\nu +5) }
\frac{
c_{3 3}{}^{4}
g_5
}{
g_3
}    
\ , \\[2mm]
c_{5 5}{}^{6}
& = 
\frac{ (\nu -4) (\nu -1) \left(\nu
   ^2-59 \nu +16\right) }{36 (\nu -16) (\nu -9) \nu }
\frac{
c_{3 3}{}^{4} c_{3 4}{}^{5} c_{3 5}{}^{6}
g_5
}{
 ( g_3 )^2
}   
\ , 
\\[2mm]
% c_{5 5}{}^{\cC(3,1)}
c_{55}{}^{ \cC_{3,1}^{W_3 W_3}  }
& = 
-\frac{10 \left(5 \nu ^3-103 \nu ^2-460 \nu -864\right) }{3 (\nu -16) (\nu -9) \nu 
   (\nu +5) }
\frac{
g_5
}{
g_3
}   
\ , 
\\[2mm]
% c_{5 5}{}^{\cC(3,3)}
c_{55}{}^{ \cC_{3,3}^{W_3 W_3} }
& = 
\frac{15 \left(3 \nu ^3-24 \nu ^2-115 \nu +64\right) }{\nu  (\nu +5) \left(\nu ^2+15
   \nu +8\right) }
\frac{
g_5
}{
g_3
}   
\ , 
\\[2mm]
% c_{5 5}{}^{\cC(4,0)_\text{I}}
c_{55}{}^{ \cC_{4,0}^{W_3 W_3} }
& = 
-\frac{8 \left(\nu ^3-13 \nu ^2+210 \nu -558\right) }{(\nu -16) (\nu -9) (\nu -7) \nu 
}
\frac{
g_5
}{
   g_3
}   
\ , 
\\[2mm]
% c_{5 5}{}^{\cC(4,0)_\text{II}}
c_{55}{}^{  \cC_{4,0}^{W_4 W_4} }
& = 
-\frac{20 (\nu +5) }{(\nu -16) (\nu +1) }
\frac{
g_5
}{
g_4
}
\ , 
\\[2mm]
% c_{5 5}{}^{\cC(4,2)_\text{I}}
c_{55}{}^{ \cC_{4,2}^{W_3 W_3}  }
& =
{
\scriptstyle
\frac{8 \left(5 \nu ^7-226 \nu ^6-2904 \nu ^5+46954 \nu ^4+338011 \nu ^3+1286280 \nu ^2+7747320 \nu
   +3959280\right) }{7 (\nu -16) (\nu -15) (\nu -9) \nu  (\nu +5)^2 \left(\nu ^2+15
   \nu +8\right) }
\frac{
g_5
}{
g_3
}   
}
\ , \\[2mm]
% c_{5 5}{}^{\cC(4,2)_\text{II}}
c_{55}{}^{ \cC_{4,2}^{W_3 W_5}  }
& = 
-\frac{ (\nu -4) (\nu -2) (\nu -1) }{(\nu
   -16) (\nu -9) \nu  }
\frac{
c_{3 3}{}^{4} c_{3 4}{}^{5}
g_5
}{
(g_3 )^2
}   
\ , 
\\[2mm]
% c_{5 5}{}^{\cC(4,2)_\text{III}}
c_{55}{}^{ \cC_{4,2}^{W_4 W_4} }
& = 
-\frac{20 (\nu +5) }{(\nu -16) (\nu +1) }
\frac{
g_5
}{
g_4
}
\ea 
\end{equation*}
\end{fleqn}
\end{minipage}
\\
\hline
    \end{tabular}
    \caption{Values of the some of the OPE coefficients in the $W_5 \times W_5$ OPE.
    The values of $c_{33}{}^4$, $c_{34}{}^5$ are given in Table \ref{ugly_tab_OPEcoeffs_part1}.
    The value of $c_{35}{}^6$ is given in Table \ref{ugly_tab_OPEcoeffs_part2}.
    \label{ugly_tab_OPEcoeffs_part8}}
\end{table}

%constraints
\begin{table}[H]
    \centering
    \begin{tabular}{| c |}
\hline  
\textbf{
\textit{
Constraints among undetermined OPE coefficients
}
}
\\ 
\hline 
\rule[-61mm]{0mm}{124mm}
\begin{minipage}{12cm}
\small
\begin{fleqn}
\begin{equation*} 
\ba 
% c_{3 7}{}^{\cC(4,4)_\text{I}}
c_{37}{}^{ \cC_{4,4}^{W_3 W_5} }
& = 
\frac{ (\nu -9) (\nu -4) (\nu
   -1) \left(\nu ^2+35 \nu +84\right) }{100 (\nu -36) (\nu -25) (\nu +1) (\nu +5)
}
\frac{c_{3 4}{}^{5} c_{4 5}{}^{7}
g_7}{
   g_4 g_5
} 
% c_{5 5}{}^{\cC(4,4)_\text{I}}
c_{55}{}^{ \cC_{4,4}^{W_3 W_5} }
   \\[2mm] 
   & -\frac{ (\nu -4)
   (\nu -1) \left(\nu ^3-42 \nu ^2+29 \nu +396\right) }{2 (\nu -36) (\nu -25) \nu 
   (\nu +5)^2 }
\frac{
c_{3 3}{}^{4} c_{4 5}{}^{7}
g_7
}{
g_3 g_5
}   
\ , \\[2mm]
% c_{3 7}{}^{\cC(4,4)_\text{II}}
c_{37}{}^{ \cC_{4,4}^{W_4 W_4}  }
& = 
\frac{  (\nu -9) (\nu -4) (\nu
   -1) \left(\nu ^2+35 \nu +84\right) }{100 (\nu -36) (\nu -25) (\nu +1) (\nu +5)
  }
\frac{
c_{3 4}{}^{5} c_{4 5}{}^{7}
g_7
}{
 g_4 g_5
}   
% c_{5 5}{}^{\cC(4,4)_\text{I}}
c_{55}{}^{  \cC_{4,4}^{W_3 W_5} }
\\[2mm]
& -\frac{2  (\nu
   -9) (\nu -4) (\nu -3) (\nu -1) }{5 (\nu -36) (\nu -25) (\nu +1)^2
   }
\frac{
c_{3 4}{}^{5} c_{4 5}{}^{7}
g_7
}{
(g_4)^2
}   
\ , \\[2mm]
c_{4 6}{}^{8}
& = 
\frac{ (\nu -4) (\nu -1) \left(\nu
   ^2+15 \nu +8\right) }{105 (\nu -25) \nu  (\nu +5) }
\frac{
c_{3 5}{}^{6} c_{4 5}{}^{7}
g_6
}{
g_3
   g_5
}   
c_{3 7}{}^{8}
\ , 
\\[2mm]
% c_{4 6}{}^{\cC(4,4)_\text{I}}
c_{46}{}^{ \cC_{4,4}^{W_3 W_5} }
& = 
\frac{28  (\nu -4) (\nu -1) (\nu +1) (\nu +2)
}{3 (\nu -25) \nu ^2 (\nu +5)^2 }
\frac{
c_{3 3}{}^{4} c_{3 5}{}^{6}
   g_4 g_6
}{
( g_3 )^2
   g_5
}   
\\[2mm]
& +\frac{
   (\nu -4) (\nu -1) \left(\nu ^2+15 \nu +8\right) }{75
   (\nu -25) \nu  (\nu +5) }
\frac{
c_{3 4}{}^{5} c_{3 5}{}^{6}
g_6
}{
g_3 g_5
}   
% c_{5 5}{}^{\cC(4,4)_\text{I}}
c_{55}{}^{  \cC_{4,4}^{W_3 W_5}  }
\ , \\[2mm]
% c_{4 6}{}^{\cC(4,4)_\text{II}}
c_{46}{}^{ \cC_{4,4}^{W_4 W_4}  }
& = 
\frac{ (\nu -4) (\nu -1)
   \left(\nu ^2+15 \nu +8\right) }{75 (\nu -25) \nu  (\nu +5) }
\frac{
c_{3 4}{}^{5} c_{3 5}{}^{6} 
g_6
}{
g_3
   g_5
}   
% c_{5 5}{}^{\cC(4,4)_\text{II}}
c_{55}{}^{ \cC_{4,4}^{W_4 W_4}  }
\\[2mm]
& -\frac{4 (\nu -4) (\nu -1) (\nu +3)
}{15 (\nu -25) \nu  (\nu +1) }
\frac{
c_{3 4}{}^{5} c_{3 5}{}^{6} 
   g_6
}{
g_3 g_4
}   
\ , \\[2mm]
c_{5 5}{}^{8}
& = 
\frac{ (\nu -4) (\nu -1) (\nu +5)
  }{84 (\nu -16) (\nu +1) }
\frac{
c_{3 4}{}^{5} c_{4 5}{}^{7}
 g_5
}{
g_3 g_4
}
c_{3 7}{}^{8}
\ea 
\end{equation*}
\end{fleqn}
\end{minipage}
\\
\hline
    \end{tabular}
    \caption{Known constraints among OPE coefficients that are not fully determined at this stage of the bootstrap.
    The values of $c_{33}{}^4$, $c_{34}{}^5$ are given in Table \ref{ugly_tab_OPEcoeffs_part1}.
    The value of $c_{35}{}^6$ is given in Table \ref{ugly_tab_OPEcoeffs_part2}.
    The value of $c_{45}{}^7$ is given in Table \ref{ugly_tab_OPEcoeffs_part4}.
    \label{ugly_tab_OPEcoeffs_part9}}
\end{table}

\newpage

\subsection{Super Virasoro covariance
and parametrization of $\mathbb W_{p_1} \times \mathbb W_{p_2}$}
\label{app_superVir}

We recall that 
$\mathbb W_{p_1} \times \mathbb W_{p_2}$ is a collective
notation 
for the OPEs
$W_{p_1} \times W_{p_2}$, 
$W_{p_1} \times G_{W_{p_2}}$,
\dots ,
$T_{W_{p_1}} \times T_{W_{p_2}}$.
The RHSs of these OPEs
are organized into contributions
of super Virasoro primaries and their descendants.
The values of $p_1$, $p_2$ determine
restrictions on the quantum
numbers $h$, $j$ of a super Virasoro
primary operator $X$ that can
potentially enter the OPE
$\mathbb W_{p_1} \times \mathbb W_{p_2}$.
We have already described in 
Section \ref{sec_construct_ansatz}
which super Virasoro primaries
are allowed for given $p_1$, $p_2$.

Let $X$ denote one the  super Virasoro primary operators
that can  contribute to $\mathbb W_{p_1} \times \mathbb W_{p_2}$
($X$ can be short or long,
composite or not).
Let $\mathbb X$ denote the
$\mathfrak{psl}(2|2)$
multiplet of $X$.
Our first task is to 
construct a basis of
super Virasoro descendants of $X$,
organized as $\mathfrak{sl}(2)_z \oplus \mathfrak{sl}(2)_y$ primaries
with weight $h$ up to $h_{\rm max}:=\tfrac 12(p_1+p_2) +1$.
(This is because the singular part of the OPE
$T_{W_{p_1}} \times T_{W_{p_2}}$
can contain 
$\mathfrak{sl}(2)_z \oplus \mathfrak{sl}(2)_y$ primaries
up to weight $h_{\rm max}$.)
To this end, we first 
compile
a list denoted
$\mathcal L'_{X, h_{\rm max}}$
and defined as follows.
$\mathcal L'_{X, h_{\rm max}}$
consists of  
all nonvanishing
$\mathfrak{sl}(2)_z \oplus \mathfrak{sl}(2)_y$ primaries
of weight $\le h_{\rm max}$
that are of the form 
\be \label{eq_nested_words}
(A_r \dots ( A_3 ( A_2  ( A_1  B   )^{j_1}_{n_1})^{j_2}_{n_2} )^{j_3}_{n_3} \dots )^{j_r}_{n_r}
\ ,  \ 
r\ge0 \ , \
A_i \in \mathbb J \ , \ 
B \in \mathbb X \ , \ 
n_i \le 0  \ . 
\ee 
The notation $(AB)^j_n$ was defined in \eqref{eq_def_slP}.
We vary the spins $j_i$   in all possible ways compatible with
$\mathfrak{sl}(2)_y$ selection
rules. The case $r=0$
corresponds  to the
operators in the 
$\mathfrak{psl}(2|2)$
multiplet
$\mathbb X$ of $X$.

The list $\mathcal L'_{X, h_{\rm max}}$ furnishes
a complete set of
$\mathfrak{sl}(2)_z \oplus \mathfrak{sl}(2)_y$ primaries
descending from $X$ with
weight $\le h_{\rm max}$.
The operators in $\mathcal L'_{X, h_{\rm max}}$, however,
are not all linear independent.
The linear relations among operators
in $\mathcal L'_{X, h_{\rm max}}$
can be solved
in terms of a subset of $\mathfrak{sl}(2)_z \oplus \mathfrak{sl}(2)_y$ primaries
(the choice of this subset is not unique; we make an arbitrary choice).
The remaining primaries
form a basis $\mathcal L_{X,h_{\rm max}}$
of 
$\mathfrak{sl}(2)_z \oplus \mathfrak{sl}(2)_y$ primaries
descending from $X$ with
weight $\le h_{\rm max}$.

We can now consider
each of the 
$\mathbb W_{p_1} \times \mathbb W_{p_2}$
OPEs
and construct an Ansatz
for the contribution
of $X$ and its super Virasoro
descendants to the singular part 
of these OPEs. 
By virtue of 
$\mathfrak{sl}(2)_z \oplus \mathfrak{sl}(2)_y$
covariance, the sought-for Ans\"atze
are all of the form
\eqref{eq_covariant_OPE}.
Therefore, they can be written
compactly as 
\be  \label{eq_AB_ansatz}
A \times B = \sum_{\substack{
C \in \cL_{X,h_{\rm max}} \\
h_C < h_A + h_B \\
j_C \in \{ |j_A - j_B| , \dots, j_A+j_B\} \\
r_C = r_A + r_B
}} c_{AB}{}^C C  \ , \qquad
A \in \mathbb W_{p_1} \ , \quad 
B \in \mathbb W_{p_2} \ , 
\ee 
where, as customary at this point,
we have omitted
the $y_{12}$, $z_{12}$
powers and the differential
operators $\cD_{h_A, h_B;h_C}$,
$\widehat \cD_{j_A, j_B;j_C}$.
The OPE coefficients $c_{AB}{}^C$
are the free parameters in these
Ans\"atze.

Finally, 
we impose the vanishing of all operators in 
\be 
\text{Jacobi}(\mathbb J , \mathbb W_{p_1} , \mathbb W_{p_2}) \ . 
\ee 
This condition implements covariance
under super Virasoro 
(and therefore in particular
$\mathfrak{psl}(2|2)$ covariance).
All OPE coefficients $c_{AB}{}^C$
in the Ans\"atze 
\eqref{eq_AB_ansatz}
are fixed in terms
of the level $k$ and of the OPE coefficient $
c_{W_{p_1} W_{p_2}}{}^X \equiv 
c_{p_1 p_2}{}^X$
governing how the super Virasoro
primary $X$ enters the OPE
$W_{p_1} \times W_{p_2}$,
\be 
W_{p_1} \times W_{p_2} =
c_{p_1 p_2}{}^{X} X + \dots   \ . 
\ee 

For illustrative purposes
we present below the results of the 
procedure outlined above
in three examples.
In all examples, the super Virasoro
primary $X$ has $r$ charge zero.

\paragraph{Contribution
of the identity operator
to $\mathbb W_3 \times \mathbb W_3$.}
In the case $X=\one$,
the operator $B$ in \eqref{eq_nested_words}
is $\one$.
Below we list 
all independent
OPEs of the form 
$\mathbb W_3 \times \mathbb W_3$
and write the contribution
of $\one$ to them.
All OPE coefficients
are determined in terms of $k$
and of the two-point function
coefficient $g_3$.
\small
\begin{flalign}
W_3 \times W_3
& = g_3 \bigg[ 
\one 
-\frac{3 J }{k}+\frac{(2 k-1)
   T}{2 k (2 k+3)}
 -\frac{6 (J  J )_0^0}{k (2 k+3)}+\frac{3
   (J  J )_0^2}{(k-1) k}  
\bigg]     \ ,  & \nonumber
\\
W_3 \times G_{W_3}
& = g_3 \bigg[ 
-\frac{3 G}{2 k}
-\frac{6 (J  G)_0^{\frac{1}{2}}}{k (2 k+3)}+\frac{3
   (J  G)_0^{\frac{3}{2}}}{(k-1) k}
\bigg] \ ,    
\nonumber
\\ 
W_3 \times \widetilde{G}_{W_3}
&=  g_3 \bigg[ 
  -\frac{3 \widetilde{G}}{2 k}
-\frac{6 (J  \widetilde{G})_0^{\frac{1}{2}}}{k (2 k+3)}+\frac{3
   (J  \widetilde{G})_0^{\frac{3}{2}}}{(k-1) k}
\bigg] \  ,   \\
\nonumber
W_3 \times T_{W_3}
&= g_3 \bigg[ 
-\frac{2 J }{k} 
   +\frac{2
   (J  J )_0^2}{(k-1) k}  
    -\frac{5 (2 k-1)
   (J  J )_{-1}^1}{2 (k-1) k (2 k+3)}  
-\frac{(4 k+1) (G \widetilde{G})_0^1}{(k-1) k (2 k+3)}-\frac{(2 k-7)
   (J  T)_0^1}{(k-1) k (2 k+3)}
\bigg] \ ,    
\end{flalign}

\begin{flalign}
G_{W_3} \times G_{W_3} & = 
g_3 \bigg[ 
-\frac{6 (G G)_0^0}{k (2 k+3)}
\bigg]  \ ,  &  \nonumber
\\
G_{W_3} \times \widetilde{G}_{W_3}
& = g_3 \bigg[ 
3 \one
   -\frac{6 J }{k}+\frac{4 (k+1) T}{k (2 k+3)}
   -\frac{6
   (J  J )_0^0}{k (2 k+3)}+\frac{3 (J  J )_0^2}{(k-1)
   k}
-\frac{(2 k-7)
   (J  J )_{-1}^1}{(k-1) k (2 k+3)}
   \nonumber
\\
&
-\frac{6 (G \widetilde{G})_0^0}{k (2 k+3)}+\frac{2 (k+4)
   (G \widetilde{G})_0^1}{(k-1) k (2 k+3)}-\frac{2 (4 k+1)
   (J  T)_0^1}{(k-1) k (2 k+3)}
\bigg] \ , 
\nonumber
\\
G_{W_3} \times T_{W_3}
& = g_3 \bigg[ 
-\frac{4 G}{k}  
-\frac{4
   (J  G)_0^{\frac{1}{2}}}{k (2 k+3)}+\frac{5
   (J  G)_0^{\frac{3}{2}}}{(k-1) k}-\frac{6
   (G T)_0^{\frac{1}{2}}}{k (2 k+3)}
-\frac{9 (G J )_{-1}^{\frac{1}{2}}}{k (2 k+3)}  
\bigg] \ ,
\end{flalign}

\begin{flalign}
\widetilde G_{W_3} \times \widetilde{G}_{W_3}
& = g_3 \bigg[
-\frac{6 (\widetilde{G} \widetilde{G})_0^0}{k (2 k+3)}
\bigg] \  , \nonumber & 
\\
\widetilde{G}_{W_3} \times T_{W_3}
& = g_3 \bigg[ 
-\frac{4 \widetilde{G}}{k}
-\frac{4
   (J  \widetilde{G})_0^{\frac{1}{2}}}{k (2 k+3)}+\frac{5
   (J  \widetilde{G})_0^{\frac{3}{2}}}{(k-1) k}
   -\frac{6
   (\widetilde{G} T)_0^{\frac{1}{2}}}{k (2 k+3)}
-\frac{9 (J  \widetilde{G})_{-1}^{\frac{1}{2}}}{k (2 k+3)}
\bigg] \ , 
\end{flalign}

\begin{flalign}
T_{W_3} \times T_{W_3}
& = g_3 \bigg[ 
-8 \one 
   +\frac{8 J }{k}-\frac{4 (10 k+13)
   T}{3 k (2 k+3)}
    +\frac{8
   (J  J )_0^0}{k (2 k+3)}
      -\frac{8 (k+4)
   (J  J )_{-1}^1}{3 (k-1) k (2 k+3)}
   & \nonumber
\\
&
+ \frac{12 (G \widetilde{G})_{-1}^0}{k (2 k+3)}-\frac{4 (7 k+13)
   (G \widetilde{G})_0^1}{3 (k-1) k (2 k+3)}+\frac{40 (k+1)
   (J  T)_0^1}{3 (k-1) k (2 k+3)}
+\frac{27
   (J  (J  J )_{-1}^1 )_0^0}{5 k (2 k+3)}-\frac{6
   (T T)_0^0}{k (2 k+3)}
   \bigg] \ . 
\end{flalign}

\normalsize

\paragraph{Contribution
of a super Virasoro primary $X$
with $h=j=\tfrac 32$
to $\mathbb W_3 \times \mathbb W_4$.}
The $\mathfrak{psl}(2|2)$ multiplet
of $X$ is the short
multiplet 
$\mathbb X = \{ X, G_X, \widetilde G_X, T_X\}$.
Below we list 
all 
OPEs of the form 
$\mathbb W_3 \times \mathbb W_4$
and write the contribution
of $X$ to them.
All OPE coefficients
are determined in terms of $k$
and $c_{34}{}^X \equiv c_{W_3 W_4}{}^X$.
(In the W-algebra studied in this work,
the only short super Virasoro
primary $X$ with $h=j=\tfrac 32$ is $X =W_3$.)
\small

\begin{flalign}
W_3 \times W_4 & =c_{34}{}^X \bigg[
X
-\frac{(k-1)
   T_X}{12 (k+4)}
+ \frac{5 (J X)_0^{\frac{1}{2}}}{3 (k+4)}-\frac{2
   (J X)_0^{\frac{3}{2}}}{3 (2 k-1)}-\frac{2
   (J X)_0^{\frac{5}{2}}}{k-3}
\bigg]  \ , 
& \nonumber \\
W_3 \times G_{W_4} & =
c_{34}{}^X \bigg[ 
\frac{2 G_X}{3}
-\frac{2 (4 k+1) (G X)_0^1}{3 (k+4) (2 k-1)}-\frac{2 k
   (G X)_0^2}{(k-3) (2 k-1)}+\frac{5
   (J G_X)_0^0}{3 (k+4)}
   \nonumber
\\ 
&  +   \frac{(2 k-7)
   (J G_X)_0^1}{3 (k+4) (2 k-1)}-\frac{2 (4 k-3)
   (J G_X)_0^2}{3 (k-3) (2 k-1)}
\bigg]    \ , 
\nonumber
\\ 
W_3 \times \widetilde G_{W_4}
& = c_{34}{}^X \bigg[ 
\frac{2 \widetilde{G}_X}{3}
-\frac{2 (4 k+1) (\widetilde{G} X)_0^1}{3 (k+4) (2 k-1)}-\frac{2 k
   (\widetilde{G} X)_0^2}{(k-3) (2 k-1)}+\frac{5
   (J \widetilde{G}_X)_0^0}{3 (k+4)}
\\ 
\nonumber
&  
   +\frac{(2 k-7)
   (J \widetilde{G}_X)_0^1}{3 (k+4) (2 k-1)}-\frac{2 (4 k-3)
   (J \widetilde{G}_X)_0^2}{3 (k-3) (2 k-1)}
\bigg] \ ,    
\nonumber
\\ 
W_3 \times T_{W_4}
& =c_{34}{}^X \bigg[ 
X+\frac{5 (3 k+13) T_X}{48 (k+4)}
  +\frac{5
   (J X)_0^{\frac{1}{2}}}{12
   (k+4)}
   -\frac{(J X)_0^{\frac{3}{2}}}{3 (2 k-1)}
   -\frac{3
   (J X)_0^{\frac{5}{2}}}{2 (k-3)}
   \nonumber
\\ 
& 
+ \frac{(4 k+1) (G \widetilde{G}_X)_0^{\frac{1}{2}}}{3 (k+4) (2 k-1)}+\frac{(2 k+3)
   (5 k-1) (G \widetilde{G}_X)_0^{\frac{3}{2}}}{6 (k-3) (k+4) (2 k-1)}-\frac{(4
   k+1) (\widetilde{G} G_X)_0^{\frac{1}{2}}}{3 (k+4) (2 k-1)}
   \nonumber
\\ 
& -\frac{(2 k+3) (5
   k-1) (\widetilde{G} G_X)_0^{\frac{3}{2}}}{6 (k-3) (k+4) (2
   k-1)}+\frac{  (38 k^2-57 k+39  )
   (J X)_{-1}^{\frac{3}{2}}}{12 (k-3) (k+4) (2 k-1)}
   +\frac{10 k
   (J X)_{-1}^{\frac{1}{2}}}{3 (k+4) (2 k-1)}
\nonumber
\\
&
-\frac{2
   (J X)_{-1}^{\frac{5}{2}}}{3 (2 k-1)}
   +\frac{2 (k-1)
   (J T_X)_0^{\frac{1}{2}}}{(k+4) (2 k-1)}
   -\frac{(k-1) (2 k+15)
   (J T_X)_0^{\frac{3}{2}}}{2 (k-3) (k+4) (2 k-1)}+\frac{  (2
   k^2-19 k-3  ) (T X)_0^{\frac{3}{2}}}{2 (k-3) (k+4) (2
   k-1)}
\bigg]    \ , \nonumber
\end{flalign}

\begin{flalign}
G_{W_3} \times W_4
& =c_{34}{}^X \bigg[
\frac{G_X}{3}
+\frac{2 (k-1) (G X)_0^1}{(k+4) (2 k-1)}-\frac{2 (k-1)
   (G X)_0^2}{(k-3) (2 k-1)}-\frac{(4 k+1)
   (J G_X)_0^1}{3 (k+4) (2 k-1)}-\frac{4 k
   (J G_X)_0^2}{3 (k-3) (2 k-1)}
 \bigg] \   , 
\nonumber  
 \\
G_{W_3} \times G_{W_4}
& = c_{34}{}^X \bigg[ 
\frac{5 (G G_X)_0^{\frac{1}{2}}}{3 (k+4)}-\frac{2
   (G G_X)_0^{\frac{3}{2}}}{3 (2 k-1)}
\bigg] \ ,  
\nonumber  \\
G_{W_3} \times \widetilde G_{W_4}
& = c_{34}{}^X \bigg[ 
2
   X-\frac{5 (k+3) T_X}{12 (k+4)}
   +\frac{5
   (J X)_0^{\frac{1}{2}}}{3 (k+4)}-\frac{2
   (J X)_0^{\frac{3}{2}}}{3 (2 k-1)}-\frac{2
   (J X)_0^{\frac{5}{2}}}{k-3}
   \nonumber  
\\
&
+\frac{  (46 k^2-419 k+213  )
   (J X)_{-1}^{\frac{3}{2}}}{36 (k-3) (k+4) (2 k-1)}+\frac{10 (k-2)
   (J X)_{-1}^{\frac{1}{2}}}{9 (k+4) (2 k-1)}-\frac{2
   (J X)_{-1}^{\frac{5}{2}}}{3 (2 k-1)}
   \nonumber  
   \\
 &  
+\frac{2 (k-1) (G \widetilde{G}_X)_0^{\frac{1}{2}}}{(k+4) (2 k-1)}-\frac{(k-1) (2
   k+15) (G \widetilde{G}_X)_0^{\frac{3}{2}}}{2 (k-3) (k+4) (2
   k-1)}+\frac{  (2 k^2+35 k+3  ) (\widetilde{G} G_X)_0^{\frac{3}{2}}}{6
   (k-3) (k+4) (2 k-1)}
 +\frac{(4 k+1) (\widetilde{G} G_X)_0^{\frac{1}{2}}}{3
   (k+4) (2 k-1)}
   \nonumber  
\\
&
+\frac{(4 k+1)
   (J T_X)_0^{\frac{1}{2}}}{3 (k+4) (2 k-1)}+\frac{(2 k+3) (5 k-1)
   (J T_X)_0^{\frac{3}{2}}}{6 (k-3) (k+4) (2 k-1)}
  +\frac{3 (k-1) (2
   k+1) (T X)_0^{\frac{3}{2}}}{2 (k-3) (k+4) (2 k-1)}
\bigg] \  ,   
\nonumber  
\\ 
G_{W_3} \times T_{W_4}
& = c_{34}{}^X \bigg[ 
\frac{5 G_X}{3}
+\frac{5 (J G_X)_0^0}{3
   (k+4)}-\frac{5 (J G_X)_0^1}{2 (k+4) (2 k-1)}-\frac{(11 k-6)
   (J G_X)_0^2}{3 (k-3) (2 k-1)}
   \nonumber  
\\
&
+\frac{5
   (J G_X)_{-1}^0}{3 (k+4)}+\frac{2 (k-1)
   (J G_X)_{-1}^1}{(k+4) (2 k-1)}-\frac{2
   (J G_X)_{-1}^2}{3 (2 k-1)}
   -\frac{5 (k+1)
   (G X)_0^1}{3 (k+4) (2 k-1)}-\frac{(7 k-3)
   (G X)_0^2}{2 (k-3) (2 k-1)}
   \nonumber  
\\
&
+\frac{4 (7 k-2) (G X)_{-1}^1}{9 (k+4) (2 k-1)}-\frac{2
   (G X)_{-1}^2}{3 (2 k-1)}
+\frac{5
   (G T_X)_0^0}{3 (k+4)}+\frac{(2 k-7)
   (G T_X)_0^1}{3 (k+4) (2 k-1)}
+\frac{2 (4 k+1)
   (T G_X)_0^1}{3 (k+4) (2 k-1)}
\bigg]  \ , 
\end{flalign}

\begin{flalign}
\widetilde G_{W_3} \times W_4
&=  c_{34}{}^X \bigg[ 
\frac{\widetilde{G}_X}{3}
+ \frac{2 (k-1) (\widetilde{G} X)_0^1}{(k+4) (2 k-1)}-\frac{2 (k-1)
   (\widetilde{G} X)_0^2}{(k-3) (2 k-1)}-\frac{(4 k+1)
   (J \widetilde{G}_X)_0^1}{3 (k+4) (2 k-1)}-\frac{4 k
   (J \widetilde{G}_X)_0^2}{3 (k-3) (2 k-1)}
\bigg] \ ,  \nonumber \\
\widetilde G_3 \times G_{W_4}
& = c_{34}{}^X \bigg[ 
-2
   X+\frac{5 (k+3) T_X}{12 (k+4)}
     -\frac{5
   (J X)_0^{\frac{1}{2}}}{3 (k+4)}+\frac{2
   (J X)_0^{\frac{3}{2}}}{3 (2 k-1)}+\frac{2
   (J X)_0^{\frac{5}{2}}}{k-3}
    \nonumber
 \\
& 
-\frac{  (46 k^2-419 k+213  )
   (J X)_{-1}^{\frac{3}{2}}}{36 (k-3) (k+4) (2 k-1)}-\frac{10 (k-2)
   (J X)_{-1}^{\frac{1}{2}}}{9 (k+4) (2 k-1)}+\frac{2
   (J X)_{-1}^{\frac{5}{2}}}{3 (2 k-1)}
    \nonumber
   \\ 
 &  
+ \frac{  (2 k^2+35 k+3  ) (G \widetilde{G}_X)_0^{\frac{3}{2}}}{6 (k-3)
   (k+4) (2 k-1)}+\frac{(4 k+1) (G \widetilde{G}_X)_0^{\frac{1}{2}}}{3 (k+4) (2
   k-1)}+\frac{2 (k-1) (\widetilde{G} G_X)_0^{\frac{1}{2}}}{(k+4) (2
   k-1)} 
-\frac{(k-1) (2 k+15) (\widetilde{G} G_X)_0^{\frac{3}{2}}}{2 (k-3) (k+4)
   (2 k-1)}
    \nonumber
\\
&
 -\frac{(4 k+1)
   (J T_X)_0^{\frac{1}{2}}}{3 (k+4) (2 k-1)}-\frac{(2 k+3) (5 k-1)
   (J T_X)_0^{\frac{3}{2}}}{6 (k-3) (k+4) (2 k-1)}
  -\frac{3 (k-1) (2
   k+1) (T X)_0^{\frac{3}{2}}}{2 (k-3) (k+4) (2 k-1)}
\bigg] \ , 
 \nonumber
 \\
\widetilde G_3 \times \widetilde G_{W_4}
& = c_{34}{}^X \bigg[ 
\frac{5 (\widetilde{G} \widetilde{G}_X)_0^{\frac{1}{2}}}{3 (k+4)}-\frac{2
   (\widetilde{G} \widetilde{G}_X)_0^{\frac{3}{2}}}{3 (2 k-1)}
\bigg] \ ,   \nonumber
\\ 
\widetilde G_{W_3} \times T_{W_4}
& = c_{34}{}^X \bigg[ 
\frac{5 \widetilde{G}_X}{3}
   -\frac{5 (k+1)
   (\widetilde{G} X)_0^1}{3 (k+4) (2 k-1)}-\frac{(7 k-3)
   (\widetilde{G} X)_0^2}{2 (k-3) (2 k-1)}
+ \frac{4 (7 k-2) (\widetilde{G} X)_{-1}^1}{9 (k+4) (2 k-1)}
-\frac{2
   (\widetilde{G} X)_{-1}^2}{3 (2 k-1)}
    \nonumber
 \\
 &
  +\frac{5 (J \widetilde{G}_X)_0^0}{3
   (k+4)}-\frac{5 (J \widetilde{G}_X)_0^1}{2 (k+4) (2 k-1)}
     -\frac{(11 k-6)
   (J \widetilde{G}_X)_0^2}{3 (k-3) (2 k-1)}
   +\frac{2 (k-1)
   (J \widetilde{G}_X)_{-1}^1}{(k+4) (2 k-1)}-\frac{2
   (J \widetilde{G}_X)_{-1}^2}{3 (2 k-1)}
    \nonumber
\\
&
+\frac{5
   (\widetilde{G} T_X)_0^0}{3 (k+4)}+\frac{(2 k-7)
   (\widetilde{G} T_X)_0^1}{3 (k+4) (2 k-1)}
+\frac{10
   f  (J (J \widetilde{G}_X)_0^1  )_0^0}{9 (k+4)}+\frac{2 (4 k+1)
   (T \widetilde{G}_X)_0^1}{3 (k+4) (2 k-1)}
\bigg] \  ,   
\end{flalign}

\be 
\ba 
T_{W_3} \times W_4
&= c_{34}{}^X \bigg[ 
\frac{2
   X}{3}
   -\frac{2
   (J X)_0^{\frac{3}{2}}}{9 (2 k-1)}-\frac{4
   (J X)_0^{\frac{5}{2}}}{3 (k-3)}
   +\frac{  (98 k^2-247 k+69  )
   (J X)_{-1}^{\frac{3}{2}}}{36 (k-3) (k+4) (2 k-1)}-\frac{2
   (J X)_{-1}^{\frac{5}{2}}}{3 (2 k-1)}
\\
&
+ \frac{(k-1) (2 k+1) (G \widetilde{G}_X)_0^{\frac{3}{2}}}{2 (k-3) (k+4) (2
   k-1)}-\frac{(k-1) (2 k+1) (\widetilde{G} G_X)_0^{\frac{3}{2}}}{2 (k-3) (k+4)
   (2 k-1)}
   \\
  &
   +\frac{  (2 k^2-19 k-3  )
   (J T_X)_0^{\frac{3}{2}}}{6 (k-3) (k+4) (2 k-1)}+\frac{(k-1) (2
   k-13) (T X)_0^{\frac{3}{2}}}{2 (k-3) (k+4) (2 k-1)}
\bigg] \ ,   \\
T_{W_3} \times G_{W_4}
& = c_{34}{}^X \bigg[
   \frac{10 G_X}{9}  
    -\frac{10 (k+1)
   (J G_X)_0^1}{9 (k+4) (2 k-1)}-\frac{4 (7 k-3)
   (J G_X)_0^2}{9 (k-3) (2 k-1)}
    +\frac{(7 k-2)
   (J G_X)_{-1}^1}{3 (k+4) (2 k-1)}-\frac{2
   (J G_X)_{-1}^2}{3 (2 k-1)}
\\
&
   +\frac{10 (k-2)
   (G X)_0^1}{9 (k+4) (2 k-1)}-\frac{2 (5 k-3)
   (G X)_0^2}{3 (k-3) (2 k-1)}
+\frac{2 (13 k-8) (G X)_{-1}^1}{9 (k+4) (2 k-1)}-\frac{2
   (G X)_{-1}^2}{3 (2 k-1)}
 \\
 &
 +\frac{(4 k+1)
   (G T_X)_0^1}{3 (k+4) (2 k-1)}
   +\frac{2 (k-1)
   (T G_X)_0^1}{(k+4) (2 k-1)}
\bigg]  \  ,   \\ 
T_{W_3} \times \widetilde G_{W_4}
& = c_{34}{}^X \bigg[ 
\frac{10 \widetilde{G}_X}{9}
   -\frac{10 (k+1)
   (J \widetilde{G}_X)_0^1}{9 (k+4) (2 k-1)}-\frac{4 (7 k-3)
   (J \widetilde{G}_X)_0^2}{9 (k-3) (2 k-1)}
   +\frac{(7 k-2)
   (J \widetilde{G}_X)_{-1}^1}{3 (k+4) (2 k-1)}-\frac{2
   (J \widetilde{G}_X)_{-1}^2}{3 (2 k-1)}
 \\
 &
 +\frac{10 (k-2)
   (\widetilde{G} X)_0^1}{9 (k+4) (2 k-1)}-\frac{2 (5 k-3)
   (\widetilde{G} X)_0^2}{3 (k-3) (2 k-1)}
+\frac{2 (13 k-8) (\widetilde{G} X)_{-1}^1}{9 (k+4) (2 k-1)}-\frac{2
   (\widetilde{G} X)_{-1}^2}{3 (2 k-1)}
 \\
 &
 +\frac{(4 k+1)
   (\widetilde{G} T_X)_0^1}{3 (k+4) (2 k-1)}
+\frac{2 (k-1)
   (T \widetilde{G}_X)_0^1}{(k+4) (2 k-1)}
\   \\
T_{W_3} \times T_{W_4}
& = c_{34}{}^X \bigg[ 
-\frac{5 X}{2}+\frac{5
   (3 k+11) T_X}{9 (k+4)}
    -\frac{20
   (J X)_0^{\frac{1}{2}}}{9 (k+4)}+\frac{5
   (J X)_0^{\frac{3}{2}}}{9 (2 k-1)}
\\
&
  +\frac{5   (46 k^2+431
   k-237  ) (J X)_{-1}^{\frac{3}{2}}}{144 (k-3) (k+4) (2
   k-1)}
   -\frac{10 (k-4)
   (J X)_{-1}^{\frac{1}{2}}}{9 (k+4) (2 k-1)}
-\frac{40 (J X)_{-2}^{\frac{1}{2}}}{9 (k+4)}+\frac{10
   (J X)_{-2}^{\frac{3}{2}}}{9 (2 k-1)}   
\\
&
+\frac{5   (46 k^2+193 k-111  ) (G \widetilde{G}_X)_0^{\frac{3}{2}}}{72
   (k-3) (k+4) (2 k-1)}
   -\frac{2
   (k-11) (G \widetilde{G}_X)_0^{\frac{1}{2}}}{9 (k+4) (2 k-1)}
-\frac{5 (G \widetilde{G}_X)_{-1}^{\frac{1}{2}}}{3
   (k+4)}+\frac{2 (G \widetilde{G}_X)_{-1}^{\frac{3}{2}}}{3 (2 k-1)}   
 \\
 &
   -\frac{5   (46
   k^2+193 k-111  ) (\widetilde{G} G_X)_0^{\frac{3}{2}}}{72 (k-3) (k+4) (2
   k-1)}
   +\frac{2 (k-11)
   (\widetilde{G} G_X)_0^{\frac{1}{2}}}{9 (k+4) (2 k-1)}
+\frac{5 (\widetilde{G} G_X)_{-1}^{\frac{1}{2}}}{3 (k+4)}-\frac{2
   (\widetilde{G} G_X)_{-1}^{\frac{3}{2}}}{3 (2 k-1)}   
  \\
 &
-\frac{5   (18 k^2+53
   k-27  ) (J T_X)_0^{\frac{3}{2}}}{24 (k-3) (k+4) (2
   k-1)}
 -\frac{2 (3 k+2)
   (J T_X)_0^{\frac{1}{2}}}{3 (k+4) (2 k-1)}   
 -\frac{2
   (J T_X)_{-1}^{\frac{3}{2}}}{3 (2 k-1)}
   +\frac{5 (J T_X)_{-1}^{\frac{1}{2}}}{6 (k+4)}
  \\
  &
-\frac{5   (22 k^2+43
   k-33  ) (T X)_0^{\frac{3}{2}}}{24 (k-3) (k+4) (2
   k-1)}
 -\frac{2 (T X)_{-1}^{\frac{3}{2}}}{3 (2 k-1)}+\frac{5
   (T T_X)_0^{\frac{1}{2}}}{3 (k+4)}
\bigg] \ .    
\ea 
\ee

\normalsize

\paragraph{Contribution
of a super Virasoro primary $X$
with $(h,j)=(3,1)$
to $\mathbb W_5 \times \mathbb W_5$.}
Among the super Virasoro
descendants of $X$ we encounter the following
$\mathfrak{psl}(2|2)$ primary operators,
\be \label{eq_some_psl22_primaries}
\ba 
\mathcal A & :=
(JX)_0^0 - \frac 15  X_{[1,-1,0]} \ , & 
(h_\cA, j_\cA) &= (4,0)
\ , \\
\mathcal B & := 
(JX)_0^1 + \frac 14 X'_{[1,0,0]} - \frac 14  X_{[1,0,0]} \ , 
& 
(h_\cB, j_\cB) &= (4,1)
\ , 
\\
\cC & :=
(JX)_0^2 - \frac 12  X_{[1,1,0]}
\ , & 
(h_\cC, j_\cC) &= (4,2) \ .
\ea
\ee
On the RHSs we are using the notation
for $\mathfrak{psl}(2|2)$ descendants
of $X$ introduced in 
Appendix \ref{app_notation}.
Below we list 
all independent
OPEs of the form 
$\mathbb W_5 \times \mathbb W_5$
and write the contribution
of $X$ to them.
We find it convenient to write the
result in terms of the super Virasoro primary $X$,
the $\mathfrak{psl}(2|2)$ primaries
$\cA$, $\cC$, and the 
$\mathfrak{psl}(2|2)$ descendants of
$X$, $\cA$, $\cC$.
The $\mathfrak{psl}(2|2)$ descendants
of $\cA$,
$\cC$ are computed
from \eqref{eq_some_psl22_primaries} using the $G^\downarrow$,
$G^\uparrow$, $\widetilde G^\downarrow$,
$\widetilde G^\uparrow$ operations.
All OPE coefficients
are determined in terms of $k$
and $c_{55}{}^X \equiv c_{W_5 W_5}{}^X$.

\small
\begin{flalign}
W_5 \times W_5
& = c_{55}{}^X \bigg[ 
X + \frac{1}{10}  X_{[1,-1,0]} 
- \frac 12 X_{[1,1,0]}
+ \frac{3 \cA }{k+3}
- \frac{3 \cC }{k-3}
\bigg] \ , & \nonumber  \\
W_5 \times G_{W_5}
& = c_{55}{}^X \bigg[ 
X_{[\frac 12 , \frac 12 , \frac 12]}
+
\frac 12 X_{[\frac 12 , - \frac 12 , \frac 12]}
- \frac{1}{20} X_{[\frac 32 , - \frac 12 , \frac 12]}
- \frac 14 X_{[\frac 32, \frac 12, \frac 12]}
\nonumber 
\\
&
+ \frac{3 \cA_{[\frac 12, \frac  12 , \frac 12]} }{k+3}
- \frac{3 \cC_{[\frac 12 , - \frac 12 , \frac 12]}}{2(k-3)}
- \frac{3 \cC_{[\frac 12 , \frac 12 , \frac 12]}}{k-3}
\bigg]
 \ , \nonumber 
 \\
W_5 \times \widetilde G_{W_5}
& = 
c_{55}{}^X \bigg[ 
X_{[\frac 12 , \frac 12 , -\frac 12]}
+
\frac 12 X_{[\frac 12 , - \frac 12 , -\frac 12]}
+ \frac{1}{20} X_{[\frac 32 , - \frac 12 , -\frac 12]}
+ \frac 14 X_{[\frac 32, \frac 12, -\frac 12]}
\nonumber 
\\
&
+ \frac{3 \cA_{[\frac 12, \frac  12 , - \frac 12]} }{k+3}
- \frac{3 \cC_{[\frac 12 , - \frac 12 , -\frac 12]}}{2(k-3)}
- \frac{3 \cC_{[\frac 12 , \frac 12 ,  -\frac 12]}}{k-3}
\bigg] 
 \ , \nonumber 
 \\ 
W_5 \times T_{W_5}
& = c_{55}{}^X 
\bigg[ 
+ \frac 25 X 
+ \frac{1}{20} X_{[2,0,0]}
+ \frac 12 X_{[1,0,0]}
- \frac 12 X'_{[1,0,0]}
- \frac 65 X_{[1,1,0]}
\nonumber 
\\
& 
- \frac{3  \cA_{[1,1,0]}}{k+3}
- \frac{6 \cC }{5(k-3)}
+ \frac{\cC_{[1,-1,0]}}{2(k-3)}
+ \frac{3 \cC_{[1,0,0]} }{2(k-3)}
- \frac{3 \cC'_{[1,0,0]} }{2(k-3)} 
+ \frac{3  \cC_{[1,1,0]}}{k-3}
\bigg]  \ , 
\end{flalign}

\begin{flalign}
G_{W_5} \times G_{W_5}
& = c_{55}{}^X \bigg[ 
X_{[1,0,1]}
+ \frac{3 \cA_{[1,0,1]}}{k+3}
- \frac{3 \cC_{[1,0,1]}}{k-3}
\bigg]
\ ,   &  \nonumber
\\
G_{W_5} \times \widetilde G_{W_5}
& = c_{55}{}^X \bigg[
2X
+ \frac 12 X_{[1,0,0]}
+ \frac 12 X'_{[1,0,0]}
+ \frac 35 X_{[1,-1,0]}
- \frac 15 X_{[2,0,0]}
- \frac 32 X_{[1,1,0]}
\nonumber
\\
&
+ \frac{3 \cA }{ k+3}
+ \frac{3 \cA_{[1,0,0]} }{k+3}
- \frac{3  \cA_{[1,1,0]}}{k+3}
\nonumber
\\
&
- \frac{3 \cC }{k-3}
- \frac{\cC_{[1,-1,0]}}{k-3}
- \frac{3 \cC_{[1,0,0]}}{2(k-3)}
- \frac{3 \cC'_{[1,0,0]}}{2(k-3)}
+ \frac{3 \cC_{[1,1,0]}}{k-3}
\bigg]  \ , 
\nonumber
\\
G_{W_5} \times T_{W_5}
& = c_{55}{}^X \bigg[
 \frac 65 X_{[\frac 12 , - \frac 12 , \frac 12]}
+ \frac 65 X_{[\frac 12, \frac 12, \frac 12]}
- \frac{27}{50} X_{[\frac 32, - \frac 12, \frac 12]}
- \frac{27}{20} X_{[\frac 32, \frac 12, \frac 12]}
\nonumber
\\
&
+ \frac{12 \cA_{[\frac 12, \frac 12, \frac 12]} }{5(k+3)}
- \frac{3  \cA_{[\frac 32,  \frac 12, \frac 12]}}{k+3}
- \frac{21 \cC_{[\frac 12, - \frac 12 , \frac 12]}  }{10(k-3)}
- \frac{6 \cC_{[\frac 12, \frac 12, \frac 12]} }{5(k-3)}
+ \frac{3 \cC_{[\frac 32, - \frac 12, \frac 12]} }{2(k-3)}
+ \frac{3 \cC_{[\frac 32, \frac 12, \frac 12
]}}{k-3}
\bigg] \ , 
\end{flalign}

\begin{flalign}
\widetilde G_{W_5} \times \widetilde G_{W_5}
& = c_{55}{}^X \bigg[ 
X_{[1,0,-1]}
+ \frac{3 \cA_{[1,0,-1]}}{k+3}
- \frac{3 \cC_{[1,0,-1]} }{k-3}
\bigg]  \ , & \nonumber   \\ 
\widetilde G_{W_5} \times
T_{W_5}
& = c_{55}{}^X \bigg[ 
 \frac 65 X_{[\frac 12, - \frac 12, - \frac 12]}
+ \frac 65 X_{[\frac 12, \frac 12, - \frac 12]}
+ \frac{27}{50} X_{[\frac 32, - \frac 12, - \frac 12]}
+ \frac{27}{20} X_{[\frac 32, \frac 12, - \frac 12]}
\nonumber
\\
&
+ \frac{12 \cA_{[\frac 12, \frac 12, - \frac 12]} }{5(k+3)}
+ \frac{3 \cA_{[\frac 32, \frac 12, - \frac 12]} }{k+3}
- \frac{21 \cC_{[\frac 12, - \frac 12, - \frac 12]} }{10(k-3)}
- \frac{6 \cC_{[\frac 12, \frac 12, - \frac 12]} }{5(k-3)}
- \frac{3 \cC_{[\frac 32, - \frac 12, - \frac 12]} }{2(k-3)}
- \frac{3 \cC_{[\frac 32, \frac 12, - \frac 12]} }{k-3}
\bigg] \ , 
\nonumber
\\
&
\end{flalign}

\begin{flalign}
T_{W_5} \times T_{W_5}
& = c_{55}{}^X \bigg[
- \frac{108}{25} X
- \frac{54}{25} X_{[1,-1,0]}
+ \frac{54}{25} X_{[1,1,0]}
+ \frac{36}{25} X_{[2,0,0]}
- \frac{24 \cA }{5(k+3)}
+ \frac{18 \cA_{[1,1,0]} }{5(k+3)}
+ \frac{3 \cA_{[2,0,0]} }{k+3}
\nonumber &
\\
&
+ \frac{84 \cC }{25(k-3)}
+ \frac{12 \cC_{[1,-1,0]} }{5(k-3)}
- \frac{6 \cC_{[1,1,0]} }{5(k-3)}
- \frac{3 \cC_{[2,0,0]} }{k-3}
\bigg] \ . 
\end{flalign}

\normalsize

\subsection{Redefinitions of $W_6$, $W_7$}
\label{app_redef}

In this appendix we study the effect of field redefinitions
of the form
\be \label{eq_W6W7redef}
W_6' = W_6 + \mu_6 \cC^{W_3 W_3}_{3,3} \ , \qquad 
W_7' = W_7 + \mu_7 \cC^{W_3 W_4}_{\frac 72, \frac 72} \ , 
\ee 
on the OPE coefficients in Table
\ref{tab_ansatz}. For simplicity, we analyze the two
redefinitions in \eqref{eq_W6W7redef}
separately in turn.

\subsubsection{Redefinition of $W_6$}

Upon redefining $W_6$ as in 
\eqref{eq_W6W7redef},
we induce a simple shift in 
all OPE coefficients of the form
$c_{pq}{}^{\cC^{W_3 W_3}_{3,3}}$
with $p,q\neq 6$,
\be 
( c_{pq}{}^{\cC^{W_3 W_3}_{3,3}} )' = 
c_{pq}{}^{\cC^{W_3 W_3}_{3,3}}
- c_{pq}{}^6 \mu_6 \ . 
\ee 
We also have a shift on OPE coefficients
of the form $c_{p6}{}^q$. More explicitly, we compute
\be 
\ba 
(c_{36}{}^{3})'   & = 
c_{36}{}^{3}+\frac{g_3 (2
   k-1) (k (2 k-17)+12) \mu _6}{(k-4) (k-2)
   (2
   k+3)} \ , 
   \\
(c_{36}{}^{5})'  & = c_{36}{}^{5}+\frac{c_{33}{}^4 c_{34}{}^5 k
   \mu _6}{2
   (k-4)} \ , 
   \\
(c_{36}{}^{7})'    & = c_{36}{}^{7} \ ,  
   \\
   (  c_{46}{}^4)'
   & = 
   c_{46}{}^4
   -\frac{24
   g_3 (k-1) \mu _6}{k^2-6
   k+8} \ , 
   \\ 
   (  c_{46}{}^6)'
   & = 
   c_{46}{}^6
   +\frac{3
   c_{34}{}^5
   c_{35}{}^6 k \mu _6}{5
   (k-4)} \ ,  
   \\
   (c_{46}{}^8)'
   & = 
   c_{46}{}^8 \ . 
\ea 
\ee 
Finally, we have shifts in the OPE coefficients of the form
$c_{p6}{}^{\text{composite}}$,
\small
\be 
\ba 
(c_{36}{}^{  \cC^{W_3 W_4}_{\frac 72, \frac 32}})'
& = 
   c_{36}{}^{  \cC^{W_3 W_4}_{\frac 72, \frac 32}}+\frac{c_{33}{}^4 (k-1) \mu _6}{3
   (k-4)} \ , 
   \\
(   c_{36}{}^{  \cC^{W_3 W_4}_{\frac 72, \frac 52} }  )'
   & = 
   c_{36}{}^{  \cC^{W_3 W_4}_{\frac 72, \frac 52} }
   -\frac{c_{33}{}^4 (k-6) \mu_6}
   {k-4} \ , 
   \\
 (c_{36}{}^{  \cC^{W_3 W_4}_{\frac 72, \frac 72} })' 
 & = c_{36}{}^{  \cC^{W_3 W_4}_{\frac 72, \frac 72} }
   +\frac
   {c_{33}{}^4 (2 k-5) \mu
   _6}{k-4} \ , 
   \\
(c_{37}{}^{  \cC^{W_3 W_3}_{3,3} })'   
   & = c_{37}{}^{  \cC^{W_3 W_3}_{3,3} }
   - c_{37}{}^6   \mu_6 \ , 
   \\ 
( c_{37}{}^{  \cC^{W_3 W_3}_{4,2} })'
   & = 
   c_{37}{}^{  \cC^{W_3 W_3}_{4,2} }
   +\frac{
   8 c_{35}{}^6
   c_{37}{}^{  \cC^{W_3 W_5}_{4,2} } k \mu
   _6}{315
   (k+7)}+\frac{c_{37}{}^{  \cC^{W_4 W_4}_{4,2} } c_{44}{}^6 k \mu _6}{35
   (k+7)} \  , 
   \\
(c_{55}{}^{  \cC^{W_3 W_3}_{3,3} })'
& = c_{55}{}^{  \cC^{W_3 W_3}_{3,3} }
   -  c_{55}{}^6   \mu_6 \ , 
   \\
(  c_{55}{}^{  \cC^{W_3 W_3}_{4,2} })' 
& = 
   c_{55}{}^{  \cC^{W_3 W_3}_{4,2} }
   +\frac{
   8 c_{35}{}^6
   c_{55}{}^{  \cC^{W_3 W_5}_{4,2} } k \mu
   _6}{315
   (k+7)}+\frac{c_{44}{}^6
   c_{55}{}^{  \cC^{W_4 W_4}_{4,2} } k \mu
   _6}{35
   (k+7)} \ , 
   \\
 (c_{46}{}^{  \cC^{W_3 W_3}_{3,3} })'   & = 
 -c_{46}{}^6 \mu
   _6+c_{46}{}^{  \cC^{W_3 W_3}_{3,3} }-\frac{1
   8 g_3 (k+4) (4 k+1) \mu
   _6}{c_{33}{}^{4} (k-1) k (2
   k+3)}+\frac{3 c_{33}{}^{4}
   c_{44}{}^{  \cC^{W_3 W_3}_{3,3} } k \mu _6}{4
   (k-4)}-\frac{3 c_{34}{}^{5}
   c_{35}{}^{6} k \mu _6^2}{5
   (k-4)} \ , 
\ea 
\ee 
\be
\ba
 (   c_{46}{}^{  \cC^{W_3 W_3}_{4,2} })'  & = 
 c_{46}{}^{ \cC^{W_3 W_3}_{4,2} }-\frac{36 g_3
   \left(20 k^4-76 k^3-1441 k^2+2917
   k+12980\right) \mu _6}{35
   c_{33}{}^{4} (k-4) (k-3) k
   (k+7) (2 k+3)}+\frac{2
   c_{33}{}^{4}
   c_{44}{}^{ 6 } k \mu _6^2}{35
   (k-4) (k+7)}
   \\
   & -\frac{2
   c_{33}{}^{4}
   c_{44}{}^{  \cC^{W_3 W_3}_{3,3} } (k-1) k \mu
   _6}{105 (k-4) (k+7)}+\frac{8
   c_{34}{}^{5}
   c_{35}{}^{6} k \mu _6^2}{525
   (k+7)}+\frac{8 c_{35}{}^{6}
   c_{46}{}^{ \cC^{W_3 W_5}_{4,2} } k \mu
   _6}{315
   (k+7)}+\frac{c_{44}{}^{ 6 }
   c_{46}{}^{ \cC^{W_4 W_4}_{4,2} } k \mu
   _6}{35 (k+7)} 
   \ , 
    \\
( c_{46}{}^{  \cC^{W_3 W_3}_{3,1} })'   
   & = 
   c_{46}{}^{  \cC^{W_3 W_3}_{3,1} }
   -\frac{12 g_3 (k-2) (2 k+1) \mu
   _6}{c_{33}{}^4 (k-4) k (2
   k+3)} \ , 
   \\
 (c_{46}{}^{  \cC^{W_3 W_5}_{4,1} })'   
   & = 
   c_{46}{}^{  \cC^{W_3 W_5}_{4,1} }
   -\frac{c_{34}{}^5 \mu_6}{5} \ , 
   \\
 ( c_{46}{}^{  \cC^{W_4 W_4}_{4,4} })'  
   & = 
   c_{46}{}^{  \cC^{W_4 W_4}_{4,4} }
   +\frac{4 c_{33}{}^4 \mu
   _6}{k-4} \ , 
   \\
(   c_{46}{}^{  \cC^{W_4 W_4}_{4,2} })'   
   & = 
   c_{46}{}^{  \cC^{W_4 W_4}_{4,2} }
   +\frac{2 c_{33}{}^4 \mu
   _6}{k-4} \ , 
   \\
(  c_{46}{}^{  \cC^{W_3 W_5}_{4,2} })'   
   & =
   c_{46}{}^{  \cC^{W_3 W_5}_{4,2} }
  + \frac{3 c_{34}{}^5
   \mu
   _6}{5}
    \ , 
    \\
(  c_{46}{}^{  \cC^{W_3 W_5}_{4,4} })'    
    & = 
    c_{46}{}^{  \cC^{W_3 W_5}_{4,4} }
    + 2
   c_{34}{}^5 \mu
   _6 \ , 
   \\
 (c_{46}{}^{  \cC^{W_3 W_5}_{4,3} })'  
   & = 
   c_{46}{}^{  \cC^{W_3 W_5}_{4,3} }
   -\frac{6
   c_{34}{}^5 \mu _6}{5} \ .
\ea 
\ee 
\normalsize
In deriving these relations, we have to take into account the fact that 
a redefinition of $W_6$ induces also a redefinition of the composite operators
$\cC^{W_4 W_4}_{4,2}$, $\cC^{W_3 W_5}_{4,2}$,
as can be checked using the expressions
reported in Appendix \ref{app_composites}.

\subsubsection{Redefinition of $W_7$}

Upon redefining $W_7$ as in 
\eqref{eq_W6W7redef},
we induce a simple shift in 
all OPE coefficients of the form
$c_{pq}{}^{\cC^{W_3 W_4}_{\frac 72, \frac 72}}$
with $p,q\neq 7$,
\be 
( c_{pq}{}^{\cC^{W_3 W_4}_{\frac 72, \frac 72}} )' = 
c_{pq}{}^{\cC^{W_3 W_4}_{\frac 72, \frac 72}}
- c_{pq}{}^7 \mu_7 \ . 
\ee 
We also have a shift on OPE coefficients
of the form $c_{p7}{}^q$. More explicitly, we compute
\be 
\ba 
(c_{37}{}^4)' & = 
c_{37}{}^4
+\frac{g_3 \left(2 k^3-39 k^2+97 k-60\right) \mu _7}{2 k^3-15 k^2+13
   k+60} \ , 
   \\
   (   c_{37}{}^6)'
   & = 
   c_{37}{}^6
   +\frac{2 c_{34}{}^5 c_{35}{}^6 k \mu _7}{5
   (k-5)} \ , \\ 
( c_{37}{}^8 )' & = c_{37}{}^8  \ . 
\ea 
\ee 
Finally, the shift of OPE coefficients
of the form $c_{p7}{}^{\text{composite}}$
are
\be 
\ba 
( c_{37}{}^{ \cC^{W_4 W_4}_{4,2} })'
& = 
  c_{37}{}^{ \cC^{W_4 W_4}_{4,2} }
+\frac{c_{33}{}^4 \mu
   _7}{6}
 \ , \qquad 
 ( c_{37}{}^{ \cC^{W_4 W_4}_{4,4} })' 
  = 
 c_{37}{}^{ \cC^{W_4 W_4}_{4,4} }
 + c_{33}{}^4 \mu_7 \ , 
 \\
 ( c_{37}{}^{ \cC^{W_3 W_5}_{4,2} })'
 & = c_{37}{}^{ \cC^{W_3 W_5}_{4,2} }
 +\frac{c_{34}{}^5 (k+5) \mu
   _7}{10 (k-5)} \ ,
   \\ 
   ( c_{37}{}^{ \cC^{W_3 W_3}_{4,2} })'
 &  = 
 c_{37}{}^{ \cC^{W_3 W_3}_{4,2} } 
{\scriptstyle  -\frac{6 g_3 \left(4 k^5-76 k^4-85 k^3+2940 k^2-10507 k+24300\right) \mu
   _7}{7 c_{33}{}^4 k \left(2 k^5-7 k^4-89 k^3+427 k^2-33
   k-1260\right)}
   }
   -\frac{c_{33}{}^4 c_{44}{}^{ \cC^{W_3 W_3}_{3,3} } (k-1) k \mu _7}{126 \left(k^2+2
   k-35\right)} \ , 
   \\
   ( c_{37}{}^{ \cC^{W_3 W_5}_{4,3} })'
   &=
   c_{37}{}^{ \cC^{W_3 W_5}_{4,3} }
   -\frac{2 c_{34}{}^5 (k-10) \mu _7}{5
   (k-5)}
 \ ,\qquad 
     ( c_{37}{}^{ \cC^{W_3 W_5}_{4,4} })'
    = 
  c_{37}{}^{ \cC^{W_3 W_5}_{4,4} }
  +\frac{c_{34}{}^5 (k-2) \mu
   _7}{k-5} \ . 
\ea 
\ee

% \be 
% \ba 
% &
% c_{37}{}^4+\frac{g_3 \left(2 k^3-39 k^2+97 k-60\right) \mu _7}{2 k^3-15 k^2+13
%    k+60} \ , \\ &c_{37}{}^6+\frac{2 c_{34}{}^5 c_{35}{}^6 k \mu _7}{5
%    (k-5)} \ , \\ 
% & 
% c_{37}{}^{ \cC^{W_3 W_3}_{3,3} }-\frac{18 g_3 \left(2 k^3-9 k^2-18 k+200\right) \mu
%    _7}{c_{33}{}^4 k (2 k+3) \left(k^2-9 k+20\right)}+\frac{c_{33}{}^4
%    c_{44}{}^{ \cC^{W_3 W_3}_{3,3} } k \mu _7}{2 (k-5)}
%    \\
% &\text{$\mathit{c}$378} \ , \\ &\frac{c_{33}{}^4 \mu
%    _7}{6}+c_{37}{}^{ \cC^{W_4 W_4}_{4,2} } \ , \\ &c_{33}{}^4 \mu
%    _7+c_{37}{}^{ \cC^{W_4 W_4}_{4,4} } \ , \\ &c_{37}{}^{ \cC^{W_3 W_5}_{4,2} }+\frac{c_{34}{}^5 (k+5) \mu
%    _7}{10 (k-5)} \ , \\ 
%  & 
%  c_{37}{}^{ \cC^{W_3 W_3}_{4,2} }-\frac{6 g_3 \left(4 k^5-76 k^4-85 k^3+2940 k^2-10507 k+24300\right) \mu
%    _7}{7 c_{33}{}^4 k \left(2 k^5-7 k^4-89 k^3+427 k^2-33
%    k-1260\right)}-\frac{c_{33}{}^4 c_{44}{}^{ \cC^{W_3 W_3}_{3,3} } (k-1) k \mu _7}{126 \left(k^2+2
%    k-35\right)}
%   \\ &c_{37}{}^{ \cC^{W_3 W_5}_{4,4} }+\frac{c_{34}{}^5 (k-2) \mu
%    _7}{k-5} \ , \\ &c_{37}{}^{ \cC^{W_3 W_5}_{4,3} }-\frac{2 c_{34}{}^5 (k-10) \mu _7}{5
%    (k-5)} \ , \\  
% \ea 
% \ee 

\subsection{Composite super Virasoro primary operators}
\label{app_composites}

In this appendix we give 
more details on the 
composite
super Virasoro primaries
introduced schematically in 
\eqref{eq_composite_schematic}.

The lightest 
super Virasoro composites 
have weight $3$
and are constructed
starting from normal ordered products of $W_3$ with itself. 
Their explicit expressions read
\be 
\ba 
\cC_{3,1}^{W_3 W_3}
& = (W_3 W_3)_0^1
+ c_{33}{}^4 \bigg[
\frac{k  T_{W_4} }{15
   (k+5)}-\frac{5 (J W_4)_0^1}{3
   (k+5)}
\bigg]   
+ g_3 \bigg[ \frac{(k-3) (2 k-1) (G \widetilde{G} )_0^1}{2 (k-1) k
   (2 k+1) (2 k+3)}
\\
&
   +\frac{\left(10 k^2-77 k-48\right)
   (J J)_{-1}^1}{5 (k-1) k (2 k+1) (2 k+3)}+\frac{(2 k-1) (3 k-2)
   (J T)_0^1}{(k-1) k (2 k+1) (2 k+3)}-\frac{3 (26 k-1)
    (J (J J)_0^0)_0^1}{5 (k-1) k (2 k+1) (2
   k+3)}\bigg]
 \ , 
 \\ 
\cC_{3,3}^{W_3 W_3}
& = 
(W_3 W_3)_0^3
+ c_{33}{}^4
\frac{ 
   (J W_4)_0^3}{k-4}
+ g_3 
\frac{ 
(J(J J)_0^2)_0^3}{(k-2) (k-1)
   k} \ . 
\ea 
\ee 
The operator $\cC_{3,1}^{W_3 W_3}$ is constructed
using the $\mathfrak{sl}(2)_z \oplus \mathfrak{sl}(2)_y$ primary
$(W_3 W_3)_0^1$
as ``seed''.
Next, we identify the super Virasoro primaries
that enter the singular part of the
$W_3 \times W_3$ OPE: they are $\one$ (with coefficient
$g_3$) and $W_4$ (with coefficient $c_{33}{}^4$).
We construct an Ansatz by enumerating
$\mathfrak{sl}(2)_z \oplus \mathfrak{sl}(2)_y$ primaries that are 
super Virasoro descendants of $\one$
and $W_4$ with the same quantum numbers 
as $(W_3 W_3)_0^1$.
The coefficients in the Ansatz are fixed
by demanding that
$\cC_{3,1}^{W_3 W_3}$ be a super Virasoro primary.
The operator $\cC_{3,3}^{W_3 W_3}$ is constructed
using the same strategy.

The same method yields
the composite super Virasoro primary
operators $\cC_{4,0}^{W_3 W_3}$,
$\cC_{4,2}^{W_3 W_3}$.
They are constructed from the ``seeds''
$(W_3 W_3)_{-1}^0$, 
$(W_3 W_3)_{-1}^2$, respectively.
More explicitly, they are given as
\small
\be 
\ba 
&
\cC_{4,0}^{W_3 W_3}
 =
(W_3 W_3)_{-1}^0
-\frac{(k-1)
   (G_{W_3} \widetilde{G}_{W_3})_0^0}{4 (k+4)}-\frac{5
   (J (W_3 W_3)_0^1 )_0^0}{2
   (k+4)}
 + c_{33}{}^4 \bigg[ \frac{25
   (J (J W_4)_0^1 )_0^0}{12 (k+4)^2}-\frac{(k-1)
   (J T_{W_4})_0^0}{6 (k+4)^2} 
   \bigg]   
   \\ 
&   
+ g_3 \bigg[ 
-\frac{ (42 k^2+137 k+20 )
   (J (\widetilde{G} G)_0^1 )_0^0}{4 k (k+4) (2 k+3) (2 k+5)
   (3 k+4)}-\frac{ (6 k^2+15 k-20 ) (G \widetilde{G})_{-1}^0}{2 k (2
   k+3) (2 k+5) (3 k+4)}
\\
& -\frac{ (33 k^2+73 k-20 )
   (J (T J)_0^1 )_0^0}{k (k+4) (2 k+3) (2 k+5) (3
   k+4)}+\frac{ (54 k^4+39 k^3+2980 k^2+11097 k+7880 )
   (J (J J)_{-1}^1 )_0^0}{40 (k-1) k (k+4) (2
   k+3) (2 k+5) (3 k+4)}
\\
&
+\frac{3  (129 k^2+341 k+20 )
   (J (J (J J)_0^0 )_0^1 )_0^0
   }{4 (k-1) k (k+4) (2 k+3) (2 k+5) (3 k+4)}+\frac{(2 k-1)  (k^2-6 k-20 )
   (T T)_0^0}{4 k (k+4) (2 k+3) (2 k+5) (3 k+4)}  \bigg] \ , 
\ea
\ee

\be 
\ba 
& 
\cC_{4,2}^{W_3 W_3}
  =
(W_3 W_3)_{-1}^2
 -\frac{  (2 k^2+3
   k-41  ) (G_{W_3} \widetilde{G}_{W_3})_0^2}{4   (k^2+5 k-17  )}
 +\frac{7
   (k+7)   (J (W_3 W_3)_0^1  )_0^2}{4   (k^2+5
   k-17  )}-\frac{63 (k-3)
     (J (W_3 W_3)_0^3  )_0^2}{20   (k^2+5
   k-17  )}
\\
& -\frac{3   (k^2+5 k-38  ) (W_3 T_{W_3})_0^2}{8
     (k^2+5 k-17  )}
+ c_{33}{}^4 \bigg[    -\frac{  (6 k^3-21 k^2+161 k-106  )
   (G \widetilde{G}_{W_4})_0^2}{32 (2 k-1) (3 k-1)   (k^2+5
   k-17  )}
\\
&
   +\frac{  (6 k^3-21 k^2+161 k-106  )
   (\widetilde{G} G_{W_4})_0^2}{32 (2 k-1) (3 k-1)   (k^2+5 k-17  )}+\frac{9
   (k-1) (2 k-31) (3 k-2) (J W_4)_{-1}^2}{64 (2 k-1) (3 k-1)
     (k^2+5 k-17  )}
 \\
   &  
     -\frac{5   (93 k^2-44 k-5  )
      (J (J W_4)_0^1  )_0^2}{24 (2 k-1) (3 k-1)
     (k^2+5 k-17  )}+\frac{9 (k-3) (11 k-13)
      (J (J W_4)_0^2  )_0^2}{32 (2 k-1) (3 k-1)
     (k^2+5 k-17  )}
 \\
 &
 +\frac{  (30 k^3+35 k^2-273 k+184  )
   (J T_{W_4})_0^2}{48 (2 k-1) (3 k-1)   (k^2+5
   k-17  )}-\frac{  (2 k^3+35 k^2-7 k+2  ) (T W_4)_0^2}{8
   (2 k-1) (3 k-1)   (k^2+5 k-17  )}  \bigg]
   \\
&
+ g_3 
      \bigg[  -\frac{  (4 k^2+27 k-271  )
     (J (\widetilde{G} G)_0^1  )_0^2}{8 (k-1) k (2 k+3)
     (k^2+5 k-17  )}+\frac{  (20 k^2+93 k-263  )
     (J (T J)_0^1  )_0^2}{4 (k-1) k (2 k+3)
     (k^2+5 k-17  )}
  \\
  &
  -\frac{(k+67) (J J)_{-2}^2}{4 (k-1) k
     (k^2+5 k-17  )}+\frac{(k+4) (5 k-53)
    (J (J J)_{-1}^1  )_0^2}{2 (k-1) k (2 k+3)
     (k^2+5 k-17  )}
     -\frac{33 (k+4)
     (J   (J (J J)_0^0  )_0^1  )_0^2
   }{2 (k-1) k (2 k+3)   (k^2+5 k-17  )}  
  \bigg] 
\ea 
\ee

\normalsize

Next, we encounter composite
super Virasoro primaries
of weight $7/2$
built from normal ordered products of $W_3$ and $W_4$. They read
\small
\begin{flalign}
\cC_{\frac 72 , \frac 12}^{W_3 W_4}
& = 
(W_3 W_4)_0^{\frac{1}{2}}
+ c_{34}{}^3 \bigg[    -\frac{(k-1) (2 k-5)
    (G \widetilde{G}_{W_3})_0^{\frac{1}{2}}}{15 (k+4) (2 k-1) (2 k+5)}+\frac{(k-1) (2
   k-5)  (\widetilde{G} G_{W_3})_0^{\frac{1}{2}}}{15 (k+4) (2 k-1) (2 k+5)}
   & \nonumber
\\
&
-\frac{2
    (6 k^2-5 k-64 )  (J W_3)_{-1}^{\frac{1}{2}}}{9 (k+4) (2
   k-1) (2 k+5)}
   +\frac{(6 k+1)
   f (J  (J W_3)_0^{\frac{1}{2}} )_0^{\frac{1}{2}}}{
   (k+4) (2 k-1) (2 k+5)}-\frac{(k-1) (42 k-5)
    (J T_{W_3})_0^{\frac{1}{2}}}{60 (k+4) (2 k-1) (2
   k+5)} \bigg] 
   \ , 
\end{flalign}
\begin{flalign}
\cC_{\frac 72 , \frac 32}^{W_3 W_4}
& = ( W_3 W_4)_0^{\frac{3}{2}}
+ c_{34}{}^3   \bigg[ 
+\frac{4 (8
   k-3)
    (J (J  W_3)_0^{\frac{1}{2}}  )_0^{\frac{3}{2}}}{
   9 (k-3) k (k+4)}-\frac{5 (13 k-15)
    (J (J  W_3)_0^{\frac{3}{2}}  )_0^{\frac{3}{2}}}{
   54 (k-3) k (2 k-1)}
 & \nonumber  \\
   &
-\frac{(k-1)   (2 k^2-17 k+12  )
   (G \widetilde{G}_{W_3})_0^{\frac{3}{2}}}{12 (k-3) k (k+4) (2 k-1)}
+\frac{(k-1)
     (2 k^2-17 k+12  ) (\widetilde{G} G_{W_3})_0^{\frac{3}{2}}}{12 (k-3) k
   (k+4) (2 k-1)}
\nonumber \\
 &
   -\frac{(k-1)   (6 k^2-19 k+24  )
   ( J T_{W_3})_0^{\frac{3}{2}}}{12 (k-3) k (k+4) (2 k-1)}
 -\frac{(k-1)
     (14 k^2-31 k-12  ) ( T W_3)_0^{\frac{3}{2}}}{12 (k-3) k
   (k+4) (2 k-1)}   
 \nonumber  \\
   &
 -\frac{  (126 k^3-2069 k^2+283 k+300  )
   (J  W_3)_{-1}^{\frac{3}{2}}}{216 (k-3) k (k+4) (2 k-1)} \bigg]    + c_{34}{}^5  \bigg[ \frac{k
   T_{W_5}}{20 (k+6)}-\frac{9
   (J W_5)_0^{\frac{3}{2}}}{5
   (k+6)} 
   \bigg] \ , 
\end{flalign}
\begin{flalign}
\cC_{\frac 72 , \frac 52}^{W_3 W_4}
& = (W_3 W_4)_0^{\frac{5}{2}}
+ c_{34}{}^3  \bigg[ 
\frac{2 (7 k-26)
   (J W_3)_{-1}^{\frac{5}{2}}}{21 (k-3) (2 k-1)}-\frac{10
    (J (J W_3)_0^{\frac{3}{2}} )_0^{\frac{5}{2}}}{
   21 (k-3) (2 k-1)} \bigg] +  c_{34}{}^5
\frac{
   (J W_5)_0^{\frac{5}{2}}}{5 (2
   k-3)} \ , 
   & 
\end{flalign}
\begin{flalign}
\cC_{\frac 72 , \frac 72}^{W_3 W_4}
& =(W_3 W_4)_0^{\frac
   {7}{2}}
   - c_{34}{}^3  
\frac{ 
(J  (J W_3)_0^{\frac{5}{2}})_0^{\frac{7}{2}}}{
   (k-4) (k-3)}
   +c_{34}{}^5
   \frac{
    (J W_5)_0^{\frac{7}{2}}}{k-5}
  \ .    &
\end{flalign} 
\normalsize
In this case  the ``seed''
is provided by the 
$\mathfrak{sl}(2)_z \oplus \mathfrak{sl}(2)_y$ primary
operators $(W_3 W_4)_0^j$
($j=\tfrac 12$, $\tfrac 32$, $\tfrac 52$, $\tfrac 72$). Suitable corrective terms are added to get super Virasoro primaries.
These terms are 
governed by the OPE
$W_3 \times W_4 = c_{34}{}^3 W_3 + c_{34}{}^5 W_5$: they are grouped into 
super Virasoro descendants
of $W_3$ and $W_5$.

The construction of composite super Virasoro
primaries proceeds with those
found in the regular parts of the
$W_3 \times W_5$ and $W_4 \times W_4$ OPEs.
We use the same strategy as above.
A  technical complication is that
the singular parts of the OPEs $W_3 \times W_5$ and $W_4 \times W_4$ contain
in turn the composites
$\cC_{3,1}^{W_3 W_3}$,
$\cC_{3,3}^{W_3 W_3}$.
For this reason, in the expressions below
we encounter terms written
with $\cC_{3,1}^{W_3 W_3}$,
$\cC_{3,3}^{W_3 W_3}$
and their super Virasoro descendants
(alongside terms constructed
with super Virasoro descendants
of $\one$, $W_4$, $W_6$).
Our notation for the 
$\mathfrak{psl}(2|2)$ descendants 
of a short or long
$\mathfrak{psl}(2|2)$ primary
was introduced in Appendix~\ref{app_notation}.

% We adopt the notation 
% \eqref{sdsds} for the $\mathfrak{psl}(2|2)$ descendants of the
% short operator
% $\cC_{3,3}^{W_3 W_3}$,
% and the notation 
% \eqref{sdsds}
% for the 
% $\mathfrak{psl}(2|2)$ descendants of the
% long operator
% $\cC_{3,1}^{W_3 W_3}$.

The composite super Virasoro primaries
of the form $W_3 W_5$
are
\small
\begin{flalign}
\cC_{4,1}^{W_3 W_5}
 & = 
(W_3 W_5)_0^1
+ c_{35}{}^4 \bigg[ 
-\frac{(k-3) (G \widetilde G_{W_4})_0^1 }{24 (k+3) (k+5)}+\frac{(k-3)
    (\widetilde G G_{W_4})_0^1  }{24 (k+3) (k+5)}
  & 
  \nonumber
 \\
 &
 -\frac{\left(7 k^2-4 k-115\right)
 (J W_4)_{-1}^1
}{16 (k-1) (k+3) (k+5)}+\frac{(3 k+1)
(J(J W_4)_0^1)_0^1
}{2 (k-1) (k+3) (k+5)}-\frac{(13 k-3)
(J T_{W_4})_0^1
  }{120 (k+3) (k+5)}
  \bigg] 
    \nonumber
\\
& 
 +
c_{35}{}^{ \cC_{3,1}^{W_3 W_3} } \bigg[ 
\frac{7 
(J \ \cC_{3,1}^{W_3 W_3})_0^1
}{2 (2 k+1)}+\frac{(k-3)
(\cC_{3,1}^{W_3 W_3})_{[1,0,0]}
}{4 (2 k+1)}-\frac{(k-3)
(\cC_{3,1}^{W_3 W_3})'_{[1,0,0]}}{4 (2 k+1)}
\bigg]     \ , 
\end{flalign}

\begin{flalign}
\widehat{\cC}\,{}_{4,2}^{W_3 W_5}
& = (W_3 W_5)_0^2
+ c_{35}{}^4 \bigg[ 
-\frac{ (18 k^4-217 k^3+530 k^2-371 k-20 ) (G \widetilde{G}_{W_4})_0^2}{40 (k-4) (k-1) (k+5) (2
   k-1) (3 k-1)}
   & 
   \nonumber
 \\
 &
 +\frac{ (18 k^4-217 k^3+530 k^2-371 k-20 ) (\widetilde{G} G_{W_4})_0^2}{40 (k-4)
   (k-1) (k+5) (2 k-1) (3 k-1)}
      \nonumber
   \\
   &
   -\frac{ (714 k^4-17251 k^3+26950 k^2-17213 k+5540 )
   (J W_4)_{-1}^2}{560 (k-4) (k-1) (k+5) (2 k-1) (3 k-1)}
      \nonumber
 \\
 &
 +\frac{ (283 k^2-249
   k-4 ) (J (J W_4)_0^1 )_0^2}{14 (k-4) (k+5) (2 k-1) (3
   k-1)}-\frac{3  (57 k^2-131 k+4 ) (J (J W_4)_0^2 )_0^2}{56
   (k-4) (k-1) (2 k-1) (3 k-1)}
      \nonumber
\\
&
-\frac{ (18 k^4-105 k^3+262 k^2-231 k-4 )
   (J T_{W_4})_0^2}{20 (k-4) (k-1) (k+5) (2 k-1) (3 k-1)}-\frac{ (38 k^4-197 k^3+210
   k^2+29 k-20 ) (T W_4)_0^2}{10 (k-4) (k-1) (k+5) (2 k-1) (3 k-1)}
   \bigg] 
      \nonumber
 \\
& + c_{35}{}^6 \bigg[ 
\frac{4 k T_{W_6}}{105 (k+7)}-\frac{28 
(J W_6)_0^2
}{15 (k+7)}
\bigg]
+ c_{35}{}^{ \cC_{3,3}^{W_3 W_3}  
} \bigg[ 
\frac{4 k T_{\cC_{3,3}^{W_3 W_3}}}{105 (k+7)}-\frac{28 
(J \ \cC_{3,3}^{W_3 W_3})_0^2
}{15 (k+7)}
\bigg]
\\
   \nonumber
& 
+ c_{35}{}^{ \cC_{3,1}^{W_3 W_3}  } \bigg[ 
\frac{2
(J \ \cC_{3,1}^{W_3 W_3})_0^2
}{k-3}+\frac{(k-5)
(\cC_{3,1}^{W_3 W_3})_{[1,1,0]}
}{2 (k-3)}
   \bigg]  \ ,
\end{flalign}
\begin{flalign}
\cC_{4,3}^{W_3 W_5}
& = (W_3 W_5)_0^3
+ c_{35}{}^4 \bigg[ 
\frac{(4 k-19) 
(J W_4)_{-1}^3
}{8 (k-4) (k-1)}
-
\frac{3
(J(J W_4)_0^2)_0^3
}{8 (k-4) (k-1)}
\bigg] 
+ c_{35}{}^6 \frac{ 
(J W_6)_0^3
}{6 (k-2)}
+ c_{35}{}^{  \cC_{3,3}^{W_3 W_3} } \frac{ 
(J \ \cC_{3,3}^{W_3 W_3})_0^3
}{6 (k-2)} \ , &
\end{flalign}
\begin{flalign}
&\cC_{4,4}^{W_3 W_5}
 = (W_3 W_5)_0^4
- c_{35}{}^4 
\frac{
(J ( J W_4)_0^3 )_0^4
}{(k-5)
   (k-4)}
   -  c_{35}{}^6  \frac{
   (J W_6)_0^4
}{k-6}
   -  c_{35}{}^{ \cC_{3,3}^{W_3 W_3} }  \frac{
   (J \ \cC_{3,3}^{W_3 W_3})_0^4
}{k-6} \ . &
\end{flalign}
\normalsize
The composite super Virasoro primaries
of the form $W_4 W_4$
are
\small 
\begin{flalign}
& 
\widehat \cC \,{}_{4,0}^{W_4 W_4}
 = (W_4 W_4)_0^0
+ g_4 \bigg[ 
-\frac{2  (54 k^2-163 k-185 ) (J (\widetilde{G} G)_0^1 )_0^0}{5 k (2 k+1)
   (2 k+3) (2 k+5) (3 k+4)}
   &
   \nonumber
 \\
 &
 +\frac{4 (k-1)  (2 k^2-19 k+20 ) (G \widetilde{G})_{-1}^0}{5 k (2
   k+1) (2 k+3) (2 k+5) (3 k+4)}+\frac{4  (98 k^2+169 k-120 )
   (J (T J)_0^1 )_0^0}{5 k (2 k+1) (2 k+3) (2 k+5) (3
   k+4)}
      \nonumber
 \\
 &
   +\frac{ (72 k^4+270 k^3-1229 k^2-3953 k-1775 )
   (J (J J)_{-1}^1 )_0^0}{10 (k-1) k (2 k+1) (2 k+3) (2 k+5) (3
   k+4)}
      \nonumber
 \\
 &
 -\frac{3  (38 k^2+127 k-18 )
   (J (J (J J)_0^0 )_0^1 )_0^0}{(k-1) k (2 k+1) (2
   k+3) (2 k+5) (3 k+4)}-\frac{(k-1)  (88 k^2+114 k-145 ) (T T)_0^0}{15 k (2
   k+1) (2 k+3) (2 k+5) (3 k+4)}
   \bigg] 
      \nonumber
\\
& + c_{44}{}^4 \bigg[ 
\frac{(k-1) 
(J T_{W_4})_0^0
}{5 (k+4) (k+5)}-\frac{5
(J(JW_4)_0^1)_0^0
}{2 (k+4) (k+5)}
\bigg]
   \nonumber
\\
& 
+ c_{44}{}^{ \cC_{3,1}^{W_3 W_3}  } 
\bigg[ 
-\frac{5 
(J \ \cC_{3,1}^{W_3 W_3})_0^0
}{2 (k+3)}-\frac{(k-2)
(\cC_{3,1}^{W_3 W_3})_{[1,-1,0]}
   }{10 (k+3)}
\bigg]    \ , 
\end{flalign}
\begin{flalign}
&
\widehat \cC \, {}_{4,2}^{W_4 W_4}
= (W_4 W_4)_0^2
+ g_4 \bigg[ 
\frac{4 (k-4) ( J ( \widetilde{G} G)_0^1 )_0^2}{(k-2) k (2 k+1) (2 k+3)}-\frac{2
    (7 k^2+7 k-122 ) ( J J)_{-2}^2}{7 (k-2) (k-1) k (2 k+1)}
    & 
    \nonumber
    \\
    &
    -\frac{4  (14
   k^2-124 k-25 ) ( J ( J J)_{-1}^1 )_0^2}{7 (k-2) (k-1) k (2 k+1)
   (2 k+3)}-\frac{4 (4 k-7) ( J ( T J)_0^1 )_0^2}{(k-2) k (2 k+1) (2
   k+3)}
    \nonumber
 \\
 &
 +\frac{6 (34 k-19)
   ( J ( J ( J J)_0^0 )_0^1 )_0^2}{7 (k-2) (k-1) k (2
   k+1) (2 k+3)}
\bigg] 
+ c_{44}{}^4 \bigg[
-\frac{3 (k-1)   (2 k^2-19 k+20  ) (G \widetilde{G}_{W_4})_0^2}{10 (k-4) (k+5) (2 k-1) (3
   k-1)}
    \nonumber
 \\
 &
 +\frac{3 (k-1)   (2 k^2-19 k+20  ) (\widetilde{G} G_{W_4})_0^2}{10 (k-4) (k+5) (2 k-1) (3
   k-1)}-\frac{27   (28 k^3-389 k^2+136 k+45  ) (J W_4)_{-1}^2}{280 (k-4) (k+5) (2
   k-1) (3 k-1)}
    \nonumber
   \\
  &
  +\frac{  (349 k^2-442 k+153  )
   (J (J W_4)_0^1  )_0^2}{14 (k-4) (k+5) (2 k-1) (3 k-1)}-\frac{9 (17
   k-19) (J (J W_4)_0^2  )_0^2}{56 (k-4) (2 k-1) (3 k-1)}
    \nonumber
 \\
 &
 -\frac{(k-1)
     (6 k^2-23 k+32  ) (J T_{W_4})_0^2}{5 (k-4) (k+5) (2 k-1) (3 k-1)}-\frac{2 (k-1)
     (11 k^2-42 k+10  ) (T W_4)_0^2}{5 (k-4) (k+5) (2 k-1) (3 k-1)}
\bigg] 
 \nonumber
\\
& + c_{44}{}^6 \bigg[ 
\frac{3 k 
T_{W_6}
}{70 (k+7)}-\frac{21
(J W_6)_0^2
}{10 (k+7)}
\bigg] 
 + c_{44}{}^{ \cC_{3,3}^{W_3 W_3}  } \bigg[ 
\frac{3 k 
T_{\cC_{3,3}^{W_3 W_3}}
}{70 (k+7)}-\frac{21
(J \ \cC_{3,3}^{W_3 W_3})_0^2
}{10 (k+7)}
\bigg] 
 \nonumber
\\
& + c_{44}{}^{  \cC_{3,1}^{W_3 W_3} } 
\bigg[ 
\frac{2 
(J \  \cC_{3,1}^{W_3 W_3})_0^2
}{k-3}+\frac{(k-5)
(\cC_{3,1}^{W_3 W_3})_{[1,1,0]}
}{2 (k-3)}
   \bigg] \ , 
\end{flalign}
\begin{flalign}
& 
\cC_{4,4}^{W_4 W_4}
= (W_4 W_4)_0^4
+ g_4 
\frac{ 
(J(J(JJ)_0^2)_0^3)_0^4
}{(k-3) (k-2)
   (k-1) k}
   -
c_{44}{}^4   
   \frac{ 
  (J(JW_4)_0^3)_0^4 
}{(k-5)
   (k-4)}
+ c_{44}{}^6 \frac{
(J W_6)_0^4
}{k-6} 
+ c_{44}{}^{  \cC_{3,3}^{W_3 W_3}  } \frac{
(J \ \cC_{3,3}^{W_3 W_3})_0^4
}{k-6}  \ , 
\end{flalign}

\normalsize

The basis of composite super Virasoro primaries
used in the main text differs slightly
from the basis constructed above.
More precisely, they are related as
\be 
\ba 
\cC_{4,2}^{W_3 W_5}
& = 
\widehat \cC\, {}_{4,2}^{W_3 W_5}
+ \bigg[ 
\frac{2(k-5)}{5(k-3)}
c_{35}{}^{  \cC_{3,1}^{W_3 W_3}   }
+ \frac{8k}{315 (k+7)}
c_{35}{}^{  \cC_{3,3}^{W_3 W_3}  }
\bigg] 
\cC_{4,2}^{W_3 W_3}
\ , \\ 
\cC_{4,0}^{W_4 W_4}
& = 
\widehat \cC \, {}_{4,0}^{W_4 W_4}
+ \frac{k-2}{5(k+3)} c_{44}{}^{ \cC_{3,1}^{W_3 W_3}  }
\cC_{4,0}^{W_3 W_3}
\ , \\ 
\cC_{4,2}^{W_4 W_4}
& = 
\widehat \cC \, {}_{4,2}^{W_4 W_4}
+ \bigg[ 
\frac{2(k-5)}{5(k-3)} c_{44}{}^{ \cC_{3,1}^{W_3 W_3}  }
+ \frac{k}{35(k+7)} c_{44}{}^{ \cC_{3,3}^{W_3 W_3}  }
\bigg] 
\cC_{4,2}^{W_3 W_3}
\  .
\ea
\ee 
The linear combinations on the RHSs
are engineered in such a way that,
when they are expanded onto $\mathfrak{sl}(2)_z \oplus
\mathfrak{sl}(2)_y$ primaries,
the operators $(W_3 W_3)_{-1}^0$,
$(W_3 W_3)_{-1}^2$
appear with coefficient 0.

\subsubsection{2-point functions of composite operators}

Let us collect here the non-trivial 2-point functions 
among the composite operators
discussed above.
At weight $h=3$ we have 
\be \label{eq_composite_norm_list_h3}
\ba 
\langle \cC^{W_3  W_3}_{3,1} 
\cC^{W_3, W_3}_{3,1} 
\rangle & = 
\frac{9 (\nu -16) (\nu -9) \nu  }{5 (\nu -11) (\nu -4) (\nu
   -2)} g_3^2
\ , 
\\
\langle \cC^{W_3  W_3}_{3,3} 
\cC^{W_3, W_3}_{3,3} 
\rangle & = 
\frac{2 \nu  \left(\nu ^2+15 \nu +8\right)  
   }{(\nu -4) (\nu +3) (\nu +7)} g_3^2
\ , 
\ea
\ee
while at weight $h=7/2$ we have
\be \label{eq_composite_norm_list_h72}
\ba 
\langle
\cC^{W_3  W_4}_{\frac 72, \frac 12} 
\cC^{W_3  W_4}_{\frac 72, \frac 12} 
\rangle & = 
\frac{2 (\nu -16) (\nu +1)  }{5 (\nu -6)
   (\nu -4)} g_3 g_4
\ , 
\\
\langle
\cC^{W_3  W_4}_{\frac 72, \frac 32} 
\cC^{W_3  W_4}_{\frac 72, \frac 32} 
\rangle & = 
\frac{6 (\nu -25) (\nu -16) (\nu +1) }{5 (\nu -13)
   (\nu -4) (\nu -1)} g_3 g_4
\ , 
\\
\langle
\cC^{W_3  W_4}_{\frac 72, \frac 52} 
\cC^{W_3  W_4}_{\frac 72, \frac 52} 
\rangle & = 
\frac{12 \nu  (\nu +1) }{7 (\nu -4) (\nu +2)} g_3 g_4
\ , 
\\
\langle
\cC^{W_3  W_4}_{\frac 72, \frac 72} 
\cC^{W_3  W_4}_{\frac 72, \frac 72} 
\rangle & = 
\frac{(\nu +1) \left(\nu ^2+35 \nu +84\right)
    }{(\nu -4) (\nu +7) (\nu +9)}  g_3 g_4
\ .
\ea 
\ee 
To avoid clutter, we have omitted the standard $z_{12}$, $y_{12}$ factors on the RHS of the above 2-point functions.

We now turn to the composites of weight $h=4$. For spins $j=1$ and $j=3$ we only have one composite. Their 2-point functions read
\be 
\ba 
\langle \cC_{4,1}^{W_3 W_5}
 \cC_{4,1}^{W_3 W_5}
 \rangle 
 & = 
 \frac{(\nu -25) (\nu +5)   }{2 (\nu -7)
   (\nu -4)} g_3 g_5
   \ , 
 \\ 
\langle \cC_{4,3}^{W_3 W_5}
 \cC_{4,3}^{W_3 W_5}
 \rangle 
 & =  
\frac{15 \nu  (\nu +5)  }{8 (\nu -4)
   (\nu +3)} g_3 g_5
\ea 
\ee 
Next, let us discuss weight $h=4$ and spin $j=0$. We compute 
\be 
\ba
\langle \cC_{4,0}^{W_3 W_3} 
\cC_{4,0}^{W_3 W_3} 
\rangle & = 
-\frac{15  (\nu -11) (\nu -7) \nu  \left(3 \nu ^2-17 \nu -10\right)}{16 (\nu -9)^2 (\nu -6) (\nu
   -4) (3 \nu -11)}
  g_3^2 
\ , 
\\
\langle \cC_{4,0}^{W_3 W_3} 
\cC_{4,0}^{W_4 W_4} 
\rangle & = 
\frac{3  (\nu -7) (\nu +1) \left(29 \nu ^3-440 \nu ^2+741 \nu +4950\right)}{4 (\nu -9)^3 (\nu -6)
   (\nu -4) (3 \nu -11)}
 g_3 g_4  
\ , 
\\
\langle \cC_{4,0}^{W_4 W_4} 
\cC_{4,0}^{W_4 W_4} 
\rangle & = 
\frac{3  (\nu +1)^2 \left(2 \nu ^6-130 \nu ^5+2915 \nu ^4-32595 \nu ^3+217053 \nu ^2-865215 \nu
   +1559250\right)}{5 (\nu -9)^4 (\nu -6) (\nu -4) \nu  (3 \nu -11)}
   g_4^2
\ .
\ea
\ee
The determinant of the Gram matrix
formed by these 2-point functions is
\be 
\det \begin{pmatrix}
    \langle \cC_{4,0}^{W_3 W_3} 
\cC_{4,0}^{W_3 W_3} 
\rangle 
& 
\langle \cC_{4,0}^{W_3 W_3} 
\cC_{4,0}^{W_4 W_4} 
\rangle
\\
\langle \cC_{4,0}^{W_3 W_3} 
\cC_{4,0}^{W_4 W_4} 
\rangle
&
\langle \cC_{4,0}^{W_4 W_4} 
\cC_{4,0}^{W_4 W_4} 
\rangle
\end{pmatrix}
= 
-\frac{9 (\nu -25) (\nu -16) (\nu -7) \nu  (\nu +1)^2}{8 (\nu -9)^3 (\nu -6) (\nu -4) (3 \nu
   -11)}
   g_3^2 g_4^2  \ .
\ee 
We proceed with weight $h=4$
and spin $j=4$. We have 
\be 
\ba 
\langle 
\cC_{4,4}^{W_3 W_5}
\cC_{4,4}^{W_3 W_5}
\rangle 
& = 
\frac{\left(\nu ^3+46 \nu ^2+109 \nu +384\right) }{(\nu -4) (\nu +9) (\nu +11)}
   g_3 g_5
\ , 
\\
\langle 
\cC_{4,4}^{W_3 W_5}
\cC_{4,4}^{W_4 W_4}
\rangle 
& = 
\frac{2  (\nu -1) (\nu +1) \left(\nu ^2+5 \nu +18\right)
}{3 (\nu -9) \nu  (\nu +9) (\nu +11)
   g_3}
c_{33}{}^4
c_{34}{}^5   
g_4 g_5
\ , 
\\
\langle 
\cC_{4,4}^{W_4 W_4}
\cC_{4,4}^{W_4 W_4}
\rangle 
& = 
\frac{2 (\nu +1)^2 \left(\nu ^4+34 \nu ^3-35 \nu ^2-1440 \nu -6480\right) } {(\nu -9) (\nu -4) \nu  (\nu +5) (\nu +9) (\nu +11)}
   g_4^2
\ .
\ea 
\ee 
The determinant of the Gram matrix
formed by these 2-point functions is
\be 
\det \begin{pmatrix}
    \langle \cC_{4,4}^{W_3 W_5} 
\cC_{4,4}^{W_3 W_5} 
\rangle 
& 
\langle \cC_{4,4}^{W_3 W_5} 
\cC_{4,4}^{W_4 W_4} 
\rangle
\\
\langle \cC_{4,4}^{W_3 W_5} 
\cC_{4,4}^{W_4 W_4} 
\rangle
&
\langle \cC_{4,4}^{W_4 W_4} 
\cC_{4,4}^{W_4 W_4} 
\rangle
\end{pmatrix}
= 
\frac{2 (\nu +1)^2 \left(\nu ^3+70 \nu ^2+469 \nu +180\right)
}{(\nu -9) (\nu -4) (\nu +9) (\nu
   +11)^2} 
 g_3 g_4^2 g_5  \ .
\ee 
Finally, we have composites with
weight $h=4$ and spin $j=2$.
\small
\be
\ba 
\langle \cC_{4,2}^{W_3 W_3}
\cC_{4,2}^{W_3 W_3}
\rangle
& = 
-\frac{63 (\nu -15) (\nu -9) (\nu +5) \left(3 \nu ^3+5 \nu ^2-138 \nu -20\right) g_3^2}{8 (\nu -4) (3 \nu -1)
   \left(\nu ^2-12 \nu -57\right)^2}
\ , 
\\ 
\langle \cC_{4,2}^{W_3 W_3}
\cC_{4,2}^{W_3 W_5}
\rangle
& = 
\frac{7 c_{33}{}^4 c_{34}{}^5 (\nu -15) (\nu -1) (\nu +5) \left(6 \nu ^6+167 \nu ^5+266 \nu ^4-6207 \nu
   ^3+66088 \nu ^2+65680 \nu +14400\right) g_5}{40 (\nu -16) \nu  (3 \nu -1) \left(\nu ^2-12 \nu -57\right)^2
   \left(\nu ^2+15 \nu +8\right)}
\ , \\
\langle \cC_{4,2}^{W_3 W_3}
\cC_{4,2}^{W_4 W_4}
\rangle
& =    
\frac{126 (\nu -15) (\nu +1) (\nu +5) \left(\nu ^5+77 \nu ^4+356 \nu ^3-6722 \nu ^2-3072 \nu +360\right) g_3
   g_4}{5 (\nu -4) \nu  (3 \nu -1) \left(\nu ^2-12 \nu -57\right)^2 \left(\nu ^2+15 \nu +8\right)}
\ , \\   
\langle \cC_{4,2}^{W_3 W_5}
\cC_{4,2}^{W_3 W_5}
\rangle
& =    
\frac{2  g_3 g_5}{35 (\nu -16) (\nu -15) (\nu -4) \nu  (3 \nu -1) \left(\nu ^2-12 \nu
   -57\right)^2 \left(\nu ^2+15 \nu +8\right)^2} \times 
   \\
  & \times \Big(75 \nu ^{13}-4225 \nu ^{12}+40586 \nu ^{11}+1586164 \nu ^{10}-29938723 \nu ^9-132686081 \nu ^8
  \\ 
  & +4461908636 \nu
   ^7+2651104202 \nu ^6-136281685134 \nu ^5-1175835464700 \nu ^4
  \\ 
   & -722010916800 \nu ^3+1328678712000 \nu ^2+1390037760000 \nu
   +342921600000\Big) 
    \ , 
\\     
\langle \cC_{4,2}^{W_3 W_5}
\cC_{4,2}^{W_4 W_4}
\rangle
& =   
-\frac{2 c_{33}{}^4 c_{34}{}^5 (\nu -1) (\nu +1) (\nu +5)  g_4 g_5}{175 (\nu -16) (\nu -15) (\nu -9) \nu ^2 (3 \nu -1)
   \left(\nu ^2-12 \nu -57\right)^2 \left(\nu ^2+15 \nu +8\right)^2 g_3}
\times 
\\ 
& \times
\Big(275 \nu ^{11}-8977 \nu ^{10}-52588 \nu
   ^9+4009341 \nu ^8-13085243 \nu ^7-495583967 \nu ^6
 \\  
   & +2173165606 \nu ^5+10317029523 \nu ^4+70708858830 \nu ^3
\\ 
& +61989055200 \nu
   ^2+8993268000 \nu -2857680000\Big) \ .
\ea 
\ee 
\normalsize
The determinant of the Gram matrix formed by these 2-point functions reads
\begin{align}
& \det \begin{pmatrix}
\langle  \cC_{4,2}^{W_3 W_3} \cC_{4,2}^{W_3 W_3}  \rangle 
& 
\langle  \cC_{4,2}^{W_3 W_3} \cC_{4,2}^{W_3 W_5}  \rangle 
& 
\langle  \cC_{4,2}^{W_3 W_3} \cC_{4,2}^{W_4 W_4}  \rangle 
\\
\langle  \cC_{4,2}^{W_3 W_3} \cC_{4,2}^{W_3 W_5}  \rangle 
& 
\langle  \cC_{4,2}^{W_3 W_5} \cC_{4,2}^{W_3 W_5}  \rangle 
& 
\langle  \cC_{4,2}^{W_3 W_5} \cC_{4,2}^{W_4 W_4}  \rangle 
\\
\langle  \cC_{4,2}^{W_3 W_3} \cC_{4,2}^{W_4 W_4}  \rangle 
&
\langle  \cC_{4,2}^{W_3 W_5} \cC_{4,2}^{W_4 W_4}  \rangle
&
\langle  \cC_{4,2}^{W_4 W_4} \cC_{4,2}^{W_4 W_4}  \rangle
\end{pmatrix}
= 
\nn \\
& = -\frac{810 (\nu -36) (\nu -25) (\nu -16) (\nu +1)^2 (\nu +5)^2 g_3^3 g_4^2 g_5}{7
   (\nu -4)^2 (3 \nu -1) \left(\nu ^2-12 \nu -57\right)^2} \ .
\end{align}

The derivation of the above 2-point function makes use of the OPEs implemented in Table 
\ref{tab_ansatz}.
For the composites of weight $h=4$
we also need some parts of the OPEs $W_5 \times W_6$ and $W_6 \times W_6$,
\be \label{eq_additional_OPEs}
W_5 \times W_6 \supset c_{56}{}^3 W_3 \ , \qquad 
W_6 \times W_6 \supset g_6 \one \ . 
\ee 
The implementation of these  OPE
contributions is performed 
with the same strategy as in 
Appendix 
\ref{app_superVir} exploiting super Virasoro covariance.
The OPE coefficient $c_{56}{}^3$ is known because
\be 
c_{56}{}^3 g_3 = \lambda_{563} = \lambda_{356} = c_{35}{}^6 g_6   \quad 
\Rightarrow \quad c_{56}{}^3 = \frac{g_6}{g_3} c_{35}{}^6 \ . 
\ee 
Here we have ``raised/lowered'' the 6 index. Thanks to \eqref{eq_gauge_fix},
this does not introduce additional terms with the composite $\cC^{W_3 W_3}_{3,3}$.

\section{Free-field realizations}
\label{app_free}

In this appendix we recall
some salient features of the
proposed free-field realization
for the VOA labelled by
the  group $A_{N-1} = \text{Weyl}({\mathfrak{su}(N)}) \cong S_N$ 
\cite{Bonetti:2018fqz} (see also \cite{Arakawa:2023cki}).
We discuss in details the cases
$N=2,3,4,5$.

\subsection{General remarks  on the free-field realization
of $\cV(A_{N-1})$}

The group $A_{N-1} = \text{Weyl}(\mathfrak{su}(N))$
is a Coxeter group of rank
$N-1$ with fundamental invariants
$2,3,4,\dots,N$.
The strong generators of the VOA $\cV(A_{N-1})$ associated to
$A_{N-1}$ 
are the generators
$\mathbb J = \{ J,G,\widetilde G, T \}$ of the small $\cN = 4$ superconformal algebra
and additional generators
organized in short $\mathfrak{psl}(2|2)$ multiplets
$\mathbb W_p = \{ W_p, G_{W_p}, \widetilde G_{W_p}, T_{W_p} \}$
with $p=3,4,\dots,N$.
The operator  $W_p$ is a Grassmann even super Virasoro primary of weight and spin  
 $h=j=p/2$.

The VOA $\cV(A_{N-1})$
admits a free-field realization in terms of the
free 
$\beta \gamma bc$ system 
spanned by $\{ \beta_{p}, \gamma_{p}, b_{p}, c_{p}\}_{p=2}^{N}$ with OPEs
\be 
\beta_{p}(z_1) \gamma_{ {p' }}(z_2) = - \frac{\delta_{p p'}}{z_{12}} + \text{reg.} \ , \quad 
b_{ p }(z_1) c_{{ p' }}(z_2)
= \frac{\delta_{p p'}}{z_{12}} + \text{reg.} \ , \quad 
p, p' = 2,\dots,N \ .
\ee 
Notice the range of the
labels $p$, $p'$,
which corresponds to the invariants of
$A_{N-1}$.
To describe the free-field realization, we introduce the notation
\be 
\ba 
J(z,y) &  = J^+(z) + J^0(z)y + J^-(z)y^2
\ , \\ 
G(z,y) &= G^+(z) + G^-(z)y \ , \\
\widetilde G(z,y) & = 
\widetilde G^+(z) + 
\widetilde G^-(z)y  \ , \\
W_p(z,y) |_{y=0} & = W_p^{\rm h.w.}(z) \ ,  \\ 
G_{W_p}(z,y) |_{y=0} & = G_{W_p}^{\rm h.w.}(z) \ .
\ea 
\ee 
In the last two expression,
we use the fact that
$W_p(z,y)$, $G_{W_p}(z,y)$
are polynomials in $y$ of degrees
$p$, $p-1$ respectively,
and we give a special name to the
$y^0$ terms in these polynomials
(h.w.~stands for highest weight).

The realizations of $J^0$,
$G^\pm$, $\widetilde G^+$, $T$,
$W_p^{\rm h.w.}$,
$G_{W_p}^{\rm h.w.}$
($p=3,\dots,N$)  are given as
\be 
\ba 
J^0 & = \sum_{p=2}^N 
\Big[ 
p \beta_p \gamma_p +(p-1) b_p c_p
\Big] \ , \\ 
G^- & = \sum_{p=2}^N b_p \gamma_p \ , \qquad 
\widetilde G^+ = 
\sum_{p=2}^N \Big[ 
p \beta_p \partial_z c_p
+ (p-1) \partial_z \beta_p c_p
\Big] \ , \\
T & = 
\sum_{p = 2}^{N} \Big[
- \frac 12   p   \beta_p \partial_z \gamma_p
+ \left ( 1 - \frac 12  p \right) \, \partial \beta_p   \gamma_p
- \frac 12 (p +1)  b_p \partial_z c_p
+ \frac 12 (1 - p )  \partial_z b_p  c_p
\Big] \ , \\
J^+ & = \beta_2 \ , \qquad 
G^+  = b_2 \ , \qquad 
W_p^\text{h.w.}   = \beta_p \ , \qquad 
G_{W_p}^\text{h.w.}   = b_p \ .
\ea 
\ee 
Above and in what follows 
monomials in free fields and their
derivatives are understood
as normal ordered.
The normal ordered product of free fields and their derivatives
is associative and graded commutative.
($\beta_p$, $\gamma_p$ are Grassmann even, $b_p$, $c_p$ are Grassmann odd.)
The central charge is given as
\be 
c = -3 \sum_{p=2}^N (2p-1) = -3(N^2-1) \ . 
\ee

The final task is the determination of the free field realization of $J^-$. Once it is identified, all other remaining
strong generators can be computed from the singular parts of the OPEs $\mathbb J \times \mathbb J$
and $\mathbb J \times \mathbb W_p$, which are completely fixed
by super Virasoro symmetry,
see Appendix \ref{app_notation}.
The operator $J^-$ has weight
$h=1$, $J^0$-charge $m=-1$,
and charge $ r = 0$ under the Cartan generator of the
$SU(2)$ outer automorphims of the
small $\cN = 4$ superconformal
algebra. The free fields $\beta_p$, $\gamma_p$, $b_p$, $c_p$ ($p=2,\dots,N$) have the following quantum numbers $h$, $m$, $r$,
\begingroup
\renewcommand{\arraystretch}{1.1}
\begin{equation}
 \begin{tabular}{| c | c  | c| c  |c || c |} 
 \hline
  & $h$ & $m$ &$h-m$ & $h+m$ & $r$ \\
 \hline\hline
 $\beta_p$  &  $\frac{1}{2} \, p $ &   $\frac{1}{2} \, p$
 &0 & $p $  & $\,\,0$ \\ 
   \hline
  $b_p$  &  $\frac{1}{2} (p + 1)  $ &   $\frac{1}{2} (p - 1)$
  &1& $p$  & $+\tfrac{1}{2}$ \\
 \hline
  $c_p$  &  $-\frac{1}{2} (p - 1)$ &   
  $-\frac{1}{2} (p - 1)$&0& $1-p$  & $-\tfrac{1}{2}$ \\
 \hline
   $\gamma_p$  &  $1 -  \frac 12 \, p $ &   $- \frac 12 \,
   p$&1 & $1-p$ & $\,\,0$  \\
 \hline
    $\partial_z$  &  $1$ &   $0$& $1$ & $1$   &  $\,\,0$ \\
    \hline
\end{tabular}
\label{betagammabc_hmr}
\end{equation}
\endgroup
One constructs the most general
Ansatz for $J^-$ compatible
with the quantum numbers $h$, $m$, $r$ and constrains it 
by demanding closure of all
OPEs $\mathbb J \times \mathbb J$,
$\mathbb J \times \mathbb W_p$,
$\mathbb W_{p_1} \times \mathbb W_{p_2}$. At the end,
there remain some undetermined
parameters in $J^-$,
which are in 1-1 correspondence
with the 2-point function
coefficients $g_p$, $p=3,4,
\dots,N$.

An important (conjectural)
feature of the free-field realization discussed above is that null operators are 
simply zero when expressed in terms of the free fields. This
property facilitates the identification of null composite operators.

We now discuss $J^-$ in greater
detail for $N=2,3,4,5$.

\subsection{The case $N=2$}

The VOA $\cV(A_1)$ coincides with
the small $\cN= 4$ superconformal algebra at central charge
$c=-9$. In this case the free-field
realization outlined above was first studied in
\cite{Adamovic:2014lra}.
The generator $J^-$
is completely determined
(compatibly with the absence of 
2-point function coefficients $g_p$, $p\ge 3$) and reads
\be 
J^- = \beta_2 \gamma_2^2
+ \gamma_2 b_2 c_2 
- \tfrac 32 \partial_z \gamma_2 \ . 
\ee 

\subsection{The case $N=3$}

This case is described in detail in section 4.2 of \cite{Bonetti:2018fqz}, so we will be brief.
One finds that $J^-$ contains one free parameter  $\Lambda$,
\be 
J^- = \beta_2 \gamma_2^2
+ 3 \beta_2 \gamma_2 \gamma_3
+ \gamma_2 b_2 c_2
+ 2 \gamma_2 b_3 c_3
+ \gamma_3 b_3 c_2
+ \Lambda \beta_2^2 \gamma_3^2 + 2 \Lambda \beta_2 \gamma_3 b_2 c_3 - 4 \partial_z \gamma_2 \ .
\ee 
The parameter $\Lambda$ is related to the 2-point function coefficient
$g_3$ as
\be 
g_3 = \tfrac{40}{3} \Lambda \  . 
\ee 
We can use the free-field realization to verify that the
composite super Virasoro primary
operator $\cC^{W_3 W_3}_{3,1}$ is null.

\subsection{The case $N=4$}

This case is described in detail in section 4.3.1 of \cite{Bonetti:2018fqz}, so we will be brief.
The expression for $J^-$ contains two free parameters
$\Lambda_{1,2}$,
\small
\be 
\ba 
 J^- & = 
5  \Lambda _1^2 \beta _2^2 \gamma _4
b_2 c_4 
-\tfrac{11}{3} \Lambda _1
 \beta _2  \gamma _3  b_3 c_4
-\tfrac{5}{3} \Lambda _1
    \beta _2 \gamma _4 b_3 c_3  
+\tfrac{7}{3} \Lambda _1
\beta _2 \gamma _4 b_4  c_4 
+2 \Lambda _1 \Lambda _2
 \beta _2 \gamma _3 
 b_2  c_3 
\\
&
-\tfrac{5}{2} \Lambda _1
 \beta _3 \gamma _3  b_2 c_4 
-\tfrac{17}{6} \Lambda _1
   \beta _3 \gamma _4  
   b_2 c_3 
+\tfrac{7}{3} \Lambda _1
\beta _4 \gamma _4 
b_2  c_4 
+ \tfrac{17}{2} \Lambda_1 \Lambda_2^{-1}
\beta_3 \gamma _4 b_3  c_4 
\\ 
&
- \Lambda _2 
\gamma _3
b_4 c_3 
+\gamma _2 b_2 c_2 
+2 \gamma _2 b_3 c_3
+3 \gamma _2 b_4 c_4 
+\gamma _3 b_3 c_2 
+ \gamma _4 b_4 c_2 
-\tfrac{1}{2} \Lambda _1 \partial_z b_2
c_4 
-\tfrac{1}{2} \Lambda _1
\partial_z \beta _2
\gamma _4  
\\
& +\tfrac{5}{3}
\Lambda _1^2
\beta _2^3 \gamma _4^2 
+\Lambda _1 \Lambda _2 
\beta_2^2 \gamma _3^2 
+\tfrac{7}{3} \Lambda _1
   \beta _4 \beta _2 \gamma _4^2 
-\tfrac{16}{3}\Lambda _1
\beta _3 \beta _2 \gamma _3 \gamma _4 
+\tfrac{7}{6} \Lambda_1 
b_2 b_3 c_3 c_4 
-\Lambda _2
\beta _4 \gamma _3^2 
\\
&
+\tfrac{17 }{4 } \Lambda _1
\Lambda _2^{-1}
\beta _3^2 \gamma _4^2 
+\beta _2\gamma _2^2
+3 \beta _3 \gamma _2 \gamma _3
+4 \beta _4 \gamma _2 \gamma _4
-\tfrac{15}{2} \partial_z \gamma _2
 \ . 
\ea 
\ee 
\normalsize
The 2-point function coefficients 
and the relevant OPE coefficients
in $W_{p_1} \times W_{p_2}$ OPEs
are given in terms of 
 $\Lambda_{1,2}$ as
\be 
\ba 
g_3 & = \tfrac{85}{2} \Lambda_1 \Lambda_2 \ , & 
g_4 & = 595 \Lambda_1^2 \ , \\
c_{33}{}^4 & = - 2 \Lambda_2 \ , &
c_{34}{}^3 & = - 28 \Lambda_1 \ , \\
c_{44}{}^4 & = \tfrac{55}{3} \Lambda_1 \ , & 
c_{44}{}^{\cC_{3,3}^{W_3 W_3}}
& = \tfrac{17}{2} \Lambda_1 \Lambda_2^{-1} \ .
\ea 
\ee 
The composite super Virasoro primary operator
$\cC_{3,1}^{W_3 W_3}$ is null,
thus does not enter
the $W_4 \times W_4$ OPE.
We can solve for $\Lambda_1$,
$\Lambda_2$ in terms of $g_3$, $g_4$. We have two solutions,
but they are physically equivalent
(they are exchanged by
the redefinition $W_4 \mapsto - W_4$ leaving $W_3$ unchanged).
Picking one solution, we find
\be 
\ba 
c_{33}{}^4 & = \tfrac{4 \sqrt 7}{\sqrt{85}} \tfrac{g_3}{\sqrt{g_4}}
\ ,  &
c_{34}{}^3 & = \tfrac{4 \sqrt 7}{\sqrt{85}} \sqrt{g_4} \ ,\\
c_{44}{}^4 & = - \tfrac{11 \sqrt 5}{3 \sqrt{119}} \sqrt{g_4} \ , &
c_{44}{}^{\cC_{3,3}^{W_3 W_3}}
& = - \tfrac{2 \sqrt 7}{\sqrt{85}} \tfrac{g_3}{\sqrt{g_4}} \ . 
\ea 
\ee

\newpage

\subsection{The case $N= 5$}
The expression for $J^-$ reads
\small
\be
\ba 
J^- & = 
\tfrac{467 }{14560}
\Lambda _1 \Lambda _2 \Lambda _3
\beta _2^4 \gamma _5^2  
+\tfrac{261 }{6400}  \Lambda_1^2
\beta _2^3
\gamma _4^2 
-\tfrac{3}{560} \Lambda _1 \Lambda _2
 \beta _2^3
 \gamma _3 \gamma _5 
+\tfrac{467 }{3640}
\Lambda _1 \Lambda _2 \Lambda _3
\beta _2^3  \gamma _5 b_2  c_5  
\\
&
+\tfrac{783 }{6400}
\Lambda _1^2
 \beta _2^2 \gamma _4  b_2 c_4 
-\tfrac{3}{280} 
\Lambda _1 \Lambda _2
  \beta _2^2 \gamma _3 
  b_2 c_5
-\tfrac{3}{560}
\Lambda _1 \Lambda _2
 \beta _2^2 \gamma _5  
    b_2 c_3
+\tfrac{209}{600} 
\Lambda _1 \Lambda_3
   \beta _2^2 \gamma _4     b_3 c_5
   \\
   &
+\tfrac{113}{800} 
\Lambda _1 \Lambda _3
 \beta _2^2 \gamma _5  
 b_3 c_4
+\tfrac{47}{96} 
\Lambda _1 \Lambda _3
\beta _2^2 \beta _3
   \gamma _4  \gamma _5  
-\tfrac{134}{273}
\Lambda _2  \Lambda _3
 \beta _2^2 \beta _4 \gamma _5^2  
-\tfrac{134}{273} 
\Lambda _2 \Lambda _3
 \beta _2^2 \gamma _5  
 b_4 c_5
 \\
 &
+\tfrac{9 }{80 } \Lambda _1 \Lambda _2 \Lambda _3^{-1}
\beta _2^2 \gamma_3^2  
+ \beta _2 \gamma _2^2 
+\tfrac{23}{18}  \Lambda _3^2
\beta _2 \beta _3^2
   \gamma _5^2  
+\tfrac{23}{9} 
\Lambda _3^2
 \beta_2 \beta _3 \gamma _5  
 b_3 c_5
+\tfrac{7}{80}
\Lambda _1
 \beta _2 \beta _4 
 \gamma _4^2 
 \\
 &
+\tfrac{17}{25} 
\Lambda _1
 \beta _2 \gamma _3 
   b_3 c_4
+\tfrac{8}{25} 
\Lambda _1
 \beta _2 \gamma _4  
 b_3 c_3
+\tfrac{7}{80} 
\Lambda _1
 \beta _2 \gamma _4 
   b_4 c_4 
-\tfrac{119}{200} 
 \Lambda _1
 \beta _2 \gamma _4 
 b_5 c_5
+   \Lambda _1
 \beta _2 \beta _3 
 \gamma _3 \gamma _4
\\
&
   -\tfrac{157}{400}
   \Lambda _1
     \beta _2 \gamma _5 
     b_5 c_4
-\tfrac{79}{80} 
\Lambda _1
\beta _2 \beta _5 \gamma _4 \gamma_5  
+ \Lambda _2
 \beta _2  \gamma _3  
 b_4 c_5
+\tfrac{2}{7} 
\Lambda _2
 \beta _2 \gamma _5 
  b_4 c_3 
+\tfrac{9}{7}
\Lambda _2
\beta _2 \beta _4
\gamma _3 \gamma _5  
\\
&
-\tfrac{6}{25} 
\Lambda _1 \Lambda _3
 \beta _2
b_2 b_3 c_4 c_5
+\tfrac{137}{300} 
\Lambda _1 \Lambda _3
 \beta_2 \beta _3 \gamma _4   
 b_2 c_5
+\tfrac{209}{400} 
\Lambda _1 \Lambda _3
 \beta _2 \beta _3 \gamma _5  
 b_2 c_4
-\tfrac{268}{273} 
\Lambda _2 \Lambda _3 
  \beta _2 \beta _4 \gamma _5
  b_2 c_5
  \\
  &
+\tfrac{9 }{40 } \Lambda _1 \Lambda _2 
\Lambda_3^{-1}
\beta_2 \gamma _3 
b_2 c_3 
+\tfrac{23}{18} 
\Lambda _3^2
\beta _3^2 \gamma _5 
b_2 c_5 
+  \gamma _2 b_2 c_2
+2 \gamma _2 b_3 c_3
+3  \gamma _2 b_4 c_4
+4 \gamma _2 b_5 c_5 
+ \gamma _3 b_3 c_2
\\
&
+3 \beta _3 \gamma _2 \gamma _3
+ \gamma _4 b_4 c_2
+4 \beta _4 \gamma _2 \gamma _4
+ \gamma _5 b_5 c_2
+5 \beta _5 \gamma _2 \gamma _5
-\tfrac{1}{5} \Lambda _1 
b_2 b_3 c_3 c_4 
+\tfrac{43}{400} \Lambda _1
b_2 b_5 c_4 c_5 
\\
&
+\tfrac{12}{25} 
\Lambda _1
 \beta _3 \gamma _3 
 b_2 c_4
+\tfrac{13}{25}
\Lambda _1
 \beta _3 \gamma _4 
 b_2 c_3
+\tfrac{7}{80}
\Lambda _1
 \beta _4 \gamma _4 
 b_2 c_4
-\tfrac{39}{80}   \Lambda _1
 \beta _5 \gamma _4
 b_2 c_5
-\tfrac{1}{2} \Lambda _1
 \beta _5 \gamma _5 
 b_2 c_4
\\
&
-\tfrac{3}{7} \Lambda _2
b_2 b_4 c_3 c_5 
+\tfrac{4}{7} \Lambda _2 
 \beta _4   \gamma _3 
 b_2 c_5
+\tfrac{5}{7} 
\Lambda _2
 \beta _4 \gamma _5 
 b_2 c_3
-\tfrac{5}{2} 
\Lambda _3
\beta _3 \beta _5 \gamma _5^2 
-\tfrac{3}{2} 
\Lambda _3
b_3 b_4 c_4 c_5 
\\
&
+\tfrac{4}{3} 
\Lambda _3
 \beta _3 \gamma _3
 b_3 c_5
+\tfrac{17}{6} 
\Lambda _3
 \beta _3 \gamma _4 
 b_4 c_5
+\tfrac{4}{3} 
 \Lambda _3
\beta _4 \gamma _4
  b_3 c_5 
+\tfrac{2}{3} 
\Lambda _3
 \beta _3 \gamma _5 
 b_3 c_3
+ \Lambda _3
 \beta _3 \gamma _5 
 b_4 c_4
 \\
 &
-\tfrac{5}{2} 
\Lambda _3
 \beta _3 \gamma _5 
 b_5 c_5
+\tfrac{5}{2}
\Lambda_3
 \beta _4 \gamma _5 
 b_3 c_4
-\tfrac{5}{2}
\Lambda _3
 \beta _5 \gamma _5 
 b_3 c_5
+ \Lambda _3
\beta _3^2 \gamma _3 \gamma _5 
+\tfrac{23}{6} 
\Lambda _3
\beta _3 \beta _4 \gamma _4 \gamma _5 
\\
&
+\tfrac{1000 }{273 } \Lambda _1^{-1}
\Lambda _2 \Lambda _3
\beta _4^2 \gamma _5^2
+\tfrac{2000 }{273}
\Lambda _1^{-1}
\Lambda _2 \Lambda _3   
 \beta _4 \gamma _5 
 b_4 c_5
+\tfrac{91 }{80 } \Lambda _2^{-1}
\Lambda _1 \Lambda _3
\beta _3^2 \gamma _4^2 
+\tfrac{91 }{40 } \Lambda _2^{-1}
\Lambda _1 \Lambda _3
   \beta _3 \gamma _4 
   b_3 c_4
  \\
  &
+\tfrac{3}{25}  \Lambda _1
  \partial_z b_2 c_4
+\tfrac{2}{3} \Lambda _3
  \partial_z b_3   c_5 
+\tfrac{3}{25} \Lambda _1 
\partial_z \beta _2    \gamma _4 
+\tfrac{2}{3} \Lambda _3
  \partial_z \beta _3
  \gamma _5
-  12 \partial_z \gamma _2   
+\tfrac{117 }{200}
\Lambda _1 \Lambda _3^{-1 }
 \gamma _3  b_5 c_4
 \\
 &
+\tfrac{39 }{100 }
   \Lambda _1
\Lambda _3^{-1}
\gamma _4
b_5 c_3 
+\tfrac{39 }{40 }
 \Lambda _1
\Lambda _3^{-1}
\beta _5 \gamma _3 \gamma _4
+\tfrac{6 }{7 }
\Lambda _2
\Lambda _3^{-1}
   \beta _4 \gamma _3^2 
+\tfrac{6 }{7 }
\Lambda _2
\Lambda _3^{-1}
\gamma _3 
b_4 c_3  \ . 
\ea
\ee
\normalsize
It contains three free parameters
$\Lambda_{1,2,3}$, consistently
with the fact that we have three 2-point function
coefficients $g_3$, $g_4$, $g_5$.

The free-field realization shows that the
following composite
super Virasoro primary operators are null,
\be \label{free_su5_nulls}
\cC_{\frac 72 , \frac 32}^{W_3 W_4}
\ , \quad 
\cC_{4,0}^{W_3 W_3}
+ \tfrac{40}{91} \tfrac{\Lambda_2}{\Lambda_1 \Lambda_3}  
\cC_{4,0}^{W_4 W_4}
\ , \quad 
\cC_{4,2}^{W_3 W_3}
+ \tfrac{ 63}{143} \tfrac{1}{\Lambda_3}
\cC_{4,2}^{W_3 W_5}
+ \tfrac{800}{1859} 
\tfrac{\Lambda_2}{\Lambda_1 \Lambda_3}
\cC_{4,2}^{W_4 W_4}
\ . 
\ee 
The OPEs of $W_p$ generators
close as
reported in the main text in 
\eqref{free_su5_OPEs},
with OPE coefficients
\be \label{eq_OPE_coeffs_with_Lambda}
\ba 
g_3 & = \tfrac{117}{10} \Lambda_1 \Lambda_2 
\Lambda_3^{-1} \ , & 
c_{33}{}^4 & = 
\tfrac{12}{7} \Lambda_2 \Lambda_3^{-1} \ , & 
c_{34}{}^3 & = \tfrac{15}{2} \Lambda_1 \ , \\
c_{34}{}^5 & = \tfrac{39}{40} \Lambda_1 \Lambda_3^{-1} \ , &
g_4 & = \tfrac{819}{16} \Lambda_1^2
 \ , 
 & 
c_{44}{}^4 & = \tfrac{67}{20} \Lambda_1 \ , \\
c_{44}{}^{\cC_{3,3}^{W_3 W_3}}
& = \tfrac{91}{40} \Lambda_1 \Lambda_3 \Lambda_2^{-1} \ , 
&
c_{44}{}^{\cC_{3,1}^{W_3 W_3}}
& = - \tfrac{273}{80} \Lambda_1 \Lambda_3
\Lambda_2^{-1} \ , &
c_{35}{}^4 & = \tfrac{78}{7} \Lambda_2 ,
\\
c_{35}{}^{\cC_{3,3}^{W_3 W_3}}
& = \Lambda_3 \ , &
c_{35}{}^{\cC_{3,1}^{W_3 W_3}}
& = - \tfrac{23}{6} \Lambda_3 \  ,
& 
c_{45}{}^3 & = \tfrac{195}{4} \Lambda_1 \Lambda_3 \ , \\
c_{45}{}^5 & = - \tfrac{261}{40} \Lambda_1 \ , &
c_{45}{}^{\cC_{\frac 72, \frac 52}^{W_3 W_4}}
& = \tfrac 32 \Lambda_3 \ , &
c_{45}{}^{\cC_{\frac 72, \frac 12}^{W_3 W_4}}
& = - \tfrac{37}{6} \Lambda_3 \ , \\
c_{45}{}^{\cC_{\frac 72, \frac 72}^{W_3 W_4}}
& = \tfrac{23}{6} \Lambda_3 \ , & 
c_{55}{}^4 & = - \tfrac{522}{7} \Lambda_2 \Lambda_3 \ , & 
c_{55}{}^{\cC_{3,1}^{W_3 W_3}}
& = - \tfrac{77}{36} \Lambda_3^2 \ , \\
c_{55}{}^{\cC_{3,3}^{W_3 W_3}}
& = \tfrac{173}{6} \Lambda_3^2 \ , &
c_{55}{}^{\cC_{4,4}^{W_3 W_5}}
& = - 5 \Lambda_3 \ , & 
c_{55}{}^{\cC_{4,4}^{W_4 W_4}}
& = \tfrac{2000}{273} \Lambda_2 \Lambda_3 \Lambda_1^{-1} \ ,
\\
\overline c_{55}{}^{\cC_{4,2}^{W_3 W_5}}
& = - \tfrac{3981}{3575} \Lambda_3 \ , &
\overline c_{55}{}^{\cC_{4,2}^{W_4 W_4}}
& = - \tfrac{928192}{117117} 
\tfrac{\Lambda_2 \Lambda_3}{\Lambda_1} \ ,
& 
\overline c_{55}{}^{\cC_{4,0}^{W_4 W_4}}
& = \tfrac{9280}{2457} \tfrac{\Lambda_2 \Lambda_3}{\Lambda_1} \ , \\
g_5 & = 585 \Lambda_1 \Lambda_2 \Lambda_3 \  .
\ea 
\ee 
It is convenient to
eliminate
$\Lambda_{1,2,3}$ in favor of $g_{3,4,5}$.
Solving for $\Lambda_{1,2,3}$ introduces
some sign ambiguities. Our choice is 
\be \label{eq_solve_for_Lambda}
\Lambda_1 = \tfrac{4}{3 \sqrt{91}} \sqrt{g_4} \ , \qquad 
\Lambda_2 = \tfrac{\sqrt{7}}{6 \sqrt{26}} \tfrac{\sqrt{g_3} \sqrt{g_5}}{\sqrt{g_4}} \ , \qquad 
\Lambda_3 = \tfrac{1}{5 \sqrt 2} \tfrac{\sqrt{g_5}} { \sqrt{g_3} }
\ee 
Different choices are physically equivalent,
since they correspond to
flipping the sign of some of the $W_p$
generators. 
Combining
\eqref{eq_solve_for_Lambda} and \eqref{eq_OPE_coeffs_with_Lambda}
we obtain the OPE coefficients
\eqref{eq_su5_coeffs} reported in the main text.

\section{The higher-spin Lie superalgebra $\hs$}
\label{app_wedge}

In this appendix we discuss how
$\hs$ is bootstrapped
and we present its full set of
(anti)commutators.

Crucial to our analysis are two
key properties of $\hs$:
$\hs$ contains
$\mathfrak{psl}(2|2)$
as a subalgebra;
the generators of $\hs$
fall into short multiplets
of $\mathfrak{psl}(2|2)$ and
are as listed in the tables
\eqref{eq_psl_table},
\eqref{eq_W_table}.

In what follows, we leverage these properties to
construct $\hs$ in a fully explicit way:
equations
\eqref{eq_psl22_compact},
\eqref{eq_JW_comm},
\eqref{eq_Wcomm},
\eqref{eq_GWcomm},
\eqref{eq_HWcomm},
\eqref{eq_TWcomm},
\eqref{eq_c_result}
 encode all
(anti)commutators of 
$\hs$ in closed form,
in the notation explained in the next subsection.

\subsection{Preliminary: $\mathfrak{sl}(2)_z \oplus \mathfrak{sl}(2)_y$ covariant (anti)commutators}

All (anti)commutators
in $\hs$
are $\mathfrak{sl}(2)_z \oplus \mathfrak{sl}(2)_y$ covariant. If $A_m$, $B_n$
are modes of 
$\mathfrak{sl}(2)_z \oplus \mathfrak{sl}(2)_y$
primary operators of
definite weights $h_A$, $h_B$ and spins $j_A$,
$j_B$,
their (anti)commutator
takes the form
\be \label{eq_sl2_commutator}
[A_m(y_1) , B_n(y_2)] 
= \sum_C \kappa_{AB}{}^C P_{h_A, h_B;h_C}(m,n) 
y_{12}^{j_A + j_B - j_C}
\widehat \cD_{j_A, j_B;j_C}(y_{12}, \partial_{y_2}) C_{m+n}(y_2) \ .
\ee 
The sum is over $\mathfrak{sl}(2)_z \oplus \mathfrak{sl}(2)_y$
primary operators $C$ with weight $h_C$
and spin $j_C$ satisfying
\be 
h_C \le h_A + h_B -1 \ , \quad 
h_C - h_A - h_B \in \mathbb Z \ , \qquad 
j_C \in \{ |j_A - j_B|, |j_A - j_B|+1, \dots, j_A + j_B \} \ . 
\ee 
The differential operators
$\widehat \cD$'s are the same
as in \eqref{eq_sl2_diff_ops}.
The $P$'s are universal polynomials
in $m$, $n$ given as
\be 
P_{h_A, h_B;h_C}(m,n)=
\sum_{k=0}^{h_{ABC}-1}
\frac{(m + h_A -1)^\downarrow_{h_{ABC}-1-k}}{(h_{ABC}-1-k)!}
(-m-n-h_C)^\downarrow_k
\frac{(h_C + h_A - h_B)_k}{k! (2h_C)_k}
 \ ,
\ee 
where $h_{ABC} := h_A + h_B - h_C$
and $(x)^\downarrow_k$ is a descending
Pochhammer symbol.
Alternative, equivalent expressions for the $P$'s can be found e.g.~in \cite{Pope:1989sr,Blumenhagen:1990jv}. 

The only piece of information in the (anti)commutator 
\eqref{eq_sl2_commutator} that is not fixed by $\mathfrak{psl}(2|2)$ covariance 
are the constants $\kappa_{AB}{}^C$. As a result, 
we find it convenient to adopt a shorthand notation for  
\eqref{eq_sl2_commutator},
in which we omit the mode indices $m$, $n$,
the universal polynomials,
the $y_i$ auxiliary variables, the differential operators 
$\widehat \cD$,
\be \label{eq_very_compact}
[A,B] = \sum_C \kappa_{AB}{}^C C \ . 
\ee 
The graded antisymmetry of the (anti)commutator
\eqref{eq_sl2_commutator}
translates into a graded symmetry property of the
$\kappa_{AB}{}^C$ constants,
\be
\kappa_{BA}{}^C = (-1)^{|A||B|}
\kappa_{AB}{}^C \ , 
\ee
where $|A|=0$ or $1$ mod 2 is the Grassmann parity of $A$, and similarly for $|B|$.

% For our purposes,
% the most interesting pieces
% of information encoded
% in the (anti)commutators
% \eqref{eq_sl2_commutator} are the constants
% $c_{AB}{}^C$ and the universal
% polynomials $P$'s.
% As a result, we find it convenient
% to adopt a more compact notation,
% in which we omit the $y$-dependence,
% the $y_{12}$ factor, and the
% differential operator
% $\widehat \cD$, which can be reinstated straightforwardly if desired.
% Thus, we write 
% \eqref{eq_sl2_commutator} as 
% \be \label{eq_compact_commutator}
% [A_m , B_n] 
% = \sum_C c_{AB}{}^C P_{h_A, h_B;h_C}(m,n) 
% C_{m+n} \ .
% \ee 

\paragraph{Observation.}
The form \eqref{eq_sl2_commutator}
is a property of any Lie algebra
with $\mathfrak{sl}(2)_y \oplus 
\mathfrak{sl}(2)_z$ symmetry. As such, it could be derived without any reference to W-algebras.
Since we have already explored the consequence of $\mathfrak{sl}(2)_y \oplus 
\mathfrak{sl}(2)_z$
covariance in W-algebras, however,
we find it convenient to
derive
\eqref{eq_sl2_commutator} 
using an argument based on  the OPE
\eqref{eq_covariant_OPE} of two
$\mathfrak{sl}(2)_z \oplus \mathfrak{sl}(2)_y$
primary operators
in the W-algebra $\WW$.
We define the Laurent modes of $A$ via 
 \be 
A(z,y) = \sum_{m} z^{-m -h_A} A_m(y) \  ,
\ee 
where the sum is over the integers
(resp.~half integers) if $h_A$ is
integer (resp.~half integer).
A similar expression
holds for $B(z,y)$.
The (anti)commutator
between the modes of $A$ and those of $B$ is computed by means of   a standard contour integral argument,
\be 
[A_m(y_1) , B_n(y_2)] = \int_0 \frac{dz_2}{2\pi i} \int_{z_2} \frac{dz_1}{2\pi i} z_1^{m + h_A -1} z_2^{n + h_B -1} A(z_1,y_1)B(z_2,y_2) \ .
\ee 
This expression, combined with \eqref{eq_covariant_OPE},
yields the functional form
\eqref{eq_sl2_commutator}. 
Let us stress that in the W-algebra
we retain all Laurent modes,
while in $\hs$ 
the modes of each field
are restricted as in tables
\eqref{eq_psl_table}, \eqref{eq_W_table}.
The functional form of 
\eqref{eq_sl2_commutator}
is the same in both scenarios.

\subsection{(Anti)commutators of $\mathfrak{psl}(2|2)$}

We first present the
(anti)commutators of
$\mathfrak{psl}(2|2)$ 
in fully explicit form
\eqref{eq_sl2_commutator},
\be \label{eq_psl_commutators_FULL}
\ba
[J_0(y_1), J_0(y_2)] &= 2 P_{1,1;1}(0,0) y_{12}
\widehat \cD_{1,1;1} J_0(y_2) 
\ , \\
[J_0(y_1), G_n(y_2)] &= P_{1,\frac 32;\frac 32}(0,n) 
y_{12} 
\widehat \cD_{1,\frac 12; 
\frac 12}
G_n(y_2) 
\ ,  \\ 
[J_0(y_1), \widetilde G_n(y_2)] &= P_{1,\frac 32;\frac 32}(0,n) 
y_{12}
\widehat \cD_{1, \frac 12 ; \frac 12}
\widetilde G_n(y_2) 
\ ,  \\ 
[J_0(y_1) , T_n] & = 0 \ , \\
[G_m(y_1), G_n(y_2)] & = 0 \ , \\
[G_m(y_1), \widetilde G_n(y_2)] &= 
2 P_{\frac 32, \frac 32 ; 1}(m,n)
\widehat \cD_{\frac 12, \frac 12 ;1}
J_{m+n}(y_2)
- P_{\frac 32, \frac 32 ; 2}(m,n)
y_{12}
\widehat \cD_{\frac 12, \frac 12;0}
T_{m+n} \ , \\ 
[G_m(y_1), T_n] & = \tfrac 32  P_{\frac 32, 2; \frac 32}(m,n) 
\widehat \cD_{\frac 12, 0;\frac 12}
G_{m+n} (y_2)\ , \\
[\widetilde G_m(y_1), \widetilde G_n (y_2)] & = 0 \ , \\ 
[ \widetilde G_m(y_1), T_n] & = \tfrac 32  P_{\frac 32, 2; \frac 32}(m,n)
\widehat \cD_{\frac 12 , 0;\frac 12}
\widetilde G_{m+n}(y_2) \ , \\
[T_m, T_n] & = 2 P_{2,2;2}(m,n) T_{m+n} 
\ .
\ea 
\ee 
On the LHSs, the modes $G_m$, $\widetilde G_m$, $T_m$
are understood to be  restricted
to the vacuum preserving modes 
\eqref{eq_psl_table}.
The universal polynomials in the above expressions are given explicitly as
\be \label{eq_some_univ_poly}
\ba  
P_{1,1;1}(m,n) & \equiv 1 \ , & 
P_{1,\frac 32;\frac 32}(m,n) & \equiv 1 \ , &    
P_{\frac 32,\frac 32;1}(m,n) & = \tfrac 12 (m-n) \ , \\  
P_{\frac 32,\frac 32;2}(m,n) & \equiv  1 \ ,
& 
P_{\frac 32,2;\frac 32}(m,n) & = \tfrac 13 (2m-n) \ , & 
P_{2,2;2}(m,n) & = \tfrac 12 (m-n) \ .
\ea 
\ee 
The algebra $\mathfrak{psl}(2|2)$
admits an $SL(2)$ outer
automorphism that rotates
the modes $G_m$, $\widetilde G_m$ as a doublet, leaving all other modes invariant.

In order to exemplify
our shorthand notation 
\eqref{eq_very_compact},
we also record 
\eqref{eq_psl_commutators_FULL}
in compact form,
\be \label{eq_psl22_compact}
\ba 
[J,J] & = 2J \ , & 
[J,G] & = G \ , &
[J, \widetilde G] & = \widetilde G \ , & 
[J,T] & = 0 \ , \\
[G, G] & = 0 \ , 
&
[G, \widetilde G]
& = 2J -T \ , &
[G,T] & = \frac 32 G \ , \\
[\widetilde G , \widetilde G] & = 0  \ , &
[\widetilde G, T] & = \frac 32 \widetilde G \ , \\
[T,T] & =  2T \ .
\ea 
\ee 
In the remainder of this appendix, all (anti)commutators will be given in compact form.

% generators read
% \be \label{eq_psl_commutators}
% \ba
% [J_0, J_0] &= 2 P_{1,1;1}(0,0) J_0 
% \ , \\
% [J_0, G_n] &= P_{1,\frac 32;\frac 32}(0,n) G_n 
% \ ,  \\ 
% [J_0, \widetilde G_n] &= P_{1,\frac 32;\frac 32}(0,n) \widetilde G_n 
% \ ,  \\ 
% [J_0 , L_n] & = 0 \ , \\
% [G_m, G_n] & = 0 \ , \\
% [G_m, \widetilde G_n] &= 
% 2 P_{\frac 32, \frac 32 ; 1}(m,n)
% J_{m+n}
% - P_{\frac 32, \frac 32 ; 2}(m,n)
% L_{m+n} \ , \\ 
% [G_m, L_n] & = \tfrac 32  P_{\frac 32, 2; \frac 32}(m,n) G_{m+n} \ , \\
% [\widetilde G_m, \widetilde G_n ] & = 0 \ , \\ 
% [ \widetilde G_m, L_n] & = \tfrac 32  P_{\frac 32, 2; \frac 32}(m,n)
% \widetilde G_{m+n} \ , \\
% [L_m, L_n] & = 2 P_{2,2;2}(m,n) L_{m+n} 
% \ .
% \ea 
% \ee 
% On the LHSs, the modes $G_m$, $\widetilde G_m$, $L_m$
% are understood to be  restricted
% to the vacuum preserving modes 
% \eqref{eq_all_vac_modesBIS}.
% The universal polynomials in the above expressions are given explicitly as
% \be \label{eq_some_univ_poly}
% \ba  
% P_{1,1;1}(m,n) & \equiv 1 \ , & 
% P_{1,\frac 32;\frac 32}(m,n) & \equiv 1 \ , &    
% P_{\frac 32,\frac 32;1}(m,n) & = \tfrac 12 (m-n) \ , \\  
% P_{\frac 32,\frac 32;2}(m,n) & \equiv  1 \ ,
% & 
% P_{\frac 32,2;\frac 32}(m,n) & = \tfrac 13 (2m-n) \ , & 
% P_{2,2;2}(m,n) & = \tfrac 12 (m-n) \ .
% \ea 
% \ee 
% Like the super Virasoro algebra,
% the algebra $\mathfrak{psl}(2|2)$
% admits an $SL(2)$ outer
% automorphism that rotates
% the modes $G_m$, $\widetilde G_m$ as a doublet, leaving all other modes invariant.

\paragraph{Observation.}
A convenient way to check
\eqref{eq_psl_commutators_FULL}
is to
 start from the OPEs
of the small $\cN = 4$ super Virasoro algebra,
reported in \eqref{eq_sVir},  \eqref{eq_SCA_compact}. 
In a first step,
we merely translate those OPEs in commutators among
all Laurent modes of $J$, $G$, $\widetilde G$, $T$.
This full set of commutators 
is of the form 
\eqref{eq_sl2_commutator}, and hence can be presented in the compact notation 
\eqref{eq_very_compact}.
We have 
\be \label{eq_translated}
\begin{array}{l}
\text{keep all}\\
\text{Laurent} \\
\text{modes}
\end{array} \; : \quad 
\begin{array}{l}
[J,J] = -k \one + 2J \ ,  \qquad 
[J,G] = G \ ,  \qquad  
[J, \widetilde G  ]
= \widetilde G \ , \qquad 
[J,T] = J \ ,  \\{}
[G,G] = 0 \ , \qquad 
[G, \widetilde G]
 = -2k \one + 2J -T \ ,\qquad 
[G,T] = \tfrac 32 G \ , \\{}
[\widetilde G , \widetilde G] = 0 \ , \qquad 
[\widetilde G, T] = \tfrac 32 \widetilde G \ , \\{} 
[T,T] = \tfrac c2 \one + 2 T \ . 
\end{array}
\ee 
Clearly, the only non-zero Laurent mode of the identity operator is
$\one_n = \delta_{n,0}$.
Next, we restrict to vacuum preserving modes.
First, the central terms drop away. This can be seen explicitly by noting that the relevant universal polynomials are
\be 
\ba 
& \text{for $[J_m,J_n] \supset \one_{m+n}$:} & P_{1,1;0}(m,n) &= -n 
\ ,  
\\
& \text{for $[G_m,\widetilde G_n]
\supset \one_{m+n}$:} & P_{\frac 32,\frac 32;0}(m,n) &= \tfrac 18 (4n^2-1) 
\ ,  
\\
& \text{for $[T_m,T_n]
\supset \one_{m+n}$:} & P_{2,2;0}(m,n) &= 
- \tfrac 16(n^2-n)
\ .
\ea 
\ee 
In the first polynomial
the only allowed value of $n$ is $n=0$ once we restrict to vacuum preserving modes. Similarly, in the second
$n=\pm \tfrac 12$, in the third
$n=0,\pm 1$. In all cases we get zero, as anticipated.

The truncation to the vacuum preserving modes kills another term in the (anti)commutators
\eqref{eq_translated}:
$[J,T]$ becomes trivial.
Once again, this is due to the form of the relevant universal
polynomial,
\be 
\ba 
& \text{for $[J_m,T_n] \supset J_{m+n}$:} & 
P_{1,2;1}(m,n) &= m  \ . 
\ea 
\ee 
Upon restricting to vacuum preserving modes, $m$ can only take the value $m=0$.

\subsection{(Anti)commutators 
of the form $[\mathbb J, \mathbb W_p]$}

The notation 
$[\mathbb J, \mathbb W_p]$
is a compact notation
for the full set of
(anti)commutators between
$J_0$, $G_m$,
$\widetilde G_m$, $T_m$
 and $(W_p)_m$,
$(G_{W_p})_m$,
$(\widetilde G_{W_p})_m$,
$(T_{W_p})_m$.
We report them in compact notation,
\be \label{eq_JW_comm} 
\ba 
[J , W_p] & = p W_p \ , 
&
[J , \widetilde G_{W_p}] & =
(p-1) \widetilde G_{W_p}
\ , 
\\
[G , W_p] & = G_{W_p}   \ , 
&
[ G , \widetilde G_{W_p}] & =
p W_p - T_{W_p} 
\ , 
\\
[ \widetilde G , W_p] & =\widetilde G_{W_p}   \ , 
&
[ \widetilde G , \widetilde G_{W_p}] & = 0
\ , 
\\
[ T , W_p] & = 
\frac p2  W_p \ , 
&
[T , \widetilde G_{W_p}] & = 
\left( \frac p2 + \frac 12 \right)
\widetilde G_{W_p}
\ , 
\\[2mm]
[J , G_{W_p} ] & = 
(p-1) G_{W_p} \ , 
&
[ J ,  T_{W_p} ] & = 
(p -2) T_{W_p}
\ , 
\\
[ G , G_{W_p} ] & = 0   \ , 
&
 [ G , T_{W_p} ] & = 
 \left( p - \frac{1}{p}\right) G_{W_p}
\ , 
\\
[ \widetilde G , G_{W_p} ] & =
- p W_p + T_{W_p} \ , 
&
[ \widetilde G ,  T_{W_p} ] & =
 \left( p - \frac{1}{p}\right)
 \widetilde G_{W_p}
\ , 
\\
[ T , G_{W_p}  ] & =
\left( \frac p2 + \frac 12
\right) G_{W_p}  \ , 
&
[ T ,   T_{W_p} ] & = 
\left( \frac p2 + 1
\right ) T_{W_p}
\ .
\\
\ea 
\ee 
These (anti)commutators
simply state that
the vaccum preserving modes
$(W_p)_m$,
$(G_{W_p})_m$,
$(\widetilde G_{W_p})_m$,
$(T_{W_p})_m$
comprise a short
multiplet of
$\mathfrak{psl}(2|2)$
with $h=j=p/2$.

\paragraph{Observation.}
The (anti)commutators 
\eqref{eq_JW_comm} could be derived
from $\mathfrak{psl}(2|2)$
representation theory,
without any reference to W-algebras.
Since we have already studied the
implications of super Virasoro covariance in W-algebras, however,
we find it convenient to 
derive \eqref{eq_JW_comm}
from
the OPEs between the generators of the small $\cN=4$ super Virasoro algebra, and a super Virasoro primary operator and its
$\mathfrak{psl}(2|2)$
descendants, already reported in 
\eqref{eq_J_short_OPEs}.
Indeed,
\eqref{eq_JW_comm} is almost 
a verbatim rewriting
of \eqref{eq_J_short_OPEs},
setting
$X=W_p$, $G_X = G_{W_p}$,
$\widetilde G_X = \widetilde G_{W_p}$,
$T_X = T_{W_p}$, 
$j_X = p/2$.
The only difference in is the commutator $[J,T_{W_p}]$. In the W-algebra,
the OPE $J \times W_p$
contains $W_p$.
This term, however,
disappears from the commutators of vacuum
preserving modes. This is because the relevant universal polynomial~is
\be 
\ba 
& \text{for $[J_m,(T_{W_p})_n] \supset (W_p)_{m+n}$:} & P_{1,\frac{p}{2} +1;\frac p2}(m,n) &= m 
\ ,  
\ea 
\ee 
but the vacuum preserving
mode of $J$ is $m=0$.

% In a first step, we translate the OPEs in (anti)commutators
% of Laurent modes, keeping all modes. In a second step,
% we restrict to the vacuum
% preserving modes.
% It is worth to point out that 
% while the $J \times T_{W_p}$
% OPE contains $W_p$,
% this term disappears from the commutators of vacuum
% preserving modes. This is because
% \be 
% P_{1,h+1;h}(m,n)=m \ , 
% \ee
% but the vacuum preserving
% mode of $J$ is $m=0$.

\subsection{Constructing the Ansatz for the $[\mathbb W_{p_1}, \mathbb W_{p_2}]$
(anti)commutators}

The notation 
$[\mathbb W_{p_1}, \mathbb W_{p_2}]$
is a compact notation
for the full set of
(anti)commutators between
$(W_{p_1})_m$,
$(G_{W_{p_1}})_m$,
$(\widetilde G_{W_{p_1}})_m$,
$(T_{W_{p_1}})_m$
 and $(W_{p_2})_m$,
$(G_{W_{p_2}})_m$,
$(\widetilde G_{W_{p_2}})_m$,
$(T_{W_{p_2}})_m$.
Below we report the most general
Ansatz for these (anti)commutators 
compatible with
$\mathfrak{psl}(2|2)$
symmetry,
with the integer vs half-integer modes of the operators involved,
with bosonic vs fermionic
statistic, and with the 
$U(1) \subset SU(2)$
outer automorphism
of $\mathfrak{psl}(2|2)$.

\be \label{eq_Wcomm}
\ba 
[W_{p_1} , W_{p_2}]
& =  \delta_{p_1,p_2} \gamma_{p_1}    \Big( J 
- \tfrac 16
T \Big)
+ \sum_q 
\kappa_{p_1 p_2}{}^{q}
\Big(
W_q -\tfrac{\left(p_1-p_2+q\right) \left(-p_1+p_2+q\right)}{4 (q-1) q (q+1)}
T_{W_q}
 \Big)
\ , \\
[W_{p_1} , G_{W_{p_2}}]
& =  
\delta_{p_1,p_2} \gamma_{p_1}  \; \tfrac 12
G  
+ \sum_q 
\kappa_{p_1 p_2}{}^{q}
\;
\tfrac{q-p_1+p_2}{2 q}
 G_{W_q}  
\ , \\
[W_{p_1} , \widetilde G_{W_{p_2}}]
& =  
\delta_{p_1,p_2} \gamma_{p_1}  \; \tfrac 12  \widetilde G  
+ \sum_q 
\kappa_{p_1 p_2}{}^{q}
\;
\tfrac{q-p_1+p_2}{2 q} \widetilde G_{W_q}  
\ , \\
[W_{p_1} , T_{W_{p_2}}]
& =  
 \sum_q 
\kappa_{p_1 p_2}{}^{q}
\Big(
-\tfrac{\left(-p_1-p_2+q\right) \left(-p_1+p_2+q\right)}{4 p_2}
W_q 
\\
&
+ 
\tfrac{\left(-p_1+p_2+q-2\right) \left(-p_1+p_2+q\right) \left(-p_1+p_2+q+2\right)
   \left(p_1+p_2+q\right)}{16 p_2 (q-1) q (q+1)}
T_{W_q}
 \Big)
\ , \\
\ea 
\ee 

\be \label{eq_GWcomm}
\ba 
[G_{W_{p_1}} , W_{p_2}]
& =  
\sum_q 
\kappa_{p_1 p_2}{}^{q}
\;
\tfrac{p_1-p_2+q}{2 q}
 G_{W_q} 
\ , \\
[G_{W_{p_1}} , G_{W_{p_2}}]
& =  0
\ , \\
[G_{W_{p_1}} , \widetilde G_{W_{p_2}}]
& =  
\delta_{p_1,p_2} \gamma_{p_1}  \Big(
(p_1-1) J 
- \tfrac 16(p_1+1)
T \Big)
\\
&+ \sum_q 
\kappa_{p_1 p_2}{}^{q}
\Big(
\tfrac{1}{2} \left(p_1+p_2-q\right)
W_q 
-\tfrac{\left(p_1-p_2+q\right) \left(-p_1+p_2+q\right) \left(p_1+p_2+q\right)}{8 (q-1) q (q+1)}
T_{W_q}
 \Big)
\ , \\
[G_{W_{p_1}} , T_{W_{p_2}}]
& =  
\delta_{p_1,p_2} \gamma_{p_1}  \; 
\tfrac{\left(p_1-1\right) \left(p_1+1\right)}{2 p_1}
G   
+ \sum_q 
\kappa_{p_1 p_2}{}^{q}
\;
\tfrac{-\left(-p_1-p_2+q\right) \left(-p_1+p_2+q\right) \left(p_1+p_2+q\right)}{8 p_2 q}
  G_{W_q} 
\ , \\
\ea 
\ee 

\be \label{eq_HWcomm}
\ba 
[\widetilde G_{W_{p_1}} , W_{p_2}]
& =   \sum_q 
\kappa_{p_1 p_2}{}^{q}
\;
\tfrac{p_1-p_2+q}{2 q}
\widetilde G_{W_q} 
\ , \\
[\widetilde G_{W_{p_1}} , G_{W_{p_2}}]
& =  
\delta_{p_1,p_2} \gamma_{p_1}  \Big( 
-(p_1-1)J 
+ \tfrac 16 (p_1+1)
T \Big)
\\
&+ \sum_q 
\kappa_{p_1 p_2}{}^{q}
\Big(
\tfrac{1}{2} \left(-p_1-p_2+q\right)
W_q + 
\tfrac{\left(p_1-p_2+q\right) \left(-p_1+p_2+q\right) \left(p_1+p_2+q\right)}{8 (q-1) q (q+1)}
T_{W_q}
 \Big)
\ , \\
[\widetilde G_{W_{p_1}} , \widetilde G_{W_{p_2}}]
& =  
0
\ , \\
[\widetilde G_{W_{p_1}} , T_{W_{p_2}}]
& =  
\delta_{p_1,p_2} \gamma_{p_1}  \;
\tfrac{\left(p_1-1\right) \left(p_1+1\right)}{2 p_1}
\widetilde G   
+ \sum_q 
\kappa_{p_1 p_2}{}^{q}
\;
\tfrac{-\left(-p_1-p_2+q\right) \left(-p_1+p_2+q\right) \left(p_1+p_2+q\right)}{8 p_2 q}
\widetilde G_{W_q} 
\ , \\
\ea 
\ee 

\be \label{eq_TWcomm}
\ba 
[T_{W_{p_1}} , W_{p_2}]
& =    \sum_q 
\kappa_{p_1 p_2}{}^{q}
\Big(
\tfrac{\left(p_1-p_2+q\right) \left(p_1+p_2-q\right)}{4 p_1}
W_q + 
\tfrac{\left(p_1-p_2+q-2\right) \left(p_1-p_2+q\right) \left(p_1-p_2+q+2\right) \left(p_1+p_2+q\right)}{16
   p_1 (q-1) q (q+1)}
T_{W_q}
 \Big)
\ , \\
[ T_{W_{p_1}} , G_{W_{p_2}}]
& =  
\delta_{p_1,p_2} \gamma_{p_1} \;
\tfrac{\left(p_1-1\right) \left(p_1+1\right)}{2 p_1}
G   
+ \sum_q 
\kappa_{p_1 p_2}{}^{q}
\;
\tfrac{-\left(-p_1-p_2+q\right) \left(p_1-p_2+q\right) \left(p_1+p_2+q\right)}{8 p_1 q}
G_{W_q}  
\ , \\
[ T_{W_{p_1}} , \widetilde G_{W_{p_2}}]
& =  
\delta_{p_1,p_2} \gamma_{p_1} \;
\tfrac{\left(p_1-1\right) \left(p_1+1\right)}{2 p_1}
   G   
+ \sum_q 
\kappa_{p_1 p_2}{}^{q}
\;
\tfrac{-\left(-p_1-p_2+q\right) \left(p_1-p_2+q\right) \left(p_1+p_2+q\right)}{8 p_1 q}
 G_{W_q}  
\ , \\
[ T_{W_{p_1}} , T_{W_{p_2}}]
& =  
\delta_{p_1,p_2} \gamma_{p_1} \Big( 
-\tfrac{\left(p_1-2\right) \left(p_1^2-1\right)}{p_1}
J + 
\tfrac{\left(p_1+2\right)
\left(p_1^2-1\right) }{6 p_1}
T \Big)
\\
&
+ \sum_q 
\kappa_{p_1 p_2}{}^{q}
\Big(
-\tfrac{\left(p_1+p_2-q-2\right) \left(p_1+p_2-q\right) \left(p_1+p_2-q+2\right)
   \left(p_1+p_2+q\right)}{16 p_1 p_2}
W_q
\\
& 
-\tfrac{\left(-p_1-p_2+q\right) \left(p_1-p_2+q\right) \left(-p_1+p_2+q\right) \left(p_1+p_2+q-2\right)
   \left(p_1+p_2+q\right) \left(p_1+p_2+q+2\right)}{64 p_1 p_2 (q-1) q (q+1)} 
T_{W_q}
 \Big)
\ , \\
\ea 
\ee 
In all above (anti)commutators, 
the range of the summation over $q$ is
\be \label{eq_q_summation_range}
\max(3, |p_1-p_2|+2) \le q \le p_1+p_2-2 \ , 
\qquad 
q - p_1 -p_2 \in 2 \mathbb Z \ . 
\ee 
% \be 
% 3 \le q \le p_1 + p_2 -2 \ , \qquad 
% q - p_1 - p_2 \in 2\mathbb Z \ . 
% \ee 
These (anti)commutators have been derived by first writing the most general Ansatz
compatible with 
$\mathfrak{sl}(2)_z \oplus \mathfrak{sl}(2)_y$
symmetry,
with the integer vs half-integer modes of the operators involved,
with bosonic vs fermionic
statistic, and with the 
$U(1) \subset SU(2)$
outer automorphism
of $\mathfrak{psl}(2|2)$.
In a second step,
we have imposed the vanishing of all super Jacobi identities of the schematic form
\be 
[\mathbb J , [\mathbb W_{p_1} , \mathbb W_{p_2}]] \pm \text{cyclic} = 0 \ , 
\ee 
in order to enforce
$\mathfrak{psl}(2|2)$
covariance.

The only free parameters
in the Ansatz are 
$\gamma_{p_1}$
and $\kappa_{p_1 p_2}{}^q$.

A remark on how $J$
can enter the $[W_{p_1}, W_{p_2}]$ commutator is in order.
A priori,
based on $\mathfrak{sl}(2)_z \oplus \mathfrak{sl}(2)_y$
selection rules only,
one might expect that $J$ could enter 
$[(W_{p_1})_m, (W_{p_2})_n]$
not only when $p_1=p_2$,
but also when $p_1$, $p_2$ differ by two units, say $p_2 = p_1+2$. If $J$ entered this commutator, it would do so as
\be 
[(W_{p_1})_m , (W_{p_1+2})_n]
\supset f_{p_1, p_1+2}{}^J 
P_{\frac{p_1}{2} , \frac{p_1}{2} +1; 1}(m,n) J_{m+n} \ . 
\ee 
The universal polynomial above is of the form
\be 
P_{h , h +1; 1}(m,n)  = \frac{\prod_{j=-h+1}^{h+1} (m+j) }{ 4^{h-1} (1)_{h-1}(\tfrac 32)_{h-1}} \ .
\ee  
We see that this polynomial vanishes 
if we set $h=p_1/2$ and we restrict $m$ to the values
$-(\tfrac{p_1}{2}-1)$, \dots, $\tfrac{p_1}{2}-1$,
which are the vacuum
preserving modes of $W_{p_1}$.

% Moreover,
% we find that $c_{p_1 p_2}{}^J$ can only be non-zero
% if $p_1 =p_2$,
% \be 
% c_{p_1 p_2}{}^J = \delta_{p_1,p_2} \gamma_{p_1} \ . 
% \ee 

We also point out
the fact that the 
$[W_{p_1}, T_{W_{p_2}}]$
commutator does not contain
$J$. Once again this is due to the universal polynomial
$P_{\frac{p_1}{2} , \frac{p_2}{2}+1 ; 1}(m,n)$,
as in the previous paragraph.
By a similar token,
$[(G_{W_{p_1}})_m , 
(W_{p_2})_n ]$
does not contain $G$,
due to the universal polynomial
$P_{\frac{p_1}{2} + \frac 12 , \frac{p_2}{2} ; 1}(m,n)$, which is of the form
\be 
P_{h+1,h;1}(m,n) = - P_{h,h+1;1}(n,m) \ .
\ee 
For the same reason
$[(\widetilde G_{W_{p_1}})_m , 
(W_{p_2})_n ]$
does not contain $\widetilde G$.

The range \eqref{eq_q_summation_range} of the sum over $q$ follows from
$\mathfrak{sl}(2)_z \oplus 
\mathfrak{sl}(2)_y$ covariance.
Clearly we must have $q\ge 3$, as this is the lowest label on a $W$ generator.
Next, standard $\mathfrak{sl}(2)_y$
selection rules imply
$|p_1 + p_2| \le q \le p_1 + p_2$,
with $q$ increasing in steps of 2.
Finally, the extremal values
$q=p_1 + p_2$ and
$q = |p_1 - p_2|$
are excluded  from the summation
by $\mathfrak{sl}(2)_z$
covariance,
because the associated universal
polynomials vanish when we restrict
$m$, $n$ to the relevant wedge modes.
More precisely, we have the relations
\be 
\ba 
P_{h_1, h_2; h_1+h_2}(m,n) &\equiv 0 \ ,  \\\quad 
P_{h_1, h_2 ; |h_1-h_2|}(m,n) &= 0 \quad  \text{ if $|m|\le h_1-1$ and $|n| \le h_2-1$} \ . 
\ea 
\ee

\subsection{Constraining the free parameters $\gamma_p$,
$\kappa_{p_1 p_2}{}^q$}

The final step in
bootstrapping
$\hs$
is to constrain
the free parameters
$\gamma_p$, 
$\kappa_{p_1 p_2}{}^q$
using the super Jacobi
identities of the form
\be 
[\mathbb W_{p_1} , [\mathbb W_{p_2} , \mathbb W_{p_3} ]]
\pm \text{cyclic} = 0 \ .
\ee 
We have not attempted to study these super Jacobi
identities analytically for
generic $p_1$, $p_2$,
$p_3$,
but we have considered
explicitly
all cases with
$p_1 + p_2 + p_3 \le 14$.

We find that 
the parameters
$\kappa_{p_1 p_2}{}^q$ are fixed
up to a common
sign. Selecting the plus sign, we have
\be  \label{eq_c_result}
\kappa_{p_1 p_2}{}^q = q \frac{\sqrt{\gamma_{p_1}} \sqrt{\gamma_{p_2}}}{
\sqrt 2 
\sqrt{\gamma_q}
} \ . 
\ee 
The value of $\gamma_p$
can be fixed by fixing the
normalization of $W_p$.
The expression
\eqref{eq_c_result} suggests
that a convenient choice is
\be 
\gamma_p = 2p^2 \ ,
\ee 
which yields the simple result
\be 
\kappa_{p_1 p_2}{}^q = p_1 p_2 \ ,
\ee 
as anticipated in the main text.

\section{Some properties of low-lying $\mathfrak{psl}(2|2)$ primaries}
\label{app_low_states}

In this appendix we present a ``catalog'' of 
$\mathfrak{psl}(2|2)$ primaries in $\WW$ for $h \le 4$,
for generic values of the central charge.
More precisely, for each $(h,j,r)$
we construct a basis of 
$\mathfrak{psl}(2|2)$ primaries
engineered to minimize the $\mathfrak R$-degree
defined by the weights for the strong generators
reported in the table in Section \ref{sec_WW_filtration} 
of $\WW$. Moreover, we choose an orthogonal basis of states
and we report the two-point functions coefficients.
In this way, we can track explicitly which states
become null as we tune $\nu$ to $N^2$ for $N=2,3, \dots$

\paragraph{The case $(h,j,r) = (1,1,0)$.}
The basis has one  element, with degree
\be 
\text{degree} = 1 \ . 
\ee 
This state is simply $J$.
The Gram matrix is 
\be 
\scriptsize
\begin{pmatrix}
\frac 12 (\nu -1)
\end{pmatrix} \ . 
\ee 

\paragraph{The case $(h,j,r) = (\tfrac 32,\tfrac 32,0)$.}
The basis has one  element, with degree
\be 
\text{degree} = \tfrac 32 \ . 
\ee 
This state is simply $W_3$.
The Gram matrix is 
\be 
\scriptsize
\begin{pmatrix}
\frac{3 (\nu -4) (\nu -1)}{\nu }
\end{pmatrix} \ . 
\ee

\paragraph{The case $(h,j,r) = (2,0,0)$.}
The basis has one  element, with degree
\be 
\text{degree} = 2 \ . 
\ee 
This state is the $\mathfrak{psl}(2|2)$ primary of the schematic form
\be 
V_{2,0} = JJ + T + J' \ .
\ee 
The Gram matrix is 
\be 
\scriptsize
\begin{pmatrix}
\frac{3}{2} (\nu -4) (\nu -1)
\end{pmatrix} \ . 
\ee

\paragraph{The case $(h,j,r) = (2,2,0)$.}
The basis has two  elements, with degrees
\be 
\text{degrees} = 2,2 \ . 
\ee 
These states are $JJ$ and $W_4$.
The Gram matrix is diagonal with entries
\be 
{
\scriptsize
\begin{pmatrix}
 \frac{1}{2} (\nu -1) (\nu +1)  \\
  \frac{4 (\nu -9) (\nu -4) (\nu -1)}{\nu  (\nu +1)} 
\end{pmatrix} \ . 
}
\ee

\paragraph{The case $(h,j,r) = (\frac 52,\frac 12,0)$.}
The basis has one  element, with degree
\be 
\text{degree} = \tfrac 52 \ . 
\ee 
This state is schematically $JW_3$
plus $\mathfrak{psl}(2|2)$ completion.
The Gram matrix is 
\be 
{
\scriptsize
\begin{pmatrix}
\frac{12 (\nu -9) (\nu -4) (\nu -1)}{\nu }
\end{pmatrix} \ . 
}
\ee 

\paragraph{The case $(h,j,r) = (\frac 52,\frac 32,0)$.}
The basis has one  element, with degree
\be 
\text{degree} = \tfrac 52 \ . 
\ee 
This state is schematically $JW_3$
plus $\mathfrak{psl}(2|2)$ completion.
The Gram matrix is 
\be 
{
\scriptsize
\begin{pmatrix}
\frac{9}{5} (\nu -4) (\nu -1)
\end{pmatrix} \ . 
}
\ee 

\paragraph{The case $(h,j,r) = (\frac 52,\frac 52,0)$.}
The basis has two  elements, with degrees
\be 
\text{degrees} = \tfrac 52 , \tfrac 52 \ . 
\ee 
The Gram matrix is diagonal with entries 
\be 
{
\scriptsize
\begin{pmatrix}
 \frac{3 (\nu -4) (\nu -1) (\nu +5)}{2 \nu }  \\
  \frac{5 (\nu -16) (\nu -9) (\nu -4) (\nu -1)}{\nu ^2 (\nu +5)}
\end{pmatrix} \ . 
}
\ee

\paragraph{The case $(h,j,r) = (3,1,0)$.}
The basis has four elements, with degrees 
\be 
\text{degrees} = 2, 3, 3, 3 \ . 
\ee 
The state with degree 2 is unique up to normalization and has the form
\be 
v^{(3,1,0)}_1= G \widetilde G + J T  + J J' + J'' \ . 
\ee 
The first state of degree 3 
can be written as
\be 
v^{(3,1,0)}_2= V_{3,1}  -\tfrac{3 (\nu -4)}{2 (3 \nu -2)} v^{(3,1,0)}_1 \ , 
\ee 
where $V_{3,1}$ is a super Virasoro descendant of $V_{2,0}$, schematically
\be 
V_{3,1} = J V_{2,0} + 
\text{$\mathfrak{psl}(2|2)$ completion} \ . 
\ee 
The second state with degree 3
is schematically $JW_4$ plus 
$\mathfrak{psl}(2|2)$ completion.
The third state with degree 3 is the composite super Virasoro primary operator
$\cC^{W_3 W_3}_{3,1}$.

The Gram matrix is diagonal with entries 
\be 
{
\scriptsize
\begin{pmatrix}
 -\frac{10}{9} (\nu -1) (3 \nu -2)   \\
  \frac{15 (\nu -4) (\nu -2) (\nu -1) (\nu +1)}{4 (3 \nu -2)}   \\
 \frac{30 (\nu -11) (\nu -9) (\nu -4) (\nu -1)}{\nu  (\nu +1)}  
   \\
 \frac{81 (\nu -16) (\nu -9) (\nu -4) (\nu -1)^2}{5 (\nu -11)
   (\nu -2) \nu } 
\end{pmatrix} \ .
}
\ee 

\paragraph{The case $(h,j,r) = (3,2,0)$.}
The basis has one element, with degree 
\be 
\text{degree} = 3 \ . 
\ee 
This state is schematically
$JW_4$ plus $\mathfrak{psl}(2|2)$
completion.
The Gram matrix is 
\be 
{
\scriptsize
\begin{pmatrix}
\frac{8 (\nu -9) (\nu -4) (\nu -1)}{3 \nu }
\end{pmatrix} \ .
}
\ee 

\paragraph{The case $(h,j,r) = (3,3,0)$.}
The basis has four elements, with degrees 
\be 
\text{degrees} = 3,3,3,3 \ . 
\ee 
The first state is of the schematic form
$JJJ + JJ' + J''$.
The second state is schematically $JW_4$.
The third state is simply $W_6$.
The fourth state is the composite super Virasoro primary $\cC_{3,3}^{W_3 W_3}$.
The Gram matrix is diagonal with entries
\be 
{
\scriptsize
\begin{pmatrix}
 \frac{3}{4} (\nu -1) (\nu +1) (\nu +3)   \\
  \frac{2 (\nu -9) (\nu -4) (\nu -1) (\nu +7)}{\nu  (\nu +1)}  \\
 \frac{6 (\nu -25) (\nu -16) (\nu -9) (\nu -4) (\nu -1)}{\nu ^2 \left(\nu
   ^2+15 \nu +8\right)}   \\
\frac{18 (\nu -4) (\nu -1)^2 \left(\nu ^2+15 \nu +8\right)}{\nu
    (\nu +3) (\nu +7)}
\end{pmatrix} \ .
}
\ee

\paragraph{The case $(h,j,r) = (\frac 72,\frac 12,0)$.}
The basis has three elements, with degrees 
\be 
\text{degrees} = \tfrac 52, \tfrac 72, \tfrac 72 \ . 
\ee 
The state with degree $\frac 52$ is unique up to normalization, of the form
\be 
v^{(\frac 72, \frac 12,0)}_1 = 
G \widetilde G_{W_3}
+ \widetilde G  G_{W_3}
+ J T_{W_3}
+ J W_3'
+ J' W_3
+ T_{W_3}'  \  .
\ee 
The first state of degree $\frac 72$ 
is a super Virasoro descendant of $W_3$ and is engineered to be orthogonal to
$v^{(\frac 72, \frac 12,0)}_1$.
The second state of degree $\frac 72$
is the composite super Virasoro primary
$\cC_{\frac 72, \frac 12}^{W_3 W_4}$.
The Gram matrix is diagonal with entries
\be 
{
\scriptsize
\begin{pmatrix}
 -\frac{105 (\nu -4) (\nu -2) (\nu -1)}{2 \nu }   \\
 \frac{15}{16} (\nu -9) (\nu -6) (\nu -4) (\nu -2) (\nu
   -1)  \\
 \frac{24 (\nu -16) (\nu -9) (\nu -4) (\nu -1)^2}{5 (\nu -6) \nu ^2}
\end{pmatrix} \ .
}
\ee

\paragraph{The case $(h,j,r) = (\frac 72,\frac 32,0)$.}
The basis has five elements, with degrees 
\be 
\text{degrees} = \tfrac 52, \tfrac 72, \tfrac 72 , \tfrac 72 \ . 
\ee 
The state of degree $\frac 52$ is unique up to normalization, of the form
\be 
v^{(\frac 72, \frac 32, 0)}_1  = 
G \widetilde G_{W_3}
+ \widetilde G  G_{W_3}
+ J T_{W_3}
+ T W_3
+ J W_3'
+ J' W_3
+ W_3'' \ . 
\ee 
The first two states of degree $\frac 72$ are super Virasoro descendants of $W_3$.
They can be chosen (non-uniquely) to form
an orthogonal basis together with
$v^{(\frac 72, \frac 32, 0)}_1$.
The third state of degree $\frac 72$
is the composite super Virasoro
primary $\cC_{\frac 72, \frac 32}^{W_3 W_4}$.
The Gram matrix is diagonal with entries
\be 
{
\scriptsize
\begin{pmatrix}
 -63 (\nu -4) (\nu -1)   \\
 \frac{9   (\nu -9) (\nu -4) (\nu -1)^3 (\nu +5)}{13 \nu
   -5}   \\
 \frac{162}{125} (\nu -4) (\nu -1) (13 \nu -5)  \\
 \frac{5 (\nu -16) (\nu -13) (\nu -9) (\nu -4) (\nu -1)}{3 \nu
   ^2 (\nu +5)}
\end{pmatrix} \ .
}
\ee

\paragraph{The case $(h,j,r) = (\frac 72,\frac 52,0)$.}
The basis has three elements, with degrees 
\be 
\text{degrees} = \tfrac 72, \tfrac 72 , \tfrac 72 \ . 
\ee 
The first state is a super Virasoro descendant of $W_3$, schematically $JJW_3$ plus $\mathfrak{psl}(2|2)$ completion.
The second state is schematically $JW_5$
plus $\mathfrak{psl}(2|2)$ completion.
The third state is the super Virasoro composite $\cC_{\frac 72, \frac 52}^{W_3 W_4}$.
The Gram matrix is diagonal with entries
\be 
{
\scriptsize
\begin{pmatrix}
 \frac{63}{50} (\nu -4) (\nu -1) (\nu +5)   \\
\frac{25 (\nu -16) (\nu -9) (\nu -4) (\nu -1) (\nu +2)}{7 \nu ^2 (\nu
   +5)}   \\
 \frac{144 (\nu -9) (\nu -4) (\nu -1)^2}{7 \nu  (\nu +2)}
\end{pmatrix} \ .
}
\ee

\paragraph{The case $(h,j,r) = (\frac 72,\frac 72,0)$.}
The basis has four elements, with degrees 
\be 
\text{degrees} = \tfrac 72, \tfrac 72 , \tfrac 72, \tfrac 72 \ . 
\ee 
The first state is a super Virasoro descendant of $W_3$.
The second state is schematically
$JW_5$ plus $\mathfrak{psl}(2|2)$ completion. The third state is the 
composite super Virasoro primary
$\cC_{\frac 72, \frac 72}^{W_3 W_4}$.
The fourth state is simply $W_7$.
The Gram matrix is diagonal with entries
\be 
{
\scriptsize
\begin{pmatrix}
  \frac{3 (\nu -4) (\nu -1) (\nu +5) (\nu +7)}{2 \nu } \\
 \frac{5 (\nu -16) (\nu -9) (\nu -4) (\nu -1) (\nu +9)}{2 \nu ^2 (\nu
   +5)}
   \\
 \frac{12 (\nu -9) (\nu -4) (\nu -1)^2 \left(\nu ^2+35 \nu +84\right)}{\nu ^2
   (\nu +7) (\nu +9)}
 \\
   \frac{7 (\nu -36) (\nu -25) (\nu -16) (\nu -9) (\nu -4) (\nu -1)}{\nu ^3
   \left(\nu ^2+35 \nu +84\right)}
\end{pmatrix} \ .
}
\ee

\paragraph{The case $(h,j,r) = (4,0,0)$.}
This case requires more care.
Firstly, we will see that, in order to
obtain one of the states of degree 3 we have to mix the super Virasoro primary
operator $\cC_{4,0}^{W_3 W_3}$
with descendants of other super Virasoro primaries. Furthermore, our choice of 
basis $\cC_{4,0}^{W_3 W_3}$, $\cC_{4,0}^{W_4 W_4}$ for the composite super Virasoro primaries with $(h,j,r) = (4,0,0)$ is not well-suited for the case $\nu = 9$. In this case, the states
$\cC_{4,0}^{W_3 W_3}$  and $\cC_{4,0}^{W_4 W_4}$ develop poles in the normalization we are using.
In particular, the state 
$\cC_{4,0}^{W_3 W_3}$ has double poles
at $\nu = 9$. We can multiply it by 
$(\nu -9)^2$. If we do so, and afterwards specialize to $\nu = 9$,
this state becomes proportional to a super Virasoro descendant of $W_4$, which is actually null for $\nu = 9$.
Despite these techincal difficulties, we can still
identify the desired basis of primaries.

The basis has seven elements, with degrees 
\be 
\text{degrees} = 2,3,3,4,4,4,4 \ . 
\ee 
The state of degree 2 is unique up to normalization, of the form
\be 
G \widetilde G' 
+ \widetilde G G' 
+ TT  + J' J' + J J''
+ T'' + J''' \ .
\ee 
The two states of degree 3 are schematically 
\small 
\be 
\ba 
&
{\widetilde G} G'+G {\widetilde G}'+G {\widetilde G} J+J^{(3)}+J^2 T+J
   J''+T J'+\left(J'\right)^2+T''+T^2 \ ,
   \\
& 
{\widetilde G} G'+G {\widetilde G}'+G {\widetilde G} J+G_{W_3} \widetilde G_{W_3}+J^{(3)}+J^2
   T+J J''+T J'+\left(J'\right)^2+W_3 W_3'+T''+T^2 \ .
\ea 
\ee 
\normalsize
The four states of degree 4 are schematically
\small
\be 
\ba 
& 
\widetilde G G'+G \widetilde G'+G \widetilde G J+G_{W_3} \widetilde G_{W_3}+J^4+J^{(3)}+J^2
   T+J J''+T J'+\left(J'\right)^2+J^2 J'+W_3
   W_3'+T''+T^2
   \ , \\ 
 &   \widetilde G G'+G \widetilde G'+G \widetilde G J+G_{W_3} \widetilde G_{W_3}+J^4+J^{(3)}+J^2
   W_4+J^2 T+J J''+W_4 J'+T
   J'
   \\
   & \qquad +\left(J'\right)^2+J^2 J'+J T_{W_4}+W_3 W_3'+T''+T^2
   \ , 
   \\
 &  
\widetilde G G'+G \widetilde G'+G \widetilde G J+G_{W_3} \widetilde G_{W_3}+J^4+J^{(3)}+J^2
   W_4+J^2 T+J J''+W_4 J'+T
   J'
   \\
   & \qquad +\left(J'\right)^2+J^2 J'+J W_3^2+J W_4'+J
   T'+J T_{W_4}+W_3 W_3'+T''+T^2
  \ , 
  \\
  &
  \widetilde G G'+G \widetilde G'+G \widetilde G J+G_{W_3} \widetilde G_{W_3}+J^4+J^{(3)}+J^2
   W_4+J^2 T+J J''+W_4 J'+T
   J'
   \\
   & \qquad +\left(J'\right)^2+J^2 J'+J W_3^2+J W_4'+J
   T'+J T_{W_4}+W_3
   W_3'+W_4''+W_4^2+W_6'+T''+T^2+T_{W_4}'
   \ .
\ea 
\ee 
\normalsize
The Gram matrix is diagonal, with diagonal entries 
\be 
\begin{pmatrix}
 \frac{35}{48} (\nu -1) (3 \nu -5) \\[2mm]
 -\frac{5 (\nu -4) (\nu -1) \left(9 \nu ^2-63 \nu +26\right)}{54 (3 \nu -5)} \\[2mm]
 -\frac{135 (\nu -9) (\nu -4) (\nu -1) \left(9 \nu ^3-63 \nu ^2+182 \nu -88\right)}{16 \nu ^2 \left(9 \nu ^2-63 \nu +26\right)} \\[2mm]
 \frac{5 (\nu -4) (\nu -1) \nu  \left(9 \nu ^4-96 \nu ^3+423 \nu ^2-684 \nu +748\right)}{24 \left(9 \nu ^3-63 \nu ^2+182 \nu
   -88\right)} \\[2mm]
 \frac{15 (\nu -9) (\nu -4) (\nu -1)^2 \left(3 \nu ^4-92 \nu ^3+1081 \nu ^2-5508 \nu +10516\right)}{2 \nu  \left(9 \nu ^4-96 \nu ^3+423
   \nu ^2-684 \nu +748\right)} \\[2mm]
 \frac{27 (\nu -16) (\nu -11) (\nu -9) (\nu -4) (\nu -1)^2 \left(3 \nu ^2-17 \nu -10\right)}{10 \nu  \left(3 \nu ^4-92 \nu ^3+1081 \nu
   ^2-5508 \nu +10516\right)} \\[2mm]
 \frac{96 (\nu -25) (\nu -16) (\nu -9) (\nu -4)^2 (\nu -1)^2}{5 (\nu -11) \nu ^2 \left(3 \nu ^2-17 \nu -10\right)}
\end{pmatrix} \ . 
\ee 
This basis of $\mathfrak{psl}(2|2)$ primaries
was constructed by first 
choosing a set of linearly independent $\mathfrak{psl}(2|2)$ primary states
with degrees 2, 3, 3, 4, 4, 4, 4.
This set is chosen in such a way that,
after expressing all OPE coefficients and 
$g_p$ 2-point function coefficients in terms of $\nu$, the resulting expressions are linear combinations of normal ordered products of 
the strong generators and their derivatives,
with coefficients that are rational functions of $\nu$. We have ensured that 
these coefficients contain no pole for $\nu = N^2$, $N = 2,3,4, \dots$.

After selecting the set of states as described above,
we have applied the Gram-Schmidt (GS) orthogonalization algorithm to this set of states, in this order.
We observe that the GS algorithm is such that any given basis element is only mixed with linear combinations of elements that precede it in the original list of states. As a result, the application of the GS algorithm does not
spoil the degree properties of the states.

\paragraph{The case $(h,j,r) = (4,1,0)$.}
The basis has five elements, with degrees 
\be 
\text{degrees} = 3,3,4,4,4 \ . 
\ee 
The states of degree 3 are of the schematic form
\be 
\ba 
& 
\widetilde G G'+G \widetilde G'+G \widetilde G J+J J''+T
   J'+\left(J'\right)^2+J^2 J'+J T'
   \ , \\ 
& 
G \widetilde G_{W_4}+G_{W_4} \widetilde G+W_4 J'+J W_4'+J T_{W_4}+T_{W_4}'  
\ . 
\ea 
\ee 
The states of degree 4 are of the schematic form
\small 
\be 
\ba 
&
G \widetilde G_{W_4}+G_{W_4} \widetilde G+J^2 W_4+W_4 J'+J W_4'+J
   T_{W_4}+T_{W_4}'
 \ ,   \\
& 
\widetilde G G'+G \widetilde G'+G \widetilde G J+G \widetilde G_{W_4}+G_{W_4}
   \widetilde G+J^{(3)}+J^2 W_4+J J''+W_4 J'+T
   J'
   \\
   & \qquad +\left(J'\right)^2+J^2 J'+J W_3^2+J W_4'+J
   T'+J T_{W_4}+W_3 W_3'+W_3 T_{W_3}+W_4''+T_{W_4}'
   \ , \\ 
   & 
   \widetilde G G'+G \widetilde G'+G \widetilde G J+G \widetilde G_{W_4}+G_{W_4}
   \widetilde G+J^{(3)}+J^2 W_4+J J''+W_4 J'+T
   J'
   \\
   & \qquad 
   +\left(J'\right)^2+J^2 J'+J W_3^2+J W_4'+J
   T'+J T_{W_4}+W_3 W_3'+W_3 W_5+W_3 T_{W_3}+W_4''+T_{W_4}' \ . 
\ea 
\ee 
\normalsize
The second state above is a descendant of the 
composite super Virasoro primary
$\cC_{3,1}^{W_3 W_3}$,
while the third state is
the composite super Virasoro primary
$\cC_{4,1}^{W_3 W_5}$.

The Gram matrix is diagonal, with diagonal entries 
\be 
\begin{pmatrix}
 -2 (\nu -4) (\nu -1) (\nu +1) \\[2mm]
 -\frac{576 (\nu -9) (\nu -4) (\nu -3) (\nu -1)}{5 \nu  (\nu +1)} \\[2mm]
 \frac{72 (\nu -11) (\nu -9) (\nu -7) (\nu -4) (\nu -1)}{125 (\nu -3) \nu } \\[2mm]
 \frac{81 (\nu -16) (\nu -9) (\nu -4) (\nu -1)^2}{10 (\nu -11) \nu } 
\\[2mm]
\frac{15 (\nu -25) (\nu -16) (\nu -9) (\nu -4) (\nu -1)^2}{2 (\nu -7) \nu ^3}
\end{pmatrix} \ . 
\ee 
This basis was obtained by the GS algorithm.
The only possible non-trivial mixing is between the two states that are super Virasoro descendants of $W_4$
(the second state of degree 3, and the first state of degree 4).

% By inspection of the  above diagonal entries of the Gram matrix we confirm the expected
% pattern of states for $\nu  = N^2$,
% \begin{center}
%     \begin{tabular}{| c c c | c 
%       | c  c c c    |}
% \hline
% $h$ & $j$ & $r$ & generic $\nu$  & 2   & 3 & 4  & 5   \\
% \hline \hline
% 4 & 1 & 0 & $ 2 \xi^3 + 3 \xi^4$ & 0 &
% $\xi^3$ &
% $2\xi^3  + \xi^4$ &
% $2\xi^3  + 2\xi^4$ 
%    \\ \hline  
% \end{tabular}
% \end{center} 

\paragraph{The case $(h,j,r) = (4,1,\pm \tfrac 12)$.}
We have a doublet of states with 
\be  
\text{degrees} = \tfrac 72, \tfrac 72  \ , 
\ee 
one with $r=1/2$, and one with $r=-1/2$.
These states are super Virasoro descendants of $W_3$, schematically 
\be 
\ba 
&
W_3 G'+G J W_3+G W_3'+G T_{W_3}+G_{W_3}''+G_{W_3}
   J^2+G_{W_3} J'+G_{W_3} T
   \ , \\ 
& W_3 \widetilde G'+\widetilde G J W_3+\widetilde G W_3'+\widetilde G T_{W_3}+\widetilde G_{W_3}''+\widetilde G_{W_3}
   J^2+\widetilde G_{W_3} J'+\widetilde G_{W_3} T   \ .
\ea 
\ee
Their mixed 2-point function coefficient reads
\be 
\begin{pmatrix}
\frac{729}{512} (\nu -9) (\nu -4) (\nu -1)
\end{pmatrix}
\ee

\paragraph{The case $(h,j,r) = (4,2,0)$.}
The basis has 11 elements, with degrees 
\be 
\text{degrees} = 3,3,3, 4,4,4,4,4,4,4,4 \ . 
\ee 
The states of degree 3 are schematically 
\be 
\ba 
v_1^{(4,2,0)} & =
G \widetilde G J+J^2 T+J J''+\left(J'\right)^2+J^2 J'
\ , \\ 
v_2^{(4,2,0)} & =
G \widetilde G_{W_4}+G_{W_4} \widetilde G+W_4 J'+J W_4'+J
   T_{W_4}+W_4''+W_4 T
   \ , \\ 
v_3^{(4,2,0)} & =  
G \widetilde G J+G \widetilde G_{W_4}+G_{W_3} \widetilde G_{W_3}+G_{W_4} \widetilde G+J^2 T+J
   J''+W_4 J'
   \\ & \qquad \qquad +\left(J'\right)^2+J^2 J'+J W_4'+J
   T_{W_4}+W_3 W_3'+W_3 T_{W_3}+W_4''+W_4 T 
    . 
\ea 
\ee 
The state $v_1^{(4,2,0)}$
is a super Virasoro descendant of $\one$,
the state $v_2^{(4,2,0)}$
is a super Virasoro descendant of $W_4$,
while the state $v_3^{(4,2,0)}$
is a linear combination of the super Virasoro primary
$\cC_{4,2}^{W_3 W_3}$ with descendants of 
$\one$, $W_4$, $\cC_{3,1}^{W_3 W_3}$, and
$\cC_{3,3}^{W_3 W_3}$.

The states of degree 4 are schematically
\small
\be 
\ba 
v_4^{(4,2,0)}
& = 
G \widetilde G J+G \widetilde G_{W_4}+G_{W_3} \widetilde G_{W_3}+G_{W_4} \widetilde G+J^4+J^2
   T+J J''+W_4 J'
   \\ & \qquad \qquad 
    +\left(J'\right)^2+J^2 J'+J
   W_4'+J T_{W_4}+W_3 W_3'+W_3 T_{W_3}+W_4''+W_4 T
\ , \\
v_5^{(4,2,0)}
& = 
G \widetilde G J+G \widetilde G_{W_4}+G_{W_3} \widetilde G_{W_3}+G_{W_4} \widetilde G+J^4+J^2
   W_4+J^2 T+J J''+W_4 J'
    \\ & \qquad \qquad 
   +\left(J'\right)^2+J^2
   J'+J W_4'+J T_{W_4}+W_3 W_3'+W_3 T_{W_3}+W_4''+W_4
   T
   \ , \\
v_6^{(4,2,0)}
& =    
G \widetilde G J+G \widetilde G_{W_4}+G_{W_3} \widetilde G_{W_3}+G_{W_4} \widetilde G+J^4+J^2
   W_4+J^2 T+J J''+W_4 J'
   \\ & \qquad \qquad 
   +\left(J'\right)^2+J^2
   J'+J W_4'+J T_{W_4}+W_3 W_3'+W_3 T_{W_3}+W_4''+W_4
   T
 \ , \\ 
v_7^{(4,2,0)}
& =    J W_6+ T_{W_6}
\ , \\ 
v_8^{(4,2,0)}
& = 
G \widetilde G J+G \widetilde G_{W_4}+G_{W_3} \widetilde G_{W_3}+G_{W_4} \widetilde G+J^4+J^2
   W_4+J^2 T+J J''+W_4 J'
    \\ & \qquad \qquad 
   +\left(J'\right)^2+J^2
   J'+J W_3^2+J W_4'+J T_{W_4}+W_3 W_3'+W_3
   T_{W_3}+W_4''+W_4 T
\ , \\
v_9^{(4,2,0)}
& = 
G \widetilde G J+G \widetilde G_{W_4}+G_{W_3} \widetilde G_{W_3}+G_{W_4} \widetilde G+J^4+J^2
   W_4+J^2 T+J J''+W_4 J'
    \\ & \qquad \qquad 
   +\left(J'\right)^2+J^2
   J'+J W_3^2+J W_4'+J T_{W_4}+W_3 W_3'+W_3
   T_{W_3}+W_4''+W_4 T
\ , \\ 
v_{10}^{(4,2,0)}
& = 
G \widetilde G J+G \widetilde G_{W_4}+G_{W_3} \widetilde G_{W_3}+G_{W_4} \widetilde G+J^4+J^2
   W_4+J^2 T+J J''+W_4 J'
    \\ & \qquad \qquad 
   +\left(J'\right)^2+J^2
   J'+J W_3^2+J W_4'+J W_6+J T_{W_4}+W_3
   W_3'+W_3 W_5
   \\ & \qquad \qquad 
   +W_3 T_{W_3}+W_4''+W_4 T+  T_{W_6}
 \ , \\ 
 v_{11}^{(4,2,0)}
 & = 
G \widetilde G J+G \widetilde G_{W_4}+G_{W_3} \widetilde G_{W_3}+G_{W_4} \widetilde G+J^4+J^2
   W_4+J^2 T+J J''+W_4 J'
    \\ & \qquad \qquad 
    +\left(J'\right)^2+J^2
   J'+J W_3^2+J W_4'+J W_6+J T_{W_4}+W_3
   W_3'+W_3 W_5
   \\ & \qquad \qquad 
   +W_3 T_{W_3}+W_4''+W_4^2+W_4
   T+W_6' +T_{W_6}  \ . 
\ea 
\ee 
% The state $v_4^{(4,2,0)}$ is a descendant of $\one$.
% The states $v_{5,6}^{(4,2,0)}$ are descendants of $W_4$.
% The state $v_7^{(4,2,0)}$ is a descendant of $W_6$.
% The state $v_8^{(4,2,0)}$ is a descendant of $\cC_{3,1}^{W_3 W_3}$.
% The state $v_9^{(4,2,0)}$ is a descendant of $\cC_{3,3}^{W_3 W_3}$.
% The state $v_{10}^{(4,2,0)}$ is 
% $\cC_{4,2}^{W_3 W_5}$
% and the state 
% $v_{11}^{(4,2,0)}$ is 
% $\cC_{4,2}^{W_4 W_4}$.

 \normalsize

The Gram matrix is diagonal with entries  
\be 
\begin{pmatrix}
 -\frac{7}{100} (\nu -1) (\nu +1) (3 \nu +2) \\[2mm]
 -\frac{1792 (\nu -9) (\nu -4) (\nu -1) (15 \nu +7)}{225 \nu  (\nu +1)} \\[2mm]
 -\frac{567 (\nu -4) (\nu -1) \left(45 \nu ^3-84 \nu ^2-343 \nu -52\right)}{8 \nu  (3 \nu +2) (15 \nu +7)} \\[2mm]
 \frac{63 (\nu -4) (\nu -1) \left(45 \nu ^5-24 \nu ^4-720 \nu ^3-108 \nu ^2+4147 \nu +468\right)}{400 \left(45 \nu ^3-84 \nu ^2-343 \nu
   -52\right)} \\[2mm]
 \frac{24 (\nu -9) (\nu -4) (\nu -1) \left(22455 \nu ^6+25023 \nu ^5-402600 \nu ^4-151284 \nu ^3+121781 \nu ^2+634517 \nu
   +12999708\right)}{\nu  \left(45 \nu ^5-24 \nu ^4-720 \nu ^3-108 \nu ^2+4147 \nu +468\right)} 
   \\[2mm]
 \frac{896 (\nu -9) (\nu -4) (\nu -1)^2 \left(15 \nu ^6-80 \nu ^5-1288 \nu ^4+4096 \nu ^3+3841 \nu ^2-32240 \nu -5544\right)}{9 \nu 
   \left(22455 \nu ^6+25023 \nu ^5-402600 \nu ^4-151284 \nu ^3+121781 \nu ^2+634517 \nu +12999708\right)} \\[2mm]
 \frac{15 (\nu -25) (\nu -16) (\nu -15) (\nu -9) (\nu -4) (\nu -1)}{7 \nu ^2 \left(\nu ^2+15 \nu +8\right)} \\[2mm]
 \frac{81 (\nu -16) (\nu -9) (\nu -4) (\nu -1)^2 \left(15 \nu ^5+211 \nu ^4+403 \nu ^3-3427 \nu ^2-9286 \nu -1260\right)}{10 \nu 
   \left(15 \nu ^6-80 \nu ^5-1288 \nu ^4+4096 \nu ^3+3841 \nu ^2-32240 \nu -5544\right)} \\[2mm]
 \frac{225 (\nu -9) (\nu -4) (\nu -1)^2 \left(\nu ^2+15 \nu +8\right) \left(3 \nu ^3+5 \nu ^2-138 \nu -20\right)}{7 \nu  \left(15 \nu
   ^5+211 \nu ^4+403 \nu ^3-3427 \nu ^2-9286 \nu -1260\right)} \\[2mm]
 \frac{150 (\nu -16) (\nu -9) (\nu -4) (\nu -1)^2 \left(3 \nu ^4-148 \nu ^3+2265 \nu ^2-4160 \nu -32400\right)}{7 (\nu -15) \nu ^3
   \left(3 \nu ^3+5 \nu ^2-138 \nu -20\right)} \\[2mm]
 \frac{1152 (\nu -36) (\nu -25) (\nu -16) (\nu -9) (\nu -4)^2 (\nu -1)^2}{7 \nu ^2 \left(3 \nu ^4-148 \nu ^3+2265 \nu ^2-4160 \nu
   -32400\right)}
\end{pmatrix} \ . 
\ee 
This basis was obtained as above using the GS algorithm.

\paragraph{The case $(h,j,r) = (4,3,0)$.}
The basis has four elements, with degrees 
\be 
\text{degrees} = 4,4,4,4 \ . 
\ee 
These states are schematically
\be 
\ba 
v_1^{(4,3,0)}
& = J^2 W_4+W_4 J'+J W_4'
\ , \\ 
v_2^{(4,3,0)}
& = 
J W_6+W_6'
\ , \\ 
v_3^{(4,3,0)}
& = 
J^2 W_4+W_4 J'+J^2 J'+J W_3^2+J W_4'+W_3
   W_3'
\ , \\
v_4^{(4,3,0)}
& = J^2 W_4+W_4 J'+J^2 J'+J W_3^2+J W_4'+J
   W_6+W_3 W_3'+W_3 W_5+W_6'
\ . 
\ea 
\ee 
The state $v_1^{(4,3,0)}$ is a descendant of $W_4$.
The state $v_2^{(4,3,0)}$ is a descendant of $W_6$.
The state $v_3^{(4,3,0)}$ is a descendant of $\cC_{3,3}^{W_3 W_3}$.
The state $v_4^{(4,3,0)}$ is $\cC_{4,3}^{W_3 W_5}$.

The Gram matrix is diagonal with entries  
\be 
\begin{pmatrix}
 \frac{16 (\nu -9) (\nu -4) (\nu -1) (\nu +7)}{9 \nu } \\[2mm]
 \frac{9 (\nu -25) (\nu -16) (\nu -9) (\nu -4) (\nu -1) (\nu +3)}{2 \nu ^2 \left(\nu ^2+15 \nu +8\right)} \\[2mm]
 \frac{27 (\nu -4) (\nu -1)^2 \left(\nu ^2+15 \nu +8\right)}{2 \nu  (\nu +7)} \\[2mm]
 \frac{225 (\nu -16) (\nu -9) (\nu -4) (\nu -1)^2}{8 \nu ^2 (\nu +3)}
\end{pmatrix} \ . 
\ee 
In this case no GS orthogonalization was necessary.

\paragraph{The case $(h,j,r) = (4,4,0)$.}
The basis has seven elements, with degrees 
\be 
\text{degrees} = 4,4,4,4,4,4,4  \ . 
\ee 
These states are schematically 
\be 
\ba 
v_1^{(4,4,0)}
& = J^4
\ , \\ 
v_2^{(4,4,0)}
& = J^2 W_4
\ , \\ 
v_3^{(4,4,0)}
& = W_8
\ , \\ 
v_4^{(4,4,0)}
& = J W_6
\ , \\ 
v_5^{(4,4,0)}
& = J W_3^3 + J^4 + J ^2 W_4
\ , \\ 
v_6^{(4,4,0)}
& = JW_6
+ J^2 W_4 + J^4
+ JW_3^3 + W_3 W_5
\ , \\ 
v_7^{(4,4,0)}
& = 
JW_6
+ J^2 W_4 + J^4
+ JW_3^3 + W_3 W_5 + W_4^2
\ , \\ 
\ea 
\ee 
The state $v_1^{(4,4,0)}$ is a descendant of $\one$.
The state $v_2{(4,4,0)}$ is a descendant of $W_4$.
The state $v_3^{(4,4,0)}$ is $W_8$.
The state $v_4{(4,4,0)}$ is a descendant of $W_6$.
The state $v_5{(4,4,0)}$ is a descendant of $\cC_{3,3}^{W_3 W_3}$.
The state $v_6{(4,4,0)}$ is 
$\cC_{4,4}^{W_3 W_5}$
and the  state $v_7{(4,4,0)}$ is
a linear combination of
$\cC_{4,4}^{W_3 W_5}$ and 
$\cC_{4,4}^{W_4 W_4}$.

The Gram matrix is diagonal with entries  
\be 
\begin{pmatrix}
  \frac{3}{2} (\nu -1) (\nu +1) (\nu +3) (\nu +5) \\[2mm]
 \frac{2 (\nu -9) (\nu -4) (\nu -1) (\nu +7) (\nu +9)}{\nu  (\nu +1)} \\[2mm]
 \frac{8 (\nu -49) (\nu -36) (\nu -25) (\nu -16) (\nu -9) (\nu -4) (\nu -1)}{\nu ^3 \left(\nu ^3+70 \nu ^2+469 \nu +180\right)} \\[2mm]
 \frac{3 (\nu -25) (\nu -16) (\nu -9) (\nu -4) (\nu -1) (\nu +11)}{\nu ^2 \left(\nu ^2+15 \nu +8\right)} \\[2mm]
 \frac{9 (\nu -4) (\nu -1)^2 (\nu +11) \left(\nu ^2+15 \nu +8\right)}{\nu  (\nu +3) (\nu +7)} \\[2mm]
 \frac{15 (\nu -16) (\nu -9) (\nu -4) (\nu -1)^2 \left(\nu ^3+46 \nu ^2+109 \nu +384\right)}{\nu ^3 (\nu +5) (\nu +9) (\nu +11)} \\[2mm]
 \frac{32 (\nu -9) (\nu -4)^2 (\nu -1)^2 \left(\nu ^3+70 \nu ^2+469 \nu +180\right)}{\nu ^2 (\nu +11) \left(\nu ^3+46 \nu ^2+109 \nu
   +384\right)}
\end{pmatrix} \ . 
\ee 
In this case we need a GS orthogonalization 
involving the two composite super Virasoro
primaries $\cC_{4,4}^{W_3 W_5}$
and $\cC_{4,4}^{W_4 W_4}$.

\newpage 

\section{Some reference results for 4d $\cN = 4$ super Yang-Mills}
In this appendix we collect some results 
for 4d $\cN = 4$ $SU(N)$ super Yang-Mills:
the Macdonald index up to $q^5$ for $N=2,\dots,10$;
the partition function of free 
4d $\cN = 4$ $SU(N)$ super Yang-Mills up to $q^4$
for $N=2,\dots,8$;
the Hilbert Series of the Hall-Littlewood chiral ring
up to $q^4$ for $N=2,\dots,8$.  

\begin{landscape}
\subsection{Macdonald indices}
\label{all_macdonald}

\tiny
\begingroup
\renewcommand{\arraystretch}{1.3}
\be 
\begin{array}{| l | l || l | l | l | l | l | l | l | l | l|}
\hline 
h & j & 2 & 3 & 4 & 5 & 6 &7 & 8 & 9 & 10 \\ \hline  \hline
  1 & 1 & \xi  & \xi  & \xi  & \xi  & \xi  & \xi  & \xi  & \xi  & \xi  \\ \hline
 \frac{3}{2} & \frac{3}{2} & 0 & \xi ^{3/2} & \xi ^{3/2} & \xi ^{3/2} & \xi ^{3/2} & \xi ^{3/2} & \xi ^{3/2} & \xi ^{3/2} & \xi ^{3/2}
   \\ \hline
 2 & 0 & 0 & \xi ^2 & \xi ^2 & \xi ^2 & \xi ^2 & \xi ^2 & \xi ^2 & \xi ^2 & \xi ^2 \\
 2 & 2 & \xi ^2 & \xi ^2 & 2 \xi ^2 & 2 \xi ^2 & 2 \xi ^2 & 2 \xi ^2 & 2 \xi ^2 & 2 \xi ^2 & 2 \xi ^2 \\ \hline
 \frac{5}{2} & \frac{1}{2} & 0 & 0 & \xi ^{5/2} & \xi ^{5/2} & \xi ^{5/2} & \xi ^{5/2} & \xi ^{5/2} & \xi ^{5/2} & \xi ^{5/2} \\
 \frac{5}{2} & \frac{3}{2} & 0 & \xi ^{5/2} & \xi ^{5/2} & \xi ^{5/2} & \xi ^{5/2} & \xi ^{5/2} & \xi ^{5/2} & \xi ^{5/2} & \xi ^{5/2}
   \\
 \frac{5}{2} & \frac{5}{2} & 0 & \xi ^{5/2} & \xi ^{5/2} & 2 \xi ^{5/2} & 2 \xi ^{5/2} & 2 \xi ^{5/2} & 2 \xi ^{5/2} & 2 \xi ^{5/2} &
   2 \xi ^{5/2} \\ \hline
 3 & 1 & \xi ^2 & \xi ^2+\xi ^3 & \xi ^2+2 \xi ^3 & \xi ^2+3 \xi ^3 & \xi ^2+3 \xi ^3 & \xi ^2+3 \xi ^3 & \xi ^2+3 \xi ^3 & \xi ^2+3
   \xi ^3 & \xi ^2+3 \xi ^3 \\
 3 & 2 & 0 & 0 & \xi ^3 & \xi ^3 & \xi ^3 & \xi ^3 & \xi ^3 & \xi ^3 & \xi ^3 \\
 3 & 3 & \xi ^3 & 2 \xi ^3 & 3 \xi ^3 & 3 \xi ^3 & 4 \xi ^3 & 4 \xi ^3 & 4 \xi ^3 & 4 \xi ^3 & 4 \xi ^3 \\
 \hline
 \frac{7}{2} & \frac{1}{2} & 0 & \xi ^{5/2} & \xi ^{5/2}+\xi ^{7/2} & \xi ^{5/2}+2 \xi ^{7/2} & \xi ^{5/2}+2 \xi ^{7/2} & \xi ^{5/2}+2
   \xi ^{7/2} & \xi ^{5/2}+2 \xi ^{7/2} & \xi ^{5/2}+2 \xi ^{7/2} & \xi ^{5/2}+2 \xi ^{7/2} \\
 \frac{7}{2} & \frac{3}{2} & 0 & \xi ^{5/2}+\xi ^{7/2} & \xi ^{5/2}+2 \xi ^{7/2} & \xi ^{5/2}+3 \xi ^{7/2} & \xi ^{5/2}+4 \xi ^{7/2} &
   \xi ^{5/2}+4 \xi ^{7/2} & \xi ^{5/2}+4 \xi ^{7/2} & \xi ^{5/2}+4 \xi ^{7/2} & \xi ^{5/2}+4 \xi ^{7/2} \\
 \frac{7}{2} & \frac{5}{2} & 0 & \xi ^{7/2} & 2 \xi ^{7/2} & 3 \xi ^{7/2} & 3 \xi ^{7/2} & 3 \xi ^{7/2} & 3 \xi ^{7/2} & 3 \xi ^{7/2}
   & 3 \xi ^{7/2} \\
 \frac{7}{2} & \frac{7}{2} & 0 & \xi ^{7/2} & 2 \xi ^{7/2} & 3 \xi ^{7/2} & 3 \xi ^{7/2} & 4 \xi ^{7/2} & 4 \xi ^{7/2} & 4 \xi ^{7/2}
   & 4 \xi ^{7/2} \\
   \hline
 4 & 0 & \xi ^2 & \xi ^2+\xi ^3+\xi ^4 & \xi ^2+2 \xi ^3+2 \xi ^4 & \xi ^2+2 \xi ^3+3 \xi ^4 & \xi ^2+2 \xi ^3+4 \xi ^4 & \xi ^2+2 \xi
   ^3+4 \xi ^4 & \xi ^2+2 \xi ^3+4 \xi ^4 & \xi ^2+2 \xi ^3+4 \xi ^4 & \xi ^2+2 \xi ^3+4 \xi ^4 \\
 4 & 1 & 0 & \xi ^3 & \xi ^3 & \xi ^3+\xi ^4 & \xi ^3+2 \xi ^4 & \xi ^3+2 \xi ^4 & \xi ^3+2 \xi ^4 & \xi ^3+2 \xi ^4 & \xi ^3+2 \xi ^4
   \\
 4 & 2 & \xi ^3 & 2 \xi ^3+\xi ^4 & 3 \xi ^3+4 \xi ^4 & 3 \xi ^3+6 \xi ^4 & 3 \xi ^3+7 \xi ^4 & 3 \xi ^3+8 \xi ^4 & 3 \xi ^3+8 \xi ^4
   & 3 \xi ^3+8 \xi ^4 & 3 \xi ^3+8 \xi ^4 \\
 4 & 3 & 0 & \xi ^4 & 2 \xi ^4 & 3 \xi ^4 & 4 \xi ^4 & 4 \xi ^4 & 4 \xi ^4 & 4 \xi ^4 & 4 \xi ^4 \\
 4 & 4 & \xi ^4 & 2 \xi ^4 & 4 \xi ^4 & 5 \xi ^4 & 6 \xi ^4 & 6 \xi ^4 & 7 \xi ^4 & 7 \xi ^4 & 7 \xi ^4 \\
 \hline
 \frac{9}{2} & \frac{1}{2} & 0 & 0 & 2 \xi ^{7/2}+\xi ^{9/2} & 2 \xi ^{7/2}+2 \xi ^{9/2} & 2 \xi ^{7/2}+3 \xi ^{9/2} & 2 \xi ^{7/2}+4
   \xi ^{9/2} & 2 \xi ^{7/2}+4 \xi ^{9/2} & 2 \xi ^{7/2}+4 \xi ^{9/2} & 2 \xi ^{7/2}+4 \xi ^{9/2} \\
 \frac{9}{2} & \frac{3}{2} & 0 & \xi ^{5/2}+2 \xi ^{7/2}+\xi ^{9/2} & \xi ^{5/2}+3 \xi ^{7/2}+2 \xi ^{9/2} & \xi ^{5/2}+3 \xi ^{7/2}+4
   \xi ^{9/2} & \xi ^{5/2}+3 \xi ^{7/2}+6 \xi ^{9/2} & \xi ^{5/2}+3 \xi ^{7/2}+7 \xi ^{9/2} & \xi ^{5/2}+3 \xi ^{7/2}+7 \xi ^{9/2} &
   \xi ^{5/2}+3 \xi ^{7/2}+7 \xi ^{9/2} & \xi ^{5/2}+3 \xi ^{7/2}+7 \xi ^{9/2} \\
 \frac{9}{2} & \frac{5}{2} & 0 & 2 \xi ^{7/2}+\xi ^{9/2} & 3 \xi ^{7/2}+4 \xi ^{9/2} & 4 \xi ^{7/2}+7 \xi ^{9/2} & 4 \xi ^{7/2}+9 \xi
   ^{9/2} & 4 \xi ^{7/2}+10 \xi ^{9/2} & 4 \xi ^{7/2}+11 \xi ^{9/2} & 4 \xi ^{7/2}+11 \xi ^{9/2} & 4 \xi ^{7/2}+11 \xi ^{9/2} \\
 \frac{9}{2} & \frac{7}{2} & 0 & \xi ^{9/2} & 3 \xi ^{9/2} & 5 \xi ^{9/2} & 6 \xi ^{9/2} & 7 \xi ^{9/2} & 7 \xi ^{9/2} & 7 \xi ^{9/2}
   & 7 \xi ^{9/2} \\
 \frac{9}{2} & \frac{9}{2} & 0 & 2 \xi ^{9/2} & 3 \xi ^{9/2} & 5 \xi ^{9/2} & 6 \xi ^{9/2} & 7 \xi ^{9/2} & 7 \xi ^{9/2} & 8 \xi
   ^{9/2} & 8 \xi ^{9/2} \\
   \hline
 5 & 0 & 0 & 0 & -\xi ^3-\xi ^4-\xi ^5 & -\xi ^3-\xi ^4-2 \xi ^5 & -\xi ^3-2 \xi ^4-2 \xi ^5 & -\xi ^3-2 \xi ^4-2 \xi ^5 & -\xi ^3-2
   \xi ^4-2 \xi ^5 & -\xi ^3-2 \xi ^4-2 \xi ^5 & -\xi ^3-2 \xi ^4-2 \xi ^5 \\
 5 & 1 & \xi ^2+\xi ^3 & \xi ^2+2 \xi ^3+\xi ^4+\xi ^5 & \xi ^2+2 \xi ^3+4 \xi ^4+3 \xi ^5 & \xi ^2+2 \xi ^3+6 \xi ^4+6 \xi ^5 & \xi
   ^2+2 \xi ^3+6 \xi ^4+9 \xi ^5 & \xi ^2+2 \xi ^3+6 \xi ^4+11 \xi ^5 & \xi ^2+2 \xi ^3+6 \xi ^4+12 \xi ^5 & \xi ^2+2 \xi ^3+6 \xi
   ^4+12 \xi ^5 & \xi ^2+2 \xi ^3+6 \xi ^4+12 \xi ^5 \\
 5 & 2 & \xi ^3 & \xi ^4 & \xi ^3+4 \xi ^4+2 \xi ^5 & \xi ^3+4 \xi ^4+4 \xi ^5 & \xi ^3+4 \xi ^4+6 \xi ^5 & \xi ^3+4 \xi ^4+8 \xi ^5 &
   \xi ^3+4 \xi ^4+9 \xi ^5 & \xi ^3+4 \xi ^4+9 \xi ^5 & \xi ^3+4 \xi ^4+9 \xi ^5 \\
 5 & 3 & \xi ^4 & 3 \xi ^4+2 \xi ^5 & 5 \xi ^4+5 \xi ^5 & 6 \xi ^4+10 \xi ^5 & 7 \xi ^4+14 \xi ^5 & 7 \xi ^4+16 \xi ^5 & 7 \xi ^4+17
   \xi ^5 & 7 \xi ^4+18 \xi ^5 & 7 \xi ^4+18 \xi ^5 \\
 5 & 4 & 0 & \xi ^5 & 4 \xi ^5 & 6 \xi ^5 & 8 \xi ^5 & 9 \xi ^5 & 10 \xi ^5 & 10 \xi ^5 & 10 \xi ^5 \\
 5 & 5 & \xi ^5 & 2 \xi ^5 & 5 \xi ^5 & 7 \xi ^5 & 9 \xi ^5 & 10 \xi ^5 & 11 \xi ^5 & 11 \xi ^5 & 12 \xi ^5 \\ \hline
\end{array} \nn 
\ee 
\endgroup
\end{landscape}

\newpage

\subsection{Counting free words}
\label{app_free_SYM_counting}

In the following tables we report the counting of
gauge-invariant $\mathfrak{psl}(2|2)$ primaries
with $h\le 4$ for $N = 2,3,\dots,8$
in free super Yang-Mills.

\begingroup

\renewcommand{\arraystretch}{1.3}
\begin{equation*}
\scriptsize
\begin{array}{| c | c ||  l | l |   }
\hline 
h & j & N=2 & N=3 
\\ \hline 
 1 & 1 & \xi  & \xi  \\
 \frac{3}{2} & \frac{3}{2} & 0 & \xi ^{3/2} \\
 2 & 0 & \xi +\xi ^2+\xi ^{3/2} \chi _1 & \xi +2 \xi ^2+\xi ^{3/2} \chi _1 \\
 2 & 2 & \xi ^2 & \xi ^2 \\
 \frac{5}{2} & \frac{1}{2} & 0 & \xi ^{3/2}+\xi ^{5/2}+\xi ^2 \chi _1 \\
 \frac{5}{2} & \frac{3}{2} & 0 & \xi ^{5/2} \\
 \frac{5}{2} & \frac{5}{2} & 0 & \xi ^{5/2} \\
 3 & 0 & 0 & \xi ^2+\xi ^3+\xi ^{5/2} \chi _1 \\
 3 & 1 & 2 \xi ^2+\xi ^3+\xi ^{5/2} \chi _1 & 3 \xi ^2+3 \xi ^3+2 \xi ^{5/2} \chi _1 \\
 3 & 2 & 0 & 0 \\
 3 & 3 & \xi ^3 & 2 \xi ^3 \\
 \frac{7}{2} & \frac{1}{2} & \xi ^{3/2}+\xi ^{5/2}+\xi ^2 \chi _1 & 2 \xi ^{3/2}+5 \xi ^{5/2}+\xi ^{7/2}+\left(3 \xi ^2+2 \xi ^3\right)
   \chi _1+\xi ^{5/2} \chi _2 \\
 \frac{7}{2} & \frac{3}{2} & 0 & 3 \xi ^{5/2}+3 \xi ^{7/2}+2 \xi ^3 \chi _1 \\
 \frac{7}{2} & \frac{5}{2} & 0 & \xi ^{7/2} \\
 \frac{7}{2} & \frac{7}{2} & 0 & \xi ^{7/2} \\
 4 & 0 & \xi +3 \xi ^2+2 \xi ^3+\xi ^4+\left(\xi ^{3/2}+\xi ^{5/2}+\xi ^{7/2}\right) \chi _1 & \xi +6 \xi ^2+8 \xi ^3+4 \xi ^4+\left(\xi
   ^{3/2}+5 \xi ^{5/2}+3 \xi ^{7/2}\right) \chi _1+\left(\xi ^2+\xi ^3\right) \chi _2+\xi ^{5/2} \chi _3 \\
 4 & 1 & \xi ^2+\xi ^3+\xi ^{5/2} \chi _1 & 2 \xi ^2+8 \xi ^3+2 \xi ^4+\left(5 \xi ^{5/2}+5 \xi ^{7/2}\right) \chi _1+3 \xi ^3 \chi _2
   \\
 4 & 2 & 2 \xi ^3+\xi ^4+\xi ^{7/2} \chi _1 & 5 \xi ^3+4 \xi ^4+3 \xi ^{7/2} \chi _1 \\
 4 & 3 & 0 & \xi ^4 \\
 4 & 4 & \xi ^4 & 2 \xi ^4 \\ \hline
\end{array}
\end{equation*}

\begin{equation*}
\scriptsize
\begin{array}{| c | c ||  l  |   }
\hline 
h & j & N=4 
\\ \hline 
 1 & 1 & \xi  \\
 \frac{3}{2} & \frac{3}{2} & \xi ^{3/2} \\
 2 & 0 & \xi +2 \xi ^2+\xi ^{3/2} \chi _1 \\
 2 & 2 & 2 \xi ^2 \\
 \frac{5}{2} & \frac{1}{2} & \xi ^{3/2}+2 \xi ^{5/2}+\xi ^2 \chi _1 \\
 \frac{5}{2} & \frac{3}{2} & \xi ^{5/2} \\
 \frac{5}{2} & \frac{5}{2} & \xi ^{5/2} \\
 3 & 0 & \xi ^2+\xi ^3+\xi ^{5/2} \chi _1 \\
 3 & 1 & 4 \xi ^2+5 \xi ^3+3 \xi ^{5/2} \chi _1 \\
 3 & 2 & \xi ^3 \\
 3 & 3 & 3 \xi ^3 \\
 \frac{7}{2} & \frac{1}{2} & 2 \xi ^{3/2}+7 \xi ^{5/2}+4 \xi ^{7/2}+\left(3 \xi ^2+4 \xi ^3\right) \chi _1+\xi ^{5/2} \chi _2 \\
 \frac{7}{2} & \frac{3}{2} & 4 \xi ^{5/2}+5 \xi ^{7/2}+3 \xi ^3 \chi _1 \\
 \frac{7}{2} & \frac{5}{2} & 2 \xi ^{7/2} \\
 \frac{7}{2} & \frac{7}{2} & 2 \xi ^{7/2} \\
 4 & 0 & \xi +7 \xi ^2+14 \xi ^3+8 \xi ^4+\left(\xi ^{3/2}+7 \xi ^{5/2}+7 \xi ^{7/2}\right) \chi _1+\left(\xi ^2+2 \xi ^3\right) \chi
   _2+\xi ^{5/2} \chi _3 \\
 4 & 1 & 3 \xi ^2+14 \xi ^3+6 \xi ^4+\left(7 \xi ^{5/2}+10 \xi ^{7/2}\right) \chi _1+4 \xi ^3 \chi _2 \\
 4 & 2 & 9 \xi ^3+10 \xi ^4+6 \xi ^{7/2} \chi _1 \\
 4 & 3 & 2 \xi ^4 \\
 4 & 4 & 4 \xi ^4
 \\ \hline
\end{array}
\end{equation*}

\begin{equation*}
\scriptsize
\begin{array}{| c | c ||  l  |   }
\hline 
h & j & N=5 
\\ \hline 
 1 & 1 & \xi  \\
 \frac{3}{2} & \frac{3}{2} & \xi ^{3/2} \\
 2 & 0 & \xi +2 \xi ^2+\xi ^{3/2} \chi _1 \\
 2 & 2 & 2 \xi ^2 \\
 \frac{5}{2} & \frac{1}{2} & \xi ^{3/2}+2 \xi ^{5/2}+\xi ^2 \chi _1 \\
 \frac{5}{2} & \frac{3}{2} & \xi ^{5/2} \\
 \frac{5}{2} & \frac{5}{2} & 2 \xi ^{5/2} \\
 3 & 0 & \xi ^2+\xi ^3+\xi ^{5/2} \chi _1 \\
 3 & 1 & 4 \xi ^2+6 \xi ^3+3 \xi ^{5/2} \chi _1 \\
 3 & 2 & \xi ^3 \\
 3 & 3 & 3 \xi ^3 \\
 \frac{7}{2} & \frac{1}{2} & 2 \xi ^{3/2}+7 \xi ^{5/2}+5 \xi ^{7/2}+\left(3 \xi ^2+4 \xi ^3\right) \chi _1+\xi ^{5/2} \chi _2 \\
 \frac{7}{2} & \frac{3}{2} & 5 \xi ^{5/2}+7 \xi ^{7/2}+4 \xi ^3 \chi _1 \\
 \frac{7}{2} & \frac{5}{2} & 3 \xi ^{7/2} \\
 \frac{7}{2} & \frac{7}{2} & 3 \xi ^{7/2} \\
 4 & 0 & \xi +7 \xi ^2+15 \xi ^3+10 \xi ^4+\left(\xi ^{3/2}+7 \xi ^{5/2}+8 \xi ^{7/2}\right) \chi _1+\left(\xi ^2+2 \xi ^3\right) \chi
   _2+\xi ^{5/2} \chi _3 \\
 4 & 1 & 3 \xi ^2+16 \xi ^3+9 \xi ^4+\left(7 \xi ^{5/2}+12 \xi ^{7/2}\right) \chi _1+4 \xi ^3 \chi _2 \\
 4 & 2 & 10 \xi ^3+13 \xi ^4+7 \xi ^{7/2} \chi _1 \\
 4 & 3 & 3 \xi ^4 \\
 4 & 4 & 5 \xi ^4
 \\ \hline
\end{array}
\end{equation*}

\begin{equation*}
\scriptsize
\begin{array}{| c | c ||  l  |   }
\hline 
h & j & N=6
\\ \hline 
 1 & 1 & \xi  \\
 \frac{3}{2} & \frac{3}{2} & \xi ^{3/2} \\
 2 & 0 & \xi +2 \xi ^2+\xi ^{3/2} \chi _1 \\
 2 & 2 & 2 \xi ^2 \\
 \frac{5}{2} & \frac{1}{2} & \xi ^{3/2}+2 \xi ^{5/2}+\xi ^2 \chi _1 \\
 \frac{5}{2} & \frac{3}{2} & \xi ^{5/2} \\
 \frac{5}{2} & \frac{5}{2} & 2 \xi ^{5/2} \\
 3 & 0 & \xi ^2+\xi ^3+\xi ^{5/2} \chi _1 \\
 3 & 1 & 4 \xi ^2+6 \xi ^3+3 \xi ^{5/2} \chi _1 \\
 3 & 2 & \xi ^3 \\
 3 & 3 & 4 \xi ^3 \\
 \frac{7}{2} & \frac{1}{2} & 2 \xi ^{3/2}+7 \xi ^{5/2}+5 \xi ^{7/2}+\left(3 \xi ^2+4 \xi ^3\right) \chi _1+\xi ^{5/2} \chi _2 \\
 \frac{7}{2} & \frac{3}{2} & 5 \xi ^{5/2}+8 \xi ^{7/2}+4 \xi ^3 \chi _1 \\
 \frac{7}{2} & \frac{5}{2} & 3 \xi ^{7/2} \\
 \frac{7}{2} & \frac{7}{2} & 3 \xi ^{7/2} \\
 4 & 0 & \xi +7 \xi ^2+15 \xi ^3+11 \xi ^4+\left(\xi ^{3/2}+7 \xi ^{5/2}+8 \xi ^{7/2}\right) \chi _1+\left(\xi ^2+2 \xi ^3\right) \chi
   _2+\xi ^{5/2} \chi _3 \\
 4 & 1 & 3 \xi ^2+16 \xi ^3+10 \xi ^4+\left(7 \xi ^{5/2}+12 \xi ^{7/2}\right) \chi _1+4 \xi ^3 \chi _2 \\
 4 & 2 & 11 \xi ^3+15 \xi ^4+8 \xi ^{7/2} \chi _1 \\
 4 & 3 & 4 \xi ^4 \\
 4 & 4 & 6 \xi ^4
 \\ \hline
\end{array}
\end{equation*}

\begin{equation*}
\scriptsize
\begin{array}{| c | c ||  l  |   }
\hline 
h & j & N=7
\\ \hline 
 1 & 1 & \xi  \\
 \frac{3}{2} & \frac{3}{2} & \xi ^{3/2} \\
 2 & 0 & \xi +2 \xi ^2+\xi ^{3/2} \chi _1 \\
 2 & 2 & 2 \xi ^2 \\
 \frac{5}{2} & \frac{1}{2} & \xi ^{3/2}+2 \xi ^{5/2}+\xi ^2 \chi _1 \\
 \frac{5}{2} & \frac{3}{2} & \xi ^{5/2} \\
 \frac{5}{2} & \frac{5}{2} & 2 \xi ^{5/2} \\
 3 & 0 & \xi ^2+\xi ^3+\xi ^{5/2} \chi _1 \\
 3 & 1 & 4 \xi ^2+6 \xi ^3+3 \xi ^{5/2} \chi _1 \\
 3 & 2 & \xi ^3 \\
 3 & 3 & 4 \xi ^3 \\
 \frac{7}{2} & \frac{1}{2} & 2 \xi ^{3/2}+7 \xi ^{5/2}+5 \xi ^{7/2}+\left(3 \xi ^2+4 \xi ^3\right) \chi _1+\xi ^{5/2} \chi _2 \\
 \frac{7}{2} & \frac{3}{2} & 5 \xi ^{5/2}+8 \xi ^{7/2}+4 \xi ^3 \chi _1 \\
 \frac{7}{2} & \frac{5}{2} & 3 \xi ^{7/2} \\
 \frac{7}{2} & \frac{7}{2} & 4 \xi ^{7/2} \\
 4 & 0 & \xi +7 \xi ^2+15 \xi ^3+11 \xi ^4+\left(\xi ^{3/2}+7 \xi ^{5/2}+8 \xi ^{7/2}\right) \chi _1+\left(\xi ^2+2 \xi ^3\right) \chi
   _2+\xi ^{5/2} \chi _3 \\
 4 & 1 & 3 \xi ^2+16 \xi ^3+10 \xi ^4+\left(7 \xi ^{5/2}+12 \xi ^{7/2}\right) \chi _1+4 \xi ^3 \chi _2 \\
 4 & 2 & 11 \xi ^3+16 \xi ^4+8 \xi ^{7/2} \chi _1 \\
 4 & 3 & 4 \xi ^4 \\
 4 & 4 & 6 \xi ^4
 \\ \hline
\end{array}
\end{equation*}

\begin{equation*}
\scriptsize
\begin{array}{| c | c ||  l  |   }
\hline 
h & j & N=8 
\\ \hline 
 1 & 1 & \xi  \\
 \frac{3}{2} & \frac{3}{2} & \xi ^{3/2} \\
 2 & 0 & \xi +2 \xi ^2+\xi ^{3/2} \chi _1 \\
 2 & 2 & 2 \xi ^2 \\
 \frac{5}{2} & \frac{1}{2} & \xi ^{3/2}+2 \xi ^{5/2}+\xi ^2 \chi _1 \\
 \frac{5}{2} & \frac{3}{2} & \xi ^{5/2} \\
 \frac{5}{2} & \frac{5}{2} & 2 \xi ^{5/2} \\
 3 & 0 & \xi ^2+\xi ^3+\xi ^{5/2} \chi _1 \\
 3 & 1 & 4 \xi ^2+6 \xi ^3+3 \xi ^{5/2} \chi _1 \\
 3 & 2 & \xi ^3 \\
 3 & 3 & 4 \xi ^3 \\
 \frac{7}{2} & \frac{1}{2} & 2 \xi ^{3/2}+7 \xi ^{5/2}+5 \xi ^{7/2}+\left(3 \xi ^2+4 \xi ^3\right) \chi _1+\xi ^{5/2} \chi _2 \\
 \frac{7}{2} & \frac{3}{2} & 5 \xi ^{5/2}+8 \xi ^{7/2}+4 \xi ^3 \chi _1 \\
 \frac{7}{2} & \frac{5}{2} & 3 \xi ^{7/2} \\
 \frac{7}{2} & \frac{7}{2} & 4 \xi ^{7/2} \\
 4 & 0 & \xi +7 \xi ^2+15 \xi ^3+11 \xi ^4+\left(\xi ^{3/2}+7 \xi ^{5/2}+8 \xi ^{7/2}\right) \chi _1+\left(\xi ^2+2 \xi ^3\right) \chi
   _2+\xi ^{5/2} \chi _3 \\
 4 & 1 & 3 \xi ^2+16 \xi ^3+10 \xi ^4+\left(7 \xi ^{5/2}+12 \xi ^{7/2}\right) \chi _1+4 \xi ^3 \chi _2 \\
 4 & 2 & 11 \xi ^3+16 \xi ^4+8 \xi ^{7/2} \chi _1 \\
 4 & 3 & 4 \xi ^4 \\
 4 & 4 & 7 \xi ^4
 \\ \hline
\end{array}
\end{equation*}

\endgroup

\subsection{Hilbert series of HL ring}
\label{app_HL}

The notation $\text{HS}_{h,j} = \text{HS}_{h,j}(\tau,a,x)$
is the contribution to the 
Hilbert series   of HL ring of a 
$\mathfrak{psl}(2|2)$ primary
with weight $h$ and spin $j$. 
If $h=j$ the $\mathfrak{psl}(2|2)$ multiplet is short.
Explicitly,
\be 
\ba 
\text{HS}_{j,j}(\tau,a,x) &= 
\tau^{2j} \chi_{2j}(a) + x \tau^{2j+1} \chi_{2j-1}(a) 
\ , \\ 
\text{HS}_{h,j}(\tau,a,x) &= 
\tau^{2h} \chi_{2j}(a)
+ x \tau^{2h + 1} \Big(
\chi_{2j-1}(a)  + \chi_{2j+1}(a) 
\Big)
+ x^2 \tau^{2h+2}  \chi_{2j}(a) \ , 
\ea 
\ee
where $h>j$ in the second line.

% The contribution of a $\mathfrak{psl}(2|2)$ primary
% of nonzero  $U(1)_r$ charge $r$
% is $x^r\text{HS}_{h,j}(q,a,x)$.

% \be 
% \ba 
% \text{HS}_{j,j}(q,a,x) &= 
% q^{j} \chi_{2j}(a) + x q^{j+\frac 12} \chi_{2j-1}(a) 
% \ , \\ 
% \text{HS}_{h,j}(q,a,x) &= 
% q^h \chi_{2j}(a)
% + x q^{h + \frac 12} \Big(
% \chi_{2j-1}(a)  + \chi_{2j+1}(a) 
% \Big)
% + x^2 q^{h+1}  \chi_{2j}(a) \ , 
% \ea 
% \ee

\paragraph{The case of $SU(2)$.}

In this case we have a simple closed formula for the Hilbert series,
\be 
1 + \sum_{n=1}^\infty \text{HS}_{n,n} \ . 
\ee 

\paragraph{The case of $SU(3)$.}
\be 
\ba 
& 1+\text{HS}_{1,1}+\text{HS}_{\frac{3}{2},\frac{3}{2}}+\text{HS}_{2,0}+\text{HS}_{2,2}+\text{HS}_{\frac{5}{2},\frac{3}{2}}+\text{HS}_{\frac{5}{2},\frac{5}{2}}
\\
&+\text{HS}_{3,1}+2
   \text{HS}_{3,3}
   +\text{HS}_{\frac{7}{2},\frac{3}{2}}+\text{HS}_{\frac{7}{2},\frac{5}{2}}+\text{HS}_{\frac{7}{2},\frac{7}{2}}
   \\
   &+\text{HS}
   _{4,0}+\text{HS}_{4,2}+\text{HS}_{4,3}+2 \text{HS}_{4,4}
   + \mathcal O(q^{9/2})
\ea
\ee 

\paragraph{The case of $SU(4)$.}

\be 
\ba 
& 1+\text{HS}_{1,1}+\text{HS}_{\frac{3}{2},\frac{3}{2}}+\text{HS}_{2,0}+2
   \text{HS}_{2,2}+\text{HS}_{\frac{5}{2},\frac{1}{2}}+\text{HS}_{\frac{5}{2},\frac{3}{2}}+\text{HS}_{\frac{5}{2},\frac{5}{2}}
   \\
   & +2
   \text{HS}_{3,1}+\text{HS}_{3,2}+3 \text{HS}_{3,3}+\text{HS}_{\frac{7}{2},\frac{1}{2}}+2 \text{HS}_{\frac{7}{2},\frac{3}{2}}+2
   \text{HS}_{\frac{7}{2},\frac{5}{2}}+2 \text{HS}_{\frac{7}{2},\frac{7}{2}}
   \\
   & +2 \text{HS}_{4,0}+(1+x) \text{HS}_{4,1}+4 \text{HS}_{4,2}+2
   \text{HS}_{4,3}+4 \text{HS}_{4,4}
      + \mathcal O(q^{9/2})
\ea 
\ee 

\paragraph{The case of $SU(5)$.}

\be 
\ba 
& 1+\text{HS}_{1,1}+\text{HS}_{\frac{3}{2},\frac{3}{2}}+\text{HS}_{2,0}+2
   \text{HS}_{2,2}+\text{HS}_{\frac{5}{2},\frac{1}{2}}+\text{HS}_{\frac{5}{2},\frac{3}{2}}+2 \text{HS}_{\frac{5}{2},\frac{5}{2}}
   \\
   &+3
   \text{HS}_{3,1}+\text{HS}_{3,2}+3 \text{HS}_{3,3}+2 \text{HS}_{\frac{7}{2},\frac{1}{2}}+3 \text{HS}_{\frac{7}{2},\frac{3}{2}}+3
   \text{HS}_{\frac{7}{2},\frac{5}{2}}+3 \text{HS}_{\frac{7}{2},\frac{7}{2}}
   \\
   &+3 \text{HS}_{4,0}+(2+x) \text{HS}_{4,1}+6 \text{HS}_{4,2}+3
   \text{HS}_{4,3}+5 \text{HS}_{4,4}
         + \mathcal O(q^{9/2})
\ea 
\ee 

\paragraph{The case of $SU(6)$.}

\be 
\ba 
& 1+\text{HS}_{1,1}+\text{HS}_{\frac{3}{2},\frac{3}{2}}+\text{HS}_{2,0}+2
   \text{HS}_{2,2}+\text{HS}_{\frac{5}{2},\frac{1}{2}}+\text{HS}_{\frac{5}{2},\frac{3}{2}}+2 \text{HS}_{\frac{5}{2},\frac{5}{2}}
   \\
   & +3
   \text{HS}_{3,1}+\text{HS}_{3,2}+4 \text{HS}_{3,3}+2 \text{HS}_{\frac{7}{2},\frac{1}{2}}+4 \text{HS}_{\frac{7}{2},\frac{3}{2}}+3
   \text{HS}_{\frac{7}{2},\frac{5}{2}}+3 \text{HS}_{\frac{7}{2},\frac{7}{2}}
   \\
   & +4 \text{HS}_{4,0}+(3+x) \text{HS}_{4,1}+7 \text{HS}_{4,2}+4
   \text{HS}_{4,3}+6 \text{HS}_{4,4}
            + \mathcal O(q^{9/2})
\ea 
\ee

\paragraph{The case of $SU(7)$.}

\be 
\ba 
& 
1+\text{HS}_{1,1}+\text{HS}_{\frac{3}{2},\frac{3}{2}}+\text{HS}_{2,0}+2
   \text{HS}_{2,2}+\text{HS}_{\frac{5}{2},\frac{1}{2}}+\text{HS}_{\frac{5}{2},\frac{3}{2}}+2 \text{HS}_{\frac{5}{2},\frac{5}{2}}
   \\
   &+3
   \text{HS}_{3,1}+\text{HS}_{3,2}+4 \text{HS}_{3,3}+2 \text{HS}_{\frac{7}{2},\frac{1}{2}}+4 \text{HS}_{\frac{7}{2},\frac{3}{2}}+3
   \text{HS}_{\frac{7}{2},\frac{5}{2}}+4 \text{HS}_{\frac{7}{2},\frac{7}{2}}
   \\
   &+4 \text{HS}_{4,0}+(3+x) \text{HS}_{4,1}+8 \text{HS}_{4,2}+4
   \text{HS}_{4,3}+6 \text{HS}_{4,4}
            + \mathcal O(q^{9/2})
\ea 
\ee 

\paragraph{The case of $SU(8)$.}

\be 
\ba 
&
1+\text{HS}_{1,1}+\text{HS}_{\frac{3}{2},\frac{3}{2}}+\text{HS}_{2,0}+2
   \text{HS}_{2,2}+\text{HS}_{\frac{5}{2},\frac{1}{2}}+\text{HS}_{\frac{5}{2},\frac{3}{2}}+2 \text{HS}_{\frac{5}{2},\frac{5}{2}}
   \\
   &+3
   \text{HS}_{3,1}+\text{HS}_{3,2}+4 \text{HS}_{3,3}+2 \text{HS}_{\frac{7}{2},\frac{1}{2}}+4 \text{HS}_{\frac{7}{2},\frac{3}{2}}+3
   \text{HS}_{\frac{7}{2},\frac{5}{2}}+4 \text{HS}_{\frac{7}{2},\frac{7}{2}}
   \\
   &+4 \text{HS}_{4,0}+(3+x) \text{HS}_{4,1}+8 \text{HS}_{4,2}+4
   \text{HS}_{4,3}+7 \text{HS}_{4,4}
             + \mathcal O(q^{9/2})
\ea 
\ee

\section{More details on the BRST construction}

\subsection{Explicit example: $(h,j)=(2,0)$ at generic $N$}

For $N$ sufficiently large, we have five $\mathfrak{psl}(2|2)$
primary states,
\be 
\ba 
\cO_1 &= {\rm Tr} (  \phi_0 \phi_1 \phi_0 \phi_1 - \phi_0 \phi_0 \phi_1 \phi_1 ) \ , \\ 
\cO_2 &= {\rm Tr} (\phi_0 \phi_1){\rm Tr} (\phi_0 \phi_1)
- {\rm Tr} (\phi_0 \phi_0){\rm Tr} (\phi_1 \phi_1) \ , \\
\cO_3 & = {\rm Tr} (b \phi_1 \phi_0 - b \phi_0 \phi_1) \ , \\
\cO_4 & = {\rm Tr} (\partial c \phi_1 \phi_0 - b \phi_0 \phi_1) \ , \\
\cO_5 & = {\rm Tr} (
\partial c b - \partial \phi_0 \phi_1 + \partial \phi_1 \phi_0
)  \ .
\ea 
\ee 
% \be 
% \ba 
% \cO_1 & = (\phi_0)^{I_1}{}_{I_2}  (\phi_1)^{I_2}{}_{I_3}
% (\phi_0)^{I_3}{}_{I_4} (\phi_1)^{I_4}{}_{I_1} 
% -  (\phi_0)^{I_1}{}_{I_2}  (\phi_0)^{I_2}{}_{I_3}
% (\phi_1)^{I_3}{}_{I_4} (\phi_1)^{I_4}{}_{I_1} 
% \ , \\ 
% \cO_2 & =  (\phi_0)^{I_1}{}_{I_2}  (\phi_1)^{I_2}{}_{I_1}
% (\phi_0)^{J_1}{}_{J_2} (\phi_1)^{J_2}{}_{J_1} 
% -  (\phi_0)^{I_1}{}_{I_2}  (\phi_0)^{I_2}{}_{I_1}
% (\phi_1)^{J_1}{}_{J_2} (\phi_1)^{J_2}{}_{J_1} 
% \ , \\ 
% \cO_3 & = b^{I_1}{}_{I_2} (\phi_1)^{I_2}{}_{I_3} (\phi_0)^{I_3}{}_{I_1}
% - b^{I_1}{}_{I_2} (\phi_0)^{I_2}{}_{I_3} (\phi_1)^{I_3}{}_{I_1}
% \ , \\ 
% \cO_4 & = \partial c^{I_1}{}_{I_2} (\phi_1)^{I_2}{}_{I_3} (\phi_0)^{I_3}{}_{I_1}
% - \partial c^{I_1}{}_{I_2} (\phi_0)^{I_2}{}_{I_3} (\phi_1)^{I_3}{}_{I_1}
% \ , \\ 
% \cO_5 & = \partial c^{I_1}{}_{I_2} b^{I_2}{}_{I_1}
% - \partial (\phi_0)^{I_1}{}_{I_2} (\phi_1)^{I_1}{}_{I_2}
% + \partial (\phi_1)^{I_1}{}_{I_2} (\phi_0)^{I_1}{}_{I_2} \ . 
% \ea 
% \ee 
The action the the BRST charge is 
\be 
\{ J_{\rm B} \cO_1 \}_1 = \sum_{j=1}^5 m_{ij} \cO_j  \ ,
\qquad 
m = \begin{pmatrix}
     0 & 0 & 0 & 3 N & 0 \\
 0 & 0 & 0 & 6 & 0 \\
 -2 & 0 & 0 & 0 & 2 N \\
 0 & 0 & 0 & 0 & 0 \\
 0 & 0 & 0 & 3 & 0 \\
\end{pmatrix} \ .
\ee 
By a similarity transformation (for generic $N$)
this can be brough to the Jordan form
\be 
\widetilde m = \left(
\begin{array}{c  cc  cc}
0 & 0 & 0 & 0 & 0  \\  
0 & 0 & 1 & 0 & 0  \\ 
0 & 0 & 0 & 0 & 0  \\  
0 & 0 & 0 & 0 & 1 \\
0 & 0 & 0 & 0 & 0 
\end{array} 
\right) \ , \qquad 
m = U \widetilde m U^{-1} \ , \qquad 
U = \begin{pmatrix}
    0 & 3N & 0 & 0 & 0 \\ 
    1 & 6 & 0 & 0 & 0 \\ 
    0 & 0 & 0 & 2N & 0 \\ 
    0 & 0 & 1 & 0 & 0 \\ 
    0 & 3 & 0 & 0 & 1 
\end{pmatrix} \ .
\ee 
We perform a change of basis for $\mathfrak{psl}(2|2)$ primaries,
\be 
\cP_i := \sum_{j=1}^5 (U^{-1} )_{ij} \cO_j  \ , \qquad 
\{ J_{\rm B} \cP_i \}_1 = \sum_{j=1}^5 \widetilde m_{ij} \cP_i \ .
\ee 
More explicitly,
\be 
\{ J_{\rm B} \cP_1 \}_1 = 0 \ , \quad 
\{ J_{\rm B} \cP_2 \}_1 = \cP_3 \ , \quad 
\{ J_{\rm B} \cP_3 \}_1 = 0 \ , \quad 
\{ J_{\rm B} \cP_4 \}_1 = \cP_5 \ , \quad 
\{ J_{\rm B} \cP_5 \}_1 = 0 \ .
\ee 
This presentation makes it manifest that the space of BRST-closed states modulo BRST-exact states is one dimensional and spanned by
$\cP_1$.
The operators $\cP_i$ are
\be 
\ba 
\cP_1 & = {\rm Tr} ( \phi_0 \phi_1 ){\rm Tr} ( \phi_0 \phi_1 )
- {\rm Tr} ( \phi_0 \phi_0 ) {\rm Tr} ( \phi_1 \phi_1 )
+ \frac 2N {\rm Tr}(\phi_0 \phi_0 \phi_1 \phi_1
-  \phi_0 \phi_1 \phi_0 \phi_1) \ , \\
\cP_2 & = \frac{1}{3N} {\rm Tr} (
\phi_0 \phi_1 \phi_0 \phi_1 
- \phi_0 \phi_0 \phi_1 \phi_1
) \ , \\ 
\cP_3 & = {\rm Tr}( \partial c \phi_1 \phi_0 
- \partial c \phi_0 \phi_1) \ , \\
\cP_4 &= \frac{1}{2N} {\rm Tr}( b \phi_1 \phi_0 
- b \phi_0 \phi_1) \ , \\ 
\cP_5 & = {\rm Tr}(\partial c b + \partial \phi_1 \phi_0
- \partial \phi_0 \phi_1)
+ \frac{1}{N} {\rm Tr} (
\phi_0 \phi_0 \phi_1 \phi_1 
- \phi_0 \phi_1 \phi_0 \phi_1 
) \ . 
\ea 
\ee 
We notice that $\cP_1$ is algebraically zero for $N=2$,
while it is non-zero for $N>2$.
The degree of $\cP_1$ is 4.
It can be checked that, even if we add to $\cP_1$
an arbitrary linear combination of $\cP_3$, $\cP_5$
(the states that are BRST-exact), we cannot lower its degree
below 4, for generic $N$.

\subsection{More examples}

Let us report some explicit bases for 
gauge singlet $\mathfrak{psl}(2|2)$ primary states. We seek bases of the form 
\be 
\{ \cZ_1, \cZ_2, \dots, \cZ_n ,\cX_1, \cY_1, \cX_2 , \cY_2, \dots, \cX_m, \cY_m \} \ , 
\ee 
such that the action of the BRST operator is 
\be 
J_{\rm B} \cdot \cZ_i = 0 \ , \quad 
i = 1,\dots,n \ ; \qquad 
J_{\rm B} \cdot \cX_i = \cY_i \ , \quad 
i = 1,\dots,m \ .
\ee 
It follows that $J_{\rm B} \cdot \cY_i=0$,
$i=1,\dots,m$.
We are free to shift any of the states
$\cZ_i$ by any linear combination of the
BRST-exact states $\cY_i$.
We exploit this freedom to get a collection of $\cZ_i$ states in which lengths are minimized. Finally, we seek a choice of 
$\cZ_i$ such that, as $N$ is lowered from the stable regime down to smaller and smaller values, the states $\cZ_i$
become zero in a sequential manner, starting from those with greater length.

% \subsubsection{The case $(h,j) = (2,0)$}
% A good choice of basis
% of gauge singlet $\mathfrak{psl}(2|2)$ primary states is
% \be 
% \ba 
% \cZ_1 & = {\rm Tr}(\phi_0 \phi_1)^2
% - {\rm Tr}(\phi_0^2){\rm Tr}(\phi_1^2)
% - \tfrac 2N {\rm Tr}(\phi_0 \phi_1 \phi_0 \phi_1)
% + \tfrac 2N {\rm Tr}(\phi_0^2 \phi_1^2)
% \ , \\
% \cX_1 & = 
% {\rm Tr}(\phi_0 \phi_1 \phi_0 \phi_1)
% - {\rm Tr}(\phi_0^2 \phi_1^2)
% \ , \\
% \cY_1 & = 3N {\rm Tr}(dc [\phi_1,\phi_0]) \ , \\
% \cX_2 & = 
% {\rm Tr}(b[\phi_1,\phi_0])
% \ , \\
% \cY_2 & = 
% 2N {\rm Tr}(dc b)
% + 2N {\rm Tr}(\phi_0 d\phi_1)
% - 2N {\rm Tr}(\phi_1 d\phi_0)
% - 2 {\rm Tr}(\phi_0 \phi_1 \phi_0 \phi_1)
% + 2 {\rm Tr}(\phi_0^2 \phi_1^2)
% \ .
% \ea 
% \ee 
% The state $\cZ_1$   is zero for $N=2$ and non-zero for $N \ge 3$.

% \bonetti{To clarify:
% Use expressions of $J$ and $T$
% in terms of BRST free fields, compute $JJ +T$ primary using
% OPE package to deal with NO's
% of BRST letters, compare result with $\cZ_1$, $\cX_1$, $\cY_2$.
% }

\subsubsection{The case $(h,j) = (2,2)$}
A good choice of basis
of gauge singlet $\mathfrak{psl}(2|2)$ primary states is
\be 
\ba 
\cZ_1 & = {\rm Tr}(\phi(y)^4)
\ , \\
\cZ_2 & = 
2 {\rm Tr}(\phi(y)^4)
- {\rm Tr}(\phi(y)^2)^2
\ .
\ea 
\ee 
The state $\cZ_2$ is identically zero for $N=2,3$, is non-zero for $N \ge 4$.
The state $\cZ_1$ is non-zero for $N \ge 2$.

% \bonetti{More suggestive basis:
% \be 
% \ba 
% \cZ_1 & = {\rm Tr}(\phi(y)^2)^2
% \ , \\
% \cZ_2 & = 
% 2 {\rm Tr}(\phi(y)^4)
% - {\rm Tr}(\phi(y)^2)^2
% \ .
% \ea 
% \ee 
% The state $\cZ_2$ is identically zero for $N=2,3$, is non-zero for $N \ge 4$.
% The state $\cZ_1$ is non-zero for $N \ge 2$.
% }

\subsubsection{The case $(h,j) = (\tfrac 52, \tfrac 12)$}
A good choice of basis
of gauge singlet $\mathfrak{psl}(2|2)$ primary states is
\be 
\ba 
\cZ_1 & =  
\tfrac 6 N {\rm Tr}(\phi_0^2 \phi_1 \phi_0 \phi_1)
- \tfrac 6N {\rm Tr}(\phi_0^3 \phi_1^2)
+ {\rm Tr}(\phi_0^2) {\rm Tr}(\phi_0 \phi_1^2)
- 2 {\rm Tr}(\phi_0 \phi_1)
{\rm Tr}(\phi_0^2 \phi_1)
+ {\rm Tr}(\phi_1^2) {\rm Tr}(\phi_0^3)
\\
& + y \Big[ 
\tfrac 6N  {\rm Tr}(\phi_0 \phi_1 \phi_0 \phi_1^2)
- \tfrac 6N {\rm Tr}(\phi_0^2 \phi_1^3)
- 2 {\rm Tr}(\phi_0 \phi_1) {\rm Tr}(\phi_0 \phi_1^2)
+ {\rm Tr}(\phi_0^2) {\rm Tr}(\phi_1^3)
+ {\rm Tr}(\phi_1^2) {\rm Tr}(\phi_0^2 \phi_1)
\Big]
\ , \\
\cX_1 & = 
{\rm Tr}(\phi_0^3 \phi_1^2)
- {\rm Tr}(\phi_0^2 \phi_1 \phi_0 \phi_1)
+ y \Big[ 
{\rm Tr}(\phi_0^2 \phi_1^3)
- {\rm Tr}(\phi_0 \phi_1 \phi_0 \phi_1^2)
\Big]  \ , \\ 
\cY_1 & = 2N {\rm Tr}(dc \phi_0^2 \phi_1)
- 2 N {\rm Tr}(dc \phi_1 \phi_0^2)
+ y \Big[ 
2N {\rm Tr}(dc \phi_0 \phi_1^2)
- 2 N {\rm Tr}(dc \phi_1^2 \phi_0)
\Big] \ , \\
\cX_2 & = 
{\rm Tr}(b \phi_0^2 \phi_1)
- {\rm Tr}(b \phi_1 \phi_0^2)
+ y \Big[ 
{\rm Tr}(b \phi_0 \phi_1^2)
- {\rm Tr}(b \phi_1^2 \phi_0)
\Big] \ , \\ 
\cY_2 & = 
2 {\rm Tr}(\phi_0^2 \phi_1 \phi_0 \phi_1)
- 2 {\rm Tr} (\phi_0^3 \phi_1^2)
\\
&
- N {\rm Tr}(dc b \phi_0)
- N {\rm Tr}(dc \phi_0 b)
- 2 N {\rm Tr}(d\phi_1 \phi_0^2)
+ N {\rm Tr}(d\phi_0 \phi_1 \phi_0)
+ N {\rm Tr}(d\phi_0 \phi_0 \phi_1)
\\
& + y \Big[ 
2 {\rm Tr}(\phi_0 \phi_1 \phi_0 \phi_1^2)
-2 {\rm Tr}(\phi_0^2 \phi_1^3)
\\
& - N {\rm Tr}(dcb \phi_1)
- N {\rm Tr}(dc \phi_1 b)
+ 2 N {\rm Tr}(d\phi_0 \phi_1^2)
- N {\rm Tr}(d\phi_1 \phi_1 \phi_0)
- N {\rm Tr}(d\phi_1 \phi_0 \phi_1)
\Big] \ . 
\ea 
\ee 
The states $\cZ_1$ is zero for $N=2,3$ and non-zero for $N \ge 4$.

\subsubsection{The case $(h,j) = (\tfrac 52, \tfrac 32)$}
A good choice of basis
of gauge singlet $\mathfrak{psl}(2|2)$ primary states is
\be 
\ba 
\cZ_1 & =  
{\rm Tr}(\phi_0^2) {\rm Tr}(\phi_0^2 \phi_1)
- {\rm Tr}(\phi_0\phi_1) {\rm Tr}(\phi_0^3)
\\
& + y 
\Big[ 
2 {\rm Tr}(\phi_0^2) {\rm Tr}(\phi_0 \phi_1^2)
- {\rm Tr}(\phi_0 \phi_1) {\rm Tr}(\phi_0^2 \phi_1)
- {\rm Tr}(\phi_1^2) {\rm Tr}(\phi_0^3 )
\Big]
\\
& + y^2 \Big[ 
{\rm Tr}(\phi_0 \phi_1)
{\rm Tr}(\phi_0 \phi_1^2)
+ {\rm Tr}(\phi_0^2) {\rm Tr}(\phi_1^3)
- 2 {\rm Tr}(\phi_1^2)  {\rm Tr}(\phi_0^2 \phi_1)
\Big]
\\
& + y^2 \Big[ 
{\rm Tr}(\phi_0 \phi_1)
{\rm Tr}(\phi_1^3)
- {\rm Tr}(\phi_1^2)
{\rm Tr}(\phi_0 \phi_1^2)
\Big] \ . 
\ea 
\ee 
The state $\cZ_1$ is zero for $N=2$
and non-zero for $N\ge 3$.

\subsubsection{The case $(h,j) = (\tfrac 52, \tfrac 52)$}
A good choice of basis
of gauge singlet $\mathfrak{psl}(2|2)$ primary states is
\be 
\ba 
\cZ_1 & =  
{\rm Tr}(\phi(y)^5)
\\\cZ_2 & = 
{\rm Tr}(\phi(y)^5) - \tfrac 56 
{\rm Tr}(\phi(y)^2)
{\rm Tr}(\phi(y)^3) \ . 
\ea  
\ee 
The state $\cZ_2$ is zero for $N=2,3,4$
and non-zero for $N\ge 5$.
The state $\cZ_1$ is non-zero
for $N\ge 2$.

\subsubsection{The case $(h,j) = (3,0)$}

In the case we find a basis of the form
$\{ \cX_1, \cY_1, \cX_2, \cY_2 \}$
so that no state survives in BRST-cohomology.
We do not report the explicit forms of 
$\{ \cX_1, \cY_1, \cX_2, \cY_2 \}$.
We notice that all these four states are
zero for $N=2$ and non-zero for $N\ge 2$.

\subsection{The case $(h,j)=(3,1)$}
We have four states of type $\cZ$
and six pairs of type $(\cX, \cY)$.
We do not report the explicit forms of the 
$(\cX, \cY)$ pairs. We report the
highest weight component of the four $\cZ$ states: 
\be
\ba 
\cZ_1  \Big|_{y=0}& = 
{\rm Tr}(\phi_0^2)
{\rm Tr}(dcb)
- 3 {\rm Tr}(\phi_0 dc)
{\rm Tr}(\phi_0 b)
- 2 {\rm Tr}(\phi_0 d\phi_1)
{\rm Tr}(\phi_0^2)
- {\rm Tr}(\phi_1 d\phi_0)
{\rm Tr}(\phi_0^2)
\\
& + 3 {\rm Tr}(\phi_0 d\phi_0)
{\rm Tr}(\phi_0 \phi_1) \ , 
\\
\cZ_2 \Big|_{y=0}
&
=
-\tfrac{5 N }{4 \left(3 N^2-2\right)}
{\rm Tr}(\phi_0^2) {\rm Tr}(\phi_0^2 \phi_1^2)
+\tfrac{5
   N }{4 \left(3
   N^2-2\right)}
   {\rm Tr}(\phi_0^2) {\rm Tr}(\phi_0  \phi_1 \phi_0 \phi_1)
   -\tfrac{\left(3 N^2+8\right) 
   }{4 \left(3
   N^2-2\right)}
   {\rm Tr}(\phi_0 ^3 \phi_1 \phi_0 \phi_1)
   \\
   &-\tfrac{(9N^2-16)
   }{4 \left(3 N^2-2\right)}
   {\rm Tr}(\phi_0 ^2 \phi_1 \phi_0^2 \phi_1)
   -\tfrac{N 
   }{4}
   {\rm Tr}(\phi_0 ^3) {\rm Tr}(\phi_0 \phi_1^2)
   +\tfrac{N 
   }{4}{\rm Tr}(\phi_0^2 \phi_1)^2+
   {\rm Tr}(\phi_0^4 \phi_1^2) \ , 
   \\
\cZ_3\Big|_{y=0}
& = 
\tfrac{1}{100} \left(3 N^2-2\right) {\rm Tr}(\phi_0^2)^2
   {\rm Tr}(\phi_1^2)-\tfrac{N \left(11
   N^2+26\right) }{20 \left(3
   N^2-2\right)}
   {\rm Tr}(\phi_0^2) 
   {\rm Tr}(\phi_0^2 \phi_1^2)
   \\
   &+\tfrac{1}{100} \left(2-3 N^2\right)
   {\rm Tr}(\phi_0^2) {\rm Tr}(\phi_0 \phi_1)^2
   +\tfrac{N \left(7 N^2+62\right) 
  }{40
   \left(3 N^2-2\right)}
    {\rm Tr}(\phi_0^2) {\rm Tr}(\phi_0  \phi_1 \phi_0 \phi_1)
   \\
   &-\tfrac{\left(13 N^2+8\right)
   }{5 \left(3 N^2-2\right)}
   {\rm Tr}(\phi_0^3 \phi_1 \phi_0 \phi_1)
   -\tfrac{2 (N-3)
   (N+3) }{5 \left(3
   N^2-2\right)}
   {\rm Tr}(\phi_0^2 \phi_1 \phi_0^2 \phi_1)
   \\
   &-\tfrac{N }{8}
   {\rm Tr}(\phi_0^4) {\rm Tr}(\phi_1^2)
   +\tfrac{N 
   }{4} {\rm Tr}(\phi_0^3 \phi_1) {\rm Tr}(\phi_0 \phi_1) +
   {\rm Tr}(\phi_0^4 \phi_1^2) \ , 
   \\
\mathcal Z_4 \Big|_{y=0}
& = 
\tfrac{1}{8} \left(3 N^2-2\right) N 
{\rm Tr}(\phi_0^2)^2 {\rm Tr}(\phi_1^2)-\tfrac{1}{8} \left(3
   N^2-2\right) N {\rm Tr}(\phi_0^2)
   {\rm Tr}(\phi_0 \phi_1)^2
   -\tfrac{1}{4} N^2 
   {\rm Tr}(\phi_0^3) {\rm Tr}(\phi_0 \phi_1^2)
 \\
 &+\tfrac{N^2 }{4} {\rm Tr}(\phi_0^2 \phi_1)^2 
 +\tfrac{\left(16 N^3-132 N^2+191
   N-792\right) N }{4 \left(3
   N^2-2\right)} {\rm Tr}(\phi_0^2)
   {\rm Tr}(\phi_0^2 \phi_1^2)
   \\
   &+\tfrac{\left(61 N^3-132 N^2-444
   N+1848\right) N }{8 \left(3
   N^2-2\right)} {\rm Tr}(\phi_0^2)
   {\rm Tr}(\phi_0 \phi_1 \phi_0 \phi_1)
   \\
   &+\tfrac{\left(215 N^3-1056 N^2+260
   N-1056\right) }{4 \left(3
   N^2-2\right)} {\rm Tr}(\phi_0^3 \phi_1 \phi_0 \phi_1)
   \\
   &+\tfrac{\left(145 N^3-528 N^2-500
   N+2112\right) }{4 \left(3 N^2-2\right)}
   {\rm Tr}(\phi_0^2 \phi_1 \phi_0^2 \phi_1)
   +\tfrac{1}{8}
   (31 N-132) N {\rm Tr}(\phi_0^4) {\rm Tr}(\phi_1^2)
   \\
   &-6 (5 N-22)
   {\rm Tr}(\phi_0^4 \phi_1^2)-\tfrac{1}{4} (31 N-132) N 
   {\rm Tr}(\phi_0^3 \phi_1) {\rm Tr}(\phi_0 \phi_1) \ , 
\ea 
\ee 
The state $\mathcal Z_1$ is non-zero for $N \ge 2$.
The state $\mathcal Z_2$ is zero for $N=2$.
The state $\mathcal Z_3$ is zero for $N=2,3$.
The state $\mathcal Z_4$ is zero for $N=2,3,4$.
The choice of $\mathcal Z$ states with this pattern of vanishing 
with $N$ is not unique, we have simply displayed one possible explicit choice.

\subsection{The case $(h,j)=(3,2)$}
We have only one state of type $\cZ$, with highest weight component
\be 
\cZ_1 \Big|_{y=0}
= {\rm Tr}(\phi_0 \phi_1) {\rm Tr}(\phi_0^2)
- {\rm Tr}(\phi_0^2) {\rm Tr}(\phi_0^4) \ . 
\ee 
This state is zero for $N=2,3$ and non-zero for $N\ge 4$.

\subsection{The case $(h,j)=(3,3)$}
We have four states of type $\cZ$,
\be
\ba 
\cZ_1
& = 
{\rm Tr}(\phi(y)^6) 
\ , \\ 
\cZ_2
& = 
{\rm Tr}(\phi(y)^6) 
- \tfrac 12 {\rm Tr}(\phi(y)^2) {\rm Tr}(\phi(y)^4) 
\ , \\ 
\cZ_3
& = 
{\rm Tr}(\phi(y)^6) 
- \tfrac 13 {\rm Tr}(\phi(y)^2) {\rm Tr}(\phi(y)^4) 
- \tfrac 14 {\rm Tr}(\phi(y)^2)^3
\ , \\ 
\cZ_4
& = 
{\rm Tr}(\phi(y)^6) 
- \tfrac 34 {\rm Tr}(\phi(y)^2) {\rm Tr}(\phi(y)^4) 
- \tfrac 134 {\rm Tr}(\phi(y)^3)^2
+ \tfrac 18 {\rm Tr}(\phi(y)^2)^3
\ .
\ea 
\ee 
The state $\cZ_1$ is non-zero for $N\ge 2$.
The state $\cZ_2$ is zero for $N=2$ and non-zero for $N\ge 3$.
The state $\cZ_3$ is zero for $N = 2,3$ and non-zero
for $N\ge 4$.
The state $\cZ_4$ is zero for $N=2,3,4,5$ and non-zero
for $N \ge 6$.

\bibliographystyle{JHEP}
\bibliography{refs}

\end{document}